%% file: tese.tex
\documentclass[a4paper,12pt,nocenter,bold,indent,oneside]{book}
\usepackage{latexsym}
\usepackage{amsmath}
\usepackage{amssymb}
\usepackage{eufrak}
\usepackage{euscript}
\usepackage{epsfig,graphics,graphicx}
\usepackage{bm}
\usepackage{color}
 \usepackage[dvips]{feynmp}
 \usepackage{slashed}
\def\k{\kappa}

\def\be{\begin{eqnarray}}
\def\ee{\end{eqnarray}}
\def\dslash{\partial\!\!\!/}

\def\bpsi{\overline{\psi}}

\def\nl{\newline}
\def\Ep{E_{{\bf p}}}
\def\Epu{E_{{\bf p}_1}}
\def\Epd{E_{{\bf p}_2}}

\def\Eq{E_{{\bf q}}}
\def\sumint{\hbox{$\sum$}\!\!\!\!\!\!\!\int}
\def\LMS{\Lambda_{\overline{\textrm{MS}}}}
\def\bs{\bar\sigma}
\def\bvp{\vec{\bar{\pi}}}
\def\bp{\bar\pi}

\makeatletter
\def\@evenhead{
               \vbox{\hbox to\hsize{\bf \thepage \hfill \sl \leftmark}
               \vspace{2pt} \hbox to\hsize{\hrulefill}}}
\def\@oddhead{
               \vbox{\hbox to\hsize{\sl \rightmark \hfill \bf \thepage}
               \vspace{2pt} \hbox to\hsize{\hrulefill}}}
\makeatother

\begin{document}
\setlength{\baselineskip}{20pt}

\pagenumbering{roman}
\bibliographystyle{unsrt}
\include{capa}

\include{agradecimentos}
\include{resumo}

\pagenumbering{arabic}
\tableofcontents
\listoffigures

\include{introducao}

\include{isospin}
\include{FS}

\include{surften}

\include{magneticQCD}
\include{OPT}

\include{relatBEC}

\include{conclusions}


\begin{appendix}
\appendix
\include{apLSM}
\include{apOPTA}
\include{apOPTB}

\include{apIsospin-Props}

\include{apIntMag}
\include{apPureGlue}
\include{apSEisospin}
\end{appendix}

\include{bibliography}

\end{document}

%% file: capa.tex
\begin{titlepage}

\vspace*{12cm}

\centerline{{\Large EXPLORING THE}} 

\vspace*{0.7cm}

\centerline{{\Large  DIFFERENT PHASE DIAGRAMS }} 

\vspace*{0.7cm}

\centerline{{\Large OF STRONG INTERACTIONS}}

\vspace*{2.cm}

\centerline{{\bf LET\'ICIA FARIA DOMINGUES PALHARES}}
\end{titlepage}

\begin{titlepage}

\includegraphics[width=2cm]{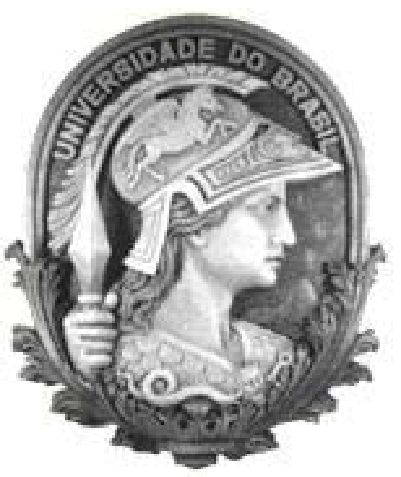}

\vspace{0.5cm}

\centerline{{\Large EXPLORING THE DIFFERENT PHASE DIAGRAMS }}
\centerline{{\Large OF STRONG INTERACTIONS}}
\vspace{2cm}
\centerline{Let\'\i cia Faria Domingues Palhares}

\vspace{3cm}

\hspace{6cm}\parbox[s]{9cm}{
PhD thesis presented to the Graduate Program in Physics 
of the Institute of Physics of the Federal University of Rio de Janeiro,
as part of the requirements for obtaining the title of Doctor in Sciences (Physics).

}

\vspace{3.1cm}

\hspace{0.85cm}\centerline{Advisor: Eduardo Souza Fraga}

\vspace{3.3cm}

{\footnotesize \centerline{Rio de Janeiro}}
{\footnotesize \centerline{february/2012}}
\end{titlepage}

%% file: agradecimentos.tex
\begin{titlepage}

\vspace*{17cm}
\hspace{8.5cm}\parbox[s]{7cm}{{\it I dedicate this thesis to my parents, Lair and Manuel,\newline from whom I have learned more than can be put on a paper, with profound admiration and respect.}}

\end{titlepage}

{\bf\large\centerline{Acknowledgements}}

\vspace{0.7cm}

First of all, I am greatly indebted to my advisor Eduardo Fraga, who has profoundly and continuously influenced my formation since early undergraduate times. For his invaluable respect, dedication and patience.  For more than guiding me through one PhD project, but rather sharing with me his everyday excitement about different physics questions; this thesis representing just a small subset of them.

I deeply acknowledge the very special work environment of the ICE (former QCD-QGP) group at the Federal University of Rio de Janeiro. I have learned a lot during the regular activities and conferences we have attended together and it was a pleasure being part of this group for many years. I would like to especially thank Takeshi Kodama for his example of integrity and wisdom, which he is always willing to share. It is an honor to have daily contact with how he sees the world.

I would like to thank the many professors of the UFRJ Physics Institute I had the opportunity to learn from. In particular, I have profited from the guidance of Carlos Arag\~ao who has introduced me (quite early) to path integrals and the imaginary time. Special thanks to Eduardo Marino for his continuous support.

My thanks go to many students with and from whom I have learned during the past years at the Physics Institute in Rio. To my officemates Ana J\'ulia Mizher, Bruno Mintz, Bruno Taketani and Daniel Kroff; I have profited a lot from daily discussions with all of you. My special thanks to Andr\'e Bessa for his help, friendship and advice, even when he was far away. To my (quasi-)group colleagues Carlos Alfonso Ball\'on Bayona, Gabriel Denicol and Philipe Mota for many interesting discussions and presentations. To Philipe Mota for friendship and computational advice always available. To my undergraduate colleagues; our early academic partnership has taught me a lot.

I acknowledge also the hospitality of the group at IPhT, CEA-Saclay, where I have spent one year as an exchange student. My special gratitude to Jean-Paul Blaizot for his patience and attention during many stimulating discussions. This whole year was an invaluable experience from which I am still learning. Je voudrais aussi remercier le group d'\'etudiants qui m'ont accueillie et, en particulier, Piotr Tourkine avec qui j'ai partag\'e un bureau et qui m'a beaucoup aid\'e pendant toute l'ann\'ee.

\newpage
I further acknowledge discussions and collaboration with Cristian Villavicencio, Gast\~ao Krein, Marcus Benghi Pinto, and Paul Sorensen. This thesis and my formation would certainly be much less rich had I not collaborated with each of you.

I would like to thank the members of the thesis committee, Jean-Paul Blaizot, Jorge Noronha, Luca Moriconi, Marcello Barbosa, Marco Moriconi, and Raimundo dos Santos for carefully reading the original version of this manuscript and for the interesting discussions during my defense.

The bureaucratic and technical help of Cas\'e, C\'esar, Cristina, Curt and Pedro at IF-UFRJ and Catherine Cataldi at IPhT were crucial in many occasions.

I would like to thank family and friends for unconditional continuous support, trust and patience.

I acknowledge partial financial support from CAPES-COFECUB, CNPq and FAPERJ.

\vspace{1cm}

\centerline{**************}

%% file: resumo.tex
\centerline{\bf ABSTRACT}
\vspace{1cm}
\centerline{{\large EXPLORING THE DIFFERENT PHASE DIAGRAMS }} 
\centerline{{\large OF STRONG INTERACTIONS}}
\vspace{1cm}
\centerline{Let\'\i cia Faria Domingues Palhares}
\vspace{0.2cm}
\centerline{Advisor: Eduardo Souza Fraga}
\vspace{1cm}

Abstract of the PhD thesis presented to the Graduate Program in Physics 
of the Physics Institute of the Federal University of Rio de Janeiro,
as part of the requirements for obtaining the title of Doctor in Sciences (Physics).

\bigskip\bigskip

In-medium field-theory is applied to different effective models and QCD to describe mass and isospin effects, finite volume corrections and magnetic fields in the phase diagram of Strong Interactions, keeping close contact with experiments and lattice results. Findings range from a technical nonperturbative solution of the hot and dense regime of a general massive Yukawa Theory and the computation of the cold and dense chiral surface tension -- a key quantity for supernovae explosions and compact star structure -- to the proposal of a novel signature for the QCD critical endpoint in heavy-ion collisions based on finite-size scaling. The behavior of the deconfining critical temperature as a function of the pion mass and the isospin chemical potential is also addressed in an effective model and the description obtained is in agreement with lattice simulations, in contrast to what is found in (Polyakov-extended) chiral models. We also discuss the thermodynamics of QCD in the presence of a (Abelian) magnetic field to two-loop order and a Functional-Renormalization-Group analysis of relativistic Bose-Einstein condensation of pions in isospin-dense media.

\vspace{0.8cm}


\vspace{0.8cm}

\noindent {\bf Related publications:}

\begin{enumerate}

\item E. S. Fraga, L. F. P., M. B. Pinto, \nl
{\it Nonperturbative Yukawa theory at finite density and temperature}, \nl
Physical Review D {\bf 79} 065026 (2009).

\item E. S. Fraga, L. F. P., C. Villavicencio, \nl
{\it Quark mass and isospin dependence of the deconfining critical temperature}, \nl
Physical Review D {\bf 79} 014021 (2009).\nl
[$+$ proceedings:  Nucl. Phys. A {\bf 820}, 287C (2009)]

\item L. F. P., E. S. Fraga, T. Kodama, \nl
{\it 
Chiral transition in a finite system and possible use of finite-size scaling in relativistic heavy ion collisions
},\nl
Journal of Physics G {\bf 38}, 085101 (2011).\nl
[$+$ proceedings:  
 PoS {\bf CPOD2009}, 011 (2009); J. Phys. {\bf G37}, 094031 (2010); Acta Phys. Polon. Supp. {\bf 4}, 715 (2011)]

  
\item L. F. P., E. S. Fraga,\nl
{\it Droplets in the cold and dense linear sigma model with quarks},\nl
Physical Review D {\bf 82}, 125018 (2010).\nl
[$+$ proceedings:  PoS {\bf FACESQCD} 2010, 014 (2010)]
%

\item  E.~S.~Fraga, L.~F.~P., P.~Sorensen,\nl
{\it Finite-size scaling as a tool in the search for the QCD critical point in
 heavy ion data}, \nl
  Physical Review C {\bf 84}, 011903 (2011) [Rapid Communications].\nl
[+ proceedings:  PoS {\bf FACESQCD} 2010, 017 (2010); Physics of Atomic Nuclei {\bf 75}, 906 (2012)]

\item  E. S. Fraga, L. F. P.,\nl
{\it Deconfinement in the presence of a strong magnetic background: an exercise within the MIT bag model},\nl
Physical Review D {\bf 86}, 016008 (2012).

\newpage

%

\item J.-P. Blaizot, E. S. Fraga, L. F. P., \nl
   {\it Thermodynamics of QCD in a strong magnetic field}.\nl
     Work in progress.

\item J.-P. Blaizot, L. F. P., \nl
   {\it Functional renormalization group analysis of relativistic Bose-Einstein condensation}.\nl
  Work in progress.

\end{enumerate} 

\vspace{14.0cm}

{\footnotesize \centerline{Rio de Janeiro}}
{\footnotesize \centerline{february/2012}}

%% file: introducao.tex
\chapter[Introduction: the genesis of the different phase diagrams]{
\label{intro}}
\chaptermark{Introduction: the genesis of the different phase ...}

\vspace{1.5cm}

{\huge \sc Introduction: the genesis of }
\vspace{0.3cm}

\noindent {\huge \sc the different phase diagrams}

\vspace{2cm}

\vspace{-12cm}
\hspace{6cm}
\includegraphics[width=8.5cm,angle=90]{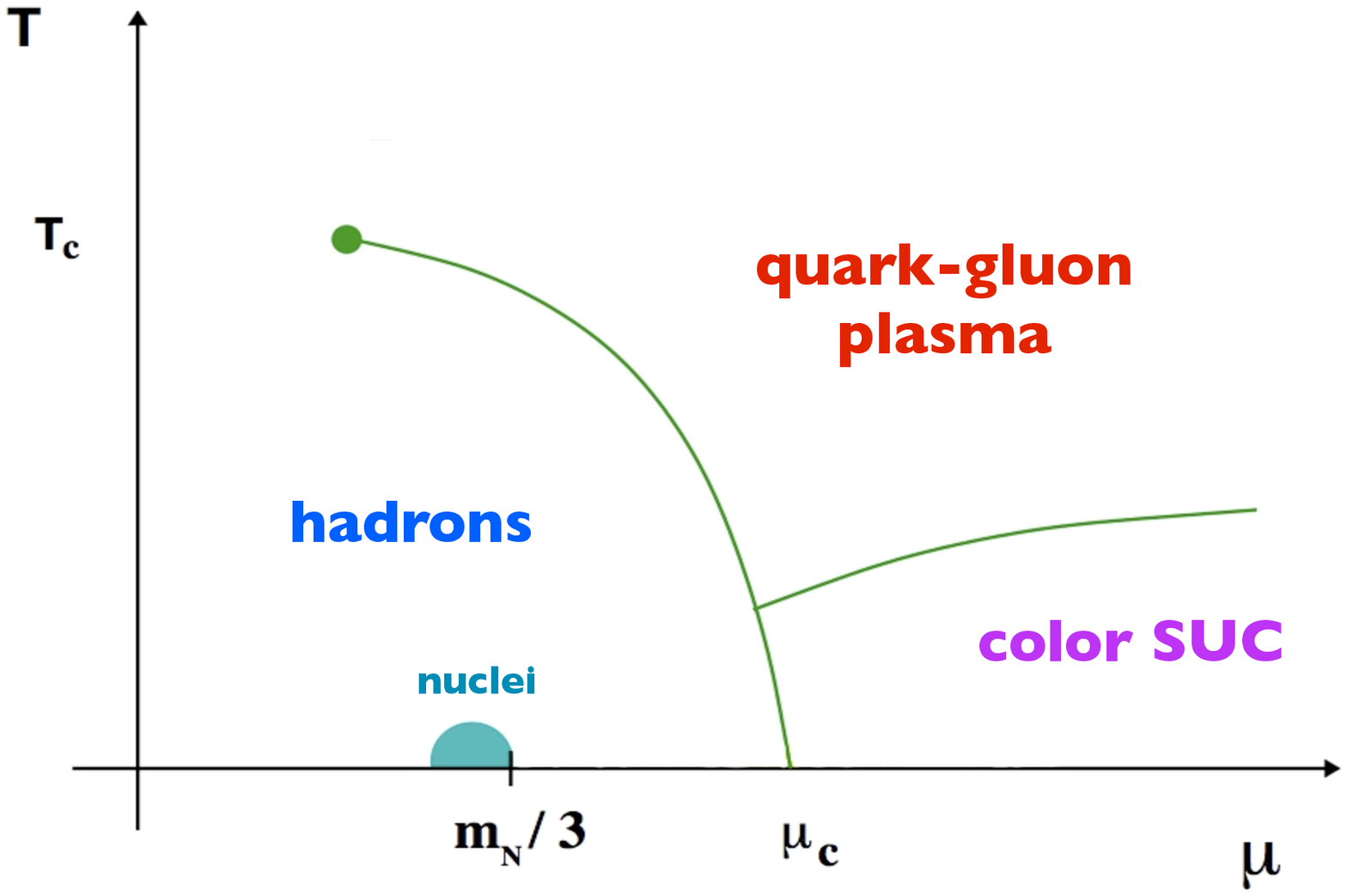}

\vspace{3.8cm}

A broad comprehension of the physical phenomena governed by a given fundamental theory requires the description of the phase structure it predicts. Since the statistical-mechanics definition of the phase of a system is intimately connected to symmetry characteristics, the first step towards understanding the phase structure of any system consists therefore of an investigation of the symmetries that play an important role in its characterization. In the case of Strong Interactions, the corresponding fundamental theory, Quantum Chromodynamics (QCD), presents a very rich structure of symmetries in the vacuum \cite{shuryak-livro}. In this vein, an intricate structure of phases is also expected.


The different phases of Strong Interactions and the transitions between them play important roles in different physical scenarios, with observational and experimental interest. In the primordial universe, for example, it is generally believed that there existed an era dominated by a hot soup of quarks and gluons at about $10^{-12}$ seconds after the Big Bang \cite{cosmo}. As the universe expanded and cooled down, a primordial confinement phase transition is thus expected.
Our universe today may also contain naturally occurring deconfined systems: 
estimates point out that the core of compact stars could accommodate high enough pressure
so that stable matter is not the conventional nuclear one, nor the hadronic dominated medium, but rather a system of deconfined quarks and gluons \cite{G_b}.
Moreover, the energy scales of the phase transitions of Strong Interactions are within the reach of laboratory experiments colliding ultra-relativistic heavy ions \cite{QGP}, such as RHIC-BNL \cite{RHIC} and LHC-CERN \cite{ALICE} and, in the near future, FAIR-GSI \cite{FAIR} and NICA-JINR \cite{NICA}.
%

Several studies and developments were accomplished in the last decades concerning the investigation of the phase structure of Strong Interactions. In special, the description 
of the phase diagram in the plane of temperature $T$ {\it versus} baryon chemical potential $\mu_B$ has received remarkable attention, with sensibly deeper advances on the temperature axis, due to the development of lattice QCD in this domain \cite{Philipsen:2011zx}. Experimentally, heavy-ion collisions at RHIC-BNL have established the existence with a new state of matter dominated by partonic degrees of freedom \cite{Adams:2005dq}.

%

Quantitatively, however, the knowledge about the phase diagram of Strong Interactions is still extremely incipient. There are even some qualitative open questions, such as e.g. the existence of the second-order critical endpoint (CEP) as the final point of a presumed first-order transition line and what concerns the connection between chiral symmetry breaking and confinement, besides the structure of the region where color superconductivity should occur \cite{Rischke:2003mt}. The phase structure of Strong Interactions is essentially, therefore, an open problem.

This is due to various difficulties encountered in the description of Strongly interacting matter under extreme conditions. First of all, it is clear that the investigation of any phase diagram requires the consideration of the theory in a broad domain of energy scales. In quantum field theory, according to the renormalization group formalism, the study of the predictions of a theory  for different energy scales must be implemented via running parameters (couplings, masses, etc). In this vein, when large enough variations of the energies are considered, non-perturbative regimes arise naturally. 

Among the non-perturbative methods to approach the fundamental gauge theory in this regime, lattice QCD stands out, since it solves in principle the full fundamental theory. There are, nevertheless, technical obstacles even for simulations of QCD on a lattice. The limitation of maximum lattice sizes and minimum lattice spacings imposed by reasonable computing times
generates infra-red and ultra-violet cutoffs for the physical phenomena being analyzed.
A deeper restriction, the notorious Sign Problem \cite{Philipsen:2011zx} renders impossible the utilization of conventional Monte Carlo methods (based on importance sampling) in the case with a finite baryon chemical potential, in which the weight of the path integral representation of the partition function becomes complex. Summing up, there exists no single method or theory that can be successfully applied to obtain the complete phase structure of Strong Interactions. The theoretical development in this area of research relies necessarily on the exploration of the complementarity of the various available methods, including different effective treatments.


%
%
%


On the experimental side, the years that follow promise significant advances in this area.
Different experiments and programs are planned to scrutinize a broad domain of intermediate chemical potentials, with the general goal of placing the first experimental point in the QCD phase diagram at high energies. Searches concentrate not only on critical correlations built up in a second-order phase transition (at the CEP), but also on possible signatures of the expected first-order line at higher baryon-antibaryon asymmetries. From the observational side, increasing improvements in astronomy recently point out to the possibility that precise data from astrophysical phenomena will soon provide constraints to ultra-dense Strongly interacting media.
This experimental and observational enterprise will certainly contribute concretely to the knowledge of how the predicted critical region of the phase diagram of Strong Interactions is in fact realized in nature.

%


The conjuncture of factors presented above implies an intense demand for theoretical studies of the phase diagram of Strong Interactions. During this challenging odyssey to overcome the obstacles in the description of the phase structure of Strongly interacting hot and dense matter, new control parameters arise naturally, springing new axes and new phase diagrams. The genesis of the different phase diagrams of Strong Interactions is therefore spontaneously triggered by the intricacy of the task of understanding Strongly interacting environments.

Axes involving the numbers of colors and flavors of the theory or the masses of the fundamental fields are examples of phase diagrams that are built due to theoretical difficulties that drive the study of ``cousin'' theories forward. Even though variations of these parameters are not realized in nature, one gains in comprehension of their role in the predictions via the study of theories which are not the original, physical one but ``next of kin'' to it. Mass effects are investigated in this thesis nonperturbatively in the context of a general Yukawa theory with massive fermions and bosons at finite temperature and density. This is a general result, with potential application in effective models of QCD and in different phenomena concerning media with interacting fermions.

On the other hand, computational technical difficulties may also motivate the analyses of new control parameters: in order to circumvent the Sign Problem when simulating dense media on the lattice, one may study imaginary chemical potentials and the isospin chemical potential, for example.
The latter should also occur in nature: the imposition of charge neutrality of compact astrophysical objects generates an asymmetry between quarks up and down in the ultra-dense core. We construct in this thesis an effective model inspired in chiral perturbation theory and perturbative QCD with the aim of investigating variations of masses and baryon and isospin chemical potentials. Findings agree with lattice QCD simulations at finite temperature and involve further predictions within the regime in which the Sign Problem is manifest, as the critical baryon chemical potential at low temperatures, which may affect compact star structure.

Moreover, extra macroscopic parameters may arise from its relevance in a specific experimental setup. In heavy-ion collisions, the energy scales associated with typical sizes of the system created are of the same order as the energies within the critical region of the phase transitions, so that it is important to investigate finite-size effects. Indeed, we show using the linear sigma model with constituent quarks that the pseudo-critical phase diagram that is currently being probed in heavy-ion collisions may differ considerably from the expectations in the thermodynamic limit. We further propose a finite-size scaling analysis as an alternative signature for the second order critical endpoint of QCD and, in collaboration with an experimentalist from STAR at the accelerator RHIC-BNL, demonstrate the viability of the application of this method to data analysis.

In Nature and in the laboratory, the QCD phase transitions take place generally out of equilibrium. The phase structure probed is therefore that of dynamical critical phenomena. In this direction, we compute the chiral surface tension in the cold and dense regime within the linear sigma model, suggesting a much lower value as compared to previous estimates, possibly allowing for observable signals of a QCD phase transition in core-collapse supernova explosions. Still in an out-of-equilibrium context, memory effects could be relevant to the time evolution of an order parameter (as can be seen in simpler models in quantum statistical mechanics \cite{Palhares:2009rq}).

The initial stage of noncentral collisions and the core of some ultra-compact astrophysical objects (the so called ``magnetars'') feature Strongly interacting media exposed to enormous magnetic fields, pointing thus to the study of background magnetic fields as a new axis of the phase diagram of Strong Interactions. We address this problem by analyzing QCD thermodynamics in the presence of a uniform and constant intense magnetic field. Questions like ``what are the adequate quasiparticles?'' and ``how does the non-Abelian interaction affects the pressure in this extreme regime?'' are investigated analytically within the well-defined context of the fundamental gauge theory. Results can be compared to recent lattice QCD simulations, and one expects in this way to gain information about different ingredients of in-medium QCD and about the behavior of perturbation theory in this domain.

With the progress of the analyses of these new phase diagrams -- initially considered as side effects of the original quest of predicting the temperature-baryon chemical potential phase structure --, it is not surprising that some of them develop independent relevance. At least academically, they are in principle part of the phase structure of the Standard Model of fundamental interactions. Besides that, such conditions may be very interesting phenomenologically, since new axes in general allow for the existence of new phases. The standard case is that of the baryon chemical potential, which gives rise to color superconducting phases \cite{Alford:2007xm}.  Nevertheless, many other examples appear in the literature. The existence of a new confined phase, ``quarkyonic'', has been proposed based on its formal prediction for the theory with infinite number of colors \cite{Andronic:2009gj}. 
More recently, the consideration of the magnetic field axis has opened the discussion on the possibility of a superconducting lowest-energy-state due to $\rho-$meson condensation at intense magnetic backgrounds \cite{Chernodub:2010qx}. Other chemical potentials (isospin/strangeness) should also generate new phases, corresponding to Bose-Einstein condensation of pions/kaons. In the final part of this thesis, pion Bose-Einstein condensation in isospin dense media is discussed. The effect of interactions has proven to be a non-trivial issue in non-relativistic Bose-Einstein condensation. Using the nonperturbative framework of the Functional renormalization Group, we address this question in the relativistic case, with the application for Strongly interacting media in mind.


Our goal in this thesis is therefore an exploration of some of these different phase diagrams through the inescapable utilization of various techniques, formal and phenomenological, involving effective models and the fundamental gauge theory, with a close connection with experiments and lattice simulations whenever possible. We hope that this study contributes to understanding the role played by different ingredients encountered in Strongly interacting environments and ultimately to
the crucial and arduous task of showing that the underlying microscopic theory of Strong Interactions and our interpretation of it are consistent with nature.

%

The thesis is organized as follows. After this initial introductory chapter, results on different phase diagrams of Strong Interactions are exposed in Chapters \ref{isospin}--\ref{relatBEC}. Mass effects are analyzed for the deconfinement phase transition in an isospin-dense QCD medium in Chapter \ref{isospin}. The pseudo-critical phase diagram of Strong Interactions, including the typical system size as an extra axis, is presented in Chapter \ref{FS}. Consequences for observables in heavy-ion experiments, especially those related to the critical endpoint, are investigated.
Chapter \ref{surften} addresses how out-of-equilibrium phenomena may influence the QCD phase transitions actually probed in observations and experiments: nucleation parameters are computed for the cold and dense chiral phase transition, with applications in astrophysics.
The thermodynamics of QCD in a magnetic background is analyzed in Chapter \ref{magneticQCD}. Nonperturbative corrections to effective models are addressed by computing the pressure of a general massive Yukawa theory within Optimized Perturbation Theory at finite temperature and baryon chemical potential in Chapter \ref{OPT}.
Chapter \ref{relatBEC} contains a Functional Renormalization Group analysis of relativistic Bose-Einstein condensation in an interacting, charged scalar field theory, as well as in a chiral model for an isospin-asymmetric pion medium. Finally, the conclusions of this thesis are presented in Chapter \ref{conclusions}. Appendices contain some technical issues tackled during the development of this work.

%% file: isospin.tex
\chapter[Phase diagrams of Strong Interactions in the presence of different chemical potentials and masses]{
\label{isospin}}
\chaptermark{Phase diagrams of Strong Interactions in the...}

\vspace{1.5cm}

{\huge \sc Phase diagrams of Strong }
\vspace{0.3cm}

\noindent {\huge \sc  Interactions in the presence of }
\vspace{0.3cm}

\noindent {\huge \sc  various chemical potentials and masses}


\vspace{-11cm}
\hspace{6cm}
\includegraphics[width=8.5cm,angle=90]{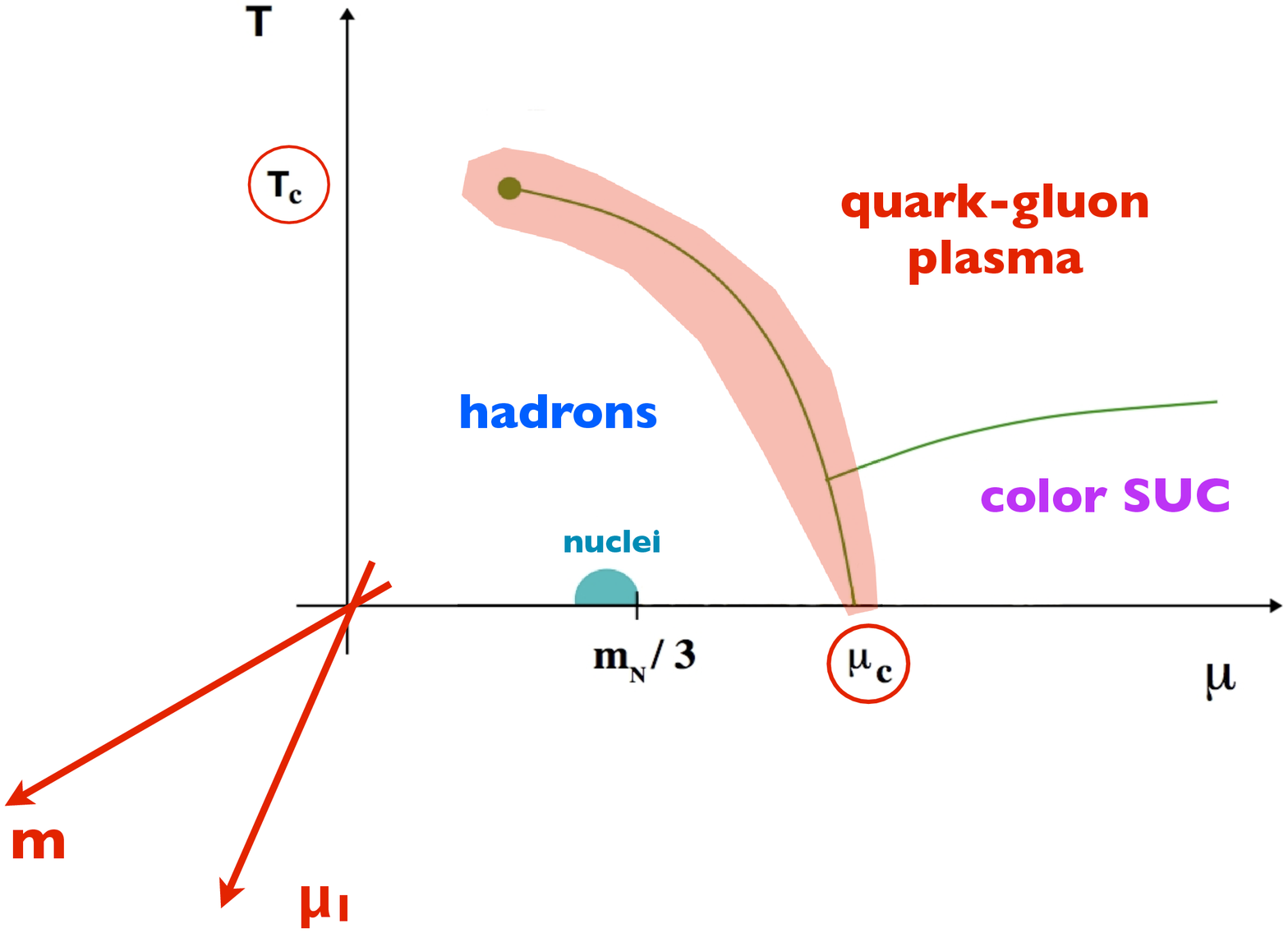}

\vspace{4.5cm}


Net baryon number is commonly considered when investigating Strongly interacting matter. Nevertheless, there are other important conserved quantum numbers involving quark degrees of freedom, such as isospin and strangeness, which are usually set to zero. In this chapter we will analyze a more general type of QCD medium including the possibility of a nonzero net isospin density and investigate its interplay with finite quark mass effects.

The specific problem we address is that of describing the deconfinement critical temperature as a multiple variable function, $T_c(\mu_B,\mu_I,m_{\pi})$ \cite{Fraga:2008be}, and its analogous at zero temperature, the critical baryon chemical potential $\mu_B^{\rm crit}(\mu_I,m_{\pi})$ \cite{Palhares:2008px}.
Due to the absence of the Sign Problem in isospin dense QCD, comparison of $T_c(\mu_B=0,\mu_I,m_{\pi})$ with lattice results is possible, providing information on the ingredients in an effective theory that suffice to capture the role played by mass and isospin effects in the critical region. The model we construct using this criterium supplies predictions also within the Sign Problem domain; the zero temperature critical baryon chemical potential  being relevant to the astrophysics of compact objects.

%

\section{Introduction}


Small quark masses and a nonzero baryon chemical potential have 
always represented major challenges for lattice simulations. Presently, 
although viable lattice sizes still prevent extensive and precise 
studies with realistic quark masses, and the Sign Problem considerably restricts 
the applicability of Monte Carlo simulations to the description of chemically 
asymmetric media, Lattice QCD is starting to provide results with smaller 
quark masses, and probing, with approximate methods, a larger domain at finite chemical 
potential \cite{Fodor:2007sy,Cheng:2007jq,Philipsen:2008gf}. For the latter, it is 
believed that much can be learned from simulations of realizations of QCD 
that avoid the Sign Problem, such as those with vanishing baryon chemical potential 
and finite isospin density, which has a positive fermionic determinant.

QCD at finite isospin density is certainly, but not only, a playground 
to test numerical approaches to the case of finite baryon density on the 
Lattice \cite{Kogut:2004zg,deForcrand:2007uz}. It is also part of the 
physical phase diagram for strong interactions, and exhibits a very rich 
phenomenology \cite{Son:2000xc}. It has been under careful study 
during the last years, both theoretically and experimentally, with a clear identification 
of certain phenomena that depend directly on the isospin asymmetry 
of nuclear matter at intermediate-energy heavy ion experiments (see Ref.  \cite{Li:2008gp} 
and references therein).

Nevertheless, theoretical and phenomenological studies often focus on the 
chiral limit of QCD, putting aside effects from finite quark masses, and 
on isospin symmetric
%
%
hot matter, mainly stimulated by the 
physical scenario found in current high-energy heavy ion collision 
experiments \cite{QM} and the quark-gluon plasma \cite{Rischke:2003mt}.
Some exceptions are, however, chiral model analyses of pion-mass dependence
of the finite temperature transition (e.g., \cite{Berges:1997eu,Dumitru:2003cf,Braun:2005fj}),
which fail completely to reproduce the lattice behavior, and results within Polyakov-loop
models that are consistent with lattice via fitting procedures. At finite isospin chemical potential,
though there are phase-diagram investigations in PNJL models \cite{Mukherjee:2006hq},
none of them has addressed the critical temperature dependence on the isospin chemical
potential in particular.
%
%

To investigate the effects of nonzero quark 
%
masses
and isospin asymmetry on the deconfining transition, we build an effective theory that 
combines ingredients from $\chi$PT in the low-energy sector with the 
phenomenological fuzzy bag model at high energy. The high-energy regime is described 
perturbatively by two-flavor QCD with massive quarks and explicit isospin symmetry breaking. 
Nonperturbative (confinement) effects at this scale are incorporated through a fuzzy bag 
description \cite{Pisarski:2006hz} with coefficients extracted from lattice data. For the 
low-energy sector, we adopt an effective action inspired by $\chi$PT that 
exhibits the same structure of symmetries contained in the high-energy theory \cite{Loewe:2002tw}.
The quasi-nucleon degrees of freedom described in this regime seem to be crucial for
understanding the pion-mass dependence of the deconfining temperature.
The definition of parameters, as well as masses, is such that variations in the deconfined sector 
are totally consistent with variations in the confined one, which guarantees the bookkeeping in 
the different degrees of freedom 
that are present in the description.

In this chapter of the thesis, after presenting the effective model, we use it to investigate the effects of finite quark masses
and isospin
number on the equation of state of hot and dense strongly interacting matter 
and on the deconfining phase transition.
%
The
setting we have proposed is simple and {\it completely} fixed by vacuum QCD properties
(measured or simulated on the lattice) and lattice simulations of finite temperature QCD.
More explicitly, there is no fitting of mass or isospin chemical potential dependence at all.
Nevertheless,
our findings for the behavior of the critical temperature as a function of
{\it both}
the pion mass and 
the isospin chemical potential are in remarkably good agreement with lattice data.

\begin{figure}[htb]
\vspace{-1cm}
\center
\includegraphics[width=10.5cm,angle=90]{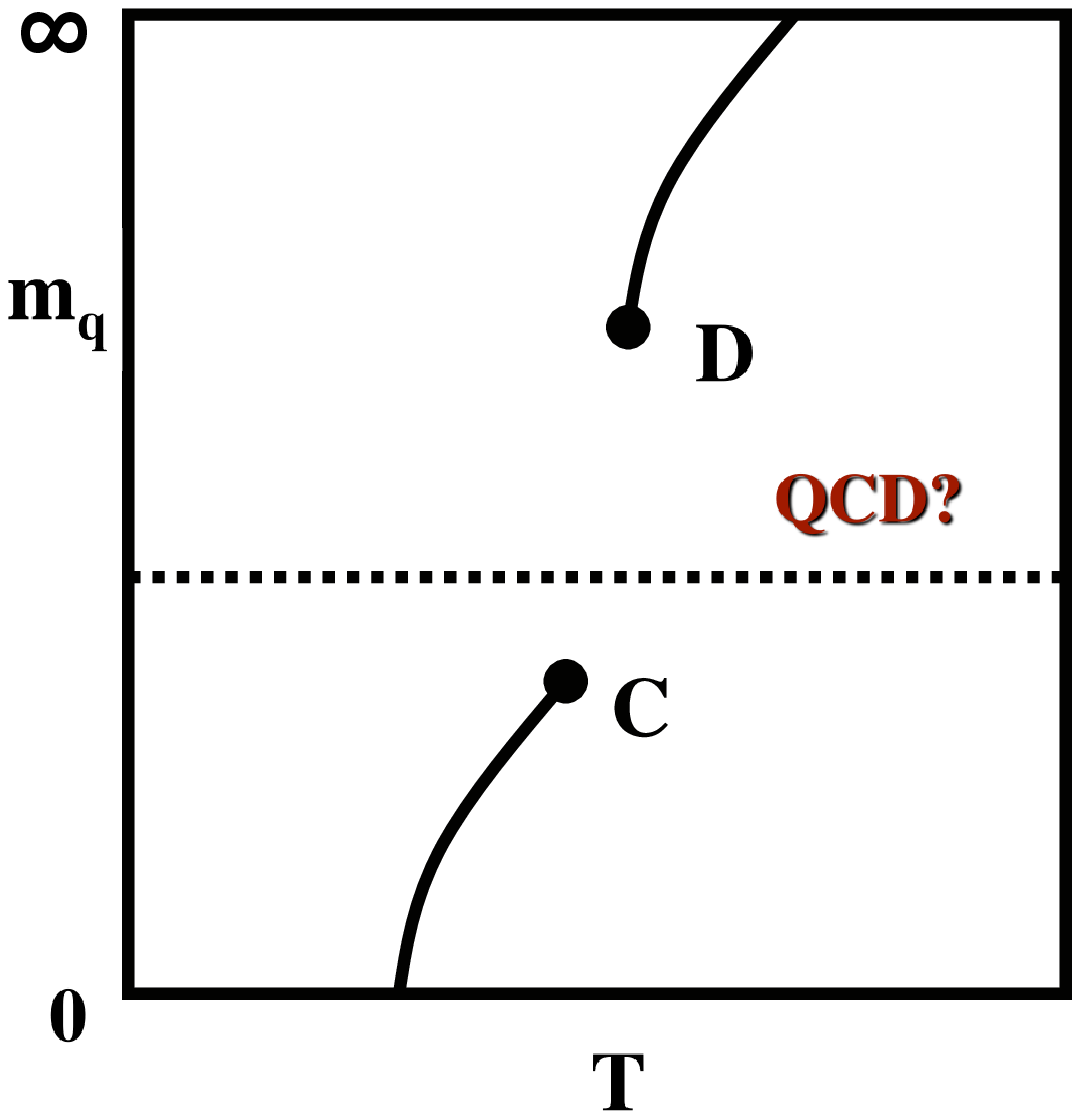}
\vspace{-2.5cm}
\caption{Cartoon of the QCD mass phase diagram.}
\label{QCD?}
\end{figure}

It is crucial to note that several detailed studies of chiral models failed to 
describe $T_c(m_{\pi})$ \cite{Berges:1997eu,Dumitru:2003cf,Braun:2005fj}, while 
Polyakov-loop models, whose results can be fitted to the lattice points for
$T_c(m_{\pi})$ \cite{Dumitru:2003cf}, cannot at the present form address isospin effects (nor any other observable related to fermionic chemical potentials).
This suggests that the functional behavior of the critical temperature is a good observable of effects from mass variations, being sensitive to the physics entering in each case. The full function $T_c\left(m_{\pi}\right)$ is probably much more sensitive to the underlying physical phenomena than the actual value of the critical temperature in Nature. Putting together the fact that first-principle lattice QCD simulations can access these``cousin'' theories, with different masses, a thorough analysis of the QCD critical temperature as a multi-variable observable might be a path worth following to identify the possible differences between chiral and deconfinement transitions and understand why they seem to be connected in real-world QCD. These issues are also related to the question of whether QCD is more influenced by the pure gauge/large $N_c$ physics or by its chiral properties, the top and bottom regions of the cartoon phase diagram in Figure \ref{QCD?}.

In our approach, one can also investigate
in a straightforward manner
the effects of quark-mass asymmetry and nonzero baryon chemical
%
potential, physical cases in which the Sign Problem develops, constraining
systematic lattice studies. 
The case of cold, dense matter is, of course, 
relevant for compact stars \cite{stars} and for the phenomenon of color 
superconductivity \cite{Alford:2007xm}.

The chapter is organized as follows. Section \ref{ETh} presents the effective model adopted, 
as well as a discussion of the approximations implied. In Section \ref{ResultsIsospin} we show our results for the effects of mass and isospin chemical potential on the critical 
temperature and critical baryon chemical potential, comparing to lattice data whenever possible. 
We also compare some of these results to those obtained in different phenomenological 
frameworks. Section \ref{ConcIsospin} contains final remarks. Some details of the technical 
derivation of the pion physical (dressed) masses are left for Appendix \ref{ApIs}.

\section{Effective theory\label{ETh}}

As discussed above, to investigate the effects of nonzero quark masses, baryon chemical potential, 
and isospin asymmetry on the deconfining transition, we build an effective theory that 
combines ingredients from chiral perturbation theory in the low-energy sector with the 
phenomenological fuzzy bag model at high energy.
In this section, the model and its degrees of freedom will be presented explicitly, 
showing thus in a concrete setting the consistent connection between the parameters and masses of the two sectors. 

\subsection{Low-energy sector}

The physical setting we adopt for the low-energy regime of strong interactions is that of a 
system of heavy nucleons in the presence of a hot gas of pions whose masses are 
already dressed by corrections from temperature, isospin chemical potential and 
quark masses. The effective Lagrangian has the form 
$\mathcal{L}_{eff} = \mathcal{L}_{N}+\mathcal{L}_{\pi N}+\mathcal{L}_{\pi}$, with
\be
\mathcal{L}_{N} &=& \overline{N} \left(
i\slashed{\partial}-M_N+\frac{1}{2}\mu_I\gamma_0 \tau^3+\frac{3}{2} \mu_B \gamma_0
\right) N
\, ,\label{LN}
\\
\mathcal{L}_{\pi N} &=&\frac{g_A}{f_{\pi}}\overline{N} i \gamma_5 \left(
\slashed{\partial}\pi-\frac{1}{2}\mu_I\gamma_0 [\tau^3,\pi] 
\right) N
\, ,
\\
\mathcal{L}_{\pi} &=&(\partial_{\mu}-ih\delta_{\mu 0})\pi^+(\partial_{\mu}+ih\delta_{\mu 0})\pi^-
-\overline{m}_{\pm}^2\pi^+\pi^- +\frac{1}{2}\left[ 
(\partial \pi^0)^2 -\overline{m}_0^2(\pi^0)^2
\right]
\, , \label{Lpi}
\ee
where $g_{A}$ is the axial vector current coefficient of the nucleon, which accounts for 
renormalization in the weak decay rate of the neutron, $f_{\pi}$ is the pion decay constant, 
and $h$ is a function of temperature and isospin chemical potential. 
Here the nucleons are represented by $N= (p,n)$ with $p,n$ being the proton and neutron
spinors, respectively, and have a mass matrix 
$M_N=\textrm{diag}(M_p,M_n)=\textrm{diag}(M-\delta M,M+\delta M)$. This is the ${\cal O}(P)$ 
nucleon chiral Lagrangian \cite{Gasser:1987rb}, but considering mass corrections at zero 
temperature and chemical potential, and coupling to the dressed pions.

The effective (dressed) masses of the pions $\pi^0=\pi^3$ and 
$\pi^{\pm}=\frac{1}{\sqrt{2}}(\pi^1\mp i\pi^2)$, which depend on temperature $T$, isospin chemical 
potential $\mu_{I}$ and mass asymmetry $\delta m=(m_d-m_u)/2$, are denoted, respectively, by 
$\overline{m}_0=\overline{m}_0(T,\mu_I,\delta m)$ and 
$\overline{m}_{\pm}=\overline{m}_{\pm}(T,\mu_I)$.
The isospin and baryonic chemical potentials can be written in terms of the chemical 
potentials for the up and down quarks. Here we adopt the following convenient 
definition\footnote{The baryonic chemical potential is also frequently defined with an 
overall factor $2/3$.}: $\mu_I=\mu_u-\mu_d$ and $\mu_B=\mu_u+\mu_d$.
We also use the customary notation $\pi\equiv \pi^a\tau^a$, with $\tau^a$ being the Pauli matrices.

The physical masses in the dressed theory defined above were computed in 
Refs.  \cite{Loewe:2002tw}.
 In the so-called first 
phase, a regime in which $|\mu_{I}|<m_{\pi}$, they have the following 
form (see Appendix \ref{ApIs} for more details):
\be
m_{\pi^0}&=& \overline{m}_0 
= m_{\pi}\left[ 1+\frac{1}{2}\alpha\sigma_{00} \right] \, ,
\label{m0}
\\
m_{\pi^{\pm}} &=&
\overline{m}_{\pm}\mp h
= m_{\pi}\left[ 1+\frac{1}{2}\alpha\sigma_1 \pm\frac{1}{2}\alpha\frac{\sigma_0}{m_{\pi}} \right]\mp \mu_I \, ,
\label{mpm}
\ee
up to first order in $\alpha=(m_{\pi}/4\pi f_{\pi})^{2}$. The functions 
$\sigma_{00}$, $\sigma_0$, $\sigma_1$ and $h$ are defined in the appendix.

In the second phase ($|\mu_{I}|>m_{\pi}$), a condensation of pions occurs, and a superfluid 
phase sets in\footnote{Here we consider the $\pi^-$ condensation taking $\mu_I=-|\mu_I|$.}. 
In this new phase, in order to reestablish the vacuum structure, a chiral rotation is produced  due to 
the isospin symmetry breaking. All this produces a pion mixing, and the nucleons also couple 
in a different way. The degrees of freedom do not correspond anymore to pions, but we can still call 
them quasi-pions since their masses in the two phases match at the transition point. 
The tree level masses do not have the shape as in the equations above. Instead, $m_{\pi^0}=|\mu_I|$, 
$m_{\pi^-}=0$, and $m_{\pi^+}=\mu_I\sqrt{1+3(m_\pi/\mu_I)^4}$ \cite{Son:2000xc}. However, it is 
possible to treat this phase with a simple approximation near the superfluid phase transition.
In the regime in which $|\mu_{I}|\gtrsim m_{\pi}$, the natural expansion parameter is given by 
$s^{2}=1-m_{\pi}^{4}/\mu_{I}^{4}$ \cite{Loewe:2002tw}, after scaling all the parameters by $|\mu_I|$. 
The result of the first terms in this expansion ($s^2=0$) provides the same equations as in the normal 
phase, just replacing $m_\pi$ by $|\mu_I|$. Strictly speaking, this is valid only for values of $|\mu_I|$ 
very close to $m_\pi$, i.e. $|\mu_I|\lesssim \sqrt{8/7} ~m_{\pi}$ (for a more detailed discussion cf.
\cite{Loewe:2002tw}). For simplicity, we also apply this results for slightly higher values of the isospin 
chemical potential\footnote{
For higher values of $|\mu_I|$ one needs to consider more terms, not only in the expansion of the 
pion Lagrangian but also in the coupling with the nucleons, due to the appearance of the pion 
condensate. These terms will also be present at the perturbative QCD level via 
the chiral rotation. The case in which $|\mu_{I}| \gg m_{\pi}$ can be explored, nevertheless, 
using $m_{\pi}^{2}/\mu_{I}^{2}$ as an expansion parameter, even though 
the validity of the whole treatment is restricted to $|\mu_{I}|$ smaller than the 
$\eta$ or $\rho$ masses \cite{Son:2000xc}. 
}.
These results in $\chi$PT are confirmed by a NJL analysis \cite{He:2005nk}.

The direct effect of the baryonic chemical potential in the pure pion quasiparticle 
gas is omitted since, without considering gluonic corrections, it appears as an anomalous 
term in the ${\cal O}(P^4)$ chiral Lagrangian \cite{AlvarezEstrada:1995mh}, and will be present 
only in two-loop corrections according to power counting. For very large values of 
$\mu_{B}$, one has to incorporate effects from the color superconductivity gap in the calculation 
of meson masses in an effective theory near the Fermi surface \cite{Alford:2007xm,Beane:2000ms}. 
For simplicity, we do not include these corrections, and restrict our analysis to more modest 
values of baryonic chemical potential.

The nucleon masses, $M_p=M-\delta M$ and $M_n=M+\delta M$, are dressed by leading-order 
contributions in zero-temperature baryon chiral perturbation theory. Using the results from 
Ref. \cite{Procura:2006bj}, for the isospin symmetric case with explicit chiral symmetry breaking, 
and Ref. \cite{Beane:2006fk}, which includes explicit isospin breaking effects, we have 
(neglecting terms $\sim m_q^2$, $\sim m_q^2\log(m_q)$, and of higher order in $m_q$):
\be
M(m)&=& M_0 +2~\gamma_1~m+2^{3/2}~\gamma_{3/2}~m^{3/2} \, ,
\\
\delta M(\delta m)&=& 2~\gamma_1^{\textrm{asymm}}~\delta m
\, ,
\ee
$M_0$ being the nucleon mass in the chiral limit, $m=(m_u+m_d)/2$ the average quark mass, 
and\footnote{To relate the pion mass (in the isospin symmetric case) with the quark masses,
we use the Gell-Mann -- Oakes -- Renner relation: 
$m_{\pi}^2f_{\pi}^2=-(m_u+m_d)\langle\bar q q\rangle=2m(-\langle\bar q q\rangle)$.}
\be
\gamma_1 &=& \frac{-4~c_1}{f_{\pi}^2}~(-\langle\bar q q\rangle) \, ,
\\
\gamma_{3/2} &=&-~\frac{3g_A^2}{32\pi f_{\pi}^5}~(-\langle\bar q q\rangle)^{3/2} \, ,
\\
\gamma_1^{\textrm{asymm}} &=&
\frac{2\bar\alpha-\bar\beta}{3}~\frac{(-\langle\bar q q\rangle)}{f_{\pi}^2}
\, .
\ee
Here, all parameters and coefficients are fixed to reproduce properties of the
QCD vacuum either measured or extracted from recent lattice simulations.
Explicitly, $\langle\bar q q\rangle = -(225~\textrm{MeV})^3$ is the (1-flavor) chiral condensate 
in the chiral limit \cite{Chiu:2003iw} and, from Table I in Ref. \cite{Procura:2006bj},
$M_0 = (0.882 \pm 0.003)~\textrm{GeV}$, 
$c_1 = (-0.93\pm0.04)~\textrm{GeV}^{-1}$, 
$g_A = 1.267$, and 
$f_{\pi} = 92.4~\textrm{MeV}$. 
From Table 3 in Ref. \cite{Beane:2006fk}, one can extract
\be
\frac{2\bar\alpha-\bar\beta}{3} &=&
\left\{
\begin{matrix}
(0.198 \pm 0.093)~(\textrm{lattice units})  & ,\; \textrm{$O(m_q)$ fit}
\\
(0.229 \pm 0.058)~(\textrm{lattice units}) & ,\; \textrm{$O(m_q^{3/2})$ fit}
\end{matrix}
\right.
\, ,
\ee
where $1~(\textrm{lattice units})=b=0.125~$fm. Therefore:
\be
\frac{2\bar\alpha-\bar\beta}{3} &=&
\left\{
\begin{matrix}
(0.12543 \pm 0.0589)~\textrm{GeV}^{-1}  & ,\; \textrm{$O(m_q)$ fit}
\\
(0.145064 \pm 0.036741)~\textrm{GeV}^{-1} & ,\; \textrm{$O(m_q^{3/2})$ fit}
\end{matrix}
\right.
\, .
\ee
Evaluating the coefficients $\gamma$ explicitly, we arrive at 
$\gamma_1 = 4.9630 \pm 0.2135$, 
$\gamma_{3/2} = -0.273424~\textrm{MeV}^{-1/2}$, and 
\be
\gamma_1^{\textrm{asymm}} &=&
\left\{
\begin{matrix}
0.16734 \pm 0.07858  & ,\; \textrm{$O(m_q)$ fit}
\\
0.19354 \pm 0.04902   & ,\; \textrm{$O(m_q^{3/2})$ fit}
\end{matrix}
\right.
\, .
\ee

Finally, since the nucleon chemical potentials are given by
\be
\mu_p &=& 2\mu_u+\mu_d= \frac{3}{2}\mu_B+\frac{1}{2}\mu_I \, ,
\\
\mu_n &=& \mu_u+2\mu_d= \frac{3}{2}\mu_B-\frac{1}{2}\mu_I
\, ,
\ee
the dispersion relation satisfied by the proton and neutron are, respectively:
\be
E_{\textrm{proton}} ({\bf p}) &=& 
\sqrt{{\bf p}^2+(M-\delta M)^2} +\frac{3}{2}\mu_B+\frac{1}{2}\mu_I \, ,
\label{Eproton}
\\
E_{\textrm{neutron}}({\bf p}) &=& 
\sqrt{{\bf p}^2+(M+\delta M)^2} +\frac{3}{2}\mu_B-\frac{1}{2}\mu_I
\, ,\label{Eneutron}
\ee
where ${\bf p}$ is the 3-momentum and the antiparticle dispersion relations are obtained
from the ones above by the substitution $\mu_i \mapsto -\mu_i$.

Using the dispersion relations for pions and nucleons, we can write the total 
pressure for the gas of free (dressed) quasiparticles as 
$P_{\textrm{chiral}} = P_{\textrm{pions}} +P_{\textrm{nucleons}}$, with
\be
P_{\textrm{pions}}  &=&
-T\int\frac{d^3{\bf k}}{(2\pi)^3}~
\Bigg\{
\log\left[
1-\textrm{e}^{-\beta [\omega(\overline{m}_{\pm})-h]}
\right]
+
\log\left[
1-\textrm{e}^{-\beta [\omega(\overline{m}_{\pm})+h]}
\right]
+
\nonumber \\
&&\quad
+
\log\left[
1-\textrm{e}^{-\beta \omega(\overline{m}_{0})}
\right]
\Bigg\} \, ,
\\ \nonumber \\
P_{\textrm{nucleons}} &=&
2T\int\frac{d^3{\bf p}}{(2\pi)^3}~
\Bigg\{
\log\left[
1+\textrm{e}^{-\beta [E(M-\delta M)+\mu_p]}
\right]
+
\log\left[
1+\textrm{e}^{-\beta [E(M-\delta M)-\mu_p]}
\right]
+
\nonumber \\ 
&&\quad
+
\log\left[
1+\textrm{e}^{-\beta [E(M+\delta M)+\mu_n]}
\right]
+
\log\left[
1+\textrm{e}^{-\beta [E(M+\delta M)-\mu_n]}
\right]
\Bigg\}
\, ,
\ee
where we have defined
$\omega(m) \equiv \sqrt{{\bf k}^2+ m^2}$ and $E(M) \equiv \sqrt{{\bf p}^2+M^2}$.

\subsection{High-energy sector}

The Fuzzy Bag Model has been proposed by Pisarski \cite{Pisarski:2006hz} as a phenomenological
parameterization of the equation of state to account for the plateau in the trace anomaly 
normalized by $T^2$, $(\epsilon-3p)/T^2$, observed in lattice results above the critical temperature.
Besides the usual MIT-type bag constant, the total pressure for QCD in this model has also a 
non-perturbative contribution\footnote{The underlying theoretical framework is that 
of an effective theory of Wilson lines and their electric 
field \cite{Pisarski:2006hz,Diakonov:2003yy,Megias:2003ui}.} $\sim T^2$:
\be
p_{\textrm{deconf}}(T) \simeq p_{\textrm{pQCD}}(T)-B_{\textrm{fuzzy}}~T^2-B_{\textrm{MIT}} \, .
\label{FuzzyTpressure}
\ee
The trace anomaly associated with this equation of state, assuming that $p_{\textrm{pQCD}}\sim T^4$, 
is then
\be
\epsilon-3p \; =\; T \frac{\partial p}{\partial T}-4p 
\; =\; 2 B_{\textrm{fuzzy}}~T^2 + 4B_{\textrm{MIT}}
\, . \label{FuzzyTraceAnomaly}
\ee

In Ref. \cite{Cheng:2007jq}, a similar parameterization has been used to fit lattice results for
the trace anomaly at high temperatures, $T>1.5~T_c \approx 300~$MeV, yielding\footnote{In the regime $T_c<T<1.5~T_c$, the authors of Ref. \cite{Cheng:2007jq} identify a 
strongly non-perturbative regime, which is not included in the fuzzy-type fit.}
\be
\left( \frac{\epsilon-3p}{T^4} \right)_{\textrm{high T}} &=&
\frac{3}{4}~b_0~g^4+\frac{b}{T^2}+\frac{c}{T^4}
\, , \label{latticefuzzyfit}
\ee
with (cf. Table VIII of Ref. \cite{Cheng:2007jq})
\be
b &=&
\left\{
\begin{matrix}
(0.101\pm0.006)~\textrm{GeV}^2 &,\, N_{\tau} = 4 \; ; \; g^2 \equiv 0
\\
(0.16\pm0.06)~\textrm{GeV}^2 &,\, N_{\tau} = 6 \; ; \; g^2 = 2.3\pm0.7
\end{matrix}
\right.
\\
c &=&
\left\{
\begin{matrix}
(0.024\pm0.001)~\textrm{GeV}^4 &,\, N_{\tau} = 4 \; ; \; g^2 \equiv 0
\\
(0.013\pm0.006)~\textrm{GeV}^4 &,\, N_{\tau} = 6 \; ; \; g^2 = 2.3\pm0.7
\end{matrix}
\right.
\ee
Notice that the first term in Eq. (\ref{latticefuzzyfit}) comes from a $O(\alpha_s^2)$ perturbative
contribution to the pressure and is important for the fit only at very high temperatures.
We neglect this term in what follows.
Comparing Eqs. (\ref{FuzzyTraceAnomaly}) and
(\ref{latticefuzzyfit}), we extract the following values for the bag coefficients:
\be
B_{\textrm{fuzzy}} &=&
\left\{
\begin{matrix}
(0.051\pm0.003)~\textrm{GeV}^2 &,\, N_{\tau} = 4 \; ; \; g^2 \equiv 0
\\
(0.08\pm0.03)~\textrm{GeV}^2 &,\, N_{\tau} = 6 \; ; \; g^2 = 2.3\pm0.7
\end{matrix}
\right.
\\
B_{\textrm{MIT}} &=&
\left\{
\begin{matrix}
(0.0060\pm0.0003)~\textrm{GeV}^4 &,\, N_{\tau} = 4 \; ; \; g^2 \equiv 0
\\
(0.0033\pm0.0015)~\textrm{GeV}^4 &,\, N_{\tau} = 6 \; ; \; g^2 = 2.3\pm0.7
\end{matrix}
\right.
\ee

In our effective theory, we adopt, phenomenologically, a simple extension of the equation of 
state in Eq. (\ref{FuzzyTpressure}) which includes the influence of finite chemical potentials in 
the perturbative pressure, neglecting for simplicity non-perturbative contributions due to the 
finite quark chemical potentials $\mu_f$, so that:
\be
p_{\textrm{deconf}}(T,{\mu_f}) \simeq p_{\textrm{pQCD}}(T,{\mu_f})-B_{\textrm{fuzzy}}~T^2-B_{\textrm{MIT}}
\, . \label{FuzzyTpressuremu}
\ee
%

\section{Results for the functional behavior of the deconfinement critical parameters\label{ResultsIsospin}}

With the explicit expressions for the pressure in the deconfined phase and in the 
low-energy (confined) phase, we can study the effects of varying the quark masses 
and the chemical potentials on the thermodynamics of the deconfining transition. In 
particular, we can study the behavior of the critical temperature, with the aim of comparing with available 
lattice results. The critical baryon chemical potential in the cold case is well within the region where the Sign Problem sets in rendering lattice QCD essentially inapplicable, so that model predictions are very welcome.

From our results for the massive free gas contribution of the pQCD pressure in the fuzzy bag model 
at finite temperature, isospin and baryon number, and the free gas pressure of quasi-pions and 
nucleons in the low-energy regime, the critical temperature and chemical potential for the deconfining 
phase transition are extracted by maximizing  the total pressure. 
The validity of our approach is, of course,
restricted by the scale of $\chi$PT: e.g. for $m_{\pi}\approx m_{\rho}\approx 770$ MeV, the expansion
parameter in $\chi$PT is $\alpha\approx 0.45$, so that the extension of the predictions to $m_{\pi}\lesssim 1$ GeV
is in principle justified\footnote{A rough estimate of the error (due to neglecting higher orders in chiral perturbation theory) in the effective masses of the low-energy sector
of our approximation for $m_{\pi}\sim 1$ GeV yields $\sim 30\%$.}.

By matching the two branches of the equations of state, corresponding to the
high and low temperature regimes, we of course obtain a first-order transition.
Systematic Lattice QCD calculations, with realistic quark masses \cite{Fodor:2007sy} (or almost realistic \cite{Cheng:2007jq}),
seem to indicate a crossover instead. On the other hand, from the experimental
standpoint a weakly first-order transition is not ruled out, and in fact corresponds
to the scenario adopted in very successful hydrodynamic calculations \cite{QM}.
Although this is a crucial question for the understanding of the phase structure
of QCD, it is essentially of no consequence to the analysis we undertake.
The value of the critical temperature we obtain is $\sim 5-10 \%$ different from
current values extracted from the Lattice \cite{Cheng:2007jq}\footnote{For many years, 
there has been no agreement between different groups
regarding the value of the critical temperature (cf. Ref. \cite{Fodor:2007sy}). This
important issue has been settled in 2010 and the results converged \cite{Cheng:2009zi} to that of Ref. \cite{Fodor:2007sy}.}, which is always the case in effective field theory approaches
and of no harm to our analysis, either. Our concern is with providing a good description of
the behavior of the critical temperature for increasing values of quark masses and the
isospin chemical potential.

%
%

\subsection{Case with $\mu_B=0$ and comparison to lattice QCD}

Let us first concentrate on the results for a system with no baryon-antibaryon asymmetry. This is the domain for which lattice QCD simulations can be applied in a controlled and systematic way, providing a reference for comparison.

\begin{figure}[htb]
\vspace{1cm}
\center
\includegraphics[width=9cm]{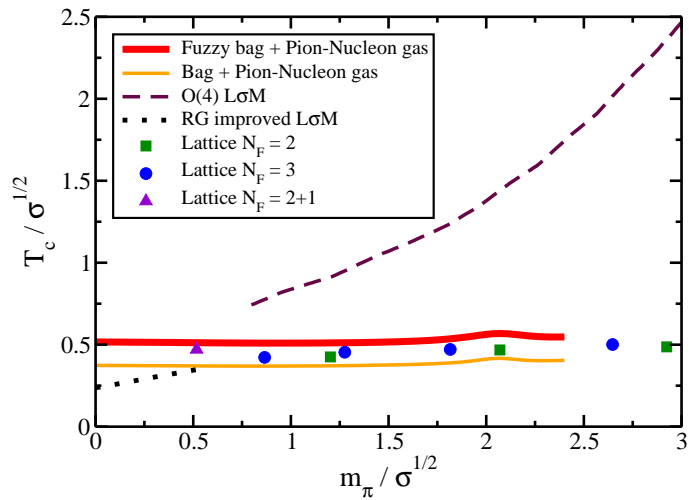}
\caption{Critical temperature as a function of $m_\pi$ normalized by 
the string tension $\sqrt{\sigma}=425~$MeV. Our results are compared to 
lattice data \cite{Karsch:2000kv,Cheng:2007jq} and other approaches \cite{Dumitru:2003cf,Berges:1997eu}.}
\label{Tc-mpi}
\end{figure}
%

The pion mass dependence, or equivalently the quark average mass dependence, of the 
critical temperature is displayed in Fig. \ref{Tc-mpi} for vanishing chemical potentials 
with the temperature and the pion mass normalized by the square root of the string tension. 
Our curves stop at a point from which our $\chi$PT approach clearly 
breaks down. The results of our framework are compared to lattice data from Refs. \cite{Karsch:2000kv} 
($N_F=2,3$) and \cite{Cheng:2007jq} ($N_F=2+1$)
and two other phenomenological treatments: the $O(4)$ linear sigma model \cite{Dumitru:2003cf} 
and a renormalization group improved computation \cite{Berges:1997eu} 
(cf. also Ref. \cite{Braun:2005fj} that discusses the quark-meson model using the proper-time 
renormalization group approach). The approximate mass 
independence observed in the lattice data is well reproduced within our framework, while 
the other descriptions tend to generate a qualitatively different behavior. 

This feature is yet another indication that the functional dependence of $T_c(m_{\pi})$
requires confinement ingredients to be reproduced, being incompatible with a phase transition dictated
by pure chiral dynamics. This argument goes in the same direction of Ref. \cite{Dumitru:2003cf},
the main difference here being the fact that we construct the mass dependence from the
$m_{quark}=0$ limit, with the heavy quasi-nucleons as the key new element at low-energies.

Moreover, our results are not strongly sensitive to the choice between the fuzzy bag model and the 
usual MIT bag model. The critical values for the MIT bag model are systematically lowered, but 
the qualitative behavior is not altered, as illustrated in Fig. \ref{Tc-mpi}. This indicates 
that a consistent treatment of the quark mass dependence connecting both perturbative regimes of 
energy is probably the essential ingredient to describe this observable.


%
\begin{figure}[htb]
\vspace{1cm}
\center
\includegraphics[width=9cm
]{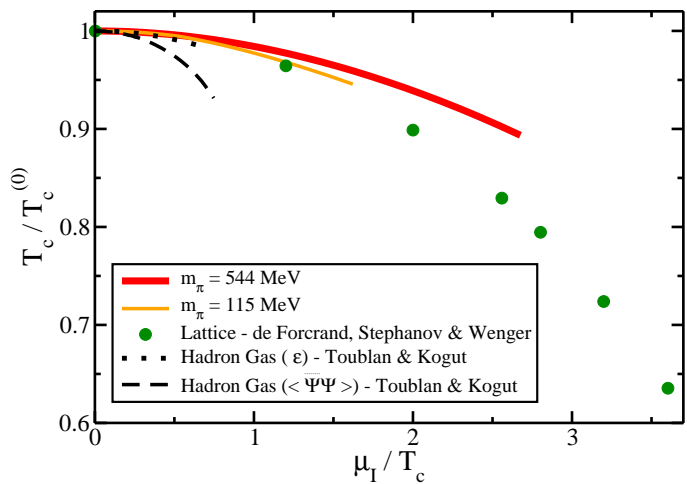}
\caption{Critical temperature as a function of $\mu_I$ for two different 
values of the pion mass. Our results are compared to lattice data \cite{deForcrand:2007uz}
and other approaches \cite{Toublan:2004ks}.}
\label{Tc-muI}
\end{figure}

In Fig. \ref{Tc-muI}, the critical temperature is plotted as a function of the isospin 
chemical potential. The critical temperature is normalized by its value in the absence 
of chemical asymmetry, whereas the isospin chemical potential is normalized by the critical 
temperature itself. The curves obtained within our framework are represented by solid lines. 
Once again our results are in good agreement with lattice computations \cite{deForcrand:2007uz}, 
even though the curve that is closer to the lattice points corresponds to a small vacuum pion mass, 
which is not the situation simulated on the lattice \cite{deForcrand:2007uz}. Our curves for different values of 
$m_{\pi}$ start to disagree more appreciably for $\mu_{I}> 2 T_{c}$. Recall that our treatment 
is valid up to isospin chemical potentials that are larger but not much larger than the 
pion mass, and contains the effects from pion condensation for $\mu_{I}> m_{\pi}$. 
Fig. \ref{Tc-muI} also exhibits, for comparison, results using the hadron resonance 
gas model \cite{Toublan:2004ks} that appear to depart from the lattice data at a much lower 
value of the isospin chemical potential. The two curves are produced by two different methods to 
determine the critical temperature: the dotted curve is obtained from the observation that the 
deconfined phase emerges at a constant energy density, whereas the dashed one uses the fact 
that the quark-antiquark condensate for the light quarks almost disappears at the quark-hadron
transition \cite{Toublan:2004ks}. Similarly to what we observe for the quark-mass dependence
of $T_c$, it is clear from Fig. \ref{Tc-muI} that plain chiral considerations render the 
largest discrepancies for the behavior of $T_c(\mu_I)$ as compared to lattice data.
It is also interesting to notice that contrary the mass-dependence of $T_c$, the behavior of $T_c$ with $\mu_I$ discriminates between the phenomenological bag treatments adopted at high energies: 
the effective model with the usual bag pressure
does not reproduce lattice data as well as the fuzzy one: even for $m_{\pi}=25~$MeV, the bag model
curve is still between the $m_{\pi}=400~$MeV and $m_{\pi}=600~$MeV fuzzy results.

\begin{figure}[h]
\vspace{1cm}
\center
\includegraphics[width=9cm]{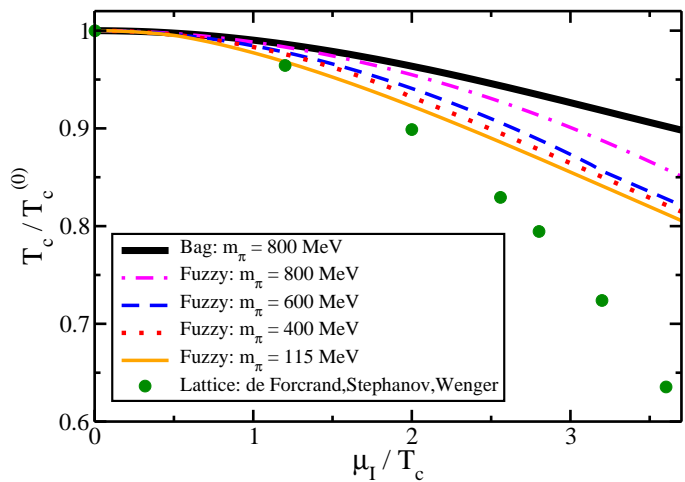}
\caption{Critical temperature as a function of $\mu_I$ for different 
values of $m_{\pi}$. Lattice data from \cite{deForcrand:2007uz}.}
\label{Tc-muI-varm}
\end{figure}

\begin{figure}[htb]
\vspace{1cm}
\center
\includegraphics[width=9cm]{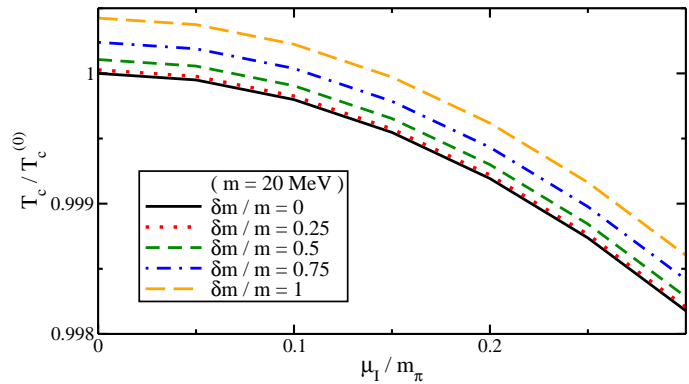}
\caption{Critical temperature as a function of $\mu_I$ for different values of the quark mass difference $\delta m$.}
\label{Tc-deltam}
\end{figure}

An up-down quark mass imbalance, characterized by  the relative difference in quark mass, 
which is in principle not much smaller than one, tends to increase 
the critical temperature, though by a quantitatively small amount, as expected. 
This result is presented in 
Fig. \ref{Tc-deltam} as a function of the isospin chemical potential. At finite $\mu_I$, the finite difference between the masses of quarks up and down spoils the cancelation of imaginary parts that caused the fermionic determinant in isospin dense QCD media to be positive definite. This situation suffers therefore from the Sign Problem even at zero baryon-chemical potential, being essentially inaccessible to current lattice algorithms.

\subsection{Including a finite baryon density}

Now that we have shown that the model yields reasonable predictions for the behavior of the critical parameters as compared to available QCD lattice data, one may explore the domain of the phase diagram in which a baryon-antibaryon asymmetry is present.

\vspace{0.65cm}

\begin{figure}[h]
\vspace{0.6cm}
\center
\includegraphics[width=9cm]{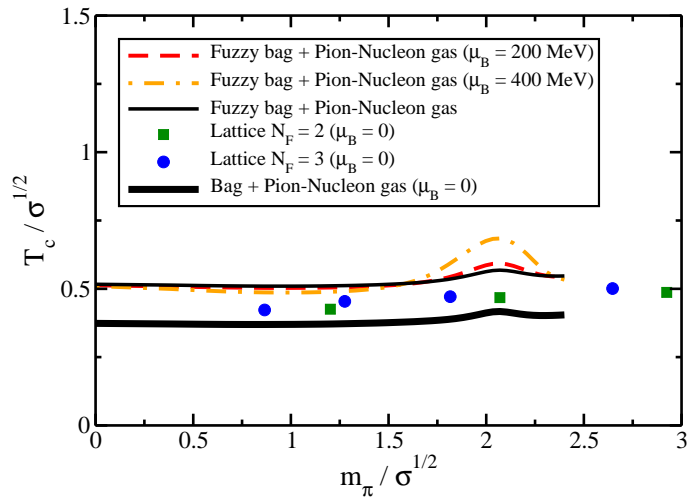}
\caption{Critical temperature as a function of $m_\pi$ normalized by 
the string tension $\sqrt{\sigma}=425~$MeV, for different $\mu_B$
($\mu_I=0$). Lattice data from \cite{Karsch:2000kv}.}
\label{Tc-mpi-varmuB}
\end{figure}

The dependence of the critical temperature on the pion mass is shown once again in Fig. \ref{Tc-mpi-varmuB}, where the influence of a finite baryon chemical potential is also displayed. 
As discussed previously, 
the approximate mass independence observed in the lattice data \cite{Karsch:2000kv} is well 
reproduced here, while previous treatments tended to generate a rather different 
behavior (cf. Fig. \ref{Tc-mpi}). The effect of a finite $\mu_B$ is
completely imperceptible for $\mu_B\le 100~$MeV. We also computed the critical 
parameters by considering the usual bag model in the high energy regime,
finding values systematically lower, but with the same qualitative behavior
as with the fuzzy bag. 


\vspace{0.4cm}

\begin{figure}[h]
\vspace{0.6cm}
\center
\includegraphics[width=9cm]{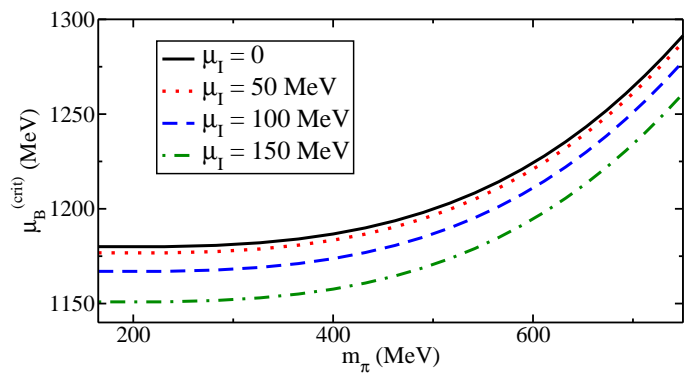}
\caption{Zero-temperature critical baryon chemical potential as a function of $m_\pi$, for different values of the isospin chemical potential.}
\label{muBc-mpi}
\end{figure}

%

Fig. \ref{muBc-mpi} displays the cold, dense case: the value of baryon chemical 
potential beyond which matter is deconfined increases with the vacuum 
pion mass. The outcome of a finite, small isospin density is to decrease 
the critical baryon chemical potential. This property could in principle 
play an important role in astrophysics.
Interestingly, the critical baryon chemical potential seems to be
significantly more sensitive to mass variations in comparison with the finite temperature
deconfinement transition. The relevance of mass effects in the thermodynamics of cold,
dense matter has already been shown in different related contexts, such as perturbative
QCD \cite{Fraga:2004gz-only,Laine:2006cp} and perturbative Yukawa theory \cite{Palhares:2008yq} (cf. also Ref. \cite{thesis}).

\section{Final Remarks\label{ConcIsospin}}

The role played by physically reasonable values for nonzero quark masses, as well as for 
finite isospin and baryonic chemical potentials becomes more and more important 
nowadays in the understanding of the phase structure of strong interactions. Present 
lattice simulations are already capable of probing the region of nonzero (small) values 
of $\mu_{B}/T$ and more realistic quark masses. Since the isospin chemical potential 
does not suffer from the Sign problem, this road is open for detailed Monte Carlo studies. 

On the experimental side, intermediate-energy experiments in nuclear physics are 
providing data and observables that are sensitive to chemical asymmetry, whereas 
high-energy heavy ion collisions to start soon at FAIR-GSI will probe a region of 
the phase diagram of QCD where effects from $\mu_{B}$ become important. Hence, 
effective theories that can encode the relevant characteristics of the rich phenomenology 
that emerges  are called for.

In this chapter we constructed an effective model profiting from results from chiral perturbation 
theory, perturbative QCD, and the fuzzy bag model to build a gas of quasiparticles that exhibits 
most of the relevant ingredients, and yet remains simple. The model implements a clear and 
consistent connection between the high and low energy parameters, allowing for a systematic 
investigation of the influence of finite quark masses and isospin symmetry breaking on the phase 
diagram for QCD at finite temperature and density. From a simple free gas calculation of the equation 
of state, we found surprisingly good agreement with different lattice data for the behavior of the 
critical temperature with masses and isospin chemical potential, indicating that the model 
captures the essential features brought about by the inclusion of those effects.
Furthermore, this model can supply information about the cold and dense regime of strongly 
interacting matter, which is not yet fully covered by lattice simulations. 

As discussed in the beginning of the chapter, to predict, using effective models, the functional behavior of the critical temperature in agreement with lattice results is not
at all a trivial task.
Several detailed studies of chiral models failed to 
describe $T_c(m_{\pi})$ \cite{Berges:1997eu,Dumitru:2003cf,Braun:2005fj}, while 
Polyakov-loop models, whose results can only be fitted to the lattice points for
$T_c(m_{\pi})$ \cite{Dumitru:2003cf}, cannot at the present form address isospin effects (nor any other observable related to fermionic chemical potentials).

Finally, the fact that different effective models seem to provide 
 distinguishable behaviors for $T_c(m_{\pi})$ together with the capability of first-principle lattice QCD simulations to assess ``cousin'' theories with various masses strongly suggests that the full function $T_c(m_{\pi},\{\mu_i\})$ is actually a good observable of the physics driving the phase boundaries. If thoroughly explored, it could even provide information on the puzzling connection between chiral and deconfinement transitions.

%% file: FS.tex
\chapter[Pseudocritical phase diagram of Strong Interactions in finite systems and CEP search]{\label{FS}}
\chaptermark{Pseudocritical phase diagram of Strong Interactions...}

\vspace{1.5cm}

{\huge \sc Pseudocritical phase diagram}
\vspace{0.3cm}

\noindent {\huge \sc  of Strong Interactions in}
\vspace{0.3cm}

\noindent {\huge \sc   finite systems and CEP search}


\vspace{-11cm}
\hspace{6cm}
\includegraphics[width=8.5cm,angle=90]{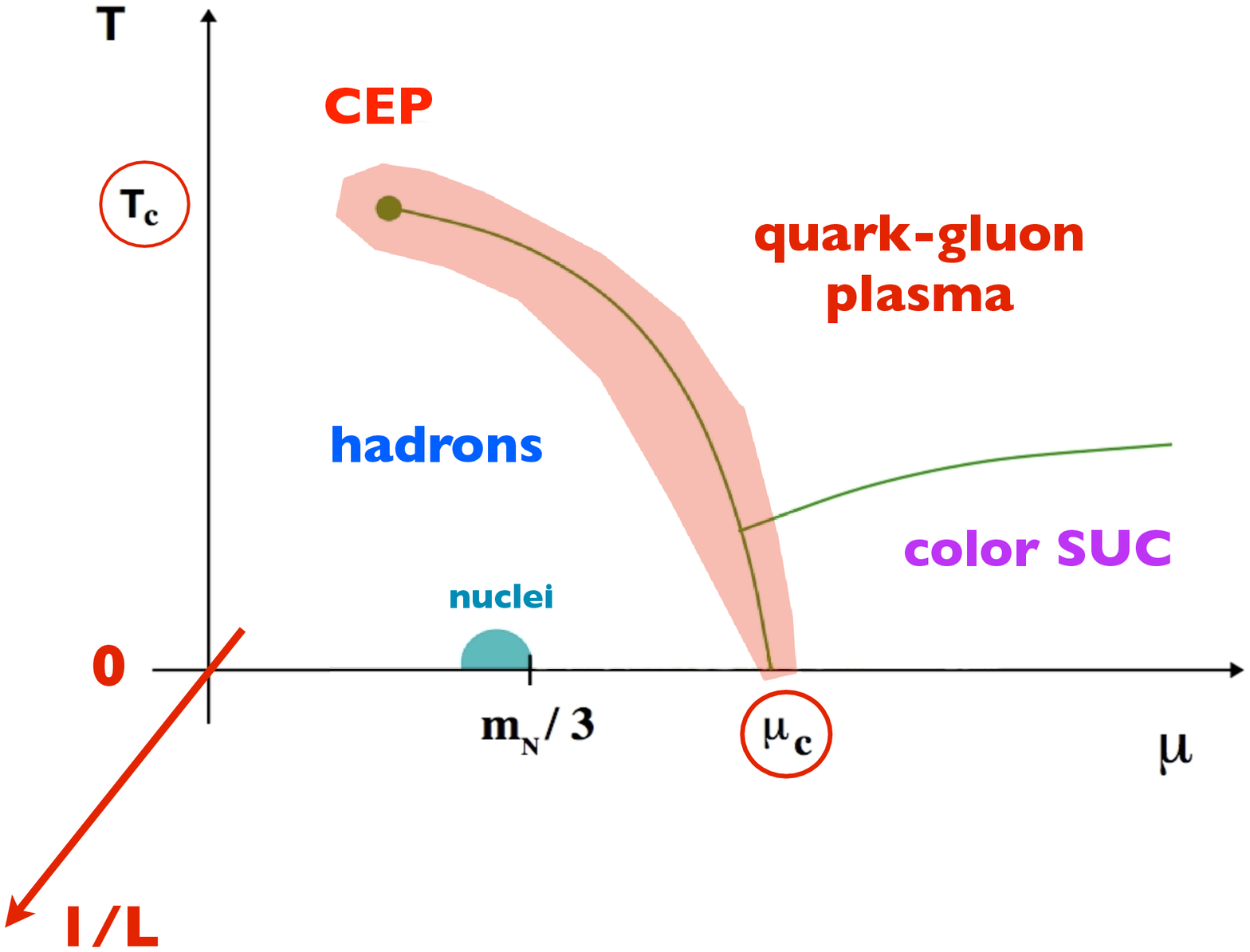}

\vspace{4.5cm}

%

In this chapter, we analyze how the QCD phase diagram is realized in real, finite systems. 
This amounts to considering a three-dimensional diagram, including the typical size of the system as an extra control parameter,
and investigating how and where in the $T-\mu_B$ plane we should expect a pseudocritical behavior to appear for each system volume.

As the discussion that follows will show, this should be an important issue when assessing the QCD phase transitions via HICs. In particular, we address different aspects of finite-size effects in the experimental search for the QCD critical endpoint in HICs, ranging from the addition of obstacles \cite{Palhares:2009tf} (further discussion in Refs. \cite{ProcsFSS1,Palhares:2011jf}) to the provision of an alternative signature \cite{FSS-RHIC} (cf. also \cite{ProcsFSS2}).

\newpage

\section{The critical endpoint in the QCD phase diagram}

Before addressing the problem of assessing the QCD phase structure through real, complicated systems as HICs, let us discuss the general picture in the thermodynamic limit and point out why we expect a second order critical point to exist in the first place. For an extended discussion and further literature the reader is referred to Ref. \cite{Stephanov:2004wx} and references therein.

The QCD phase transitions at high energies are usually described in terms of chiral and/or confinement properties. The symmetries associated with both of them (namely, $SU(N_f)\times SU(N_f)$ and $Z(N_c)$) are explicitly broken in Nature by the presence of quarks and their nonzero current masses, respectively. No exact order parameter can be thus defined in real-world QCD. Nevertheless, one clearly expects a drastic change in the medium of QCD at very high energies, in which a much larger number of degrees of freedom should be dynamical. This expectation has been intensively tested in lattice QCD simulations that indeed find a smooth but sensible increase in thermodynamical properties such as pressure and entropy for very hot media \cite{Laermann:2003cv}. At zero baryon chemical potential, first-principle lattice results strongly indicate a crossover transition between hadron-chirally broken and quark-chirally (approximately) restored phases. At large $\mu_B$, as discussed in the last chapter, the lattice QCD reference is lost\footnote{Some approximate lattice approaches to the baryon-dense domain also predict critical point \cite{FodorCEP} (cf., however, Refs. \cite{deForcrand:2007rq}).} and one has to resort to effective approaches. The great majority of (chiral or mixed chiral-confinement\footnote{It should be noted that currently known pure confinement models cannot describe baryon-antibaryon asymmetric media in a simple way.}) models predicts a first-order phase transition for cold and dense systems. In order to compatibilize these theoretical expectations for the phase diagram, a second-order critical endpoint has to be assumed at a finite $\mu_B$, below which the transition becomes a crossover.

If we concentrate on chiral properties and the up-down sector, this CEP ``topology'' of the phase diagram may be viewed as reminiscent of the chiral limit, in which the small explicit symmetry breaking due to the current quark masses is neglected. In this case, with $m_{\rm quark}=0$, chiral symmetry is exactly preserved by the lagrangian and the chiral condensate is a well-defined order parameter. There is thus a sharp distinction between the two phases and a singularity (or finite discontinuity) must be reached when passing from one of them to the other; no crossover transition is allowed. Indeed, two-flavor chiral QCD belongs to the same universality class as the Ising Model, featuring a second order phase transition at finite temperature (as also verified on the lattice\footnote{Actually, there is still some controversy in the literature regarding this result. In Ref. \cite{latt-controv}, for instance, the authors argue that the transition may be dominated by confinement properties, rendering it of first order instead. This is yet another indication of the subtleties sometimes involved in extracting continuum results from lattice simulations.}; cf. e.g. \cite{Laermann:2003cv}). This second order line turns into a first-order one at higher $\mu_B$'s; the intersection being a tricritical endpoint. With this chiral picture of the phase diagram in mind, one can turn on the current quark masses $m_{\rm quark}$. The clear distinction between the phases is lost and all transitions get smeared: while the second-order one becomes a smooth crossover, the first-order transition remains, but with a reduced discontinuity. The CEP is the smoothed version of the tricritical point present in the chiral limit.
%


Of course, these theoretical indications are not proof of existence. Any experimental guidance would be invaluable in this quest for mapping the true QCD phase diagram. That is the opportunity provided by the current and near-future heavy-ion collision experiments.

After more than a decade of experiments colliding heavy nuclei to investigate the properties of matter under extreme conditions of temperature and density, a clear signal of collectivity at the partonic level observed at RHIC-BNL together with other indications point to the formation of a new state of matter, even though its properties seem to be more intricate than those of the originally predicted quark-gluon plasma \cite{Adams:2005dq}. Current and near-future colliders stretch the currently probed frontiers in two directions. The higher energies attained at the LHC provide the closest conditions to the primordial universe plasma and the rich statistics will allow for better studies of fluctuations and the properties of the medium. In the other direction, different experimental programs are dedicated to probe ultra-dense media with higher baryon-antibaryon asymmetry: the on-going Beam Energy Scan (BES) program \cite{starbes} at RHIC-BNL and the future colliders at FAIR-GSI and NICA-JINR. This region of the phase diagram is exactly the one expected to contain the richer part of the phase structure of strong interactions.

In particular, the establishment of the existence and location of the critical endpoint of QCD (CEP) is a crucial step in the mapping of the phase diagram. The investigation of the experimental signatures of the CEP in the specific setup of heavy-ion collisions (HICs) is therefore of great importance. In this chapter, we discuss some of the proposed signatures for the CEP in HICs, concentrating on the possible role played by the finiteness of the system created in these collisions, as proposed in Refs. \cite{Palhares:2009tf,FSS-RHIC}.

%

\section{From the thermodynamic limit to real systems}

In the idealized case of an equilibrium system in the thermodynamic limit, a critical endpoint is always associated with unambiguous, sharp signals due to the second order phase transition that characterizes it. Large fluctuations are expected at all length scales and the correlation length diverges, leading to the conformal invariance present in second-order phase transitions (cf. e.g. Ref. \cite{goldenfeld}). Direct consequences and signals are then divergences in correlation functions of the order parameter $\sigma$ (e.g. susceptibilities), that scale with the correlation $\xi$
\begin{eqnarray}
\langle\sigma^n \rangle\sim \xi^{p_n}
\end{eqnarray}
where $p_n$ is a positive exponent related to the specific correlation function and to universal aspects of the system under investigation.

\begin{figure}
%
  \begin{center}
    \includegraphics[width=.45\textwidth]{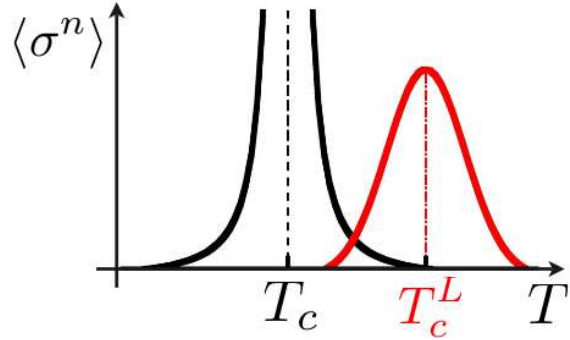}
  \end{center}
  \caption{The $n$-point correlation function of the order parameter $\sigma$ is shown as a function of the external parameter $T$. The singular (black) curve corresponds to the thermodynamic limit, while the behavior in a finite system is illustrated by the shifted peak (red curve). }\label{pseudocorr}
\end{figure}

In any real system, however, even if we can assume equilibrium, the volume is finite, so that the partition function (and hence all other thermodynamic quantities) is always well-defined and nonsingular. Similarly, the correlation length is trivially bounded by the finite volume of the system\footnote{The correlation length is also bounded, through causality,  by the finite lifetime of the system. Actually, in principle, one has to consider the growth rate of correlated domains in a fully dynamical approach, the speed of light (i.e. causality) being the ultimate bound in a relativistic system.} and no criticality can be defined. Instead of the divergent critical behavior expected for correlation functions in the thermodynamic limit, the observables measured will show {\it pseudocritical} peaks, corresponding to a smoothening of the associated critical singularity. As illustrated in Figure \ref{pseudocorr}, these peaks are actually shifted from the position of the original singularity by a size-dependent amount \cite{goldenfeld,FSS-books}.

In this case, the only way to precisely establish the presence of a second-order phase transition is to verify that the correlations in the system satisfy a specific behavior as a function of the system size leading to the conformal invariance that characterizes the second-order phase transition in the thermodynamic limit. Such specific trend is the well-known Finite-Size Scaling (FSS) \cite{FSS-books}.

\section{CEP signatures in HICs}

\begin{figure}[h]
\vspace{-4cm}
\center
\includegraphics[width=12.5cm]{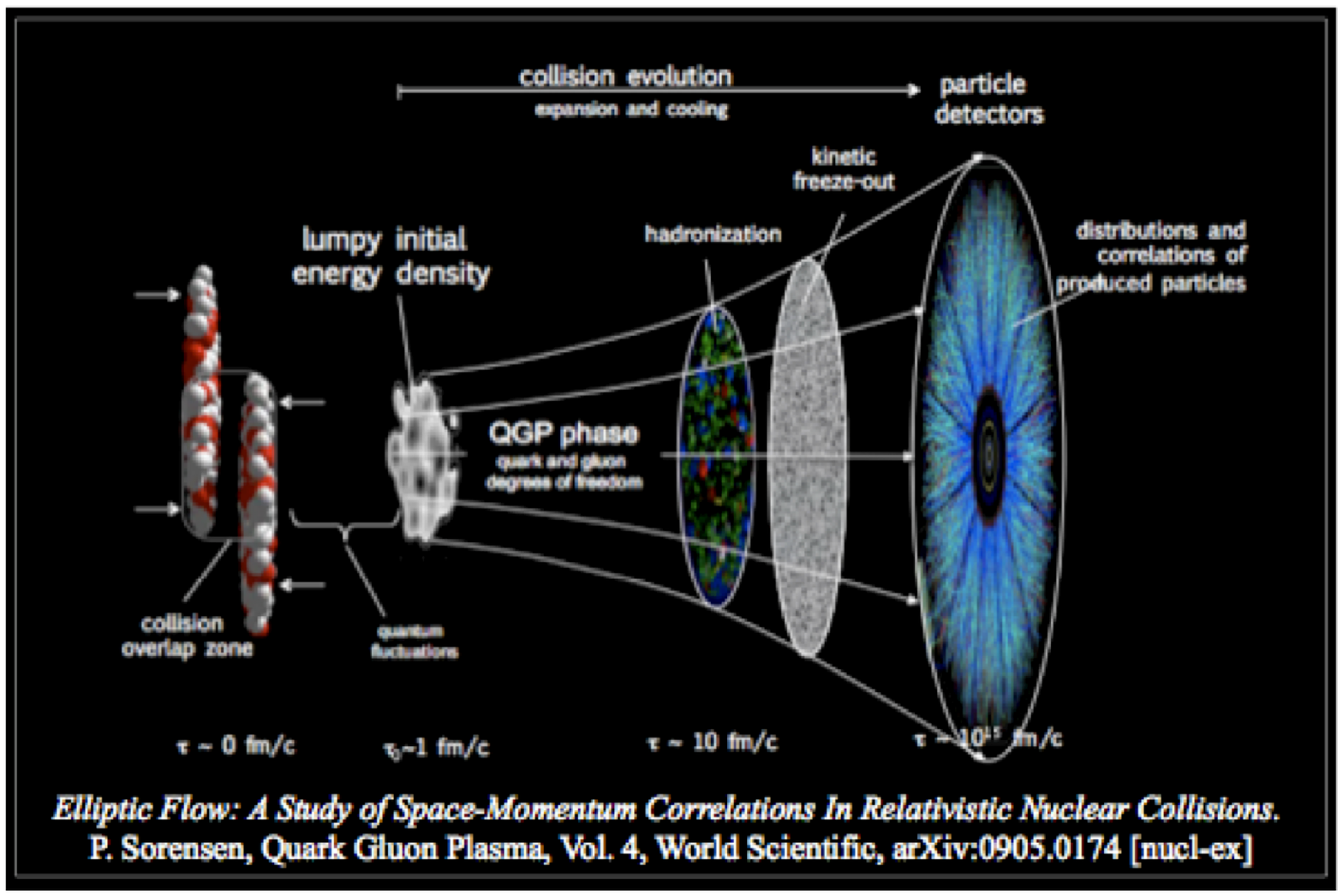}
\vspace{-5cm}
\caption{[extracted from Ref. \cite{Paul-Review}] Schematic cartoon of the stages occuring during a heavy-ion collision.}
\label{HICsPaul}
\end{figure}

The connection between CEP phenomena in the thermodynamic limit and the asymptotic particles measured in HIC detectors is even more indirect. The medium created in such experiments is an extremely complicated and fastly evolving system that, as illustrated in Figure \ref{HICsPaul}, goes through different stages that cannot in general be disentangled in the final observed spectrum. The QCD phase transition is only one of these stages, being followed by an interacting hadronic medium before the freeze-out that ultimately defines the final particle distributions detected. Moreover, it is probable that the phase conversion process occurs actually far from equilibrium, which could change completely the correlations in the system and even manifest features which are not consistent with the ones expected from the phase structure in the thermodynamic limit \cite{Hohenberg-Halperin}.

The problem of defining a clear and unambiguous signature for the CEP in HICs is therefore a nontrivial task that should be pursued with caution, always having the various caveats resulting from the specific experimental setup in mind. In what follows, we will disregard completely several of these caveats (e.g. out-of-equilibrium effects, critical slowing down near the CEP \cite{Berdnikov:1999ph}, hadronic medium modification, etc) and concentrate on the role played by finite-size effects. Besides discussing the caveats brought about by the finiteness of the system, we show how the fact that HICs provide data from media of different sizes may actually allow for a complementary alternative signature of the CEP based on FSS \cite{Palhares:2009tf}.

As discussed above, the CEP in a real system manifests as peaks in the cumulants of the distribution of the order parameter as functions of the distance to the criticality in the external parameter space. In this vein, it is natural to try to construct signatures of the CEP in HICs related to non-monotonic behavior in observables connected to the order parameter of the QCD transitions as one passes through or nearby the critical temperature $T_c$ and baryon chemical potential $\mu_{B c}$. In HICs, however, one does not have direct experimental access to neither of the (approximate) order parameters for the chiral and deconfinement transitions, namely the chiral condensate $\sigma$ and the expectation value of the Polyakov Loop, respectively.
Moreover, even if we assume that the system is equilibrated, direct measurements of $T$ and $\mu_B$ are not available.

\begin{figure}[h]
\center
\includegraphics[width=4cm,angle=90]{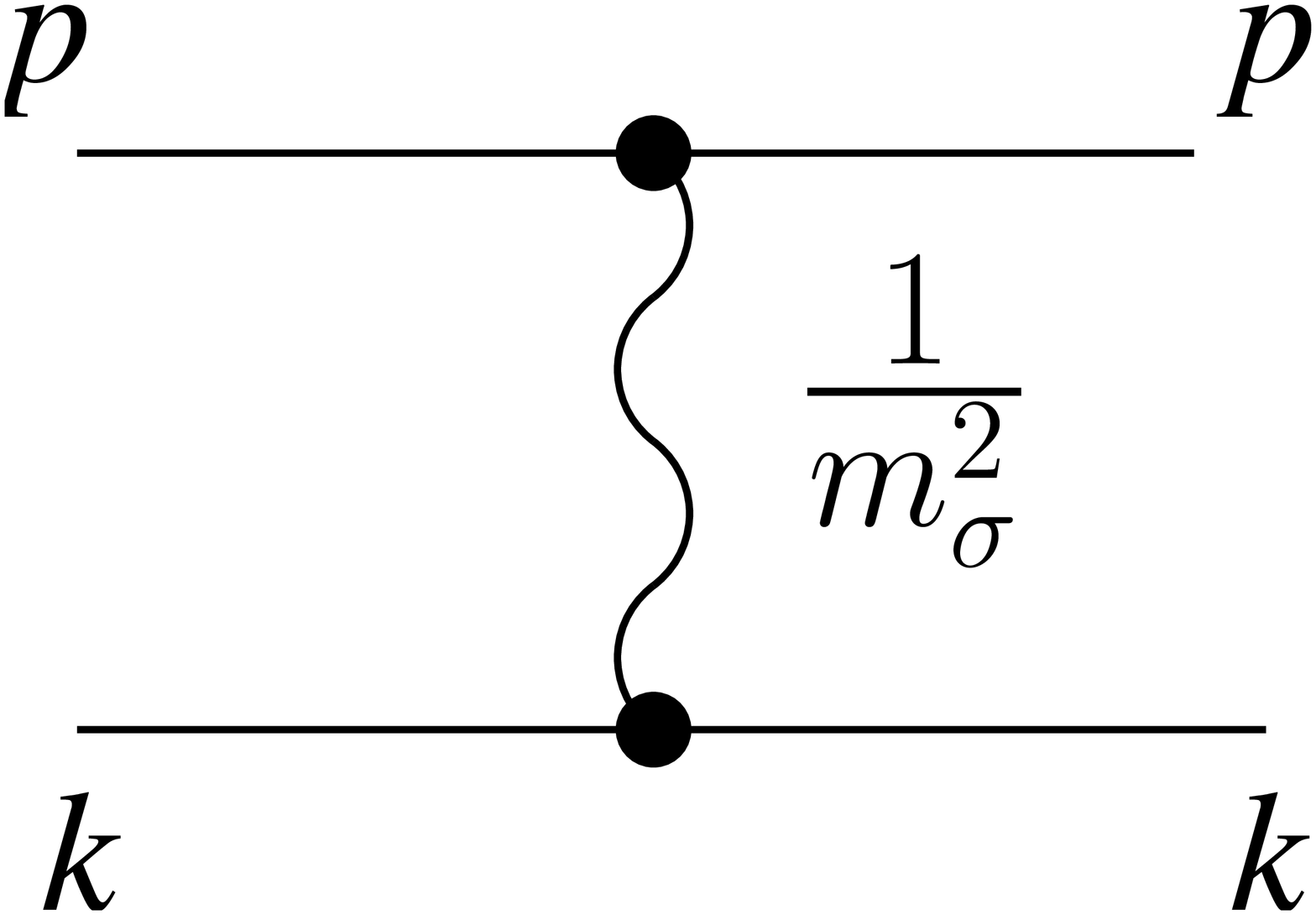}
\caption{Diagram of the leading contribution to two-point pion correlations featuring the transfer of critical correlations of the order parameter field $\sigma$ to pions.}
\label{diagSteph}
\end{figure}

The first concrete proposal of signature that tries to address these issues is the one by Stephanov {\it et al} \cite{Stephanov:1998dy}. The idea is based on the assumption that the critical correlations of the chiral order parameter will be transmitted to particles in the final HIC spectra that can effectively couple to the chiral condensate field $\sigma$, such as pions (via the interaction $G\sigma\pi^+\pi^-$) and nucleons ($g_N\sigma\overline{N}N$). Within this effective description, the critical contribution to the correlation of fluctuations in the particle numbers can be written in terms of the correlation length $\xi$ (or the inverse of the $\sigma$ mass), that should drastically increase (or diverge in the thermodynamic limit) at the CEP. The leading term in two-particle correlations comes from the exchange of a $\sigma$, as depicted diagrammatically in Figure \ref{diagSteph}, giving (for more details and notation cf. e.g. Ref. \cite{Athanasiou:2010kw}): $\langle\delta n_p\delta n_k\rangle\sim 1/m_{\sigma}^2\sim \xi^2$. Thus, a peak in this observable is expected in the vicinity of the CEP. Its height is related to the maximum correlation length, which in turn is limited by various factors: the size of the nuclei overlapping region (in practice, the effective system size can be considerably small, as for example is the case in core-corona scenarios \cite{Werner:2007bf}),  the finite lifetime of the medium, critical slowing down \cite{Berdnikov:1999ph}, etc.

Besides the critical contribution to the particle number fluctuations, there are also different noncritical contributions (e.g. those coming from thermal decays) that form a background to the CEP signature which could consequently be hidden. With the aim of increasing the sensitivity with respect to the critical behavior near the CEP, different observables that couple to higher cumulants of the order parameter have been proposed. Respectively connected to the three- and four-particle correlators, the critical contributions to skewness $\omega_3\sim \xi^{9/2}$ and kurtosis $\omega_4\sim \xi^{7}$ have stronger dependence on the correlation length, yielding a higher chance that these peaks surmount the noncritical background \cite{Stephanov:2008qz}.

\begin{figure}[h]
\center
\begin{minipage}[t]{54mm}
\vspace{-2cm}
\includegraphics[width=6cm]{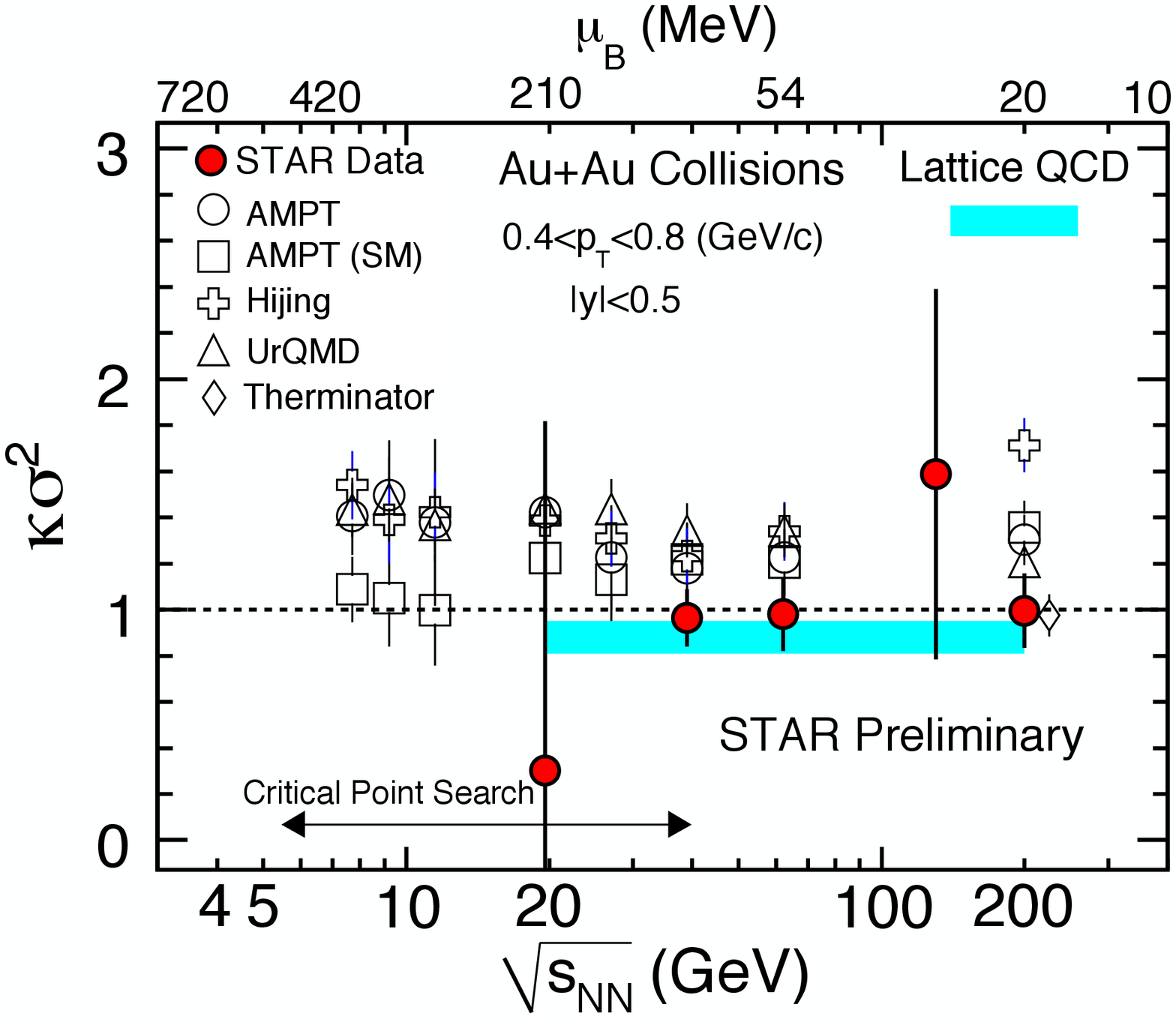}
\end{minipage}
\hspace{.4cm} 
\begin{minipage}[t]{70mm}
 \center
\vspace{-3.5cm}
\includegraphics[width=8cm]{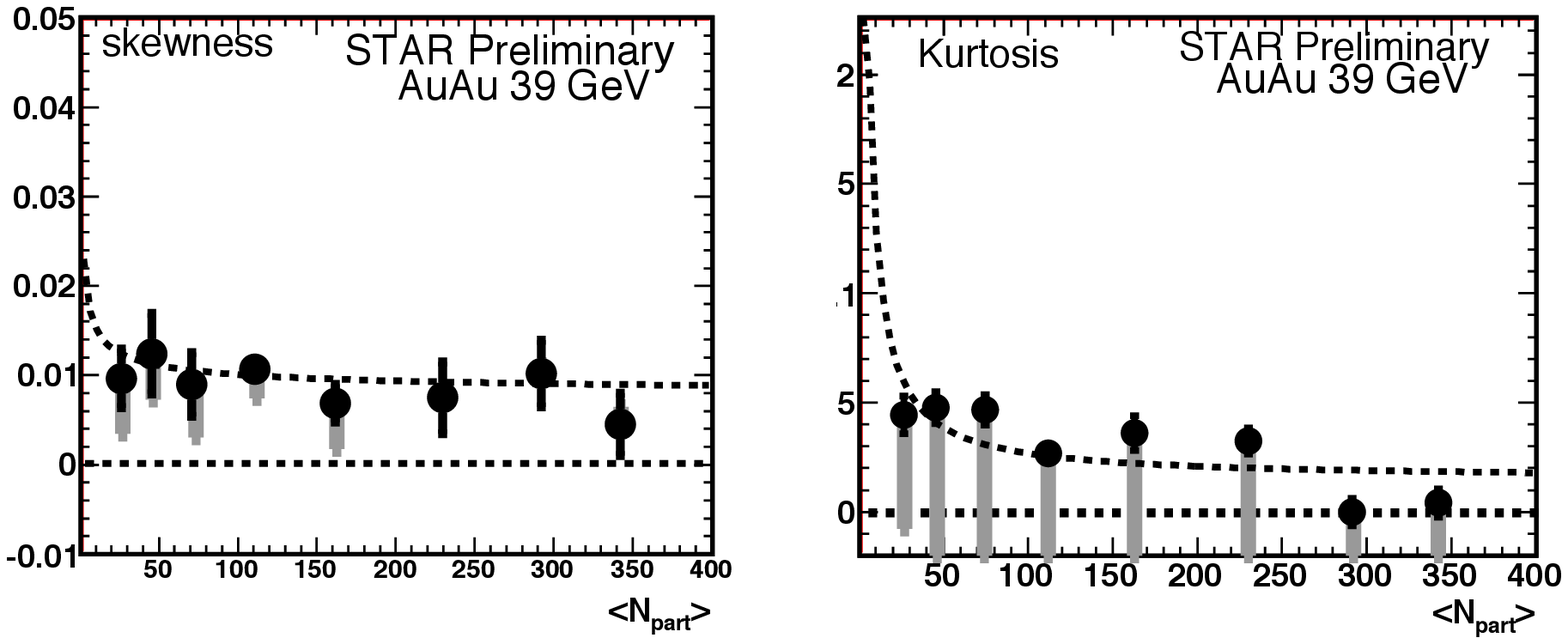}
\end{minipage}
\noindent
\vspace{-3.2cm}
\caption{
Measurements of second, third and fourth cumulants of fluctuations. Preliminary data from the Beam Energy Scan program at RHIC.}
\label{nonmon-exp}
\end{figure}

Specific dimensionless ratios of these higher cumulants that are believed to be less model-dependent \cite{Athanasiou:2010kw} have been compared to current available HIC data and, very recently, to the analysis of the first run of the RHIC BES program. Some results are shown in Figure \ref{nonmon-exp}. Up to now no clear indication of criticality has been observed in this type of signatures. Instead of the expected increase in the fluctuations at lower energies, one has actually found a decreasing trend in the kurtosis observable. Recent work has been put forward that interprets this apparent drop as the behavior in the direction of negative kurtosis, which is claimed to be a signature of the vicinity of the CEP from the crossover side \cite{Stephanov:2011pb}. Nevertheless, this is not a sufficient condition for the CEP and can be a result of other constraints of the medium created in HIC, such as baryon number conservation \cite{Schuster:2009jv}.

\section{Role of finite-size effects in HICs and CEP search}

First of all, the finiteness of the system will only play a significant role if it is comparable to the relevant scales associated with the QCD phase transitions under investigation. The typical critical temperatures for deconfinement and chiral symmetry restoration are predicted by lattice QCD to be $T_c\sim 150~$MeV \cite{Laermann:2003cv}, which gives a ``critical size scale'' $L_c\sim 1.3~$fm. If we consider the early universe during the epoch of the primordial quark-hadron phase transition, one can estimate roughly its radius as given by the particle horizon in a Robertson-Walker space-time (in which the scale factor grows as $a(t)\sim t^n$). Using a simple approximation for the equation of state ($3P\approx\epsilon\approx(\pi^2/30) N T^4$, with $N$ being the total number of thermal degrees of freedom), one gets:
\begin{eqnarray}
L_{\rm univ}(T)\approx \frac{1}{4\pi}\left(\frac{1}{1-n}\right)\left(\frac{45}{\pi N}\right)^{1/2}\frac{M_{\rm Planck}}{T^2}
\approx\frac{1.45\times 10^{18}}{(T/{\rm GeV})^2\sqrt{N}} ~{\rm fm}\,,
\end{eqnarray}
or $L_{\rm univ}\sim 10^{18}\,$fm for QCD ($n=1/2$, $N\sim 50\,$ and $T_c \sim 150\,$MeV).
The primordial universe corresponds thus to a system of typical size several orders of magnitude bigger than the critical size scale $L_c$, being essentially in the thermodynamic limit as far as the QCD phase transitions are concerned.

 Although the process of phase conversion of hot hadronic matter
that presumably happens in ultra-relativistic heavy-ion collision
experiments is often compared to the cosmological quark-hadron transition in
the early universe, the relevant space-time scales differ by almost twenty
orders of magnitude.
In HIC experiments, the volume of the system is not only finite but also centrality-dependent \cite{Adams:2005dq}. The typical linear sizes $L$ are actually quite small ($L\lesssim d_{\rm ion}$, with the ion diameter $d_{\rm ion} \approx 10-15~{\rm fm}$), corresponding to energies $L^{-1}\gtrsim 13.1-19.7~ {\rm MeV}$, not negligible in comparison with the expected scales in the critical region of the phase diagram $T_c\lesssim 150~{\rm MeV}$, especially in non-central collisions. This suggests that finite-volume effects may affect significantly the different physical phenomena occuring in a heavy-ion collision.

In what follows, we address the role played by finite-size effects in the QCD (pseudocritical) phase diagram as probed by HICs and in the context of the signatures of the CEP. In the next subsection, we show, using a well-defined and established effective model for the chiral transition, that indeed the expectation put forward by the rough estimates above is verified: for system sizes realized in current HICs, the (pseudo)critical region of the chiral phase diagram differs sensibly from the result in the thermodynamic limit \cite{Palhares:2009tf}. Subsection \ref{caveatsfromFS} presents then the consequences for the proposed observable signatures of the CEP in HICs; finite-volume effects representing in general an extra obstacle. In Subsection \ref{FSSHICs}, we argue that the presence of large finite-size corrections in HICs data allows for an alternative signature for the CEP search \cite{Palhares:2009tf}. The practical implementation of the method of data analysis is demonstrated \cite{FSS-RHIC} in Subsection \ref{FSSPaul}.

\subsection{Volume dependence of the chiral pseudocritical phase diagram\label{TmuL}}

Let us consider the pseudocritical phase diagram for the chiral transition in Strong Interactions, including an extra axis for the system size. 

Finite-size effects in the study of phase transitions via lattice
simulations, seen as an inevitable drawback of the method, and the necessary
extrapolation to the thermodynamic limit have been thoroughly studied for
decades \cite{goldenfeld}, the ultimate answer to the problem being given by the
method of FSS 
\cite{FSS-books}. Systematic calculations of finite-volume corrections
can then be computed not only near the criticality of continuous
(second-order) transitions, but also for first-order phase transitions 
\cite{binder-landau}, so that the thermodynamic limit can be taken in the
calculation of properties of the phase diagram.

The study to be presented here goes in the opposite direction:
 for natural systems that are truly small, one should study the
modifications caused by its finiteness in the phase diagram before comparing
to experimental observables. As discussed above, this is the situation in the study of the
chiral transition in 
HICs. Nevertheless,
contrasting to the large number of studies of finite-volume corrections in the computation of the
chiral condensate and related quantities within chiral perturbation theory
(see e.g. \cite{Damgaard:2008zs} and references therein), this issue is
often overlooked in the case of the quark-gluon plasma. Among the
exceptions, there is a lattice estimate of finite-size effects in the
process of formation of the quark-gluon plasma \cite{Gopie:1998qn}, a few
studies within the Nambu--Jona-Lasinio model \cite{finite-NJL}, and an
analysis of color superconductivity in dense QCD in a finite volume \cite{Yamamoto:2009ey}, besides investigations of finite-size effects on the
dynamics\footnote{For simplicity, we completely ignore the dynamics of the phase conversion.
Nevertheless, it was shown in Ref. \cite{Fraga:2003mu} that finite-size
effects will play a relevant role in the case of 
HICs.} of the plasma \cite{Spieles:1997ab,Fraga:2003mu}. The effect of
finite volume on the pion mass and chiral symmetry restoration at finite
temperature and zero density was investigated by Braun \textit{et al.} using
the proper-time renormalization group \cite{Braun:2004yk}.

In this subsection, we present results \cite{Palhares:2009tf} (cf. also Refs. \cite{ProcsFSS1}) for the modification of the
pseudocritical line in the temperature-chemical potential phase diagram, as
well as for isentropic trajectories, employing an effective chiral model: the linear
sigma model coupled to quarks with two flavors, $N_{f}=2$ \cite{GellMann:1960np}, which is presented in detail in Appendix \ref{LSM}. This effective theory has been widely used to describe
different aspects of the chiral transition, such as thermodynamic properties 
\cite{quarks-chiral,Scavenius:2000qd,Scavenius:2001bb,Taketani:2006zg} and
the nonequilibrium phase conversion process \cite{Fraga:2004hp}, as well as
combined to other models in order to include effects from confinement 
\cite{polyakov}. We focus on the chiral phase transition, avoiding the inclusion
of confinement ingredients (although the linear sigma model coupled to
quarks contains mesons and quarks that alternate dominance in the two
different phases), and show that the volume-dependence of the phase diagram
of the model in the regime of energy scales probed by current heavy ion
experiments can be large. Even though the output numbers in an effective
model should not be taken as accurate predictions, the quantitative question
that matters here, and that can be answered within this approach, is how big
the dislocations of the pseudocritical line and critical point, isentropic
trajectories, etc are \textit{relative} to the same phase diagram in the
thermodynamic limit. As mentioned above they turn out to be non-negligible.

Strictly speaking, only systems that are infinite in volume exhibit
``true'' spontaneous symmetry breaking 
\cite{weinberg}. Nevertheless, for finite systems of volume $V$ the barrier
penetration factor is already suppressed as $\exp (-aV)$, $a$ being a
positive constant determined by the microscopic features of each system. Of
course, if one waits long enough, the system will be 
``distributed'' between the two minima. If this suppression
is very effective, the two minima are physically well separated and
one can consider them as two almost different
phases of the system, even if there is no true phase transition.

Keeping this caveat in mind, our phenomenological analysis is motivated by
the fact that heavy ion experiments strongly suggest a new phase of
deconfined matter \cite{Adams:2005dq} and the system is obviously finite,
and by the enormous success of this methods in the description of (\textit{%
pseudo})phase transitions in small systems \cite{goldenfeld,FSS-books}. For
simplicity, we disregard inhomogeneity corrections. Even though we are aware
that these are relevant (see Refs. \cite{Taketani:2006zg,Nickel:2009wj}),
they should not modify the general picture we present or diminish the
significance of the (large) finite size effects we obtain in this simpler
description.

The effective theory adopted is described in Appendix \ref{LSM}.
In the case of a finite system of linear size $L$, the momentum integral
from the one-loop quark contribution to the effective potential (cf. Eq.(\ref{TH-OmegaMed1})) is
substituted by a sum 
\begin{equation}
\frac{V_{q}}{T^{4}}=\frac{2N_{f}N_{c}}{(LT)^{3}}\sum_{\mathbf{k}}\left[ \log
\left( 1+e^{-(E_{\mathbf{k}}-\mu )/T}\right) +\log \left( 1+e^{-(E_{\mathbf{k%
}}+\mu )/T}\right) \right] \;,
\end{equation}
where $E_{\mathbf{k}}=\sqrt{\mathbf{k}^{2}+m_{eff}^{2}}$, and 
$m_{eff}=g|\sigma |$ is the effective mass of the 
quarks\footnote{In general, the effects of finite size are not fully accounted for by the
replacement of continuum integrals by discrete sums over momentum states. In
certain cases, the importance of inhomogeneities can be enhanced in a small
system, especially near its surface. For simplicity, we neglect these
effects here.}. It is clear that finite-size effects in a quantum field
theory are, in a way, very similar to thermal effects with $1/L$ playing the
role of the temperature \cite{ZinnJustin:2000dr}. However, in the latter
boundary conditions are determined by the spin-statistics theorem, whereas
in the former there is no clear guidance (see, e.g., discussions in Refs. 
\cite{finite-NJL,Elze:1986db}). Choosing periodic boundary conditions (PBC)
are useful to focus on bulk properties of the model undisturbed by surface
effects \cite{ZinnJustin:2000dr,Brezin:1981gm,Brezin:1985xx}, but one is
always free to choose anti-periodic boundary conditions (APC) or any other,
instead. Depending on the physical situation of interest, a boundary
condition may be more realistic than others. To illustrate the dependence of
the pseudocritical phase diagram on the boundary conditions, we show results
for PBC and APC\footnote{Lattice studies \cite{Bazavov:2007zz} of the pure-gauge SU(3) deconfining
transition in a box with wall-type boundary conditions show sensibly larger
effects as compared to the analogous case with PBC. We also do not
investigate the influence of shape variations, choosing a cubic system.}.
For PBC, e.g., the components of the momentum assume the discretized values 
$k_{i}=2\pi \ell _{i}/L$, $\ell _{i}$ being integers, and there is a zero
mode (absent for APC, for which $k_{i}=\pi (2\ell _{i}+1)/L$). At zero
temperature, this zero mode in the case with PBC will modify the classical
potential, generating size-dependent effective couplings and masses.

\begin{figure}
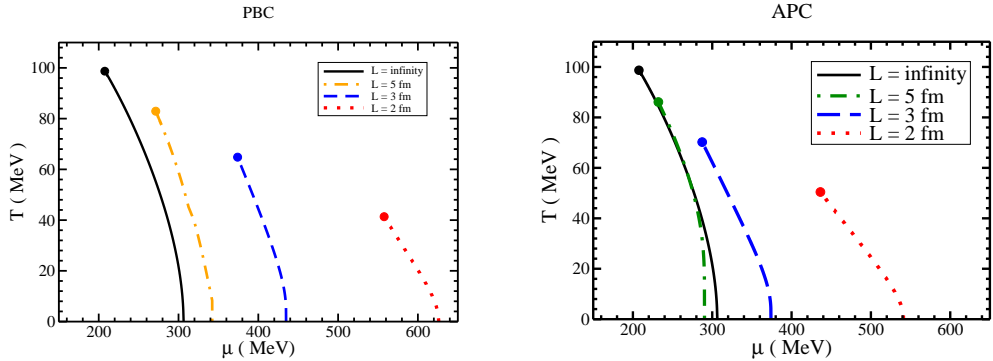

\center
\begin{minipage}[t]{64mm}
\includegraphics[width=6cm]{Plots/PhDiagram-PBC.eps}
\end{minipage}
\hspace{.4cm} 
\begin{minipage}[t]{70mm}
\includegraphics[width=6cm]{Plots/PhDiagram-APC.eps}
\end{minipage}
\noindent\caption{Pseudocritical transition lines and endpoints for different system
sizes within the linear $\protect\sigma$ model with periodic (left) and
antiperiodic (right) boundary conditions.}
\label{PhDiags}
\end{figure}

\begin{figure}[tbh]
\vspace{0.5cm}
\center
\begin{minipage}[t]{66.5mm}
\includegraphics[width=6.3cm]{Plots/CEP.eps}
\caption{Displacement of the pseudocritical endpoint in the $T-\mu$ plane as the system size is
decreased for different boundary conditions.}
\label{CEP}
\end{minipage}
\hspace{.3cm} 
\begin{minipage}[t]{66.5mm}
\includegraphics[width=6.45cm]{Plots/Tcrossover.eps}
\label{T-crossover}
\caption{Normalized crossover temperature at $\mu=0$ as a function of the inverse size $1/L$ for the cases with PBC and APC.}
\end{minipage}
\end{figure}


\begin{figure}[tbh]
\vspace{0.5cm}
\par
\center
\begin{minipage}[t]{64mm}
\includegraphics[width=6cm]{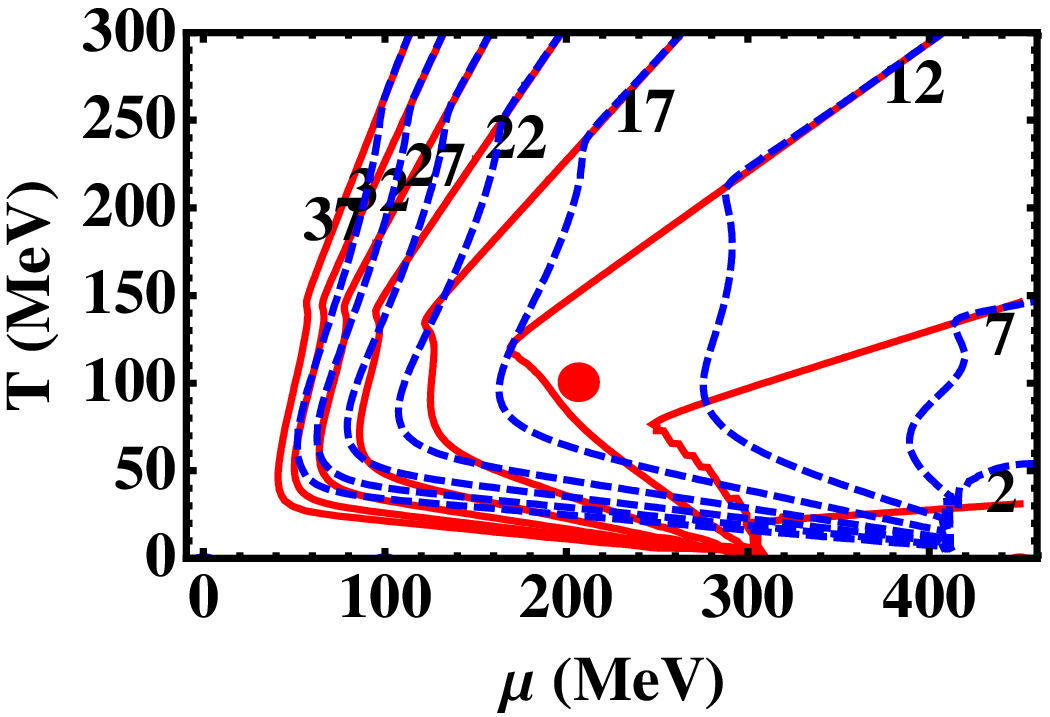}
\end{minipage}
\hspace{.5cm} 
\begin{minipage}[t]{64mm}
\includegraphics[width=6cm]{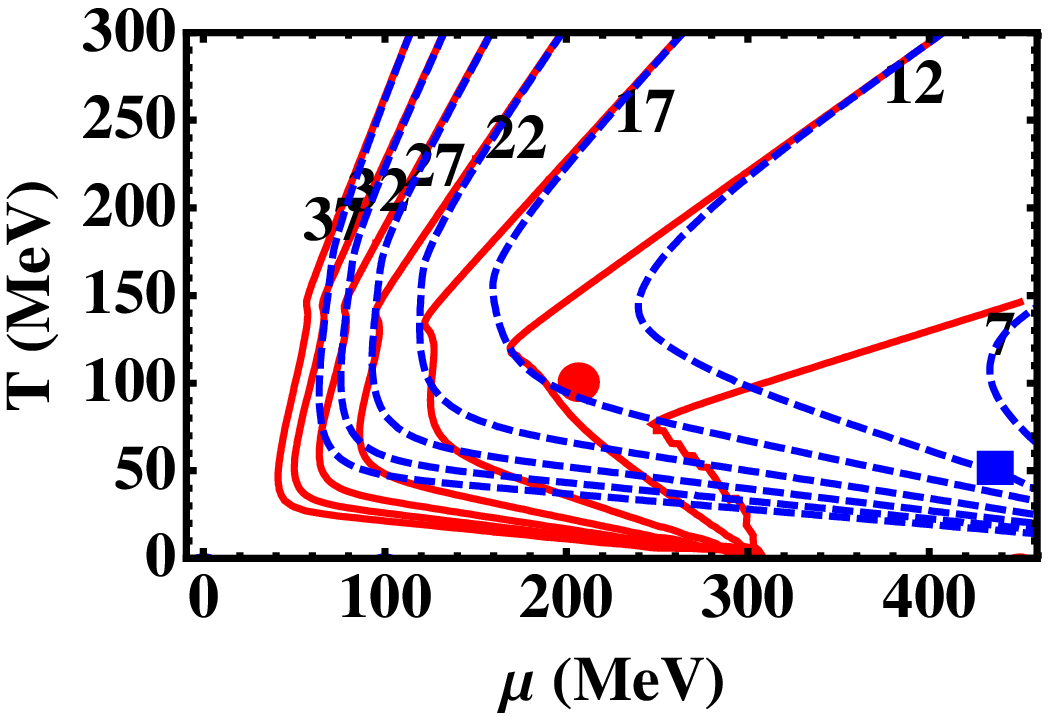}
\end{minipage}
\caption{Isentropic trajectories, labeled by the respective value of entropy
per baryon number, in the thermodynamic limit (solid, red lines) and for $%
L=2 $ fm with PBC (left) and APC (right). The red dot is the genuine
critical endpoint, while the square is pseudocritical one.}
\label{Isentropics}
\end{figure}

In what follows, we analyze systems of sizes\footnote{%
Our range of values for the linear size $L$ is motivated by the estimated
plasma size presumably formed in high-energy heavy ion collisions at RHIC 
\cite{Adams:2005dq}. The upper limit is essentially geometrical, provided by
the radius of the nuclei involved, whereas the lower limit is an estimate
for the smallest plasma observed.} between $10$ fm and $2$ fm which relate
to the typical linear dimensions involved in central and most peripheral
collisions of $Au$ or $Pb$ ions at RHIC and LHC, respectively. Figure \ref%
{PhDiags} displays the shift of the pseudocritical transition lines and
their respective endpoints as the size of the system is decreased. The
transition lines represent pseudo-first-order transitions, characterized by
a discontinuity in the approximate order parameter, the chiral condensate $%
\sigma$, and the production of latent heat through the process of phase
conversion. We find that those lines are displaced to the region of higher $%
\mu$ and shrinked by finite-size corrections. The former effect is sensibly
larger when PBC are considered, indicating that the presence of the spatial
zero mode tends to shift the transition region to the regime of larger
chemical potentials. Both boundary conditions reproduce the infinite-volume
limit for $L\gtrsim10$ fm. Figure \ref{CEP} shows the corresponding
displacement of the pseudocritical endpoint, comparing PBC and APC: both
coordinates of the critical point are significantly modified, and $\mu_{%
\mathrm{CEP}}$ is about $30\%$ larger for PBC. For $\mu=0$, the crossover
transition is also affected by finite-size corrections, increasing as the
system decreases, as shown in Figure 3.7.
Again, PBC generate
larger effects: up to $\sim80\%$ increase in the crossover transition
temperature at $\mu=0$ when $L=2$ fm.

Results for the isentropic trajectories are shown in Figure \ref{Isentropics}%
, comparing the infinite-volume limit with the finite system with $L=2$ fm
in the cases with PBC and APC. For sufficiently high temperatures, the
isentropic lines in the thermodynamic limit are reproduced, while large
discrepancies are found around and below the transition region. The
remarkable variations at low temperatures are due to the shift of the
pseudocritical line and, for PBC, to the zero-mode-induced modifications of
the vacuum properties, especially the vacuum constituent quark mass.
Although there is no strong focusing effect around the critical endpoint (as
observed previously for this chiral model \cite{Scavenius:2000qd} and
similar ones \cite{Fukushima:2009dx}), it is clear that the density of
isentropic trajectories around the pseudocritical endpoint is sensibly
higher than in the thermodynamic limit.

\subsection{Consequences for signatures of the CEP\label{caveatsfromFS}}

\begin{figure}[h]
%
  \begin{center}
  \includegraphics[width=0.5\linewidth]{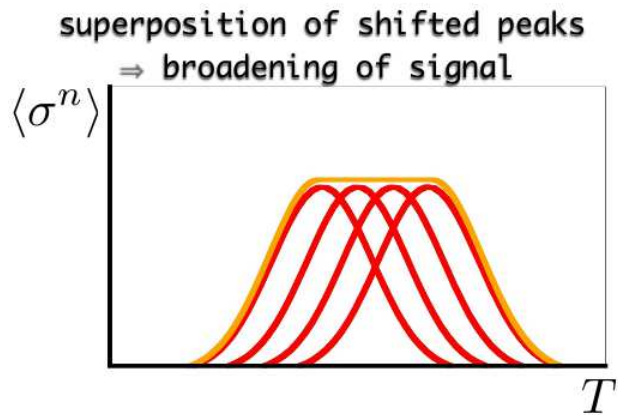}
  \end{center}
\caption{Cartoon illustrating the broadening of a pseudocritical signal due to the averaging over a centrality interval.}\label{broadening}
\end{figure}

As already stated, the finiteness of the system  created in HICs imposes a natural limitation on the growth of the correlation length. Besides this trivial constraint that smoothens out CEP divergences into peaks, the finiteness of the system created in HICs also guarantees that non-monotonic signals will actually probe pseudocritical quantities which might be significantly shifted from the genuine criticality, as illustrated in Fig. \ref{pseudocorr}. Our results for these shifts in a chiral model clearly indicate that the displacement of the (pseudo-)CEP and the \newline (pseudo-)first-order lines can be large for the size scales present in current HIC experiments \cite{Palhares:2009tf} (cf. Fig. \ref{PhDiags}). If this is indeed the case in QCD, the
non-monotonic behavior signaled by the fluctuation observables will not only be blurred in a region and the effects from criticality severely smoothened, but also this region will be shifted with respect to most of the effective-model and lattice predictions, that are done in general in the thermodynamic limit. Moreover, once finite-size corrections are important, the observables will be also sensitive to boundary effects. Smoothening of the fluctuation peaks will be further increased by the averaging procedure within not sufficiently small centrality (i.e. system size) bins, as illustrated in Figure \ref{broadening}, tending to hide them underneath the background.  Of course, the width of the centrality bins in data analysis is also bounded from below so that the statistics is enough to guarantee reasonable statistical errors. Figure \ref{Isentropics} indicates further that signatures based on the focussing of isentropic trajectories could probe a quantitatively different scenario due to the finiteness of the systems created at HICs.

\subsection{Finite-size scaling as a tool for the CEP search\label{FSSHICs}}

Now that we have argued that finite-size corrections to critical phenomena in the context of HICs can be sizable, we can approach this issue from a different perspective with the aim of taking advantage of this feature in the experimental search for the CEP.

As is well-known \cite{goldenfeld,amit}, the second-order transition that occurs at the CEP in the thermodynamic limit is characterized by a divergent correlation length and by the property of scale invariance. These constraints in the thermodynamic limit are translated into real, finite systems as the existence of finite-size scaling (FSS) \cite{FSS-books,Brezin:1981gm,Brezin:1985xx} in the vicinity of criticality. The phenomenon of FSS can be rigorously proved through a renormalization group analysis and is extensively studied and successful in condensed matter physics (cf. Ref. \cite{sldq} for an example within the context of spin glass transitions in disordered Ising systems).
Since the conformal invariance expected at the CEP in the thermodynamic limit leads to strong constraints for the behavior of finite systems, large finite-size effects as those encountered in Subsection \ref{TmuL} may, instead of only bringing obstacles for the CEP search, be turned into a tool.  

FSS is a 
powerful statistical mechanics technique that prescinds from the knowledge of the details of a 
given system; instead, it provides information about its criticality based solely on very general 
characteristics. Furthermore, it is valid also when one is not so close to the
critical point\footnote{In a system with size $L$ and temperature $T$, the region where the scaling
becomes important is characterized by $L\sim \xi _{\infty }(T)=(\mathrm{%
length~scale})t^{-\nu }$.}, provided that we use full scaling plots. This
technique is predictive even for tiny systems in statistical mechanics (see,
e.g., Ref. \cite{sldq}), and, since it can be used in various different
regions of the phase diagram not far from the critical point, it can in
principle provide enough statistics for data analysis.

And since the thermal environment corresponding to the region of quark-gluon 
plasma formed in heavy ion collisions can be classified according to the centrality of the collision, 
events can be separated according to the size of the plasma that is created. Heavy ion collisions indeed provide an ensemble of differently-sized systems, calling for a finite-size scaling analysis.  In the
following, we suggest an application of this method to
search for the critical point of QCD (and possibly determining  its
universality class), assuming that we can build out of heavy ion data a set
of systems of different temperature and sizes via the incident
energies and the distribution of collisional centrality.

The FSS hypothesis \cite{FSS-books} was conjectured before the development of the renormalization 
group (RG). However, it can be derived quite naturally by applying RG techniques to critical 
phenomena \cite{Brezin:1985xx,amit}. For a system with typical linear dimension $L$, the singular 
part of the free energy density scales as
\begin{equation}
f_{s}(\{g\},L^{-1})=\ell^{-d}f_{s}(\{g'\},\ell L^{-1}) \; ,
\label{eq-fs1}
\end{equation}
when the lengths are reduced by a factor $\ell$ in a given RG transformation. $\{g\}$ are couplings 
and $d$ is the space dimension. The fact that the system is finite is irrelevant for this implementation, 
since the RG transformations are local. Near a RG fixed point, one can write (\ref{eq-fs1}) in terms of 
the right eigenvectors of the linearized RG transformation:
\begin{equation}
f_{s}(t,r,...,L^{-1})=\ell^{-d}f_{s}(t\ell^{y_{t}},r\ell^{y_{r}}...,\ell L^{-1})\;,
\label{eq-fs2}
\end{equation}
so that one notices that $L^{-1}$ behaves like a relevant eigenvector with eigenvalue 
$\Lambda_{L}=\ell$, and $y_{L}=1$. Here, $t=(T-T_{c})/T_{c}$ is the reduced temperature, 
$T_{c}$ being the temperature associated with the critical point. The reduced temperature 
$t$ is a dimensionless measure of the distance to the critical point when no other external 
parameter is considered. In the presence of other external parameters, such as the baryonic 
chemical potential $\mu$, one should redefine this distance accordingly
to include the dependence on $r=(\mu-\mu_{c})/\mu_{c}$, which plays a role analogous to the magnetic field variable in a Ising system.
$T_{c}$ and $\mu_{c}$ are defined in the thermodynamic limit, and 
the true second-order phase transition occurs for vanishing $t$, $r$ and $L^{-1}$ (and other 
couplings reaching their value at the fixed point, $\{g=g^{*}\}$). 

For finite $L$, crossover effects become important. If the correlation length diverges as 
$\xi_{\infty}\sim t^{-\nu}$ at criticality, where $\nu$ is the corresponding 
critical exponent, in the case of $L^{-1} t^{-\nu} \gg 1$ the system is no longer governed by the 
critical fixed point and  $L$ limits the growth of the correlation length, rounding all 
singularities \cite{goldenfeld}. If $L$ is finite, $\xi$ is analytic in the limit $t \to 0$, etc and one can 
draw scaling plots of $L/\xi$ vs. a given coupling $g$ for different values of $L$ to find that all curves 
cross at $g=g^{*}$ in this limit, which is a way to determine $g^{*}$. The critical temperature can 
also be determined in this fashion, since the curves will also cross at $t=t_{c}$.

This scaling plot technique can be extended, taken to its full power for other quantities, such as 
correlation functions. An observable $X$ in a finite thermal system can be written, 
in the neighborhood of criticality, in the following form \cite{Brezin:1985xx}:
\begin{equation}
X(t,\{g\};\ell;L)=L^{\gamma_{x}/\nu} f(t L^{1/\nu}) \;,  \label{scaling}
\end{equation}
where $\gamma_{x}$ is the bulk (dimension) exponent of $X$ and $\{g\}$ dimensionless coupling 
constants. The function $f(y)$ is universal up to scale fixing, and the critical exponents are sensitive 
essentially to dimensionality and internal symmetry, which will give rise to the different universality 
classes \cite{goldenfeld,amit}. To simplify the discussion, we ignore for the moment the chemical potential 
variable $r$. In principle, $f$ is a function of two scaling variables, though. Again, if one plots the 
scaled observable versus $t$, the curves should cross at $t=t_{c}$. Moreover, using the appropriate 
scaling variable, instead of $t$, all curves should collapse into one single curve if one is not far 
from the critical point. So, this technique can be applied to the analysis of observables  that are 
directly related to the correlation function of the order parameter of the transition, such as fluctuations 
of the multiplicity of soft pions \cite{Stephanov:1998dy}. 

This is the idea of using full scaling plots
to search for the critical endpoint of QCD in heavy ion data. Here, as in statistical mechanics searches 
for critical points, one can treat $T_{c}$, $\mu_{c}$ and the critical exponents as parameters in a scaling 
plot fit, the system size being provided by different centrality bins or by the number of participants in the collision. In fact, the scaling variable is defined up to $L$- and $t$-independent multiplicative factors, 
so that the knowledge of the actual size of the system is not needed.


\begin{figure}[h]
%
  \begin{center}
  \includegraphics[width=0.5\linewidth]{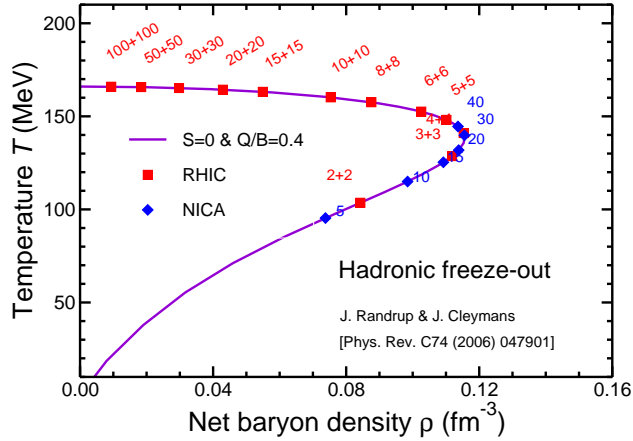}
  \end{center}
\caption{[extracted from Ref. \cite{Randrup:2009ch}] Values of temperature and net baryon density at chemical freeze-out obtained with statistical model fits to hadron abundances.}\label{freezeoutcurve}
\end{figure}

As mentioned previously, the correct scaling variable should measure the distance from the 
critical point, thereby involving both temperature and chemical potential, or their reduced versions 
$t$ and $r$. This produces a two-dimensional scaling function and makes the analysis of heavy 
ion data highly nontrivial.  
%
Phenomenologically, we adopt a simplification motivated by results 
from thermal models for the freeze-out region, connecting temperature and chemical potential. These models, which fit ratios of particle multiplicities surprisingly well, 
suggest that the temperature and chemical potential characterizing the data measured in a heavy-ion collision are constrained to a ``freeze-out curve'', as depicted in Figure \ref{freezeoutcurve}.
Since this curve is parameterized by the center-of-mass energy of the collision ($\sqrt{s_{NN}}$), it is reasonable to build our one-dimensional 
scaling variable from this quantity and the size of the system. 

Actually, the reduction in the number of dimensions in
the external parameter space seems very natural also from the standpoint of hydrodynamical descriptions of the evolution of the system.
As suggested by the
success of ideal hydrodynamic models, the expansion of the system may well
be considered adiabatic. Therefore, we expect that the point of the emission
of soft pions should be on this adiabatic line, characterized by the initial
energy density and naturally parameterized  in the $\left(
T-\mu \right) $ plane by $\sqrt{s_{NN}}$.


The range of sizes that can be accessed in heavy ion collisions over which FSS can be tested is limited. It is reasonable to assume that for a locally thermalized quark gluon plasma to form, the system needs to be several times the hadronic size (several fm). The largest possible system will have a diameter of approximately $15$ fm. Estimates from an analysis of HBT data from STAR indicate that from peripheral to central collisions, the system size changes by a factor of $3-4$~\cite{hbtsize}. We plot the scaled observable vs $\left[(\mu-\mu_c)/\mu_c\right]L^{1/\nu}$.  The value of $L$ for the various centralities can be estimated from a Glauber Monte-Carlo model. For our study, the overlap area is calculated from $S_{\perp}=R_{Au}^2(\Theta-\sin\Theta)$ where $\Theta=2\cos^{-1}(b/2R_{Au})$, where $R_{Au}$ is the nuclear radius for Au and $b$ is the collision impact parameter. Then the length is taken to be $L=2\sqrt{S_{\perp}/\pi}$.

A system of size $L_1$ can be compared to a system of size $L_2$ only when both settings correspond to the same value of the scaling variable, i.e.
\begin{equation}
  \mu_1-\mu_c=(\mu_2-\mu_c)\left( \frac{L_2}{L_1} \right)^{1/\nu}.
\end{equation}
This constrains which $\mu$ values can be compared when testing for scaling by directly comparing data rather than by extrapolations. Since a central $Au+Au$ collision is about four times larger than a peripheral one ($L_{2}=4L_{1}$) and taking $\nu=2/3$, we find that a measurement in peripheral Au+Au collisions at $\mu_1$ can be compared to a central collision at $\mu_2$ when $\mu_1-\mu_c\approx8(\mu_2-\mu_c)$.

\subsection{FSS analysis of preliminary results from RHIC Beam Energy Scan\label{FSSPaul}}

Data on various fluctuations have been measured at RHIC and the SPS over an energy range 
from $\sqrt{s_{NN}}=5$ GeV to $200$ GeV and the centrality dependence of $p_T$ fluctuations 
has been measured at RHIC for energies from $19.6$ GeV to $200$ GeV. These data provide an 
opportunity to look for evidence of finite-size scaling. 

To search for scaling, we consider the correlation measure $\sigma_{p_T}/\langle p_T\rangle$~\cite{Adams:2005ka} 
scaled by $L^{-\gamma_x/\nu}$, according to Eq. \ref{scaling}. We consider the $p_T$ fluctuations 
$\sigma_{p_T}$ scaled by $\langle p_T\rangle$ to obtain a dimensionless variable \footnote{
The STAR collaboration has also published $p_T$ fluctuations using an alternative variable 
$\Delta\sigma_{p_T}$~\cite{Adams:2006sg}. The results found using this variable are qualitatively equivalent to the ones for 
$\sigma_{p_T}/\langle p_T\rangle$.}. We use the correlation data measured in bins corresponding to the 0-5\%, 5-10\%, 10-20\%, 20-30\%, 30-40\%, 40-50\%, 50-60\%, and 60-70\% most central collisions. We estimate the corresponding lengths $L$ to be 12.4, 11.1, 9.6, 8.0, 6.8, 5.6, 4.5, and 3.4 fm.
%
%
The exponent $\nu = 2/3$ is determined by the Ising universality 
class of QCD and we consider values of $\gamma_{x}$ around $1$ (ignoring small anomalous 
dimension corrections). 
We also varied the value of $\gamma_{x}$ from $0.5$ to $2.0$ and found that changing $\gamma_{x}$ 
within this range does not improve the scaling behavior. Using a smaller value of $\gamma_{x}$ 
simply causes $\sigma_{p_T}/\langle p_T\rangle$ scaled by $L^{-\gamma_x/\nu}$ to drop more 
slowly with increasing centrality and increasing $\gamma_x$ causes this drop to happen more 
quickly.

\begin{figure}[htb]
\centering\mbox{
\includegraphics[width=0.75\textwidth]{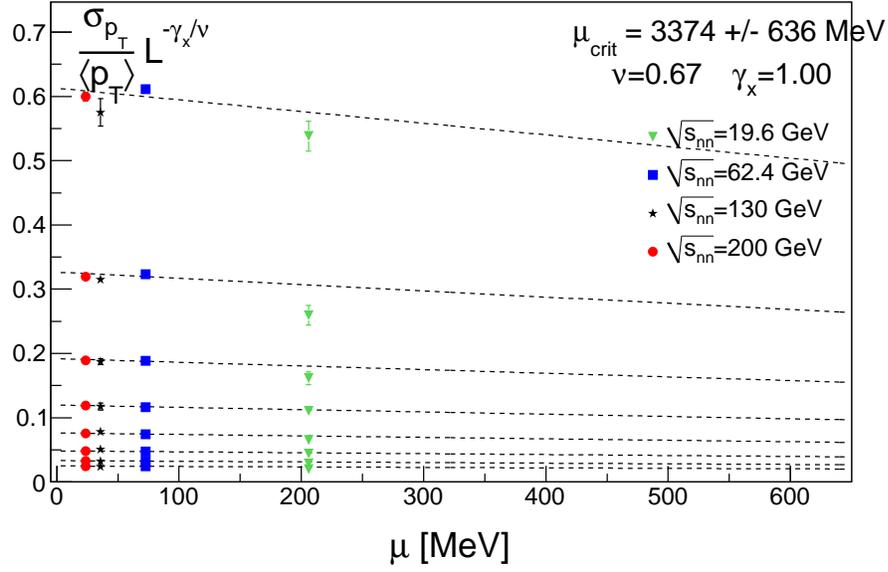}}
\caption{ Scaled $\sigma_{p_T}/\langle p_T\rangle$ vs $\mu$ for different system sizes, and 
with $\nu=2/3$ and $\gamma_{x}=1$. Data extracted from RHIC collisions at energies 
$\sqrt{s_{NN}}=19.6, 62.4, 130$, and $200$ GeV (linear fit, see text). }
\label{scaling-linear}
\end{figure}

\begin{figure}[htb]
\centering\mbox{
\includegraphics[width=0.75\textwidth]{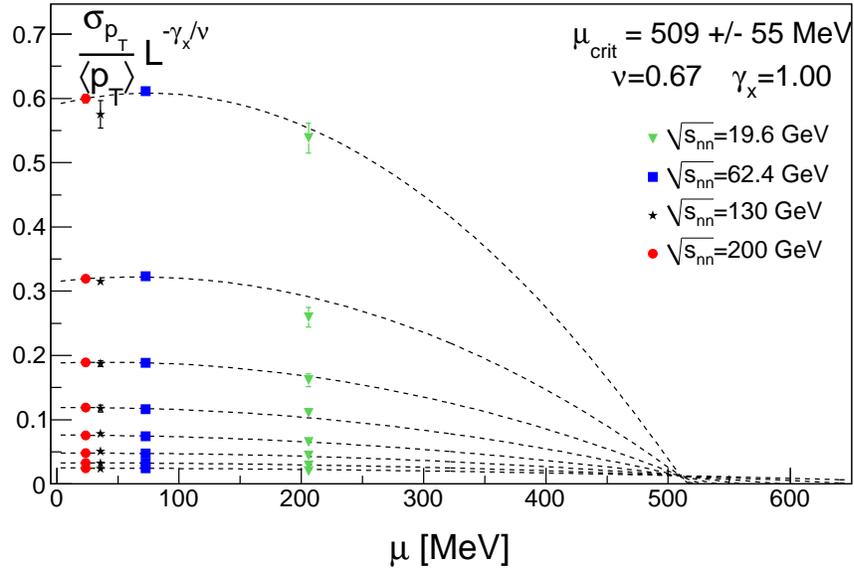}}
\caption{ Scaled $\sigma_{p_T}/\langle p_T\rangle$ vs $\mu$ 
for different system sizes, assuming $\mu_{\mathrm{crit}}=509$ MeV, which corresponds to a critical point at 
$\sqrt{s_{NN}}=5.75$ GeV. Again, $\nu=2/3$ and $\gamma_{x}=1$.
Data extracted from RHIC collisions at energies $\sqrt{s_{NN}}=19.6, 62.4, 130$, and $200$ GeV 
(second order polynomial fit, see text).}
\label{scaling-pol2}
\end{figure}

In Figs.~\ref{scaling-linear} and \ref{scaling-pol2} we plot $\sigma_{p_T}/\langle p_T\rangle$ scaled by 
$L^{-\gamma_x/\nu}$ vs $\mu$ for different system sizes, using data extracted from collisions at 
$\sqrt{s_{NN}}= 19.6, 62.4, 130, 200$ GeV. We use the parameterization from Ref.~\cite{conversion} to convert from $\sqrt{s_{NN}}$ to $\mu$. If there is a critical point at $\mu=\mu_{\mathrm{crit}}$, the 
curves for different sizes of the system should cross at this value of $\mu$. However, since the 
currently available data is restricted to not so large values of the chemical potential, one has to 
perform extrapolations using fits. The scaling function $f$ in Eq. \ref{scaling} is expected to be 
smoothly varying around the critical point, so we fit the data corresponding to a given linear size $L$ to 
a polynomial, but constraining the polynomials to enforce the condition that all the curves cross at 
some $\mu=\mu_{\mathrm{crit}}$, where $\mu_{\mathrm{crit}}$ is an adjustable parameter in the fit. This clearly assumes 
the existence of a critical point. In Fig.~\ref{scaling-linear} we use a linear fit. The approximate energy independence of $\sigma_{p_T}/\langle p_T\rangle$ along with the linear fit, leads to a very large $\mu$ value where the curves can cross ($\mu\sim 3~$GeV).  There is no reason however to assume a linear fit function, so in Fig.~\ref{scaling-pol2} we also try a second order polynomial. Using a second order polynomial function for $f$ allows the curves from different system sizes to cross at a much smaller value of $\mu$. Based on this fit, we find that the data is consistent with a critical point at $\mu\sim510~$MeV corresponding to a $\sqrt{s_{NN}}$ of 5.75 GeV.

The value estimated for $\mu_{\mathrm{crit}}$ based on finite-size scaling of current data is highly dependent on the assumed functional form of $f$. The approximate energy independence of $\sigma_{p_T}/\langle p_T\rangle$ for a given $L$, however, already indicates within the finite-size scaling assumption that the critical point should be at $\mu$ values well above those currently available. Based on finite-size scaling of $\sigma_{p_T}/\langle p_T\rangle$, one would not expect a critical point at $\mu<400~$MeV. Having data for lower values of $\sqrt{s}$, i.e. higher values of chemical potential, as expected from the analysis of the Beam Energy Scan program at RHIC~\cite{starbes}, one should be able to study full scaling plots of $\sigma_{p_T}/\langle p_T\rangle$ scaled by $L^{-\gamma_x/\nu}$ vs. $\frac{\mu-\mu_{\mathrm{crit}}}{\mu_{\mathrm{crit}}}L^{1/\nu}$ without the need of long extrapolations. For the current set of data, these full scaling plots would not be very enlightening.

\begin{figure}[htb]
\centering\mbox{
\includegraphics[width=0.75\textwidth]{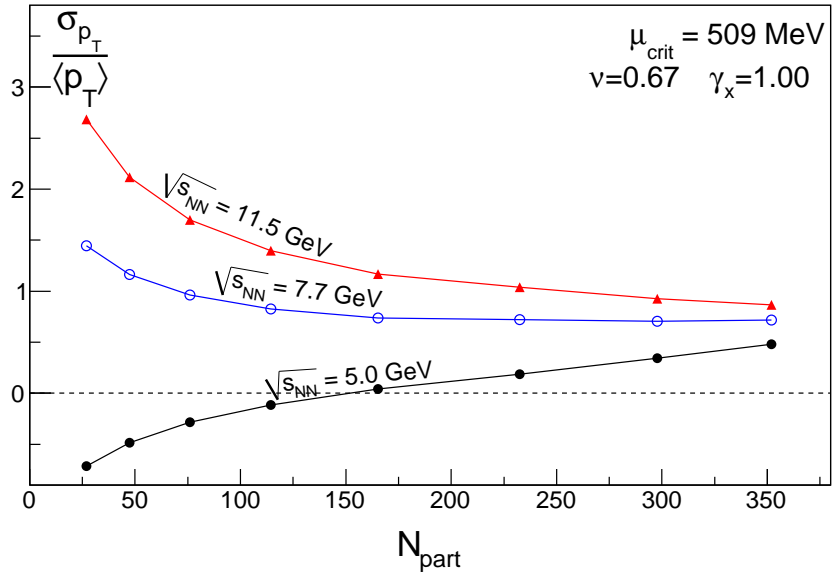}}
\caption{ The expected measurement of $\sigma_{p_T}/\langle p_T\rangle$ as a function 
of the number of participants at lower energies assuming the critical point is at $509$ MeV 
as extracted from the quadratic polynomial fit of STAR data.}
\label{predict}
\end{figure}

RHIC has also run at lower energies in order to search for a critical point in the Beam Energy 
Scan program.  That data is currently being analyzed. Here we use the quadratic polynomial fit 
of STAR data (Fig. \ref{scaling-pol2}) and assume the critical point is at $509$ MeV to make 
predictions for $\sigma_{p_T}/\langle p_T\rangle$ at lower energies. The finite-size scaling 
scenario along with currently available data, allows us to predict the energy and system size 
dependence of fluctuations for any given values of $\mu_{\mathrm{crit}}$, $\gamma_x$ and $\nu$.  
We show this expectation as a function of the number of participants, $N_{\mathrm{part}}$, for three 
proposed beam energies: $11.5, 7.7$ and $5$ GeV in Fig.~\ref{predict}. Notice that the 
centrality dependence changes once one moves to the other side of the critical point.
This is a condition enforced by finite-size scaling which provides a generic signal for having reached the first-order phase transition side of the critical point.

%


\section{Final remarks}

Searching for the the critical endpoint in the phase diagram for strong interactions is a remarkably 
challenging task. On one hand, lattice simulations can not provide the sort of guidance that is possible 
at zero density, where there is no sign problem, and effective models point to its existence, qualitatively, 
but yield very different quantitative predictions. On the other hand, heavy ion collision experiments 
probe limited regions of the phase diagram. Besides, they come with non-trivial background contributions 
that tend to blur the signatures provided by the non-monotonic behavior of observables built from 
correlation functions of the order parameter.

In this chapter, we point out that this task should be pursued within an extended phase diagram of Strong Interactions: the one containing an extra axis associated with the volume of the system under consideration. Finite-size effects most probably play an important role in HIC experiments and particularly in the QCD CEP search. As a caveat: restricting the access of non-monotonic signatures to pseudo-critical quantities that are severely smoothened and might be significantly shifted with respect to the genuine CEP (in the thermodynamic limit). But also as a complementary tool: FSS analysis is perfectly feasible and simple to be implemented in the context of HIC’s, yielding predictions for comparison of data for different center-of-mass energies.

The fact that finite-size scaling prescinds from the knowledge of the details of the system under 
consideration, providing information about its criticality based solely on its most general features, 
makes it a very powerful and pragmatic tool for data analysis in the heavy ion collision search for 
the critical point. From a very limited data set in energy spam, we have used FSS to exclude the 
presence of a critical point at small values of the baryonic chemical potential, below $450~$MeV. 
We have also used the scaling function to predict the behavior of data with system size at lower 
energies. We are looking forward to compare our predictions to the outcome of data analysis from 
the Beam Energy Scan program at RHIC.

%% file: surften.tex
\chapter[Nonequilibrium phenomena and phase structure in the cold and dense domain]{\label{surften}}
\chaptermark{Nonequilibrium phenomena and phase...}

\vspace{1.5cm}

{\huge \sc Nonequilibrium phenomena and}
\vspace{0.3cm}

\noindent {\huge \sc  phase structure in the cold}
\vspace{0.3cm}

\noindent {\huge \sc   and dense domain}


\vspace{-11.3cm}
\hspace{6cm}
\includegraphics[width=9cm,angle=90]{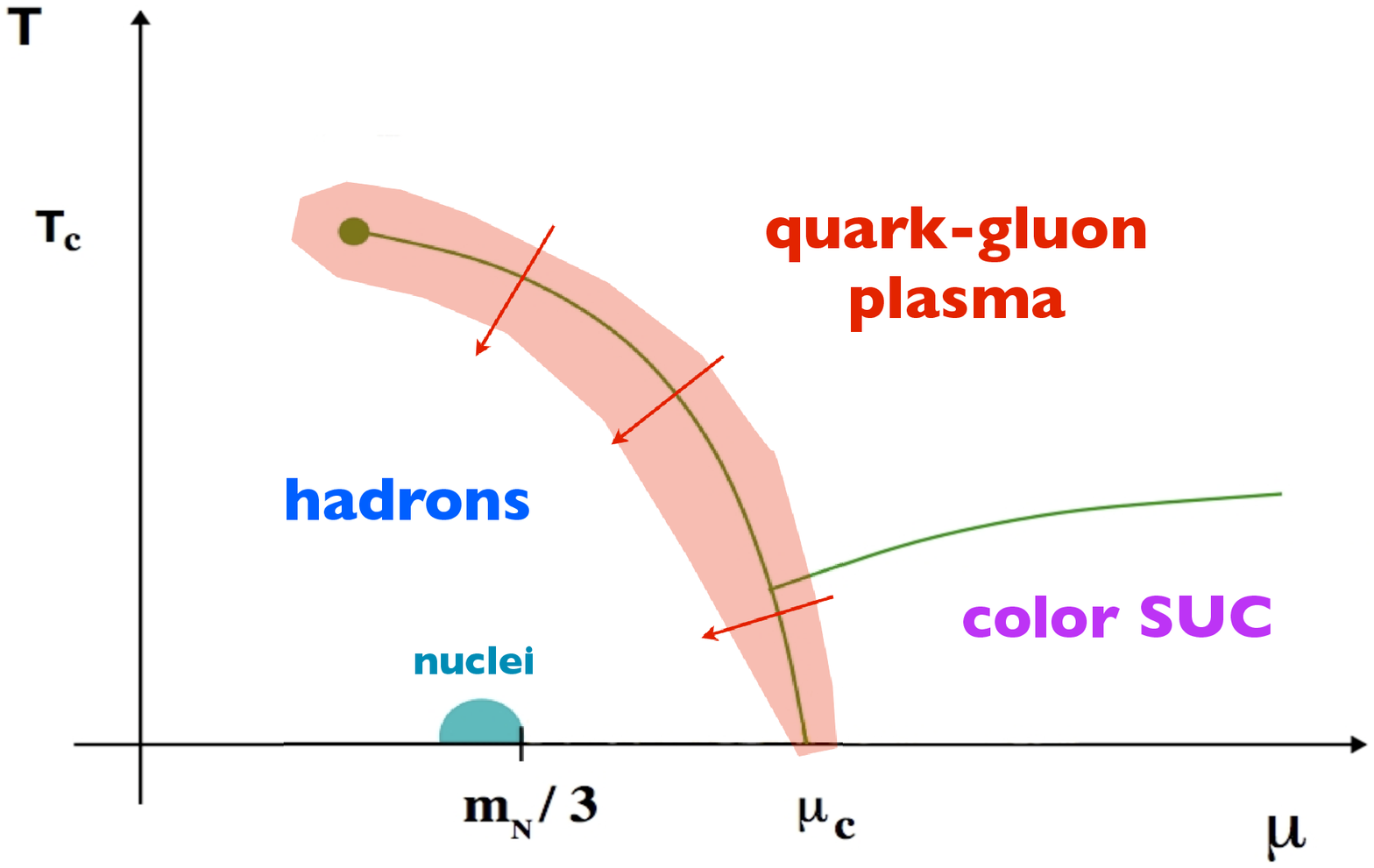}

\vspace{4.5cm}

In this chapter we consider the phase diagram of Strong Interactions from a dynamical point of view. Although a very challenging enterprise, this is crucial when investigating phase transitions at high energies, due to the current lack of well-controlled experiments in this domain. In Nature or in the laboratory, the structure of phases of Strong Interactions is generally probed via indirect, out-of-equilibrium conversion processes.

The particular problem being addressed here is the dynamics of nucleation for the chiral phase transition in the cold and dense regime. Estimates of relevant time and energy scales are computed within the linear sigma model with quarks \cite{Palhares:2010be} and the presentation focus is on the possible consequences of these results for different astrophysical applications \cite{Palhares:2010be,Procs-ST}.

\section{QCD matter and astrophysical phenomena}

The thermodynamics of strong interactions for cold matter under extremely high 
densities is of utmost importance for the understanding of the structure of compact 
stars \cite{stars}. Since quantum chromodynamics (QCD) is asymptotically free, it is 
believed that for high enough densities quarks will be in a deconfined state, the 
quark-gluon  plasma \cite{Rischke:2003mt}. Moreover, due to the approximately chiral 
nature of the QCD action, one also expects quarks to be essentially massless above a 
sufficiently high value of the chemical 
potential\footnote{Nevertheless, these features 
and other thermodynamic properties of dense media were also shown to be significantly 
affected by nonzero quark masses \cite{Fraga:mass,Kurkela:2009gj}.}. 
Depending on the location of the critical 
density, one might find several sorts of condensates and even deconfined quark matter 
in the core of neutron stars \cite{Alford:2006vz}. Furthermore, the order and strength of 
the chiral transition are crucial features in establishing the existence of a new class of 
compact stars \cite{Fraga:2001id,SchaffnerBielich:2004ch}. 

On the other hand, going in the opposite direction, in which observations provide insight to theory,
astrophysical phenomena might shed some light in a mostly obscure region of the QCD phase diagram: the domain of low temperatures and high baryon-antibaryon asymmetry.
This region is exactly where one expects 
the richest structure in the high energy QCD phase diagram to be and where theorists have less guidance for what QCD predicts, since first-principle lattice calculations are still not reliable (due to the Sign Problem, cf. e.g. Ref. \cite{Laermann:2003cv}).
Moreover, despite the great experimental efforts to create matter with a finite baryon chemical potential,  accelerators cannot reach the domain of very low temperature and high chemical potential.

\begin{figure}
\center
\includegraphics[width=8.5cm]{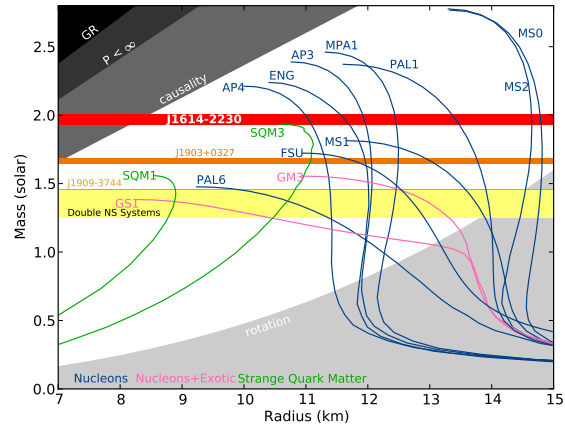}
\caption{[extracted from Ref. \cite{Demorest:2010bx}] Neutron star mass-radius diagram. The horizontal red band shows the observational constraint from the J1614−2230 mass measurement. Any equation of state whose line that does not intersect it is ruled out by this measurement.}
\label{shapiro}
\end{figure}

 Since extremely high densities and relatively low temperatures are believed to exist in the core of ultracompact objects, it is probable that the QCD phase transitions play an important role in the structure and dynamics of formation of such systems. Even though in general observations are much less controlable than laboratory experiments, the astronomical measurement techniques and instrumentation have much developed in the last decade and this might be the dawn of a precision era in which astrophysics really constrains nonperturbative QCD phenomena. One concrete example of astronomical observation that has reached the capability of constraining models for strong interactions in the nonperturbative regime is that of Ref. \cite{Demorest:2010bx}, shown in Figure \ref{shapiro}, in which the error bars of less than $5\%$ in the $2$ solar masses measurement rule out several equations of state.

\subsection{QCD surface tension and observable implications in astrophysics}

\begin{figure}[h!]
\center
\includegraphics[width=7.5cm]{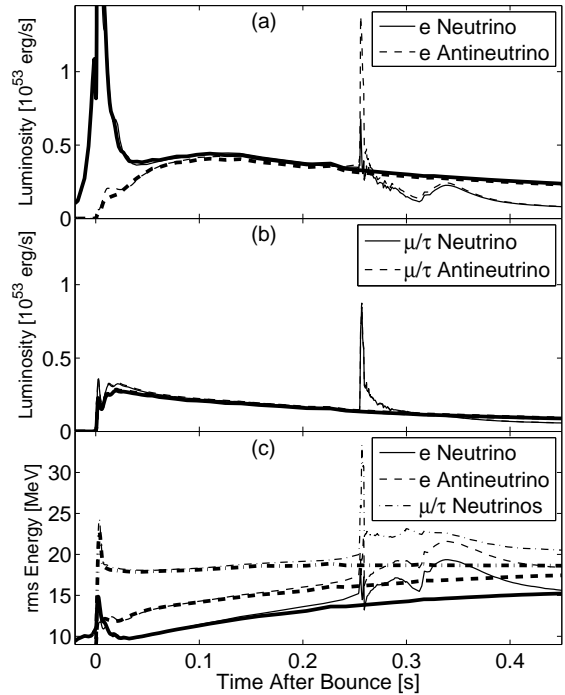}
\caption{[extracted from Ref. \cite{Sagert:2008ka}] Neutrino luminosities obtained in a core-collapse simulation assuming a hadron-quark transition.}
\label{antinuburst}
\end{figure}

The surface tension between partonic and hadronic phases of Strong Interactions seems to be a key quantity in connecting microscopic modeling of cold and dense QCD with astrophysical applications: recently it has been put forward that a low enough surface tension could allow for possible observable consequences in different phenomena related to compact objects.
It was shown for instance that deconfinement can happen during the early post-bounce accretion stage of a core collapse supernova event, which could result not only in a delayed explosion but also in an antineutrino burst that could provide a signal of the presence of quark matter in compact stars \cite{Sagert:2008ka}. In Figure \ref{antinuburst}, the output of a core-collapse simulation shows the emission of the extra antineutrino burst more than $0.2$ second after the standard neutrino signal of supernovae explosions. 


\begin{figure}[h]
\center
\includegraphics[width=7.5cm]{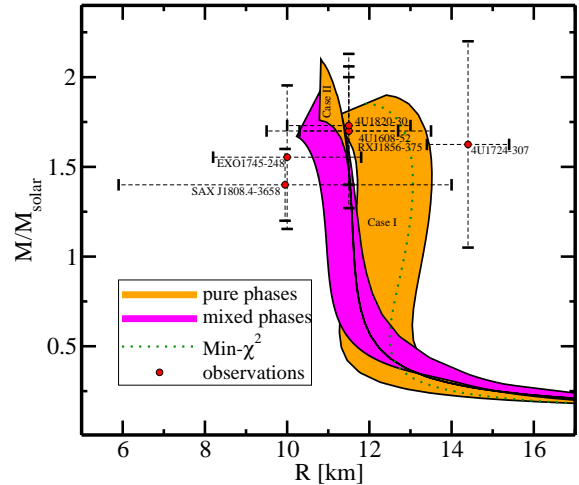}
\caption{[extracted from Ref. \cite{Kurkela:2010yk}] Mass-radius relation obtained from equations of state using perturbative QCD results at high energies are compared to different observations. The left-most magenta band is the result in the presence of mixed phases, energetically viable only if the surface tension is low enough.}
\label{mixedstars}
\end{figure}

However, as was discussed in detail in Ref. \cite{Mintz:2009ay, Mintz-thesis} (see also Ref. \cite{Bombaci:2009jt} )  those possibilities depend on the actual dynamics of phase conversion, more specifically on the time scales that emerge. In a first-order phase transition, as is expected to be the case in QCD at very low temperatures, the process is guided by bubble nucleation (usually slow) or spinodal decomposition (``explosive'' due to the vanishing barrier between phases), depending on how fast the system reaches the spinodal instability as compared to the nucleation rate \cite{reviews}. Nucleation in relatively high-density, cold strongly interacting matter, with chemical potential of the order of the temperature, can also play an important role in the scenario proposed in Ref. \cite{Boeckel:2009ej}, where a second (little) inflation at the time of the primordial quark-hadron transition could account for the dilution of an initially high ratio of baryon to photon numbers. Moreover, as shown in Figure \ref{mixedstars}, significantly different compact star structures are obtained if one considers the possibility of a layer of a quark-hadron mixed phase in the core of compact stars \cite{Kurkela:2010yk}. A key ingredient in all these scenarios is, of course, the surface tension, since it represents the price in energy one has to pay for the mere existence of an interface between quark and hadron phases. If the cost in energy is too high, quark-hadron mixed phases will not be favourable and similarly nucleation time scales will be too long for these astrophysical phenomena to take place.

\vspace{0.8cm}

However, the surface tension magnitude and the whole process of phase conversion via nucleation 
in a cold and dense environment are not known for strong interactions. In fact, the 
mapping of this sector of the (equilibrium) phase diagram is still in its infancy \cite{Stephanov:2007fk}, 
let alone dynamical processes of phase conversion. 

Computing the surface tension for the QCD transitions with first-principle methods is a prohibitively complicated task.
It is a genuine nonperturbative quantity, in the sense that 
it cannot be obtained by a na\"ive expansion around one minimum of the effective action, and is also inaccessible by current first-principle lattice QCD simulations, since the Sign Problem at finite chemical 
potential brings about major technical difficulties for performing Monte Carlo lattice simulations. One has therefore to resort to effective models. 
Estimates for the surface tension between a quark phase and hadron matter were 
considered previously in different contexts. In a study of the minimal interface between a 
color-flavor locked phase and nuclear matter in a first order transition, the authors of 
Ref. \cite{Alford:2001zr} use dimensional analysis and obtain $\Sigma\sim 300~$MeV/fm$^{2}$ 
assuming that the transition occurs within a fermi in thickness. Taking into account the effects 
from charge screening and structured mixed phases, the authors of Ref. \cite{Voskresensky:2002hu} 
provide estimates in the range of $50-150~$MeV/fm$^{2}$ but do not exclude smaller or larger 
values. These values of the surface tension are too high, rendering the nucleation of quark matter bubbles improbable in the context of compact objects.

In what follows, we present our computation \cite{Palhares:2010be} of the surface tension for the cold and dense chiral phase transition within a well-established chiral model (namely, the linear sigma model with constituent quarks (LSMq)) and assuming the homogeneous nucleation scenario \cite{Langer:1967ax} (cf. also \cite{reviews}). Results indicate that the surface tension in the cold and dense regime can be considerably smaller than previous estimates, possibly allowing for the interesting observable astrophysical phenomena described above.

The remainder of this chapter is organized as follows.
 In Section \ref{homnucl}, we delineate how, within the framework of homogeneous nucleation, it is possible to derive the nucleation parameters starting from an effective potential for the chiral order parameter $V_{\rm eff}(\sigma)$. 
Using the results for the effective potential $V_{\rm eff}(\sigma)$ of the linear sigma model with quarks from Appendix \ref{LSM}, we compute the nucleation parameters for three different cases, applying the method described in Section \ref{methodfit}. The cold and dense result is presented in Section \ref{ResCold}, while Section \ref{ResTVac} contains the discussion on the competition between thermal effects and vacuum logarithmic corrections. Application of the results to the context of supernovae explosions is discussed in Section \ref{Conseq}. Final remarks are found in Section \ref{Conc}.


\section{How to estimate the surface tension?}\label{homnucl}

Let us now consider the formation of droplets of quark matter at high density 
that happens via homogeneous nucleation \cite{reviews}. Dynamically this process 
will occur either via thermal activation of droplets (thermal nucleation) or quantum 
nucleation. The physical setting we have in mind is the one that gives the best chances 
for thermal nucleation of quark droplets in cold hadronic matter found in ``hot'' protoneutron 
stars. As discussed in Ref. \cite{Mintz:2009ay,Mintz-thesis}, and previously 
in Refs. \cite{Horvath:1992wq,Olesen:1993ek}, that corresponds to temperatures of the order 
of $10-20~$MeV \cite{refs-temp}. At these temperatures, and in the presence of a barrier in the effective 
potential, thermal nucleation dominates over quantum nucleation. As soon as the barrier 
disappears, the spinodal instability is reached and the mechanism that takes over is the explosive 
spinodal decomposition. The range of temperatures under consideration is, then, high enough 
to allow for thermal nucleation (quantum nucleation being comparatively 
negligible \cite{Mintz:2009ay,Mintz-thesis,Bombaci:2009jt,Iida:1997ay}) and low enough to justify 
the use of the zero-temperature effective 
potential. Temperatures of a couple of tens of MeV will not modify appreciably 
the equation of state, bringing corrections $O(T^{2}/\mu^{2})\sim 1\%$ for the typical values of 
chemical potential for the system under consideration. Nevertheless, as will be shown later,
thermal corrections can be important for the process of nucleation.

As stated above, with the aim of quantifying the surface tension associated with the cold and dense chiral phase transition in QCD, we will adopt the framework of homogeneous nucleation proposed by Langer \cite{Langer:1967ax}. It is based on the construction of a coarse-grained free energy functional $\mathcal{F}[\sigma]$ for the configuration of the order parameter field $\sigma$ from a given effective potential $V_{\rm eff}$ as follows:
\begin{eqnarray}
\mathcal{F}[\sigma]
&=&
4\pi\int r^2 dr\left[\frac{1}{2}\left(\frac{d\sigma}{dr}\right)^2+V_{\rm eff}(\sigma)\right]\,,
\label{Ffunc}
\end{eqnarray}
where spherical symmetry has been assumed.

\begin{figure}[h]
    \centering
        \includegraphics[width=0.35\textwidth]{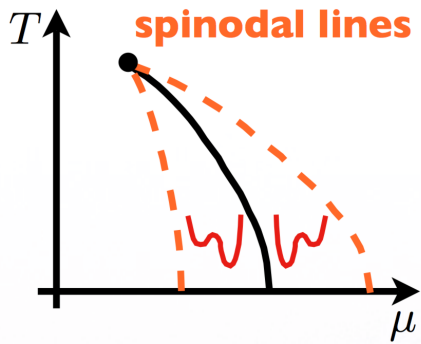}
\caption{Cartoon of the temperature {\it versus} chemical potential phase diagram with a first order critical line (solid) and the spinodal curves (dashed) defining the metastable region.}
\label{spinodals}
\end{figure}

\begin{figure}
\center
\begin{minipage}[h]{62mm}
\includegraphics[width=5.5cm]{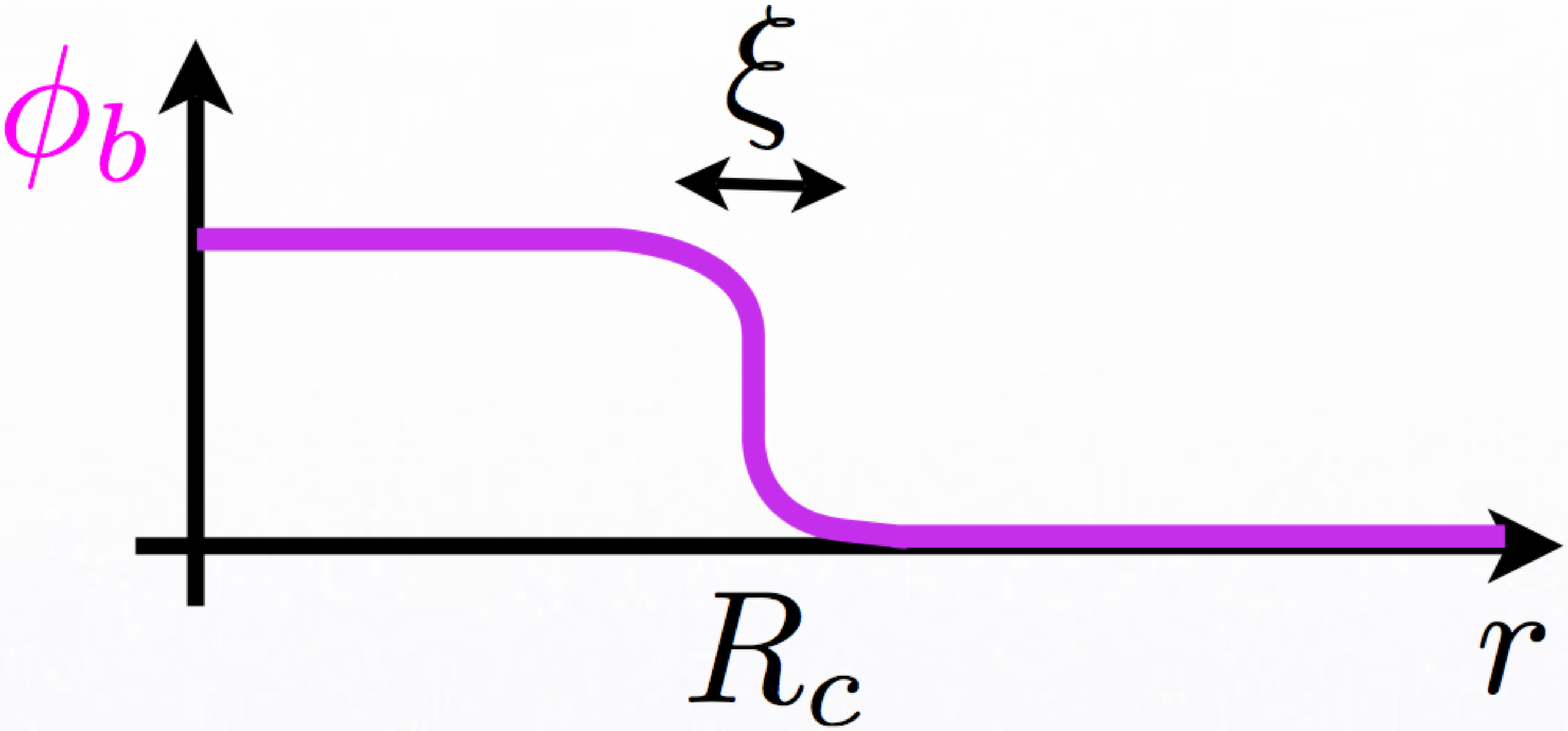}
\caption{Example of bubble profile.}
\label{bubble}
\end{minipage}
\hspace{.7cm} 
\begin{minipage}[h]{62mm}
\includegraphics[width=5.5cm]{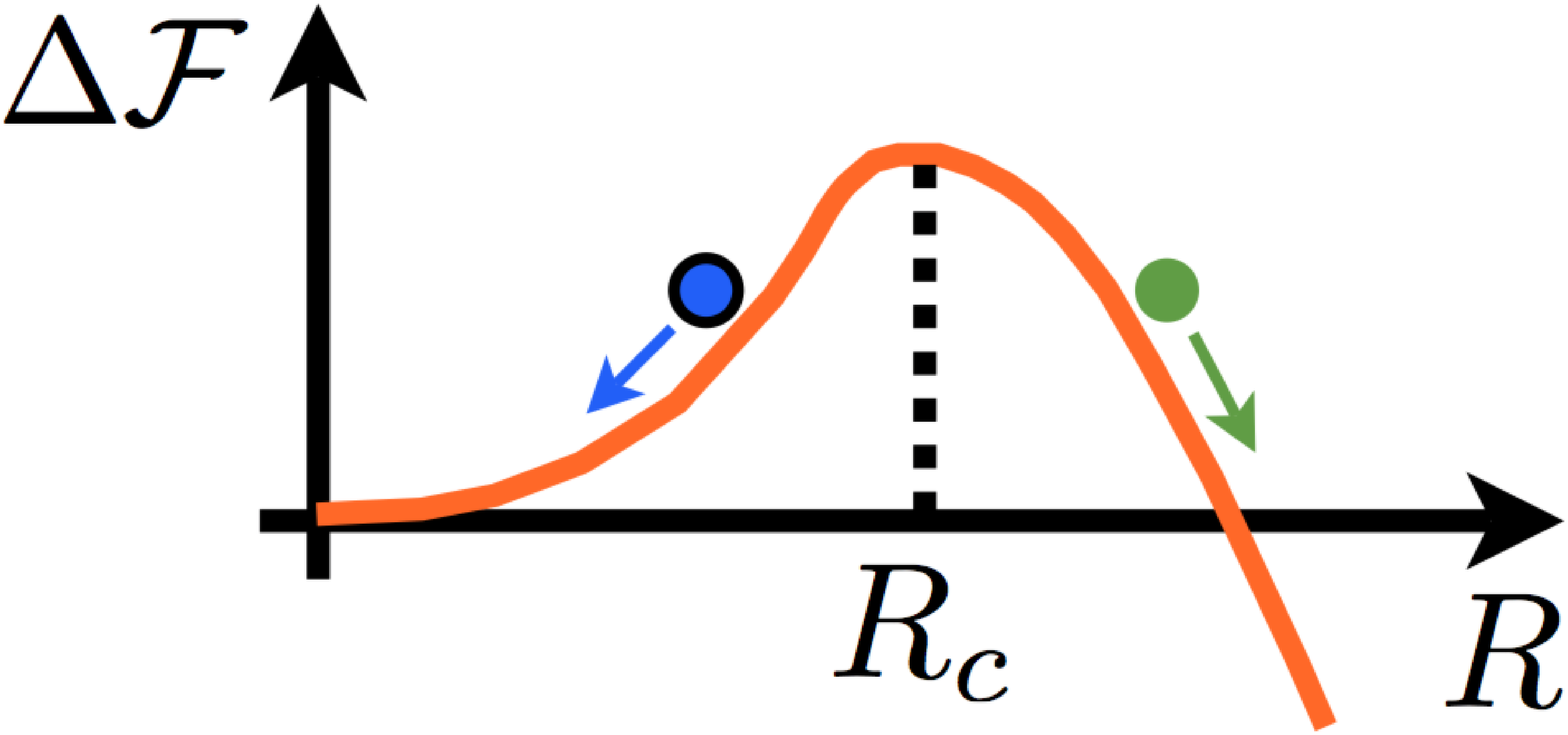}
\caption{Sketch of the difference of coarse-grained free energy between the configurations with and without a bubble of radius $R$.}
\label{DeltaF}
\end{minipage}
\end{figure}

Around the domain of temperature $T$ and chemical potential $\mu$ in which a first-order phase transition takes place, a metastable phase appears. More precisely, metastability manifests as the existence of (at least) two minima of the effective potential with a maximum in between inside the region defined by the two spinodal lines, as depicted in Fig. \ref{spinodals}. Within the so-called metastable region, the coarse-grained free energy functional in Eq. (\ref{Ffunc}) develops a nonuniform extremum configuration of the bubble type $\phi_b(r; R=R_c)$, connecting the metastable minimum to the absolute one (for an illustration cf. Fig. \ref{bubble}). This unstable nonuniform extremum is called the critical bubble and is intimately related to the dynamics of phase conversion via bubble nucleation. Differently from the uniform configuration sitting at the absolute minimum, the critical bubble is unstable, corresponding actually to a maximum of the free-energy $\mathcal{F}$.

If the system goes through an out-of-equilibrium stage in which the external parameters $T$ and $\mu$ are modified fast enough as compared to the time scales of the microscopic dynamics, then the system may get trapped in the metastable phase.
In this scenario, fluctuations (brought about by the interaction with a thermal or particle {\it reservoir})  may generate bubbles of the absolute minimum inside the metastable medium. These fluctuation bubbles will then evolve according to the competition between volume and surface terms, as dictated by the free-energy difference between the configurations with and without a bubble of radius $R$:
\begin{eqnarray}
\Delta\mathcal{F}
&\equiv&\mathcal{F}[\phi_b(r;R)]-\mathcal{F}[\phi_b(r;R=0)]
=
-\frac{4\pi}{3}R^3\Delta P+4\pi R^2\Sigma
\,,
\label{eqDeltaF}\end{eqnarray}
whose general form is sketched in Fig. \ref{DeltaF}. The volume term encodes the gain in energy resulting from the conversion to the true equilibrium state and is proportional to the pressure difference $\Delta P>0$ between the phases. On the other hand, the surface term accounts for the 
energy spent in the construction of a surface between the phases, being given essentially by the surface tension $\Sigma$. 
%

As illustrated in Fig. \ref{DeltaF}, the free-energy difference $\Delta \mathcal{F}$ shows a maximum associated with the critical bubble ($R=R_c$) and fluctuation-generated configurations containing bubbles with radii $R<R_c$ will tend to the uniform metastable vacuum in order to minimize the energy. For radii bigger than the critical radius, the minimization of energy implies that the droplet will grow and eventually complete the phase conversion. The critical radius goes to zero at the spinodal curves, where the metastable false vacuum becomes unstable (the barrier disappears) and the phase conversion occurs explosively via the spinodal decomposition process.

Finally, from Eq. (\ref{eqDeltaF}) it is clear that the surface tension can be obtained via the knowledge of the effective potential $V_{\rm eff}$ and the critical bubble profile $\phi_b(r;R=R_c)$. Since the critical bubble itself is an unstable extremum of the coarse-grained action $\mathcal{F}$ in Eq. (\ref{Ffunc}), it is ultimately obtained from the effective potential $V_{\rm eff}$ as well. In the thin wall approximation (valid when the critical bubble has a thin interface region as compared to its radius, i.e. $\xi\ll R_c$; cf. Fig. \ref{bubble}) for a quartic effective potential, the solution $\phi_b(r;R=R_c)$ can be written analytically in terms of the Taylor coefficients of the potential.

\subsection{Extracting nucleation parameters from the effective potential}\label{methodfit}

Since our framework is an effective model, we can only aim for reasonable estimates and 
functional behavior, not numerical precision. Therefore, it is convenient to work with approximate 
analytic relations by fitting the relevant region of the effective potential by a quartic polynomial 
and in the thin-wall limit approximation for bubble nucleation. Following 
Refs. \cite{Scavenius:2001bb,Taketani:2006zg}, we can express the effective potential over the range 
between the critical chemical potential, $\mu_{c}$, and the spinodal, $\mu_{sp}$, in the familiar 
Landau-Ginzburg form
\begin{equation}
V_{\rm eff} \approx \sum_{n=0}^4 a_n \, \phi^n \;.
\label{LandGinz}
\end{equation}
Although this approximation is obviously incapable of reproducing all three 
minima of $V_{\rm eff}$, this polynomial form is found to provide a good quantitative 
description of this function in the region of interest for nucleation, i.e. where 
the minima for the symmetric and broken phases, as well as the barrier 
between them, are located. 

A quartic potential such as Eq. (\ref{LandGinz}) can always be rewritten in the form
\begin{equation}
{\mathcal V}(\varphi)=\alpha ~ (\varphi^2-a^2)^2+j\varphi \;,
\label{www}
\end{equation}
with the coefficients above defined as follows:
\begin{eqnarray}
\alpha &=& a_4\quad, \\
a^2 &=& \frac{1}{2}\left[ -\frac{a_2}{a_4}
+\frac{3}{8}\left(\frac{a_3}{a_4}\right)^2  \right]\;, \\
j &=& a_4\left[ \frac{a_1}{a_4}-
\frac{1}{2}\frac{a_2}{a_4}\frac{a_3}{a_4}+
\frac{1}{8}\left(\frac{a_3}{a_4}\right)^3   \right]\;, \\
\varphi&=&\phi + \frac{1}{4}\frac{a_3}{a_4} \; . 
\end{eqnarray}
The new potential ${\mathcal V}(\varphi)$ reproduces the original $V_{\rm eff} ( \phi )$ 
up to a shift in the zero of energy.  We are interested in the effective 
potential only between $\mu_c$ and $\mu_{sp}$.  At $\mu_c$, we will have two 
distinct minima of equal depth.  This clearly corresponds to the choice $j = 0$
in Eq.\,(\ref{www}) so that ${\mathcal V}$ has minima at $\varphi = \pm a$ and a maximum 
at $\varphi = 0$.  The minimum at $\varphi = -a$ and the maximum move closer 
together as the chemical potential is shifted and merge at $\mu_{sp}$.  Thus, 
the spinodal requires $j/\alpha a^3 = -8/3\sqrt{3}$ in Eq.\,(\ref{www}). 
The parameter $j/\alpha a^3$ falls roughly 
linearly from $0$, at $\mu=\mu_c$, to $-8/3\sqrt{3}$ at the spinodal.

The explicit form of the critical bubble in the thin-wall limit is then given 
by~\cite{Fraga:1994xd}
\begin{equation}
\varphi_b (r;\xi,R_c)=\varphi_f + \frac{1}{\xi\sqrt{2\alpha}}
\left[ 1-\tanh \left( \frac{r-R_c}{\xi} \right) \right] \;,
\label{bubproftw}
\end{equation}
where $\varphi_f$ is the new false vacuum, $R_c$ is the radius of the 
critical bubble, and $\xi=2/m$, with $m^2\equiv {\mathcal V}''(\varphi_f)$, is a measure 
of the wall thickness.  The thin-wall limit corresponds to $\xi/R_c\ll 1$, which can be rewritten as $(3|j|/8\alpha a^3)\ll 1$ \cite{Fraga:1994xd}.  
Nevertheless, it was shown in \cite{Scavenius:2001bb,Bessa:2008nw} for 
the case of zero density and finite temperature that the thin-wall limit becomes 
very imprecise as one approaches the spinodal (This is actually a very general 
feature of this description \cite{reviews}). In this vein, also as remarked above, 
the analysis presented below is to be regarded as semi-quantitative, and we aim 
for estimates. 

In terms of the parameters $\alpha$, $a$, and $j$ defined above, we 
find \cite{Scavenius:2001bb,Taketani:2006zg}
\begin{eqnarray}
\varphi_{t,f} &\approx& \pm a - \frac{j}{8\alpha a^2}  \quad, \\
\xi &=& \left[ \frac{1}{\alpha (3\varphi_f^2-a^2)} \right]^{1/2}
\label{twcorlength}
\end{eqnarray}
in the thin-wall limit. The surface tension, $\Sigma$, is given by
\begin{equation}
\Sigma\equiv \int_0^{\infty}{\rm d}r~\left( \frac{{\rm d}
\varphi_b}{{\rm d}r} \right)^2 
\approx \frac{2}{3\alpha\xi^3} \ ,
\end{equation}
and the critical radius is obtained from $R_c = (2\Sigma/\Delta V)$, where
$\Delta V \equiv V(\phi_f)-V(\phi_t) \approx 2 a | j |$. Finally, the free energy of a 
critical bubble is given by $F_b=(4\pi\Sigma/3)R_c^2$, and 
from knowledge of $F_b$ one can evaluate the nucleation rate 
$\Gamma \sim e^{-F_b/T}$. In calculating thin-wall properties, we shall 
use the approximate forms for $\phi_t$, $\phi_f$, $\Sigma$, and $\Delta V$ 
for all values of the potential parameters.

\section{Results for nucleation in the cold and dense LSM}\label{ResCold}

In what follows, we consider the LSM with quarks in the absence of vacuum corrections.
The effective potential up to 1-loop order is then given by Eq. (\ref{Veff0}) with 
$\Omega_{\xi}^{\rm ren}=\Omega_{\rm med}^{(1)}$. This corresponds to the standard case, 
adopted frequently in the literature (cf. e.g. Ref. \cite{Scavenius:2000qd}).

Using the method described in the previous section, we characterize quantitatively
the nucleation process predicted within this case, calculating different nucleation
parameters. In this section, we present results for both metastable regions, above and below 
the critical chemical potential $\mu_{\rm crit}=305.03~$MeV, which correspond respectively to 
the nucleation of quark droplets in a hadronic environment and the formation of hadronic bubbles 
in a partonic medium.

\vspace{0.3cm}
\begin{figure}[!hbt]
\begin{center}
\vskip 0.2 cm
\includegraphics[width=8cm,angle=0]{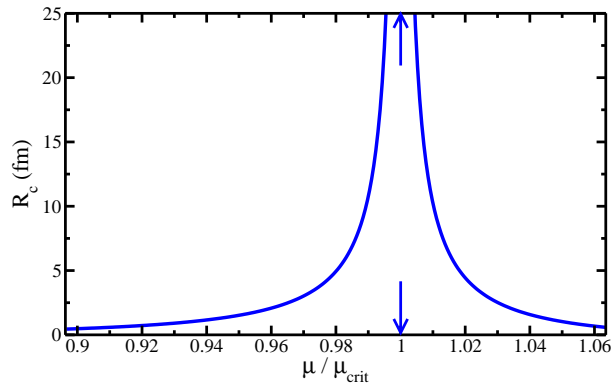}
\end{center}
\caption{Critical radius as a function of the quark chemical potential in the two metastable regions (between the spinodals).
The arrows indicate the critical chemical potential.}
\label{Rc-fig}
\end{figure}

In Fig. \ref{Rc-fig},
the critical radius, namely the radius of the critical bubble, is displayed. 
Any bubble created in the system via external influences or thermal 
fluctuations will either grow or shrink unless
its radius equals the critical value, which settles the threshold between suppressed and favored bubbles.
For radii bigger than the critical radius, the minimization of energy implies that the droplet will grow
and eventually complete the phase conversion. The critical radius goes to zero at the spinodal curves, where
the metastable false vacuum becomes unstable (the barrier disappears) and the phase conversion occurs explosively via the spinodal decomposition 
process\footnote{In the thin-wall approximation, which is a poor description of the regions close to the 
spinodals, the critical radius does not vanish, only becomes very small. The same is true for other 
quantities, as the surface tension.}. 
On the critical line the vacua become degenerate so that both of them are stable
yielding no nucleation, or equivalently, a divergent critical radius. In the case of the chiral transition
in the LSM with quarks, we obtain that the critical radius is $R_c<10~$fm over about $\sim 90\%$ of the metastable
regions.

\begin{figure}[!hbt]
\begin{center}
\vskip 0.2 cm
\includegraphics[width=8cm,angle=0]{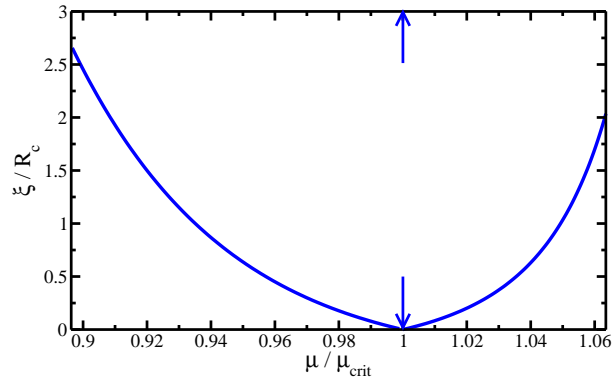}
\end{center}
\caption{Ratio between the correlation length $\xi$ and the critical radius $R_c$ as a function of the quark chemical potential in the two metastable regions.}
\label{Ratio-fig}
\end{figure}

The correlation length $\xi$ which provides a measure of the size of the bubble wall is plotted
in Fig. \ref{Ratio-fig}, in units of the critical radius.
As discussed in the previous section, the thin-wall approximation relies on the assumption that the 
ratio between the surface correlation length $\xi$ and the radius of the critical bubble is small.
Fig. \ref{Ratio-fig}
shows clearly that this is a reasonable condition in the vicinity of the critical
line, away from the spinodals.

\begin{figure}[!hbt]
\begin{center}
\vskip 0.2 cm
\includegraphics[width=8cm,angle=0]{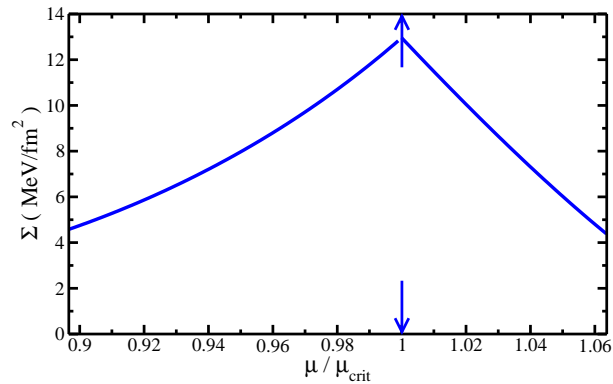}
\end{center}
\caption{Surface tension as a function of the quark chemical potential in the two metastable regions.}
\label{Sigma-fig}
\end{figure}

The surface tension is a key parameter in quantifying the nucleation process in a given medium
since it measures the amount of energy per unit of area that is spent in the construction of
a surface between the phases. In Fig. \ref{Sigma-fig}
we show our result for the surface tension, $\Sigma$,
for the chiral transition in the cold and dense LSM with quarks. The surface tension assumes values between 
$\sim 4$ and $\sim 13~$ MeV/fm$^2$, being, throughout the whole metastable region, of the order of magnitude that renders the formation of
quark matter viable during a supernova explosion, according to Ref. \cite{Sagert:2008ka}.
The biggest values occur near criticality,
since this domain is characterized by large barriers and a small free energy difference between the true and false vacua.
It should be noticed
that those values are well bellow previous estimates of this quantity, as will be discussed in the conclusion of this chapter.

\vspace{0.3cm}

\begin{figure}[!hbt]
\begin{center}
\vskip 0.2 cm
\includegraphics[width=8cm,angle=0]{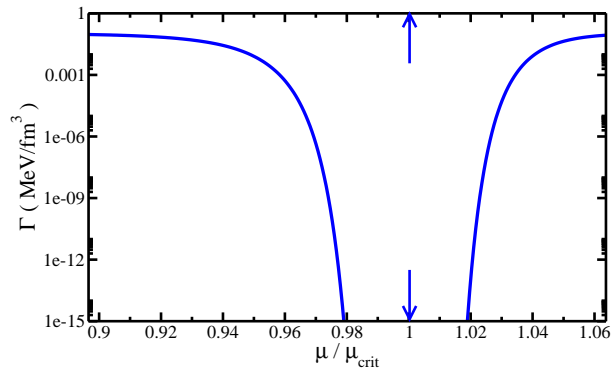}
\end{center}
\caption{Nucleation rate as a function of the quark chemical potential in the two metastable regions.}
\label{Gamma-fig}
\end{figure}

We have also estimated the nucleation rate for homogeneous nucleation \cite{Langer:1967ax,Csernai:1992tj} 
as $\Gamma\sim T_f^4\, {\rm e}^{-F_b/T_f}$, where $F_b$ is the free energy
of the critical bubble configuration and $T_f=30~$MeV is an {\it ad hoc} temperature. Applying the expression above for the nucleation rate in this case
means that we are neglecting the temperature dependence of the critical-bubble free energy in the exponent. The results are shown in 
Fig. \ref{Gamma-fig} for both metastable regions. The nucleation rate falls abruptly as the chemical potential
approaches its critical value.

It is interesting to point out that the difference in the size of the metastable regions above and below criticality might play
an important role in determining whether nucleation is a viable process of phase conversion for a given system, as compared
to spinodal decomposition scenarios or even if the lifetime of the system represents enough time for nucleation to take place.
It is clear in Fig. \ref{Gamma-fig}, for instance, that the domain in chemical potential for which the nucleation rate
assumes sizable values is larger at the metastable region corresponding to bubble formation of hadronic matter in a partonic medium.

\vspace{0.3cm}

\section{Effects from vacuum terms versus thermal corrections}\label{ResTVac}

Having presented in the section above the important parameters for nucleation and
discussed their results for the zero-temperature chiral transition in the LSM,
let us now analyze the influence of quantum logarithmic terms in the vacuum effective potential
as well as thermal corrections.

In this section we focus on the metastable region above the critical chemical potential, aiming
for thermal nucleation of quark matter droplets in compact objects and especially in supernovae explosions.
In this vein, we also keep relatively low temperatures when including thermal corrections, 
up to $T=30~$MeV. The systematic inclusion of thermal corrections provides a measure of the
validity of the zero-temperature approximation for the thermodynamics and phase structure 
in low-temperature scenarios such as those found in compact objects. We find that results for the nucleation parameters
up to $T=10~$MeV present variations within $\sim 10\%$ of the zero-temperature values, 
displayed in the previous section.

\begin{figure}[!h]
\begin{center}
\vskip 0.2 cm
\includegraphics[width=8cm,angle=0]{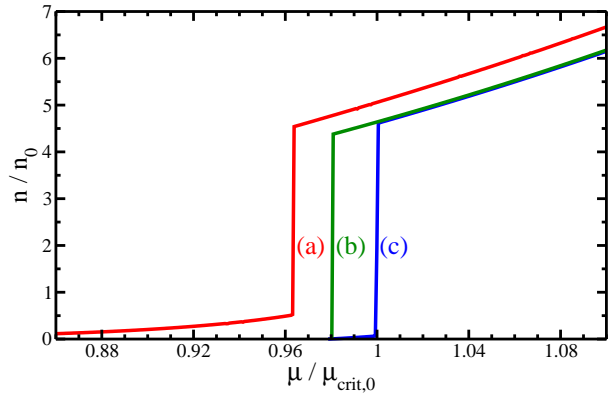}
\end{center}
\caption{Density in units of the nuclear saturation density $n_0=0.16~$fm$^{-3}$ as a function of the quark chemical potential in the cold and dense LSM (curve $(c)$), including vacuum terms (curve $(b)$) and thermal corrections (curve $(a)$).}
\label{All-Density-fig}
\end{figure}

\begin{figure}[h!]
\vspace{0.3cm}
\begin{minipage}[t]{75mm}
\includegraphics[width=8cm]{Plots/All-Rcvsmu-SpA.eps}
\end{minipage}
\hspace{0.7cm}
\begin{minipage}[t]{75mm}
\includegraphics[width=7.8cm]{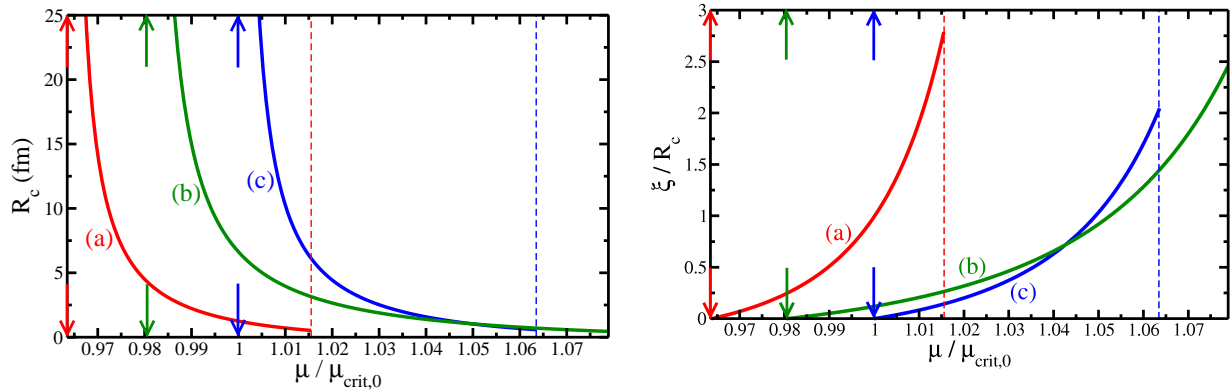}
\end{minipage}
\caption{Critical radius (left) and correlation length (right) in the nucleation region of quark droplets as a function of the quark chemical potential in the cold and dense LSM (curve $(c)$), including vacuum terms (curve $(b)$) and thermal corrections (curve $(a)$).}
\label{All-Rc-fig}
\end{figure}

\begin{figure}[htb!]
\vspace{0.3cm}
\begin{center}
\vskip 0.2 cm
\includegraphics[width=9cm,angle=0]{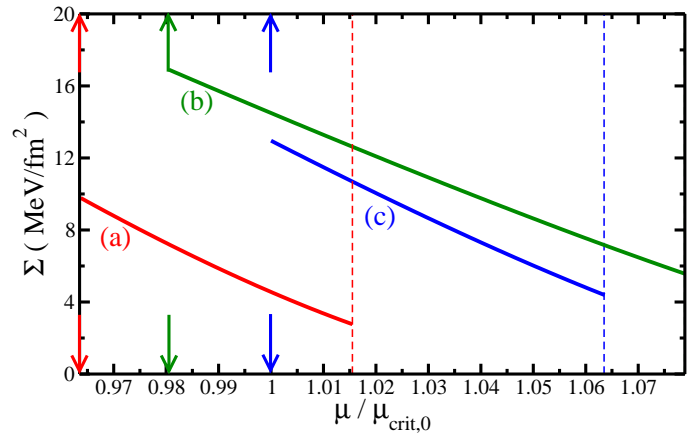}
\end{center}
\caption{Surface tension in the nucleation region of quark droplets as a function of the quark chemical potential in the cold and dense LSM (curve $(c)$), including vacuum terms (curve $(b)$) and thermal corrections (curve $(a)$).}
\label{All-Sigma-fig}
\end{figure}

\begin{figure}[hbt]
\begin{center}
\vskip 0.2 cm
\includegraphics[width=8.5cm,angle=0]{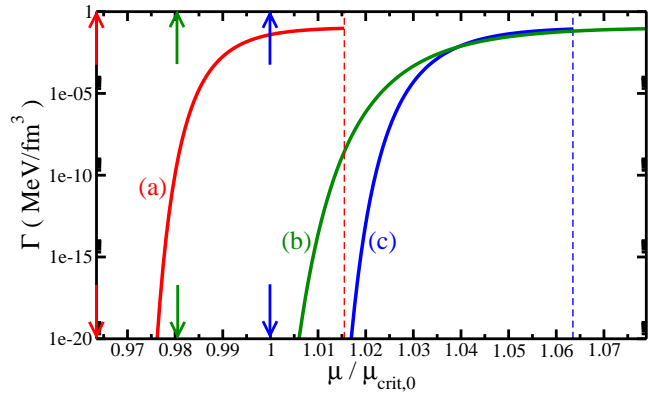}
\end{center}
\caption{Nucleation rate in the nucleation region of quark droplets as a function of the quark chemical potential in the cold and dense LSM (curve $(c)$), including vacuum terms (curve $(b)$) and thermal corrections (curve $(a)$).}
\label{All-Gamma-fig}
\end{figure}

In Figures \ref{All-Density-fig}--\ref{All-Gamma-fig}, the role played by temperature corrections and quantum vacuum effects
is presented. More specifically, we compare results for the three following situations: $(a)$ classical vacuum
effective potential with thermal corrections for $T=30~$MeV, $(b)$ quantum vacuum effective potential
at zero temperature, $(c)$ classical vacuum effective potential at zero temperature, i.e. the
results shown in the last subsection. Each case is associated with a different critical chemical potential (indicated by arrows
in the plots; the normalization $\mu_{{\rm crit},0}=305.03~$MeV corresponds to the critical chemical potential of the case $(c)$, of the last subsection) and also
a different spinodal chemical potential (denoted by the vertical, dashed lines). 

Concerning the modification of the metastable regions, 
we show that the quantum vacuum corrections increase significantly the domain in chemical potential in which metastability persists; the metastable region
in the presence of quantum vacuum terms is $\sim 40\%$ bigger than the one associated with the classical vacuum effective potential.
This feature indicates that such corrections might generate large differences in the dynamic evolution of the system, even though 
the absolute value of the critical chemical potential itself shifts only $\sim 2\%$.

\vspace{0.3cm}

During the dynamical process of phase conversion, the density is increased (e.g. via gravitational pressure in the supernova explosion scenario)
and therefore the free energy required for including a new quark in the system also augments. 
The mapping between density and quark chemical potential within the cases we consider is plotted in Figure \ref{All-Density-fig}.
The discontinuities signal the first-order phase transition that restores chiral symmetry at high energies.
As the evolving system reaches a high enough density, it penetrates the metastable region above the critical chemical potential, 
where higher densities are associated with higher supercompression.

It is clear that the values of density predicted below the transition are irrealistically low.
This is a consequence of the fact that the LSMq fails to describe nuclear matter. Being an effetive theory constructed with the purpose
of describing the chiral symmetry features of strongly interacting matter, 
the LSMq does not predict the low-energy nuclear matter properties nor the gas-liquid transition that forms it.
However, since our ultimate interest is investigating the formation of quark matter in astrophysical processes involving ultra-compact objects,
the framework most suited is exactly one that describes the high-energy transitions of strong interactions, namely chiral restoration
and/or deconfinement. We choose therefore the LSMq and analyze the nucleation of chirally-symmetric droplets in a strongly interacting medium.

In order to interpret the results for the nucleation parameters, one should keep in mind the
variation of the metastable regions and its implication on the value of chemical potential that needs to be reached above the respective criticality,
i.e. the amount of supercompression.


%

%

In Figure \ref{All-Rc-fig}, the left panel shows the variation of the radius of the critical bubble,
while the right one displays the correlation length (or equivalently
the approximate width of the bubble wall) for the three cases.
A direct consequence of the shifting of the metastable domain is the fact that, for 
fixed chemical potential, the critical radius decreases significantly both when 
one includes quantum vacuum corrections (curve $(b)$) or in the presence of 
thermal effects (curve $(a)$). However, the amount of supercompression needed for reaching a
certain viable value of critical radius clearly increases when one includes quantum corrections
and decreases in the case with thermal corrections. Therefore, there is a nontrivial
competition between those effects: while quark matter nucleation is 
disfavored by the inclusion of quantum vacuum corrections, it is favored by thermal corrections to the nucleation
parameters.

The competition between vacuum and thermal corrections is more explicitly shown by our findings for the surface tension, 
displayed in Figure \ref{All-Sigma-fig}. As compared to the zero-temperature, classical
computation (curve $(c)$), the establishment of a interface between quark matter droplets and
the hadronic medium costs more with the incorporation of quantum corrections in the effective potential.
On the other hand, finite temperature terms tend to push the surface tension down, reaching a minimum 
of only $\sim 3~$MeV/fm$^2$, facilitating the surface formation. Therefore, if the environment
in the core of a compact object is cold enough
and if the effective theory including quantum corrections in the vacuum is the most suited for describing
the chiral transition, then the surface tension can be as high as $\sim 17.5~$MeV/fm$^2$ near criticality
rendering quark matter nucleation much slower and even an improbable phenomenon.

Analogous results for the nucleation rate $\Gamma$ can be seen in Figure \ref{All-Gamma-fig}. 
Once again, the amount of supercompression associated with a given chemical potential should 
be kept in mind. The nucleation rate falls abruptly for chemical potentials approaching the respective 
critical values. In the vicinity of the spinodals, the nucleation rate is dominated by the pre-exponential 
factor and reaches sizable values, around $\sim 0.1~$MeV/fm$^3$.

\vspace{1cm}

\section{Consequences for the quark-matter-induced supernova explosion scenario}\label{Conseq}

To contribute to the investigation of the possibility of nucleating quark matter droplets during the early
post-bounce stage of core collapse supernovae, we follow Ref. \cite{Mintz:2009ay,Mintz-thesis} and define the
nucleation time as being the time it takes for the nucleation of a single critical bubble inside a volume
of $1$km$^3$, which is typical of the core of a protoneutron star, i.e.
\begin{equation}
\tau_{nucl}\equiv \left( \frac{1}{1 {\rm km}^3} \right) \frac{1}{\Gamma} \;.
\end{equation}

\vspace{0.3cm}
\begin{figure}[!h]
\begin{center}
\vskip 0.2 cm
\includegraphics[width=8.5cm,angle=0]{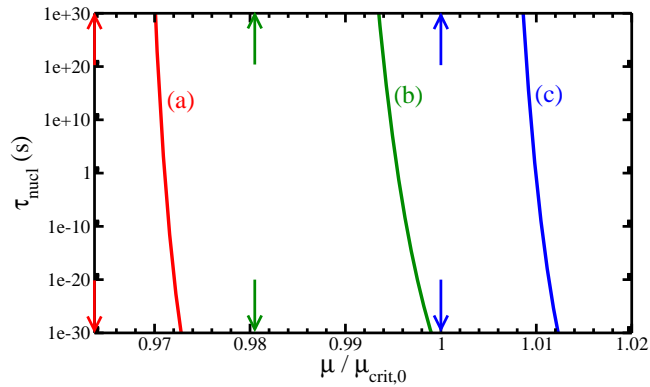}
\end{center}
\caption{Nucleation time defined for a typical volume of $1$ km$^3$ in the cold and dense LSM (curve $(c)$), including vacuum terms (curve $(b)$) and thermal corrections (curve $(a)$).}
\label{All-tau-fig}
\end{figure}

Fig. \ref{All-tau-fig} shows this quantity as a function of the normalized chemical potential for the three scenarios
considered in this analysis. The relevant time scale to compare is the time interval the system takes
from the critical chemical potential to the spinodal during the supernova event, if it ever reaches such
high densities in practice. Implicit in the definition above is the approximation of constant density and
temperature over the core, which is fine for an estimate, since density profiles are quite flat in this region
of the star. The typical time scale for the early post-bounce phase is of the order of a fraction of a second,
so that the time within the metastable region is smaller, in the ballpark of milliseconds.

Keeping these simplifications in mind, together with the caveat that the LSMq does not describe nuclear
matter besides the chiral properties of strong interactions, we show in Fig. \ref{All-tau-fig} that vacuum corrections tend
to increase the density depth required for efficient nucleation whereas thermal corrections push in the
opposite direction, consistently with the results discussed in the previous subsection. Our results for the
nucleation time, together with those for the surface tension, tend to favor the best setting for nucleation
of quark matter in the supernova explosion scenario considered in Ref. \cite{Mintz:2009ay,Mintz-thesis}, especially when
thermal corrections with physical temperatures are included. To a great measure, this happens because
the nucleation time, and the whole process of phase conversion, depends very strongly on the surface
tension, since it enters cubed in the Boltzmann exponential of the rate $\Gamma$.

\section{Final remarks}\label{Conc}

In this chapter we have explored the interplay between the dynamics of cold and dense phase transitions of Strong Interactions and astrophysical phenomena involving ultra-compact objects. Since these environments contain QCD matter under extreme pressures, much beyond any current laboratory can achieve, their understanding requires a theoretical description of Strongly interacting cold and dense media. For the same reason, with the development of observational techniques, astrophysical observations may become an invaluable experimental channel to assess the low-temperature high-density domain of the QCD phase diagram as realized in Nature. 
Taking thorough advantage of this channel demands quite often the description of dynamical critical phenomena within Strong Interactions.

As emphasized throughout the chapter, a striking example of this connection of out-of-equilibrium QCD phase transitions with observable astrophysical phenomena is the surface tension between the hadronic and the chirally restored deconfined phases in the cold and dense regime. If its value is small enough, quark-matter formation could occur during core-collapse supernovae explosions, providing an alternative dynamics and even an observable signal of the QCD phase transition within compact objects. Moreover, a reasonably small surface tension between partonic and hadronic phases could also contribute to allow for different compact star structures including mixed phases.

Modelling the chiral phase transition via the linear sigma model with quarks, we have studied 
homogeneous nucleation in a framework that allowed for analytic calculations, given by 
a potential fit by a quartic polynomial and the thin-wall approximation. All the relevant 
quantities were computed as functions of the chemical potential (or the baryonic density), 
the key function being the surface tension. 

 In particular, 
we have showed that the LSMq predicts a surface tension of $\Sigma \sim 5$--$15~$MeV/fm$^{2}$, 
much lower than previous estimates (e.g. \cite{Alford:2001zr} and \cite{Voskresensky:2002hu} ) and possibly allowing for the interesting observable phenomena described above.
Including temperature effects and vacuum logarithmic corrections, we find a clear competition between these
ingredients in characterizing the dynamics of the chiral phase conversion, so that if the temperature is low enough
the consistent inclusion of vacuum corrections could help preventing the nucleation of quark matter during
the early post-bounce stage of core collapse supernovae.

As discussed in Subsection \ref{Int}, effects from interactions between the sigma field and the quarks that come about at two-loop order
could in principle contribute to this competition as a third sizable modification and should be investigated
as well.

The linear sigma model, however, does not contain essential ingredients to describe nuclear 
matter, e.g. it does not reproduce features such as the saturation density and the binding energy. 
Therefore, the results obtained in this chapter should be considered with caution when applied to 
compact stars or the early universe. It is an effective theory for a first-order chiral phase transition 
in cold and dense strongly interacting matter, and allows for a clean calculation of the physical 
quantities that are relevant for homogeneous nucleation in the process of phase conversion. In 
the spirit of an effective model description, our results should be viewed as estimates that indicate 
that the surface tension is reasonably low and falls with baryon density, as one increases the 
supercompression. First-principle calculations 
in QCD in this domain are probably out of reach in the near future. Therefore, estimates within 
other effective models would be very welcome.

%% file: magneticQCD.tex
\chapter[Thermodynamics of QCD in a magnetic background]{
\label{magneticQCD}}

\vspace{1.5cm}

{\huge \sc Thermodynamics of QCD}
\vspace{0.3cm}

\noindent {\huge \sc in a magnetic background}

\vspace{2cm}

\vspace{-12cm}
\hspace{6cm}
\includegraphics[width=8.5cm,angle=90]{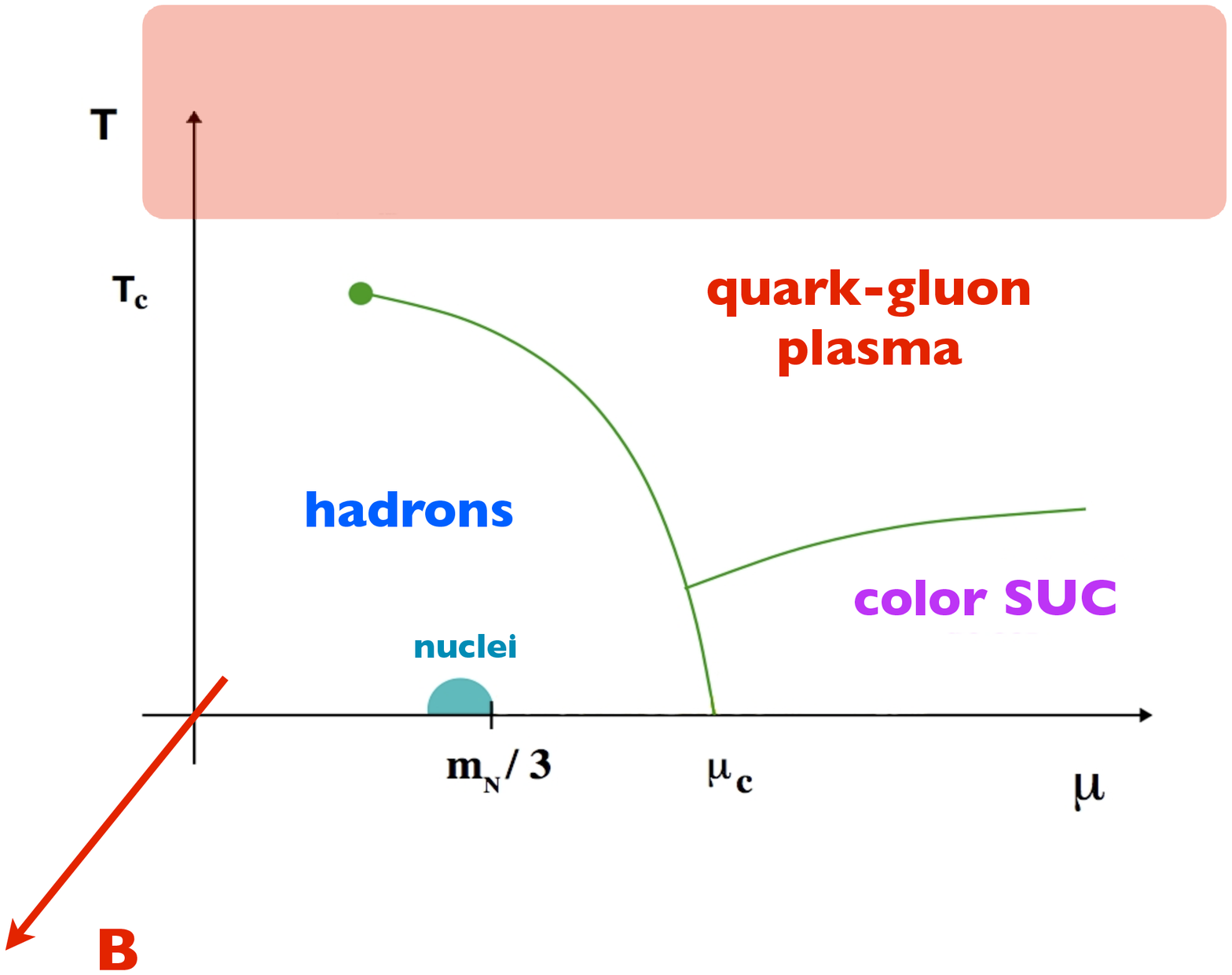}

\vspace{3.8cm}

In this chapter, a hot QCD medium exposed to a constant and uniform external magnetic background is considered. The magnetic field becomes therefore a new axis in the phase diagram of Strong Interactions to be explored, giving rise, for example, to the temperature {\it versus} magnetic field plane, which we address in particular.

Some interesting phenomena have been predicted and explored recently regarding Strongly interacting matter under extremely intense magnetic fields, but we will be interested here in posing a well-defined question within the fundamental gauge theory: what happens to the thermodynamics of QCD in the presence of such an intense magnetic background? 
At sufficiently high energies, the answer to this question should be given by perturbative QCD in a nonperturbative magnetic background. That is the approach we adopt in this on-going project \cite{MagneticQCD-wip} to compute the QCD pressure and the thermal self-energies, on the path of identifying the adequate quasiparticles in this domain of the QCD phase diagram.

\section{Motivation and framework}

The area of in-medium Strong Interactions under extreme magnetic fields has gone through a fast development due to the combination of three features realized recently: experimental relevance, capability of being investigated on the lattice and a rich phenomenology.

Hot QCD matter exposed to sufficiently high magnetic fields is currently being generated in laboratory experiments. The setup of ultra-relativistic heavy ion collisions is such that extremely intense magnetic fields are produced in peripheral collisions, with a nonzero impact parameter. A beam of heavy ions like Au accelerated to velocities very close to that of the light creates an enormous electric current. Parallel, non-coinciding beams going in opposite directions induce, through a basic classical electrodynamics process, a maximum magnetic field at the center, exactly where the medium forms in such collisions. At RHIC the intensity of the magnetic fields created may reach $6\,m_{\pi}^2\approx 10^4$--$10^5~$MeV${}^2\approx10^{19}~$Gauss \cite{magnetic-HIC}.

There have been many developments in the direction of using this extreme field to produce signatures to assess the properties of the new state of matter created in these experiments. For instance, the Chiral Magnetic Effect \cite{Kharzeev:2007jp} takes advantage of the capability of  magnetic fields of spatially separating different electric charges to render measurable the possible formation of sphaleron-induced CP-odd domains in the medium. 
Currently under investigation are also the modifications of the QCD phase structure in the presence of this new external parameter. Sufficiently large magnetic fields could allow for drastic phenomena, from affecting the chiral and the deconfinement phase transitions \cite{Agasian:2008tb,Fraga:2008qn,Boomsma:2009yk,Fukushima:2010fe,Ana-thesis} to transforming vacuum in a superconducting medium via $\rho$-meson condensation \cite{Chernodub:2010qx}. Although complemented by lattice results \cite{lattice-maxim,lattice-delia,Braguta:2011hq,Bali:2011qj}, most of these analyses were done within effective frameworks, as one is usually obliged to do when treating the many ingredients that are relevant to approach experimental conditions or describing phase transitions. 

It is important, however, to address the high-temperature, high-magnetic field regime of the QCD phase diagram (semi-)analytically from the point of view of the fundamental gauge theory. Since the $T-B$ plane of the phase diagram is amenable to lattice QCD simulations, it represents a new open channel for comparison between theory and numerical nonperturbative approaches. Our aim in this on-going project \cite{MagneticQCD-wip} is to make proper use of this opportunity by providing perturbative QCD predictions for the thermodynamics of a hot environment immersed in an intense magnetic field. In this way, one gains further analytical understanding of lattice results on one side and tests the validity of the perturbative expansion in this new regime on the other. Furthermore, this first-principle investigation within the fundamental gauge theory shall also furnish guidance for effective-model building to go beyond the domain of applicability of perturbation theory.

Interesting phenomena in Strongly interacting media arise only when magnetic fields are sufficiently large as compared to relevant scales such as $m_{\pi}^2$. Therefore, the specific physical system we address in this chapter is hot QCD matter under an intense constant and uniform magnetic background at very high energies. Asymptotic freedom then suggests the applicability of perturbation theory, while the (classical)  magnetic sector must be treated nonperturbatively. In this vein, the framework adopted
is described by the QCD Lagrangian in the presence of a classical electromagnetic vector potential $\vec{A}$ ($\nabla\times \vec{A}=\vec{B}=B \hat z$; $A_{\rm cl}=(0,\vec{A})$):
\be
\mathcal{L}_{\rm QCD}
&=&
-\frac{1}{4}{\rm Tr}\left[G^{\mu\nu}G_{\mu\nu}\right]
+\sum_f\overline{\Psi}_f\left[i\slashed{\partial}-m_f-q_f\slashed{A}_{cl}\right]\Psi_f
-g\sum_f\overline{\Psi}_f\,\slashed{G}\,\Psi_f
\,,
\ee
where $G_{\mu}=G_{\mu}^a\lambda^a$ is the gluon field ($\lambda^a$ are the Gell-Mann matrices) and $G^a_{\mu\nu}=\partial_{\mu}G^a_{\nu}-\partial_{\nu}G^a_{\mu}+gf^{abc}G^b_{\mu}G^c_{\nu}$ is the gluon field tensor in a standard $SU(N_c)$ gauge theory.
Here, the quark-gluon interaction is treated perturbatively up to $O(g^2)$, while the magnetic coupling is considered as a nonperturbative dressing of the fermion propagator. In this vein, our description is that of a medium of magnetically-dressed quarks perturbed by the coupling with gluons.

In what follows, the framework is concretely built by the derivation of the {\it exact} fermion propagator in a magnetic background in Section \ref{MagProp} and then used to compute the one-loop gluon selfenergy in the vacuum (Section \ref{GluonFuk}) and the QCD thermodynamical potential to two-loop order (Section \ref{Pressure}). Remarks and perspectives of this work in progress are summarized in the end.

\section{Exact Fermion propagator in a magnetic background\label{MagProp}}

The most crucial nontrivial ingredient of our approach to QCD in an intense magnetic background is the nonperturbative description of the propagator of the fermions dressed by magnetic effects. This propagator corresponds to the Green's function associated with the Dirac equation in the presence of a classical electromagnetic potential $A_{\rm cl}^{\mu}(x)=(0,\vec{A}(x))$, such that the magnetic field is produced $\vec{B}=B\hat z=\nabla\times \vec{A}$:
\be
\big[i(\slashed{\partial}+iq\slashed{A}_{\rm cl}(x))-m_f\big]S_0(x,y)=(2\pi)^4\delta^4(x-y)
\,,
\ee
or, in general operator form, using $\Pi^{\mu}(\hat p,\hat x)=(\hat p^{\mu}-qA^{\mu}_{\rm cl}(\hat x))$ and $\langle x|S_0|y\rangle=S_0(x,y)$,
\be
\big[\slashed{\Pi}(\hat p,\hat x)-m_f\big]S_0=1\!\!1
\,,
\label{MagD-eq}
\ee
where it is clear that the presence of a space-dependent classical potential $A_{\rm cl}$ (essential for generating the constant and uniform magnetic field) renders the Dirac operator a mixture of momentum and coordinate operators. The matrices in both of these representations are therefore nondiagonal and infinite-dimensional, so that their diagonalization is an intricate task, as well as computing their determinant.

One may still formally write the solution of this equation as
\be
S_0&=&\frac{1}{\slashed{\Pi}(\hat p,\hat x)-m_f}=\frac{1}{[\slashed{\Pi}(\hat p,\hat x)]^2-m_f^2}\,[\slashed{\Pi}(\hat p,\hat x)+m_f]
\,.
\label{S0-form}
\ee
In practice, one needs to transform this formal solution into an explicit matrix element (e.g. in coordinate space: $\langle x|S_0|y\rangle=S_0(x,y)$) which can be used directly in diagrammatic computations. Originally, this task was accomplished by Schwinger \cite{Schwinger:1951nm}, who introduced the now called Schwinger formalism to find an {\it exact} explicit solution which we shall sketch in what follows.

First, it is convenient to re-write the inverse operator appearing in Eq. (\ref{S0-form}) as an integral of a unitary operator, getting the following Schwinger integral form for the propagator:
\be
S_0&=&i\int_0^{\infty} ds \exp\left\{
-is\big[\slashed{\Pi}(\hat p,\hat x)]^2-m_f^2\big]
\right\}
[\slashed{\Pi}(\hat p,\hat x)+m_f]
\,.
\label{S0-1}
\ee
We may therefore re-interpret the unitary operator $U(s)\equiv\exp[-is\big[\slashed{\Pi}(\hat p,\hat x)]^2]$ as an evolution operator in an auxiliary time $s$ governed by an auxiliary hamiltonian $\mathcal{H}\equiv \slashed{\Pi}(\hat p,\hat x)]^2$. Within this picture, the propagator becomes an average in the time $s$ of the operator $[\slashed{\Pi}(\hat p,\hat x)+m_f]$ evolved according to the auxiliary hamiltonian $\mathcal{H}$:
\be
S_0&=&i\int_0^{\infty} ds 
~{\rm e}^{ism_f^2}
~
U(s)[\slashed{\Pi}(\hat p,\hat x)+m_f]
\,,
\label{S0-2}
\ee
or, in coordinate space,
\be
S_0(x,y)&=&i\int_0^{\infty} ds 
~{\rm e}^{ism_f^2}
~
\langle x|U(s)[\slashed{\Pi}(\hat p,\hat x)+m_f]
|y\rangle
\\
&=&
i\int_0^{\infty} ds 
~{\rm e}^{ism_f^2}
~
\Big\{
\langle x=x(s)|\slashed{\Pi}(\hat p,\hat x)
|y=x(0)\rangle
+m_f  \langle x=x(s) |y=x(0)\rangle
\Big\}
\,.
\label{S0-3}
\ee
Solving the evolution in the auxiliary time $s$ is therefore equivalent to solving the operatorial Dirac equation in the presence of a classical external field $A_{\rm cl}^{\mu}(\hat x)$.

The detailed computation of the evolved matrix elements $\langle x=x(s) |y=x(0)\rangle$
and $\langle x=x(s)|\slashed{\Pi}(\hat p,\hat x)
|y=x(0)\rangle$ can be found in Ref. \cite{Schwinger:1951nm} and the results are in Eqs. (3.14) and (3.20) of this reference. They have the form:
\be
\langle x=x(s)|\slashed{\Pi}(\hat p,\hat x)
|y=x(0)\rangle
&=&f_1(x-y)\,\langle x=x(s) |y=x(0)\rangle
\,,
\nonumber\\
\langle x=x(s) |y=x(0)\rangle
&=&
f_2(x-y)\, \Phi(x,y)
\,,
\label{MatEle}
\ee
where $\Phi(x,y)$ is the only contribution that breaks translational invariance, being known as the Schwinger phase:
\be
\Phi(x,y)&\equiv&
\exp\left[
iq\int_x^ydx'_{\mu}A_{\rm cl}^{\mu}(x')
\right]
\,.
\ee
Ultimately, this phase is responsible for the non-invariant form of the  fermion propagator dressed by magnetic effects:
\be
S_0(x,y)&=&
\Phi(x,y)~\overline{S}_0(x-y)
\\
&=&
\Phi(x,y)~ 
i\int_0^{\infty} ds 
~{\rm e}^{ism_f^2}
~
f_2(x-y)\, 
~\big[
f_1(x-y)
+m_f
\big]
\,.
\label{S0-4}
\ee
Notice that this apparent breaking of translational invariance is not a physical property (all points in space-time are completely equivalent), since the propagator itself is not physical, being gauge dependent as well. In fact, this phase evaluates to $1$ in a specific nontrivial gauge choice \cite{Chodos:1990vv}.

Choosing the electromagnetic gauge as $A_{cl}=(0,0,Bx,0)$, the final form for the {\it exact} fermion propagator in a uniform and constant magnetic background ($\vec{B}=B~\hat z$)
is \cite{Schwinger:1951nm}
 (cf. also \cite{Chodos:1990vv,Gusynin:1995nb}):
\begin{eqnarray}
S_0(x,y)&=&{\rm exp}\left\{\frac{iq}{2}[x^{\mu}-y^{\mu}]A_{\mu}^{\rm ext}(x+y)\right\}
~\overline{S}_0(x-y)
\\
&=&
{\rm exp}\left\{\frac{iq}{2}[x^{\mu}-y^{\mu}]A_{\mu}^{\rm ext}(x+y)\right\}
~\int\frac{d^dP}{(2\pi)^d}
{\rm e}^{-iP\cdot(x-y)}
~\overline{S}_0(P)
\\
&=&
{\rm exp}\left\{\frac{iq}{2}[x^{\mu}-y^{\mu}]A_{\mu}^{\rm ext}(x+y)\right\}
~\int\frac{d^dP}{(2\pi)^d}
{\rm e}^{-iP\cdot(x-y)}
\times
\nonumber\\
&&\quad
\times
\int_0^{\infty}
ds~{\rm exp}\left[is\left({\bf p}_L^2-{\bf p}_T^2\frac{\tan(qBs)}{qBs}-m_f^2\right)\right]
\times
\nonumber\\
&&\quad\quad
\times
\Big\{
[{\bf p}_L\cdot\gamma_L+m_f][1\!\!1+\gamma^1\gamma^2\tan(qBS)]
-{\bf p}_T\cdot{\bf \gamma}_T[1+\tan^2(qBs)]
\Big\}
\,,
\nonumber\\
\label{S0-exact}
\end{eqnarray}
where we have used a compact notation for the transverse and longitudinal quantities: ${\bf p}_T=(p_1,p_2)$, ${\bf \gamma}_T=(\gamma^1,\gamma^2)$, ${\bf p}_L=(p_0,p_3)$, and ${\bf \gamma}_L=(\gamma^0,\gamma^3)$.

This propagator presents consistently some desired properties. First, the free Dirac propagator is recovered in the absence of magnetic fields. Indeed, this is obtained directly from $\lim_{B\to 0}\tan(qBs)=qBs\to0$ and the basic property of the Schwinger integral (with a regularization of the exponential implicitly assumed):
\be
\int_0^{\infty}
ds~{\rm exp}\left[isF\right]
&=&
\frac{i}{F}
\,.
\ee

Moreover, it can be shown that the poles in Eq. (\ref{S0-exact}) occur at (and only at) the energies for which the Landau level dispersion relation is satisfied: $p_0^2=m_f^2+p_3^2+2nB$ ($n=0,1,2,\cdots$). The demonstration \cite{Chodos:1990vv} (cf. also \cite{Gusynin:1995nb}) goes through rewriting the Schwinger integral into a sum over Landau levels. Using
($z\equiv [-\exp(-2iqBs)]$)
\be
i\tan(qBs)&=&
1-2\frac{z}{z-1}
\\
(1-z)^{-(\alpha+1)}
\exp\left[\frac{xz}{z-1}\right]
&=&
\sum_{n=0}^{\infty}L_n^{\alpha}(x)\,z^n
\,,\label{maracutaia}
\ee
where $L_n^{\alpha}(x)$ are the generalized Laguerre polynomials,
the exact result for the propagator, Eq. (\ref{S0-exact}), becomes a sum over Landau levels\footnote{
This is nevertheless a subtle transformation that relies on assuming a nonzero $B$ (so that, in the final expression, the free Dirac result cannot be recovered) and regularization of the Schwinger integral (and of the sums appearing in the relation in Eq. (\ref{maracutaia})).}:
\be
\overline{S}_0(P)
&=&
i \exp\left[-\,\frac{{\bf p}_T^2}{qB}\right]
\sum_{n=0}^{\infty}(-1)^n
\frac{D_n(qB,P)}{{\bf p}_L^2-m_f^2-2nqB}
\,,
\ee
with (defs.: $L_n\equiv L_n^0$ and $L_{-1}^{\alpha}=0$)
\be
D_n(qB,P)
&=&
({\bf k}_L\cdot\gamma_L+m)
\Big[
(1+i\gamma^1\gamma^2)L_n\left(2\frac{{\bf p}_T^2}{qB}\right)
-(1-i\gamma^1\gamma^2)L_{n-1}\left(2\frac{{\bf p}_T^2}{qB}\right)
\Big]
\nonumber\\&&
+4({\bf k}_T\cdot\gamma_T)L_{n-1}^1
\left(2\frac{{\bf p}_T^2}{qB}\right)
\,.
\label{S0-exact-LLsum}
\ee

\subsection{Lowest Landau Level approximation\label{LLL-sub}}

This expansion over Landau levels is particularly convenient when one is interested in phenomena at very intense magnetic fields, because large fields should guarantee the fast convergence of this series, allowing for approximations that take into account only the first levels.

When all mass and momentum scales are indeed much smaller than the considered magnetic field, one may in principle take the Lowest Landau Level (LLL) approximation\footnote{From the form of the propagator in Eq. (\ref{S0-exact-LLsum}), and the behavior of the Laguerre polynomials when their argument vanishes, the connection between the LLL approximation and the large $B$ limit becomes clear. This equivalence between the leading-order $1/B$ expansion and the LLL approximation is {\it not} always true (cf. the computation of the free pressure in Subsection \ref{FreePressure}).}, which is equivalent to substituting the sum over Landau levels for its value at $n=0$. Up to regularization differences, the LLL propagator reads:
\begin{eqnarray}
S_0^{\rm LLL}(x,y)
&=&
{\rm exp}\left\{\frac{iq}{2}[x^{\mu}-y^{\mu}]A_{\mu}^{\rm ext}(x+y)\right\}
\int\frac{d^dP}{(2\pi)^d}
{\rm e}^{-iP\cdot(x-y)}
i {\rm exp}\left(-\frac{{\bf p}_T^2}{|qB|}\right)
~\frac{1+i\gamma^1\gamma^2}{{\bf p}_L\cdot\gamma_L-m_f}
\,,
\nonumber\\
\end{eqnarray}
which is equivalent to the result used in Ref. \cite{Fukushima:2011nu}, obtained via a method of constructing the projectors for the different Landau levels from the exact solution of the Dirac equation \cite{Fukushima-projectors}:
\begin{eqnarray}
S_0^{\rm LLL}(x,y)
&=&
\int\frac{dp_0dp_2dp_3}{(2\pi)^{d-1}}
\sqrt{\frac{qB}{\pi}}
{\rm exp}\Big\{
-ip_0(x_0-y_0)
+ip_2(x_2-y_2)
+ip_3(x_3-y_3)
\Big\}
\times
\nonumber\\&&\quad
\times
{\rm exp}\Big\{
-\frac{eB}{2}\big[
\left(x_1-\frac{p_2}{qB}\right)^2
+\left(y_1-\frac{p_2}{qB}\right)^2
\big]
\Big\}
~\mathcal{P}_0
\frac{i}{{\bf p}_L\cdot\gamma_L-m_f}
\label{PropLLL-Fuk}
\end{eqnarray}
with $\mathcal{P}_0={\rm diag}(1,0,1,0)=(1\!\!1+i\gamma^1\gamma^2)/2$ being the projector over the spin states which are present physically in the LLL. The peculiarity of the LLL approximation should be noted: this propagator is still a $4\times 4$ matrix, but it describes only two physical propagating modes. Two eigenvalues of $S_{0}^{\rm LLL}$ go to zero in the vicinity of the $n=0$ Landau level pole ($p_0^2=m_f^2+p_3^2$), posing problems for the computation of the free pressure in this approximation (cf. Section \ref{FreePressure}).


\section{Extreme magnetic fields and gauge boson selfenergy\label{GluonFuk}}

As a first step towards understanding the quasiparticle content of a Strongly interacting system exposed to an intense constant and uniform magnetic field, let us compute the self-energy of the gluon corrected by one loop of dressed quarks in the vacuum.

This is a simple calculation that shall illustrate basic features of perturbation theory over the nonperturbatively dressed magnetic fermions at work, such as the effective reduction  to $(1+1)$ dimensions for very intense backgrounds.
In Ref. \cite{Fukushima:2011nu} the analogous QED case in the LLL approximation was considered and complemented by an approximate spectral construction to show the appearance of a peculiar mode of the corrected photon propagation, similar to the {\it zero sound} mode in Fermi liquid theory \cite{abrikosov-gorkov}.

The actual computation for the one-loop gluon selfenergy is very similar; the only difference resting on the extra color tensor structure in the vertex ($e\gamma^{\mu}$ in QED is mapped into $g\gamma^{\mu}\lambda^a$ in QCD, where $\lambda^a$ are the Gell-Mann matrices). Indeed, in the general situation, in which translational invariance is not assumed, the selfenergies for the photon $\Pi_{\rm photon}^{\mu\nu}(k,q)$ and the gluon $\Pi^{\mu\nu}(k,q)$ may be written as:
\be
i\Pi_{\rm photon}^{\mu\nu}(k,q)
&=&
\int d^4xd^4y 
{\rm e}^{iq\cdot x + ik\cdot y}
i\Pi^{\mu\nu}(x,y)
\nonumber\\ 
&=&
\int d^4xd^4y 
{\rm e}^{iq\cdot x + ik\cdot y}
e^2{\rm Tr}\left[
\gamma^{\mu}S_0(x,y)\gamma^{\nu}
S_0(y,x)
\right]\,,
\\
i\Pi^{\mu\nu}(k,k')
&=&
\int d^4xd^4y 
{\rm e}^{ik\cdot x + ik'\cdot y}
i\Pi^{\mu\nu}(x,y)
\label{Pidef0}
\\ 
&=&
\int d^4xd^4y 
{\rm e}^{ik\cdot x + ik'\cdot y}
g^2{\rm Tr}\left[
\gamma^{\mu}\lambda^aS_0(x,y)\gamma^{\nu}\lambda^a
S_0(y,x)
\right]
\label{Pidef}
\,,
\ee
so that the $\Pi_{\rm photon}$ maps into the gluon selfenergy if the following substitution is implemented\footnote{In this section, we shall consider the contribution to the selfenergy coming from only one fermion with electric charge $e$ in both QED and QCD cases. Nevertheless, in the full QCD gluon selfenergy quarks carry color and flavor indices and their charge is flavor-dependent, so that $\Pi^{\mu\nu}_{\rm QCD}=N_c\sum_f\Pi^{\mu\nu}(q_f)$.}:
$e^2\mapsto g^2 {\rm Tr}_{\rm color}\lambda^a\lambda^a=g^2 (N_c^2-1)/2$. Therefore, the presentation of this section follows closely that of Ref. \cite{Fukushima:2011nu}.

In accordance with the framework adopted,
 Eq. (\ref{Pidef}) is exactly the perturbative expression for the gluon selfenergy at one loop, but with the fermion propagator being dressed by (nonperturbative) magnetic effects.

As discussed above, we adopt the LLL approximation, valid for extremely intense fields, much larger than any other mass or momentum scale in the problem.
Taking the LLL quark propagator, Eq. (\ref{PropLLL-Fuk}), into the definition of the gluon selfenergy (Eq. (\ref{Pidef})), one gets:
\be
i\Pi^{\mu\nu}(k,k')
&=&
-g^2\frac{N_c^2-1}{2}
\int\frac{dp_0dq_0dp_2dq_2dp_3dq_3}{(2\pi)^{6}}
\nonumber\\&&
\phi_{xy}
{\rm Tr}\Big[
\gamma^{\mu}~\mathcal{P}_0
\frac{1}{{\bf p}_L\cdot\gamma_L-m_f}~
\gamma^{\nu}
~\mathcal{P}_0
\frac{1}{{\bf q}_L\cdot\gamma_L-m_f}
\Big]
\label{Pi-0}
\,,
\ee
where the phase $\phi_{xy}$ carries the $x$ and $y$-integrals from the Fourier transform,
\be
\phi_{xy}&\equiv&
\int d^4xd^4y 
{\rm e}^{ik\cdot x + ik'\cdot y}
{\rm exp}\Big\{
-i(p_0-q_0)(x_0-y_0)
+i(p_2-q_2)(x_2-y_2)
+i(p_3-q_3)(x_3-y_3)
\Big\}
\nonumber\\&&
\frac{eB}{\pi}{\rm exp}\Big\{
-\frac{eB}{2}\big[
\left(x_1-\frac{p_2}{eB}\right)^2
+\left(y_1-\frac{p_2}{eB}\right)^2
+\left(x_1-\frac{q_2}{eB}\right)^2
+\left(y_1-\frac{q_2}{eB}\right)^2
\big]
\Big\}
\,,\label{defphi}
\ee
which is solved in Appendix \ref{apIntMag}, yielding Eq. (\ref{resphixy})). The delta functions from this phase are then used to perform the $q_i$ ($i=0,2,3$) integrals, furnishing momentum conservation for the external legs in directions $i=0,2,3$:
\be
i\Pi^{\mu\nu}(k,k')
&=&
-g^2\frac{N_c^2-1}{2}
(2\pi)^3\delta^{(0,2,3)}(k'+k)
\int\frac{dp_0dp_2dp_3}{(2\pi)^3}
\nonumber\\&&
{\rm exp}\Big\{
-\frac{1}{4eB}
\left[
{k'}_1^2+k_1^2+2i(2p_2+k_2)(k'_1+k_1)+2(k_2)^2
\right]
\Big\}
\nonumber\\&&
{\rm Tr}\Big[
\gamma^{\mu}~\mathcal{P}_0
\frac{1}{{\bf p}_L\cdot\gamma_L-m_f}~
\gamma^{\nu}
~\mathcal{P}_0
\frac{1}{({\bf p}_L+{\bf k}_L)\cdot\gamma_L-m_f}
\Big]
\label{Pi-1}
\,,
\ee
The last integral that does not involve the trace in Eq. (\ref{Pi-1}) is then the integral over $p_2$:
\be
\int \frac{dp_2}{2\pi}
{\rm exp}\Big\{
-\frac{1}{4eB}
\left[
2i(2p_2+k_2)(k'_1+k_1)
\right]
\Big\}
&=&
eB\delta(k'_1+k_1)
\,,
\ee
where we have absorbed $k_2$ via a shift before integrating.

Finally, the expression for the gluon selfenergy showing explicitly the momentum conservation of the external legs reads:
\be
i\Pi^{\mu\nu}(k,k')
&=&
(2\pi)^4\delta^{4}(k'+k)
\frac{eB}{2\pi}
{\rm exp}\Big\{
-\frac{k_1^2+k_2^2}{2eB}
\Big\}
(-g^2)\frac{N_c^2-1}{2}
\int\frac{dp_0dp_3}{(2\pi)^2}
\nonumber\\&&
{\rm Tr}\Big[
\gamma^{\mu}~\mathcal{P}_0
\frac{1}{{\bf p}_L\cdot\gamma_L-m_f}~
\gamma^{\nu}
~\mathcal{P}_0
\frac{1}{({\bf p}_L+{\bf k}_L)\cdot\gamma_L-m_f}
\Big]
\label{Pi-2}
\,.
\ee

It is interesting to note that the effect of an extremely intense magnetic field on the gluon (or photon) propagation is two-fold: (i) the $B$ dependence is restricted to an overall decaying exponential factor on the transverse momentum and (ii) the nontrivial dynamical contribution is exactly the vacuum gluon selfenergy in $(1+1)$ dimensions. 

To verify this latter property it is necessary to examine further the projected trace appearing above, i.e.
\be
T_4&\equiv&
{\rm Tr}\Big[
\gamma^{\mu}~\mathcal{P}_0
\frac{1}{{\bf p}_L\cdot\gamma_L-m_f}~
\gamma^{\nu}
~\mathcal{P}_0
\frac{1}{({\bf p}_L+{\bf k}_L)\cdot\gamma_L-m_f}
\Big]
\,,
\ee
where $\mathcal{P}_0=(1\!\!1+i\gamma^1\gamma^2)/2$ is the projector over the degrees of freedom actually present in the LLL (only one spin state). Its properties $\gamma^{\mu}\mathcal{P}_0=\mathcal{P}_0\gamma^{\mu}\stackrel{\mu=1,2}{=}0$
guarantee for example that $\Pi^{\mu\nu}=0$ whenever $\mu=1,2$ or $\nu=1,2$. Thus, effectively the Dirac indices run only through $(1+1)$ dimensions ($\mu=0,3$). One can show that (cf. Appendix \ref{apIntMag}):
\be
{\rm Tr}\left[\mathcal{P}_0\mathcal{M}\right]
&=&\frac{1}{2}{\rm Tr}\left[\mathcal{M}\right]
\,,
\ee
for any given matrix $\mathcal{M}$ that is a linear combination of $1\!\!1,\gamma^{\mu}$ and $\gamma^{\mu}\gamma^{\nu}$, with $\mu,\nu\ne 1,2$. The trace $T_4$ entering the gluon selfenergy computation is therefore equivalent to a trace in dimension $\bar d=2$, in which the Dirac indices assume only two values (but the matrices are still $4\times4$):
\be
T_4&=&
\frac{1}{2}
{\rm Tr}_{\bar d=2}\Big[
\gamma^{\mu}~\mathcal{P}_0
\frac{1}{{\bf p}_L\cdot\gamma_L-m_f}~
\gamma^{\nu}
~\mathcal{P}_0
\frac{1}{({\bf p}_L+{\bf k}_L)\cdot\gamma_L-m_f}
\Big]
\,.
\ee
Using this result, the nonzero components of the one-loop gluon selfenergy in a magnetic background within the LLL approximation may be rewritten in terms of the vacuum gluon selfenergy in dimension $(1+1)$:
\be
i\Pi^{\mu\nu}(k,k')
&\stackrel{\mu,\nu\ne 1,2}{=}&
(2\pi)^4\delta^{4}(k'+k)
\frac{eB}{2\pi}
{\rm exp}\Big\{
-\frac{k_1^2+k_2^2}{2eB}
\Big\}
\frac{1}{2}
i\Pi_{\bar d =2}^{\mu\nu}(k_0,k_3)
\,,\label{Pi-3}
\ee
with
\be
\Pi_{\bar d =2}^{\mu\nu}(k_0,k_3)
=
ig^2\frac{N_c^2-1}{2}
\int\frac{dp_0dp_3}{(2\pi)^2}
{\rm Tr}_{\bar d =2}\Big[
\gamma^{\mu}
\frac{1}{{\bf p}_L\cdot\gamma_L-m_f}
\gamma^{\nu}
\frac{1}{({\bf p}_L+{\bf k}_L)\cdot\gamma_L-m_f}
\Big]
\label{Pid2}
\,,
\ee
or, solving the Dirac trace in the standard way \cite{Peskin:1995ev},
\be
i\Pi_{\bar d =2}^{\mu\nu}(k_0,k_3)
&=&
(-g^2)\frac{N_c^2-1}{2}
\int\frac{dp_0dp_3}{(2\pi)^2}
4
\Big\{
[m_f^2-{\bf p}_L\cdot({\bf p}_L+{\bf k}_L)]g^{\mu\nu}
+p^{\mu}(k+p)^{\nu}+
\nonumber\\
&&\quad+p^{\nu}(k+p)^{\mu}
\Big\}
\,.
\label{Pid2-2}
\ee

Notice that the theory in dimension 2 is in general much better behaved in the UV as compared to $d=4$, requiring no renormalization procedure. The integral above, however, is not convergent, but, as we shall see below, has no poles in $1/\epsilon$ in dimensional regularization, being thus of the form: $\int d^{\bar d}p\,p^n$. The absence of poles in dimensional regularization guarantees that no nontrivial subtraction is needed, so that the final result does not depend on a renormalization scale at all, as expected.

Let us now obtain an explicit form for $\Pi^{\mu\nu}(k,k')$, Eq. (\ref{Pi-3}).
The result for the one-loop vacuum selfenergy in $\bar d$ dimensions is of course well-known and can be obtained straightforwardly using dimensional regularization and Feynman parameters \cite{Peskin:1995ev}:
\be
i\Pi_{\bar d =2}^{\mu\nu}(k_0,k_3)
&=&
(-g^2)\frac{N_c^2-1}{2}
\int\frac{dp_0dp_3}{(2\pi)^2}
4
\Big\{
[m_f^2-{\bf p}_L\cdot({\bf p}_L+{\bf k}_L)]g^{\mu\nu}
+p^{\mu}(k+p)^{\nu}+
\nonumber\\
&&\quad+p^{\nu}(k+p)^{\mu}
\Big\}
\nonumber\\
&=&
(-g^2)\frac{N_c^2-1}{2}
\left(\frac{-i}{4\pi}\right)
\left(\frac{{\rm e}^{\gamma}\Lambda}{\Delta}\right)^{\frac{\epsilon}{2}}
\Gamma(1+\epsilon/2)
8\left\{
\left[g^{\mu\nu}-\frac{k^{\mu}k^{\nu}}{{\bf k}_L^2}\right]f\left(\frac{{\bf k}_L^2}{m_f^2}\right)
\right\}
\nonumber\\
&\stackrel{\epsilon\to 0}{=}&
g^2\frac{N_c^2-1}{2}
\left(\frac{2i}{\pi}\right)
\left\{
\left[g^{\mu\nu}-\frac{k^{\mu}k^{\nu}}{{\bf k}_L^2}\right]f\left(\frac{{\bf k}_L^2}{m_f^2}\right)
\right\}
\,,
\label{Pid2-3}
\ee
with
\be
f(x)
&\equiv&
\int_0^1 dy\frac{y(1-y)}{y(1-y)-x^{-1}}
\,.
\ee

Therefore, the final result for the gluon selfenergy corrected by a loop of fermions\footnote{In the case of quarks, the electric charge is flavor-dependent: $e\mapsto q_f$.} dressed by the magnetic field in the LLL approximation is:
\be
\Pi^{\mu\nu}(k,k')
&\stackrel{\mu,\nu= 1,2}{=}&0
\\
\Pi^{\mu\nu}(k,k')
&\stackrel{\mu,\nu\ne 1,2}{=}&
(2\pi)^4\delta^{4}(k'+k)
\frac{eB}{2\pi}
{\rm exp}\Big\{
-\frac{k_1^2+k_2^2}{2eB}
\Big\}
\frac{g^2}{\pi}\frac{N_c^2-1}{2}
\left[g^{\mu\nu}-\frac{k^{\mu}k^{\nu}}{{\bf k}_L^2}\right]f\left(\frac{{\bf k}_L^2}{m_f^2}\right)
\,,
\nonumber\\
\ee
reproducing the analogous photon result \cite{Fukushima:2011nu} for $g^2(N_c^2-1)/2\mapsto e^2$. 

As discussed in the QED context, the function $f(x)$ presents a divergence at the threshold for decay of the photon into an electron-positron pair and an imaginary part signaling this instability for the high momentum modes. Since the position of the threshold is of course dependent on the fermion mass $m_f$, the chiral limit is subtle and is not interchangeable with the zero-momentum limit. This feature may be problematic for the thermal version of this computation, since the integrals over $p_0$ and $\vec{p}$ are naturally separated by the discretization of the zeroth component into Matsubara frequencies. The computation of the thermal one-loop gluon selfenergy in the presence of an intense magnetic background is work in progress.

\section{QCD pressure\label{Pressure}}

After the experience acquired in the computation of the gluon selfenergy in our framework of perturbative QCD in a nonperturbative magnetic background, let us address the thermodynamics. It is interesting to recall that these results may be directly compared to high-quality lattice QCD simulations with a magnetic field that are currently underway, with some results already reported in Ref. \cite{Bali:2011qj}.

Once again, we shall use perturbative QCD with the dressed fermion propagator from Section \ref{MagProp}. 
The standard diagrammatic expansion for the thermodynamic potential of QCD reads:
\vspace{0.2cm}
\begin{fmffile}{fmftese}
\begin{eqnarray}
\Omega_{QCD}&\equiv& -~\frac{1}{\beta V}~\ln Z_{QCD}
\nonumber \\ \nonumber \\
&=& -~\frac{1}{\beta V}~~
\parbox{10mm}{
\begin{fmfgraph*}(35,35)\fmfkeep{bolhagluon}
\fmfpen{0.8thick}
\fmfleft{i} \fmfright{o}
\fmf{gluon,left,tension=.08}{i,o}
\fmf{gluon,left,tension=.08}{o,i}
\end{fmfgraph*}}
\quad
+~\frac{1}{\beta V}~~
\parbox{10mm}{
\begin{fmfgraph*}(35,35)\fmfkeep{bolhaghost}
\fmfpen{0.8thick}
\fmfleft{i} \fmfright{o}
\fmf{dashes,left,tension=.08}{i,o}
\fmf{dashes,left,tension=.08}{o,i}
\end{fmfgraph*}}
\quad
+\frac{1}{\beta V}~\sum_{f}~
\parbox{10mm}{
\begin{fmfgraph*}(35,35)\fmfkeep{bolhaquark}
\fmfpen{0.8thick}
\fmfleft{i} \fmfright{o}
\fmf{fermion,left,tension=.08,label=$\noexpand\psi_f$}{i,o}
\fmf{fermion,left,tension=.08}{o,i}
\end{fmfgraph*}}
\quad
+\nonumber \\ \nonumber \\ \nonumber\\
&&
+ ~\frac{1}{2}~\frac{1}{\beta V}~\sum_{f}~~
\parbox{10mm}{
\begin{fmfgraph*}(35,35)\fmfkeep{exchange}
\fmfpen{0.8thick}
\fmfleft{i} \fmfright{o}
\fmf{fermion,left,tension=.08,label=$\noexpand\psi_f$}{i,o,i}
\fmf{gluon}{i,o}
\fmfdot{i,o}
\end{fmfgraph*}}
~~~
+ ~\frac{1}{2}~\frac{1}{\beta V}~~
\parbox{10mm}{
\begin{fmfgraph*}(35,35)\fmfkeep{exchange-ghost}
\fmfpen{0.8thick}
\fmfleft{i} \fmfright{o}
\fmf{dashes,left,tension=.08}{i,o,i}
\fmf{gluon}{i,o}
\fmfdot{i,o}
\end{fmfgraph*}}
~~~
- ~\frac{1}{2}~\frac{1}{\beta V}\frac{1}{6}~~
\parbox{10mm}{
\begin{fmfgraph*}(35,35)\fmfkeep{exchange-3g}
\fmfpen{0.8thick}
\fmfleft{i} \fmfright{o}
\fmf{gluon,left,tension=.08}{i,o,i}
\fmf{gluon}{i,o}
\fmfdot{i,o}
\end{fmfgraph*}}
~~~
- 
\nonumber \\ \nonumber\\ \nonumber \\
&& -\frac{1}{2}~\frac{1}{\beta V}\frac{1}{8}~~~~~
\parbox{10mm}{
\begin{fmfgraph*}(35,35)\fmfkeep{db-4g}
\fmfpen{0.8thick}
\fmfbottom{i} \fmftop{o}
\fmf{phantom}{i,v}
\fmf{gluon,tension=.5}{v,v}
\fmf{phantom}{v,o}
\fmf{gluon,left=90,tension=.5}{v,v}
\fmfdot{v}
\end{fmfgraph*}}
~~~ +
\nonumber \\ \nonumber \\
&& +~ [diagrams ~with ~counterterms]~
+~O(3~loops) , \label{OmegaY}
\end{eqnarray}
\vspace{-0.4cm}

\noindent where full lines are fermions, dressed by the magnetic field, curly lines are gluons and dashed lines represent ghosts that will cancel the spurious degrees of freedom in the gluonic pressure.

To investigate the role of color interactions, our computation will go up to two-loop order.
In this approximation, the thermodynamic potential may be separated into a fermionic part ($\Omega_{QCD}^{F}(B)$), that carries all the magnetic-field dependence, and a pure glue contribution
($\Omega_{QCD}^{G}$) in the following way:
\be
\Omega_{QCD} &=& \Omega_{QCD}^{F}(B)+\Omega_{QCD}^{G}+~ [diagrams ~with ~counterterms]~
+O(3~loops) 
 \, ,
\ee
where
\be
\Omega_{QCD}^{F}(B) &=&
\frac{1}{\beta V}~\sum_{f}~
\parbox{10mm}{
\fmfreuse{bolhaquark}
}
\quad
+ ~\frac{1}{2}~\frac{1}{\beta V}~\sum_{f}~~
\parbox{10mm}{
\fmfreuse{exchange}
}
~~~
\, .\label{OmegaQCDF}
\ee

\be
\Omega_{QCD}^{G} &=&
 -~\frac{1}{\beta V}~~
\parbox{10mm}{
\fmfreuse{bolhagluon}
}
\quad
+~\frac{1}{\beta V}~~
\parbox{10mm}{
\fmfreuse{bolhaghost}
}
~~~
+ ~\frac{1}{2}~\frac{1}{\beta V}~~
\parbox{10mm}{
\fmfreuse{exchange-ghost}
}
~~~
- ~\frac{1}{2}~\frac{1}{\beta V}\frac{1}{6}~~
\parbox{10mm}{
\fmfreuse{exchange-3g}
}
~~~
- 
\nonumber \\ \nonumber\\ 
&& -~\frac{1}{2}~\frac{1}{\beta V}\frac{1}{8}~~~~~
\parbox{10mm}{
\fmfreuse{db-4g}
}
~~~
\, . \label{OmegaQCDG}
\ee

The gluonic part is equivalent to the usual hot perturbative QCD result and is, therefore, well-known (cf. Appendix \ref{apPureGlue} for references and details on the collection of renormalized results):
\be
\Omega_{QCD}^G
&=&
-2(N_c^2-1)\frac{\pi^2T^4}{90}
+(N_c^2-1)N_c~g^2T^4~\frac{1}{144}
\, .
\label{OmegaGQCD}
\ee

In the presence of a constant and uniform magnetic field, however, one needs to compute the nontrivial contribution $\Omega_{QCD}^{F}(B)$, which contains two terms: $O(g^0)$ and $O(g^2)$.
 
\subsection{Pressure of free magnetically-dressed quarks\label{FreePressure}}

The first diagram in Eq. (\ref{OmegaQCDF}) corresponds to the free Fermi gas pressure in the presence of the magnetic background. In the free theory, the action becomes quadratic, so that the partition function can in principle be solved exactly, in terms of the determinant of the inverse propagator which in this case is the exact fermion propagator including the magnetic dressing. The computation of this determinant, however, turns out to be prohibitively complicated due to the absence of translational invariance of the propagator. 

For the LLL approximation, the problem is worsened by the presence of vanishing eigenvalues associated with the spurious degrees of freedom with spin anti-aligned with the magnetic field, which are physically absent from the LLL. It is crucial to realize, however, that the LLL approximation for the free gas pressure is not equivalent to the leading order of a large magnetic field expansion at all. As it shall become clear in what follows, for the zero-temperature, finite-$B$ term of the pressure, the LLL is the energy level which less contributes in the limit of large $B$; the result being dominated by high values of $n$. Nevertheless, the equivalence between the LLL approximation and the large $B$ limit remains valid for the temperature dependent part of the free pressure.

This free contribution to the fermionic pressure has been considered in different contexts (usually, in effective field theories \cite{Fraga:2008qn,Boomsma:2009yk,Ebert:2003yk,Menezes:2008qt,Andersen:2011ip}) and computed from the direct knowledge of the energy levels of the system, i.e. the Landau levels\footnote{It is equivalent to describe the Landau levels in terms of a orbital number $l=0,1,2,\cdots$ and the spin quantum number $s=\pm 1$ as $E^2(l,s,p_3)=p_3^2+m_f^2+q_fB(2l+s+1)$. In this representation the degeneracy is the same for every state $\eta_{l,s}=q_fB/(2\pi)$.}
\be
E^2(n,p_3)=p_3^2+m_f^2+2q_fBn\,,
\ee
and their degeneracies $q_fB/(2\pi)$ for $n=0$ and $q_fB/(\pi)$ for $n=1,2,\cdots$. The exact result, including all Landau levels, is:

\be
\parbox{10mm}{
\fmfreuse{bolhaquark}
}
\quad
&=&
-N_c V \sum_{n,f}\frac{q_fB}{\pi}(1-\delta_{n0}/2)
\int \frac{dp_3}{2\pi} \bigg\{ 
\beta\, E(n,p_3)
+\ln\left( 1+e^{-\beta[E(n,p_3)-\mu_f]} \right)
+\nonumber\\ &&\quad
+\ln\left( 1+e^{-\beta[E(n,p_3)+\mu_f]} \right)
\bigg\}
\, ,\label{bolhaquark}
\ee
where the first term is a clearly divergent zero-point energy term and the other two are particle and anti-particle finite-temperature contributions. Since $E(n,p_3)$ grows with $B$, the largest the $n$ labeling the Landau level considered the larger the zero-point energy term becomes, being minimal for the LLL. Thus, in the limit of large $B$, the LLL approximation is inadequate. The decaying exponential dependence of the finite-temperature term on the energies $E(n,p_3)$ guarantees, however, that the LLL dominates indeed the result for intense magnetic fields. 

To obtain a good approximation for the large $B$ limit of the free pressure, we choose to treat the full exact result and take the leading order of a $m_f^2/q_fB$ expansion in the final expression. Let us then discuss the treatment of the divergent zero-point term.
Despite being a zero-temperature contribution, the first term in Eq. (\ref{bolhaquark}) cannot be fully subtracted because it carries the modification to the pressure brought about by the magnetic dressing of the quarks. One needs therefore to address its renormalization. Using dimensional regularization and the zeta-function representation\footnote{Notice that this is also a type of regularization.} for the sums over Landau levels and subtracting the vacuum term in $(3+1)$ dimensions, one gets (def.: $x_f\equiv m_f^2/(2q_fB)$):
\be
-N_c \beta V \sum_{n,f}\frac{q_fB}{\pi}(1-\delta_{n0}/2)
\int \frac{dp_3}{2\pi}
\, E(n,p_3)
&=&
-\frac{N_c}{2\pi^2} \beta V\sum_f (q_fB)^2\Big[
\zeta'\left(-1,x_f\right)
+\nonumber\\&&
+\frac{1}{2}(x_f-x_f^2)\ln x_f+\frac{x_f^2}{4}-
\nonumber\\&&-
\frac{1}{12}\big(
2/\epsilon+\ln (\Lambda^2/2q_fB)+1\big)
\Big]
\,,
\ee
where a pole $\sim (q_fB)^2[2/\epsilon]$ still remains. This infinite contribution that survived the vacuum subtraction can be interpreted as a pure magnetic pressure coming from the artificial scenario adopted, with a constant and uniform $B$ field covering the whole universe. In this vein, one may neglect all terms $(\sim q_f B^2)$ and independent of masses and other couplings (as done, e.g. in Refs. \cite{Fraga:2008qn,Menezes:2008qt,Andersen:2011ip}), concentrating on the modification of the pressure of the QCD matter under investigation. 

The final exact result for the free pressure of magnetically dressed quarks is therefore (pressure $=-(1/\beta V)$[diagram]):
\be
\frac{P_{\rm free}^{F}}{N_c}&=&
\sum_f \frac{(q_fB)^2}{2\pi^2}\Big[
\zeta'\left(-1,x_f\right)-\zeta'\left(-1,0\right)
+\frac{1}{2}(x_f-x_f^2)\ln x_f+\frac{x_f^2}{4}
\Big]
\nonumber\\&
+&T \sum_{n,f}\frac{q_fB}{\pi}(1-\delta_{n0}/2)
\int \frac{dp_3}{2\pi} \bigg\{ 
\ln\left( 1+e^{-\beta[E(n,p_3)-\mu_f]} \right)
+\ln\left( 1+e^{-\beta[E(n,p_3)+\mu_f]} \right)
\bigg\}
\,.
\nonumber\\
\ee
In the limit of large magnetic field (i.e. $x_f=m_f^2/(2q_fB)\to 0$), one obtains:
\be
\frac{P_{\rm free}^{F}}{N_c}&\stackrel{{\rm large}~ B}{=}&
\sum_f \frac{(q_fB)^2}{2\pi^2}\Big[
x_f\ln \sqrt{x_f}
\Big]
\nonumber\\&
+&T \sum_{f}\frac{q_fB}{2\pi}
\int \frac{dp_3}{2\pi} \bigg\{ 
\ln\left( 1+e^{-\beta[E(0,p_3)-\mu_f]} \right)
+\ln\left( 1+e^{-\beta[E(0,p_3)+\mu_f]} \right)
\bigg\}
\,.
\nonumber\\
\ee

\begin{figure}[h!]
\center
\includegraphics[width=10cm]{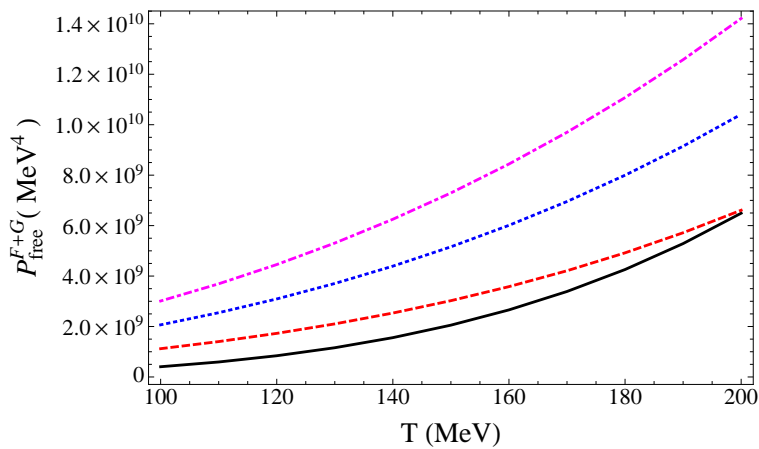}
\caption{
Free pressure of a gas of gluons and magnetically-dressed up and down quarks as a function of temperature. Different curves represent different magnetic background intensities:
$eB=\big\{0$ (black, solid, bottom), $20\,m_{\pi}^2,$ $40\,m_{\pi}^2,$ $60\,m_{\pi}^2$ (magenta, dash-dotted, top)$\big\}$, where $m_{\pi}=138~$MeV is the vacuum pion mass.}
\label{MagFreepressure}
\end{figure}

Adding the $O(g^0)$ piece of the gluonic contribution in Eq. (\ref{OmegaGQCD}), the free pressure of the quark-gluon system in the presence of an intense magnetic field reads:
\be
P_{\rm free}^{F+G}
&=&
2(N_c^2-1)\frac{\pi^2T^4}{90}
+\sum_f P_{\rm free}^{F}\left(T,x_f\right)
\,.
\ee
Figure \ref{MagFreepressure} illustrates the behavior of $P_{\rm free}^{F+G}(T,B)$ for two-flavor QCD with $m_{u}=m_d=5~$MeV.

%

\subsection{Turning on quark-gluon interaction: the exchange diagram in a magnetic field}

The second diagram in Eq. (\ref{OmegaQCDF}) is referred to as the ``exchange energy'' and corresponds here to the first nontrivial contribution that perturbs the decoupled system of gluons and free dressed quarks by turning on the $SU(N_c)$ gauge interaction. Its computation and analysis is our main goal in this on-going project.

In terms of the propagators in coordinate space (notice that ${\rm log}Z_{\rm exch}=(-1/2)[{\rm diagram}]$), the exchange diagram reads:

\begin{eqnarray}
\parbox{10mm}{
\fmfreuse{exchange}
}
\quad &=& 
\beta V~g^2~N_c(\lambda_a\lambda_a)~
\int\frac{d^dxd^dy}{\beta V}\int\frac{d^dK}{(2\pi)^d}
\frac{{\rm e}^{-iK\cdot(y-x)}}{K^2}
{\rm Tr}
\big[
\gamma_{\mu}S_0(x,y)\gamma^{\mu}S_0(y,x)
\big]
\, , \nonumber\\ \label{excRF}
\end{eqnarray}

\noindent where $\lambda_a$ are Gell-Mann matrices, with $\lambda_a \lambda_a=\frac{N_c^2-1}{2}$,
the trace $\textrm{Tr}$ acts over Dirac indices  and the 4-momentum is given in terms of the
Matsubara frequencies ($\omega_l^B=2l\pi T$) 
and of the 3-momentum as: $K=\left( k^0=i\omega_l^{B}\, ,\, {\bf k} \right)$.

Notice that if we assume translational invariance and the free Dirac propagator for the fermions (with $P=\left( p^0=i\omega_{n}^{F}+\mu_f \, ,\,  {\bf p}\right)$ and $\omega_n^F=(2n+1)\pi T$), 
\begin{eqnarray}
S_0(x,y)&=&
\int\frac{d^dP}{(2\pi)^d}
{\rm e}^{-iP\cdot(x-y)}
~S_0(P)
=\int\frac{d^dP}{(2\pi)^d}
\frac{{\rm e}^{-iP\cdot(x-y)}}{\slashed{P}-m_f}
\,,
\end{eqnarray}
this expression reduces to the usual one (cf. e.g. Eqs. (5.36) and (5.49) in Ref. \cite{kapusta-gale}):

\begin{eqnarray}
\parbox{10mm}{
\fmfreuse{exchange}
}
\quad &=& 
\beta V~g^2~\nu_c~
\sumint_{P_1,P_2,K} (2\pi)^3\beta 
~\delta^{(4)}(K-P_1+P_2)
\textrm{Tr}\left[ \gamma^{\mu}\frac{1}{(\slashed{P}_1-m_f)K^2}\gamma_{\mu}\frac{1}{\slashed{P}_2-m_f} \right]
\, , \nonumber\\ \label{excRFB=0}
\end{eqnarray}
where we have defined $\nu_c\equiv N_c(N_c^2-1)/2$.

In the presence of a uniform and constant magnetic background (${\bf B}=B~\hat z$), however, the fermion propagator becomes dependent on $x$ and $y$ in a nontrivial way due to the Schwinger phase, as discussed in Section \ref{MagProp}.

Introducing the LLL form of the fermionic propagators, Eq. (\ref{PropLLL-Fuk}), in the exchange contribution, Eq. (\ref{excRF}), we obtain:

\begin{eqnarray}
\parbox{10mm}{
\fmfreuse{exchange}
}^{\quad\rm LLL}
\,&=& 
\beta V~g^2~\nu_c~
\int\frac{d^dxd^dy}{\beta V}\int\frac{d^dK}{(2\pi)^d}
\frac{{\rm e}^{-iK\cdot(y-x)}}{K^2}
\frac{q_fB}{\pi}
\int\frac{dp_0dp_2dp_3}{(2\pi)^{d-1}}
\int\frac{dq_0dq_2dq_3}{(2\pi)^{d-1}}
\nonumber\\&&
{\rm exp}\Big\{
-i(p_0-q_0)(x_0-y_0)
+i(p_2-q_2)(x_2-y_2)
+i(p_3-q_3)(x_3-y_3)
\Big\}
\nonumber\\&&
{\rm exp}\Big\{
-\frac{q_fB}{2}\big[
\left(x_1-\frac{p_2}{q_fB}\right)^2
+\left(x_1-\frac{q_2}{q_fB}\right)^2
+\left(y_1-\frac{p_2}{q_fB}\right)^2
+\left(y_1-\frac{q_2}{q_fB}\right)^2
\big]
\Big\}
\nonumber\\&&
{\rm Tr}
\Bigg[
\gamma_{\mu}
\frac{\mathcal{P}_0}{{\bf p}_L\cdot\gamma_L-m_f}
\gamma^{\mu}
\frac{\mathcal{P}_0}{{\bf q}_L\cdot\gamma_L-m_f}
\Bigg]
\, .  \label{excRF-LLL}
\end{eqnarray}
%
%

\vspace{0.5cm}

As argued in Appendix \ref{apIntMag}, from the properties of Dirac gamma matrices and the definition of the projector $\mathcal{P}_0$, one concludes that the trace appearing in Eq.(\ref{excRF-LLL}) is essentially occurring in a (1+1)-dimensional subspace ($\mu=0,3$). Using $\bar d=2$, Eqs. (\ref{P0Dirac}) and (\ref{TrP0}) and the standard relations for Dirac traces ($\textrm{Tr}[(\gamma_{\mu})^n]=0$, with $n$ odd; $g_{~\mu}^{\mu}=\bar d$; 
 $\textrm{Tr}[\gamma_{\mu}\gamma_{\nu}]=4~g_{\mu\nu}$), we get (defining the 2-dimensional slash as: $\slashed{P}\equiv{\bf p}_L\cdot\gamma_L$):
\begin{eqnarray}
\textrm{Tr}\left[ \gamma^{\mu}\mathcal{P}_0(\slashed{P}+m_f)\gamma_{\mu}\mathcal{P}_0(\slashed{Q}+m_f) \right]
&=&
\textrm{Tr}\left[\mathcal{P}_0 \gamma^{\mu}(\slashed{P}+m_f)\gamma_{\mu}(\slashed{Q}+m_f) \right]
\nonumber \\
&=&
 \frac{1}{2}\textrm{Tr}_{\bar d=2}\left[ \gamma^{\mu}(\slashed{P}+m_f)\gamma_{\mu}(\slashed{Q}+m_f) \right]
\nonumber \\
&=&
2\left( 2-\bar d\right)~{\bf p}_L\cdot {\bf q}_L+2\bar d~m_f^2
\, ,
\end{eqnarray}
so that the exchange contribution for the LLL becomes


%
\begin{eqnarray}
\parbox{10mm}{
\fmfreuse{exchange}
}^{\quad\rm LLL}
\quad &=& 
\beta V~g^2~\nu_c~
\int\frac{d^dK}{(2\pi)^d}\int\frac{d^{\bar d}Pd^{\bar d}Q}{(2\pi)^{2\bar d}}
\frac{2\left( 2-\bar d\right)~{\bf p}_L\cdot {\bf q}_L+2\bar d~m_f^2}{K^2[{\bf p}_L^2-m_f^2][{\bf q}_L^2-m_f^2]}
\frac{q_fB}{\pi}
\int\frac{dp_2dq_2}{(2\pi)^2}
\nonumber\\
&&
\int\frac{d^dxd^dy}{\beta V}
{\rm e}^{-iK\cdot(y-x)
-i(p_0-q_0)(x_0-y_0)
+i(p_2-q_2)(x_2-y_2)
+i(p_3-q_3)(x_3-y_3)
}
\nonumber\\
&&
{\rm exp}\Big\{
-\frac{q_fB}{2}\big[
\left(x_1-\frac{p_2}{q_fB}\right)^2
+\left(x_1-\frac{q_2}{q_fB}\right)^2
+\nonumber\\
&&\quad\quad\quad
+\left(y_1-\frac{p_2}{q_fB}\right)^2
+\left(y_1-\frac{q_2}{q_fB}\right)^2
\big]
\Big\}
\, . \label{excRF-LLL-1}
\end{eqnarray}

The integrals over $x$ and $y$, in the second line above, give (def.: $\beta V\equiv \beta L_1L_2L_3$):
\begin{eqnarray}
&&\int\frac{d^dxd^dy}{\beta V}
{\rm e}^{-iK\cdot(y-x)
-i(p_0-q_0)(x_0-y_0)
+i(p_2-q_2)(x_2-y_2)
+i(p_3-q_3)(x_3-y_3)
}=
\int \frac{dx_1dy_1}{\beta V}
{\rm e}^{ik_1(y_1-x_1)}\times
\nonumber\\
&&\times
\int dx_0dy_0dx_2dy_2dx_3dy_3
{\rm e}^{
-i(p_0-q_0-k_0)(x_0-y_0)
+i(p_2-q_2-k_2)(x_2-y_2)
+i(p_3-q_3-k_3)(x_3-y_3)
}
\nonumber\\
&&=\frac{\beta V}{L_1}
(2\pi)\delta(p_0-q_0-k_0)(2\pi)\delta(p_2-q_2-k_2)
(2\pi)\delta(p_3-q_3-k_3)\int \frac{dx_1dy_1}{\beta V}
{\rm e}^{ik_1(y_1-x_1)}
\nonumber\\
&&=\frac{\beta V}{L_1}
(2\pi)^{\bar d}\delta({\bf p}_L-{\bf q}_L-{\bf k}_L)(2\pi)\delta(p_2-q_2-k_2)
\int\frac{dx_1dy_1}{\beta V}
{\rm e}^{ik_1(y_1-x_1)}
\,.
\end{eqnarray}

We are then left with:


\begin{eqnarray}
\parbox{10mm}{
\fmfreuse{exchange}
}^{\quad\rm LLL}
&=& 
\beta Vg^2\nu_c
\int\frac{d^dK}{(2\pi)^d}\int\frac{d^{\bar d}Pd^{\bar d}Q}{(2\pi)^{2\bar d}}
\frac{2\left( 2-\bar d\right)~{\bf p}_L\cdot {\bf q}_L+2\bar d~m_f^2}{K^2[{\bf p}_L^2-m_f^2][{\bf q}_L^2-m_f^2]}
(2\pi)^{\bar d}\delta({\bf p}_L-{\bf q}_L-{\bf k}_L)
\nonumber\\
&&\!\!\!\!\!\!
\frac{\beta V}{L_1}
\int\frac{dx_1dy_1}{\beta V}
{\rm e}^{ik_1(y_1-x_1)}
\frac{q_fB}{\pi}
\int\frac{dp_2dq_2}{(2\pi)^2}
(2\pi)\delta(p_2-q_2-k_2)
\nonumber\\
&&\!\!\!\!\!\!
{\rm exp}\Big\{
-\frac{q_fB}{2}\big[
\left(x_1-\frac{p_2}{q_fB}\right)^2
+\left(x_1-\frac{q_2}{q_fB}\right)^2
+\left(y_1-\frac{p_2}{q_fB}\right)^2
+\left(y_1-\frac{q_2}{q_fB}\right)^2
\big]
\Big\}
\nonumber\\
&=&
\beta Vg^2\nu_c
\int\frac{dk_1dk_2}{(2\pi)^2}\nonumber\\
&&\!\!\!\!\!\!
\int\frac{d^{\bar d}Kd^{\bar d}Pd^{\bar d}Q}{(2\pi)^{3\bar d}}
\frac{2\left( 2-\bar d\right)~{\bf p}_L\cdot {\bf q}_L+2\bar d~m_f^2}{K^2[{\bf p}_L^2-m_f^2][{\bf q}_L^2-m_f^2]}
(2\pi)^{\bar d}\delta({\bf p}_L-{\bf q}_L-{\bf k}_L)
\label{line2}\\
&&\!\!\!\!\!\!
\int\frac{dx_1dy_1}{L_1}
{\rm e}^{ik_1(y_1-x_1)}
\frac{q_fB}{\pi}
\int\frac{dp_2dq_2}{(2\pi)^2}
(2\pi)\delta(p_2-q_2-k_2)
\nonumber\\
&&\!\!\!\!\!\!
{\rm exp}\Big\{
-\frac{q_fB}{2}\big[
\left(x_1-\frac{p_2}{q_fB}\right)^2
+\left(x_1-\frac{q_2}{q_fB}\right)^2
+\left(y_1-\frac{p_2}{q_fB}\right)^2
+\left(y_1-\frac{q_2}{q_fB}\right)^2
\big]
\Big\}
\, . \nonumber\\ \label{excRF-LLL-2}
\end{eqnarray}

\noindent Defining the function that comes from the integrations related to the transverse directions as
\begin{eqnarray}
I(k_1,k_2,q_fB)
&\equiv&
\int\frac{dx_1dy_1}{L_1}
{\rm e}^{ik_1(y_1-x_1)}
\frac{q_fB}{\pi}
\int\frac{dp_2dq_2}{(2\pi)^2}
(2\pi)\delta(p_2-q_2-k_2)
\nonumber\\
&\times&
{\rm exp}\Big\{
-\frac{q_fB}{2}\big[
\left(x_1-\frac{p_2}{q_fB}\right)^2
+\left(x_1-\frac{q_2}{q_fB}\right)^2
+\left(y_1-\frac{p_2}{q_fB}\right)^2
+\left(y_1-\frac{q_2}{q_fB}\right)^2
\big]
\Big\}\nonumber\\
\label{Ik}
\end{eqnarray}
and realizing that the line labeled as Eq.(\ref{line2}) is nothing but the 2-dimensional result for the exchange contribution including a ``massive'' gluon ($K^2={\bf k}_L^2-m_k^2$; $m_k^2=k_1^2+k_2^2$), we have:

\begin{eqnarray}
\parbox{10mm}{
\fmfreuse{exchange}
}^{\quad\rm LLL}
 &=& 
\int\frac{dk_1dk_2}{(2\pi)^2}~I(k_1,k_2,q_fB)~
\mathcal{G}\left(k_1^2+k_2^2,m_f^2\right)
\, . \nonumber\\ \label{excRF-LLL-3}
\end{eqnarray}
with $\mathcal{G}\left(m_k^2,m_f^2\right)$ being the 2-dimensional exchange diagram:
\begin{eqnarray}
\mathcal{G}\left(m_k^2,m_f^2\right)
 &=& 
\parbox{10mm}{
\fmfreuse{exchange}
}^{\quad  \bar d=2;\; m_k^2=k_1^2+k_2^2}
\nonumber\\
&&
\nonumber\\
&&
\nonumber\\
\!\!\!\!\!\!\!
&&\hspace{-2cm}=
\beta Vg^2\nu_c
\int\frac{d^{\bar d}Kd^{\bar d}Pd^{\bar d}Q}{(2\pi)^{3\bar d}}
\frac{2\left( 2-\bar d\right)~{\bf p}_L\cdot {\bf q}_L+2\bar d~m_f^2}{[{\bf k}_L^2-m_k^2][{\bf p}_L^2-m_f^2][{\bf q}_L^2-m_f^2]}
(2\pi)^{\bar d}\delta({\bf p}_L-{\bf q}_L-{\bf k}_L)
\, . \nonumber\\ \label{excRF-d2}
\end{eqnarray}

Finally, taking the result in Eq. (\ref{Ik-3}) for the integral over the transverse directions $I(k_1,k_2,q_fB)$ (Eq. (\ref{Ik})) obtained in Appendix \ref{apIntMag},
\begin{eqnarray}
I(k_1,k_2,q_fB)
&=&
\frac{q_fB}{2\pi}
{\rm e}^{
-\frac{k_1^2+k_2^2}{2q_fB}}
\,.
\label{Ik-3}
\end{eqnarray}
into Eq. (\ref{excRF-LLL-3}) for the exchange contribution to the QCD pressure, one obtains:

\begin{eqnarray}
\parbox{10mm}{
\fmfreuse{exchange}
}^{\quad\rm LLL}
 &=& 
\left(
\frac{q_fB}{2\pi}\right)
\int\frac{dk_1dk_2}{(2\pi)^2}~{\rm e}^{
-\frac{k_1^2+k_2^2}{2q_fB}}~
\mathcal{G}\left(k_1^2+k_2^2,m_f^2\right)
\\
\nonumber\\
 &=& 
\left(
\frac{q_fB}{2\pi}\right)
\int\frac{dk_1dk_2}{(2\pi)^2}~{\rm e}^{
-\frac{k_1^2+k_2^2}{2q_fB}}\quad
\parbox{10mm}{
\fmfreuse{exchange}
}^{\quad  \bar d=2;\; m_k^2=k_1^2+k_2^2}
\, .
 \label{excRF-LLL-4}
\end{eqnarray}

\vspace{.5cm}

This expression realizes concretely the intuitive expectation that the nontrivial dynamics in an extremely intense magnetic field should be restricted to $(1+1)$ dimensions. Since the gluons do not couple directly to the magnetic field, their dispersion relation maintains its $(3+1)$-dimensional character ($\omega^2=k_1^2+k_2^2+k_3^2$), what effectively results in a ``massive'' gluon ($m_k^2=k_1^2+k_2^2$) in the reduced $(1+1)$-dimensional diagram. In the end the exchange contribution to the QCD pressure in the lowest-Landau level approximation for the fermion propagation is essentially an average over the effective gluon transverse mass $m_k^2=k_1^2+k_2^2$ of the exchange diagram in $(1+1)$-dimensions with the Gaussian weight $(q_fB/2\pi) \exp[-m_k^2/2q_fB]$.

\newpage

Let us now compute the 2-dimensional massive exchange diagram. Departing from Eq. (\ref{excRF-d2}),
\begin{eqnarray}
\mathcal{G}\left(m_k^2,m_f^2\right)
 &=& 
\beta Vg^2\nu_c
\int\frac{d^{\bar d}Kd^{\bar d}Pd^{\bar d}Q}{(2\pi)^{2\bar d}}
\delta({\bf p}_L-{\bf q}_L-{\bf k}_L)
\frac{2\left( 2-\bar d\right)~{\bf p}_L\cdot {\bf q}_L+2\bar d~m_f^2}{[{\bf k}_L^2-m_k^2][{\bf p}_L^2-m_f^2][{\bf q}_L^2-m_f^2]}
\nonumber
\\
&=&
\beta Vg^2\nu_c
\int\frac{d^{\bar d-1}Kd^{\bar d-1}Pd^{\bar d-1}Q}{(2\pi)^{3(\bar d-1)}}
(2\pi)^{\bar d-1}\delta({\bf p}_L-{\bf q}_L-{\bf k}_L)
\nonumber\\
&&
T\sum_l~T\sum_{n_ 1}~T\sum_{n_2}~
\beta  \delta_{n_1\, ,\, n_2+l}
\frac{2\left( 2-\bar d\right)~{\bf p}_L\cdot {\bf q}_L+2\bar d~m_f^2}{[{\bf k}_L^2-m_k^2][{\bf p}_L^2-m_f^2][{\bf q}_L^2-m_f^2]}
\, . \nonumber\\ \label{excRF-d2-1}
\end{eqnarray}

\noindent Apart from the tensorial structure, this diagram corresponds to the 2-dimensional version of the exchange diagram  of in-medium Yukawa theory with both fermions and bosons massive that was computed originally in Ref. \cite{Palhares:2008yq} (cf. also \cite{thesis}) and is also used in Chapter \ref{OPT}. In dimension $(1+1)$, however, UV divergences are much less severe, being absent in this case, and one expects renormalization to be trivial. Our concern here will be directed towards the IR domain, which should be subtle as proved to be the case for the selfenergy above. 

In principle the diagram computation in dimension 2 is fully convergent so that one may set $\bar d =2$:
\begin{eqnarray}
\mathcal{G}\left(m_k^2,m_f^2\right)
&=&
\beta Vg^2\nu_c
\int\frac{dk_zdp_zdq_z}{(2\pi)^{3}}
(2\pi)\delta(p_z-q_z-k_z)
\nonumber\\
&&
T\sum_l~T\sum_{n_ 1}~T\sum_{n_2}~
\beta  \delta_{n_1\, ,\, n_2+l}
\frac{4~m_f^2}{[{\bf k}_L^2-m_k^2][{\bf p}_L^2-m_f^2][{\bf q}_L^2-m_f^2]}
\, . \nonumber\\ \label{excRF-d2-2}
\end{eqnarray}
It is interesting to note from this result that in dimension 2 the exchange contribution to the QCD pressure is exactly zero in the chiral limit $m_f\to 0$. The magnetic QCD case is, however, more intricate and the extra $k_1,k_2$ integrals in Eq. (\ref{excRF-LLL-4}) could effectively produce a different result for the chiral limit or even bring IR divergences that should be treated carefully. It is not clear, for example, whether one should evaluate the Matsubara sums before the momentum integrations or not (as discussed in Ref. \cite{Fukushima:2011nu}).

Let us proceed in the usual way, evaluating the Matsubara sums first, via the standard contour integration procedure \cite{kapusta-gale}.

Using Kapusta's representation for the Kronecker Delta \cite{Kapusta:1979fh}\footnote{This substitution actually allows for the computation of the sums in any desired order, solving a subtle problem related to the absence of unicity of the analytical continuation of a function defined in a discrete domain \cite{Norton:1974bm}.}, 
\be
\beta \delta_{n_1\, ,\, n_2+l} &=& \int_0^{\beta}d\theta~ \textrm{e}^{\theta\left[
2\pi i T~(n_1-n_2-l)\right]}
=
\int_0^{\beta}d\theta~ \textrm{e}^{\theta\left[
p_1^0-p_2^0-k^0\right]}
\nonumber \\
&=& \frac{\textrm{e}^{\beta\left[
k^0+p_2^0-\mu_f \right]}-\textrm{e}^{\beta\left[
p_1^0-\mu_f \right]}}{p_1^0-p_2^0-k^0}
\, ,\label{RepDelta}
\ee
we can rewrite Eq. (\ref{excRF-d2-1}) as:

\begin{eqnarray}
\mathcal{G}\left(m_k^2,m_f^2\right)
 &=& 
\beta Vg^2\nu_c
\int\frac{d^{\bar d-1}Kd^{\bar d-1}Pd^{\bar d-1}Q}{(2\pi)^{3(\bar d-1)}}
(2\pi)^{\bar d-1}\delta({\bf p}_z-{\bf q}_z-{\bf k}_z)
~\mathcal{S}({\bf p}_z,{\bf q}_z,{\bf k}_z)
\, . \nonumber\\ \label{excRF-d2-3}
\end{eqnarray}
%
%
%
%
%
%
%
\noindent where
\be
\mathcal{S}({\bf p}_z,{\bf q}_z,{\bf k}_z)&=&
T\sum_l
~T\sum_{n_ 1}~T\sum_{n_2}~\frac{s_0(p_0,p_0,k_0)}{[{\bf k}_L^2-m_k^2][{\bf p}_L^2-m_f^2][{\bf q}_L^2-m_f^2]}
\\ \nonumber \\
s_0(p_0,q_0,k_0)&=& 
\left[2\left( 2-\bar d\right)~{\bf p}_L\cdot {\bf q}_L+2\bar d~m_f^2\right]
~\frac{\textrm{e}^{\beta\left[
k_0+q_0-\mu_f \right]}-\textrm{e}^{\beta\left[
p_0-\mu_f \right]}}{p_0-q_0-k_0}
\, . \label{s0}
\ee

\noindent Notice that $s_0(p_0,q_0,k_0)$ is an analytic function, even in the limit
$p_0 \to (q_0+k_0)$:
\be
\lim_{p_0 \to (q_0+k_0)} s_0(p_0,q_0,k_0)&\propto&
\textrm{e}^{\beta\left[
k_0+q_0-\mu_f \right]}~
\frac{1-[1+\beta(p_0-q_0-k_0)]}{p_0-q_0-k_0}
\propto\textrm{e}^{\beta\left[
k_0+q_0-\mu_f \right]}
\, .
\ee
%


Therefore, solving the Matsubara sums using the standard results
\be
T\sum_l \frac{g(k^0)}{m_k^2-K^2}&=& \frac{1}{2\omega}\left.\left\{ g(k^0) n_B(k^0) \right\}\right|^{k^0=\omega}_{k^0=-\omega}
\\
T\sum_n \frac{g(p^0)}{P^2-m_f^2}&=& 
\frac{1}{2\Ep}\left.\left\{ g(p^0) n_F(p^0-\mu_f) \right\}\right|^{k^0=\Ep}_{k^0=-\Ep}
\ee

\noindent with g(x) analytic, $K^2=-(\omega_l^B)^2-{\bf k}_z^2$, $P^2=(i\omega_{n_1}^F+\mu_f)^2-{\bf p}_z^2$, $Q^2=(i\omega_{n_2}^F+\mu_f)^2-{\bf q}_z^2$
$\omega^2=m_k^2+{\bf k}_z^2$, $\Ep^2={\bf p}_z^2+m_f^2$,
$n_B(\omega)=[\textrm{e}^{\beta\omega}-1]^{-1}$ and $n_F(E)=[\textrm{e}^{\beta E}+1]^{-1}$,
we arrive at:

\be
\mathcal{S}({\bf p}_z,{\bf q}_z,{\bf k}_z)&=&
T\sum_l
~T\sum_{n_1}~T\sum_{n_2}~\frac{s_0(p_0,q_0,k_0)}{(P^2-m_f^2)(K^2-m_k^2)(Q^2-m_f^2)}
\nonumber \\
&=&
-T\sum_l ~\frac{1}{m_k^2-K^2}
~T\sum_{n_1} ~\frac{1}{P^2-m_f^2}
~T\sum_{n_2}~\frac{s_0(p_0,q_0,k_0)}{Q^2-m_f^2}
\nonumber \\
&=&
-\frac{1}{8\omega\Ep\Eq}\left.\Bigg\{
s_0(p_0,q_0,k_0)~n_F(p_0-\mu_f )
~n_F(q_0-\mu_f )
~n_B(k_0)
\Bigg\}\right|_{(p_0,q_0,k_0)=(-\Ep,-\Eq,-\omega)}^{(p_0,q_0,k_0)=(\Ep,\Eq,\omega)}
\, , \nonumber \\
\label{Ssolv1}
\ee
where we adopt the notation 
$\left.\left\{ f(x,y,z) \right\}\right|^{(x,y,z)=(a,b,c)}_{(x,y,z)=(-a,-b,-c)}\equiv
\left.\left\{ \left.\left[ \left.\left( 
f(x,y,z)
\right)\right|_{x=-a}^{x=a}\right]\right|_{y=-b}^{y=b}\right\}\right|_{z=-c}^{z=c}$.

In order to further simplify Eq. (\ref{Ssolv1}), we can rewrite the function $s_0(p_0,q_0,k_0)$
through the identity
\be
\textrm{e}^{\beta\left[
k_0+q_0-\mu_f \right]}-\textrm{e}^{\beta\left[
p_0-\mu_f \right]}
=
\frac{1}{n_F(q_0-\mu_f )n_B(k_0)}-\frac{1}{n_B(k_0)}+\frac{1}{n_F(q_0-\mu_f )}-\frac{1}{n_F(p_0-\mu_f )}
\, .
\nonumber\\
\ee

Finally, using these results in Eq. (\ref{excRF-d2-3}), after a long but straightforward algebra, we obtain:
\begin{eqnarray}
\mathcal{G}(m_k^2,m_f^2)
&=&
-
\beta V~g^2N_c\left( \frac{N_c^2-1}{2} \right)
\int \frac{d^{\bar d-1}{\bf p}_z d^{\bar d-1}{\bf q}_z d^{\bar d-1}{\bf k}_z}{(2\pi)^{3(\bar d-1)}}
~(2\pi)^{\bar d-1}\delta({\bf k}_z-{\bf p}_z+{\bf q}_z)
\nonumber\\
&&
\frac{4}{8\omega\Ep\Eq}\Bigg\{
\mathcal{J}_{+}~\omega~\Sigma_1
+
\mathcal{J}_{-}~\omega~\Sigma_2
+2\Big[ \mathcal{J}_{-}~E_+-\mathcal{J}_{+}~E_- \Big]~
n_B(\omega)~N_F(1)
-\nonumber \\
&&
-\Big[ \mathcal{J}_+ (E_-+\omega)-\mathcal{J}_- (E_+-\omega) \Big]~N_F(1)
-2~\mathcal{J}_-~E_+~n_B(\omega)
+\nonumber \\
&&
-\mathcal{J}_-(E_+-\omega)
\Bigg\}
%
%
\, , \label{excRes}
\end{eqnarray}

\noindent where 
\be
{\mathcal{J}}_{\pm} &=& -2~\frac{\frac{\bar{d}}{2}~m_f^2-
\left( \frac{\bar d}{2}-1\right)~(\pm\Ep \Eq-{\bf p}_z\cdot{\bf q}_z)}{E_{\mp}^2-\omega^2}
~\stackrel{\bar d\to 2}{\longrightarrow}~ -~\frac{2m_f^2}{E_{\mp}^2-\omega^2}
\,,
\label{barJpm}
\ee
and

\be
\omega &=& \sqrt{{\bf k}_z^2+m_k^2}=\sqrt{k_1^2+k_2^2+k_3^2}\,,
\\
E_{\pm}&\equiv& \Ep\pm\Eq=\sqrt{{\bf p}_z^2+m_f^2}\pm\sqrt{{\bf q}_z^2+m_f^2}\,,
\\
N_F(1) &\equiv& n_F(\Ep+\mu_f ) +n_F(\Ep-\mu_f ) \, ,
\label{Nfs}
\\  
\Sigma_1 &\equiv& n_F(\Ep+\mu_f )~n_F(\Eq+\mu_f )+n_F(\Ep-\mu_f )~n_F(\Eq-\mu_f ) \, ,
\\
\Sigma_2 &\equiv& n_F(\Ep+\mu_f )~n_F(\Eq-\mu_f )+n_F(\Ep-\mu_f )~n_F(\Eq+\mu_f ) \, .
\ee

In $\bar d=2$ explicitly, we have:
\begin{eqnarray}
\mathcal{G}(m_k^2,m_f^2)
&=&
\beta V~g^2N_c\left( \frac{N_c^2-1}{2} \right)
~
m_f^2~
\int \frac{d p_3 d q_3 d  k_3}{(2\pi)^{3}}
~(2\pi)\delta(k_3-p_3+q_3)
\times
\nonumber\\
&\times&
\frac{1}{\omega\Ep\Eq}\Bigg\{
\frac{\omega~\Sigma_1}{E_-^2-\omega^2}
+
\frac{\omega~\Sigma_2}{E_+^2-\omega^2}
+2
\left[\frac{E_+}{E_+^2-\omega^2}-\frac{E_-}{E_-^2-\omega^2}\right]
~
n_B(\omega)~N_F(1)
-\nonumber \\
&&
-
\left[
\frac{2(\Eq+\omega)}{(E_--\omega)(E_++\omega)}
\right]
~N_F(1)
-2~
\frac{E_+}{E_+^2-\omega^2}
~n_B(\omega)
-
\frac{1}{E_++\omega}
\Bigg\}
\, . \label{excRes-d=2}
\end{eqnarray}

Finally, our preliminary result for the exchange contribution to the pressure of perturbative QCD in a magnetic background is\footnote{For the full result in QCD, summation over the flavor index $f$ is still required.}:

\vspace{0.3cm}

\begin{eqnarray}
\parbox{10mm}{
\fmfreuse{exchange}
}^{\quad\rm LLL}
 &=& 
\beta V~g^2N_c\left( \frac{N_c^2-1}{2} \right)
~
m_f^2
\nonumber\\&&
\left(
\frac{q_fB}{2\pi}\right)
\int\frac{dk_1dk_2}{(2\pi)^2}~{\rm e}^{
-\frac{k_1^2+k_2^2}{2q_fB}}
~
\int \frac{d p_3 d q_3 d  k_3}{(2\pi)^{3}}
~(2\pi)\delta(k_3-p_3+q_3)
\nonumber\\
&&
\frac{1}{\omega\Ep\Eq}\Bigg\{
\frac{\omega~\Sigma_1}{E_-^2-\omega^2}
+
\frac{\omega~\Sigma_2}{E_+^2-\omega^2}
+2
\left[\frac{E_+}{E_+^2-\omega^2}-\frac{E_-}{E_-^2-\omega^2}\right]
~
n_B(\omega)~N_F(1)
-\nonumber \\
&&
-
\left[
\frac{2(\Eq+\omega)}{(E_--\omega)(E_++\omega)}
\right]
~N_F(1)
-2~
\frac{E_+}{E_+^2-\omega^2}
~n_B(\omega)
-
\frac{1}{E_++\omega}
\Bigg\}
\,.
 \label{exc-LLL-final}
\end{eqnarray}

\vspace{0.3cm}

Interestingly, the result seems to be proportional to the quark mass, although the remaining momentum integrations may still bring $1/m_f^2$ factors. If not, this means that in the chiral limit the gauge interaction does not correct the pressure up to $O(g^2)$.

To obtain a final result, the momentum integrations need to be carried out explicitly. However, the possibility of divergences, especially in the IR domain, is not excluded and a regularization procedure may be required, as usually happens in the description of a magnetic background.

\end{fmffile}
\section{Remarks and perspectives}

In this chapter, we have considered an external magnetic field as an extra axis in the phase diagram of hot QCD matter. As usual, the new axis opens up novel channels of comparison between different theoretical approaches, but in this case also between theory and experiment. As discussed in the first section, several recent developments have shown that hot QCD matter under intense magnetic fields has an exciting phenomenology, closely connected with on-going heavy-ion collision experiments at RHIC and LHC, and is amenable to first-principle lattice QCD simulations. 

Our approach here \cite{MagneticQCD-wip} goes in the direction of furnishing predictions for the thermodynamics of QCD in a magnetic background within the fundamental gauge theory. In this vein, we adopt an approach that combines perturbative QCD with a nonperturbative treatment of the coupling of quarks to the intense magnetic field. After the building blocks of the framework were constructed, the application to the computation of the one-loop vacuum selfenergy for the gluon and the preliminary results for the thermodynamics have revealed interesting features and subtleties intrinsic to media under extreme magnetic backgrounds.

The effective dimensional reduction to $(1+1)$ dimensions was verified in different contexts, in agreement with the intuitive physical picture of Landau orbits in the limit of intense magnetic fields. Although the dimensional reduction tends to soften possible UV divergences, we have seen in several examples that regularizations are still required. The computation of the free pressure of magnetically-dressed quarks revealed an (infinite) pure magnetic contribution due to the artificial picture of an omnipresent constant and uniform field. Moreover, we have shown that the connection between the LLL approximation and the large $B$ limit is not universal. A lot, however, is still to be understood and the subtleties encountered so far point out that the remaining momentum integrations in the exchange contribution, Eq. (\ref{exc-LLL-final}), should be carefully analyzed.
Near future perspectives of this on-going work \cite{MagneticQCD-wip} include as well the computation of the one-loop quark and gluon thermal selfenergies within this framework and comparison to lattice data.

%% file: OPT.tex
\chapter[Masses and condensates in nonperturbative thermodynamics of Strongly interacting matter]{\label{OPT}}
\chaptermark{Masses and condensates in nonperturbative...}

\vspace{1.5cm}

{\huge \sc Masses and condensates in}
\vspace{0.3cm}

\noindent {\huge \sc  nonperturbative thermodynamics of}
\vspace{0.3cm}

\noindent {\huge \sc Strongly interacting matter}


\vspace{-11cm}
\hspace{6cm}
\includegraphics[width=8.5cm,angle=90]{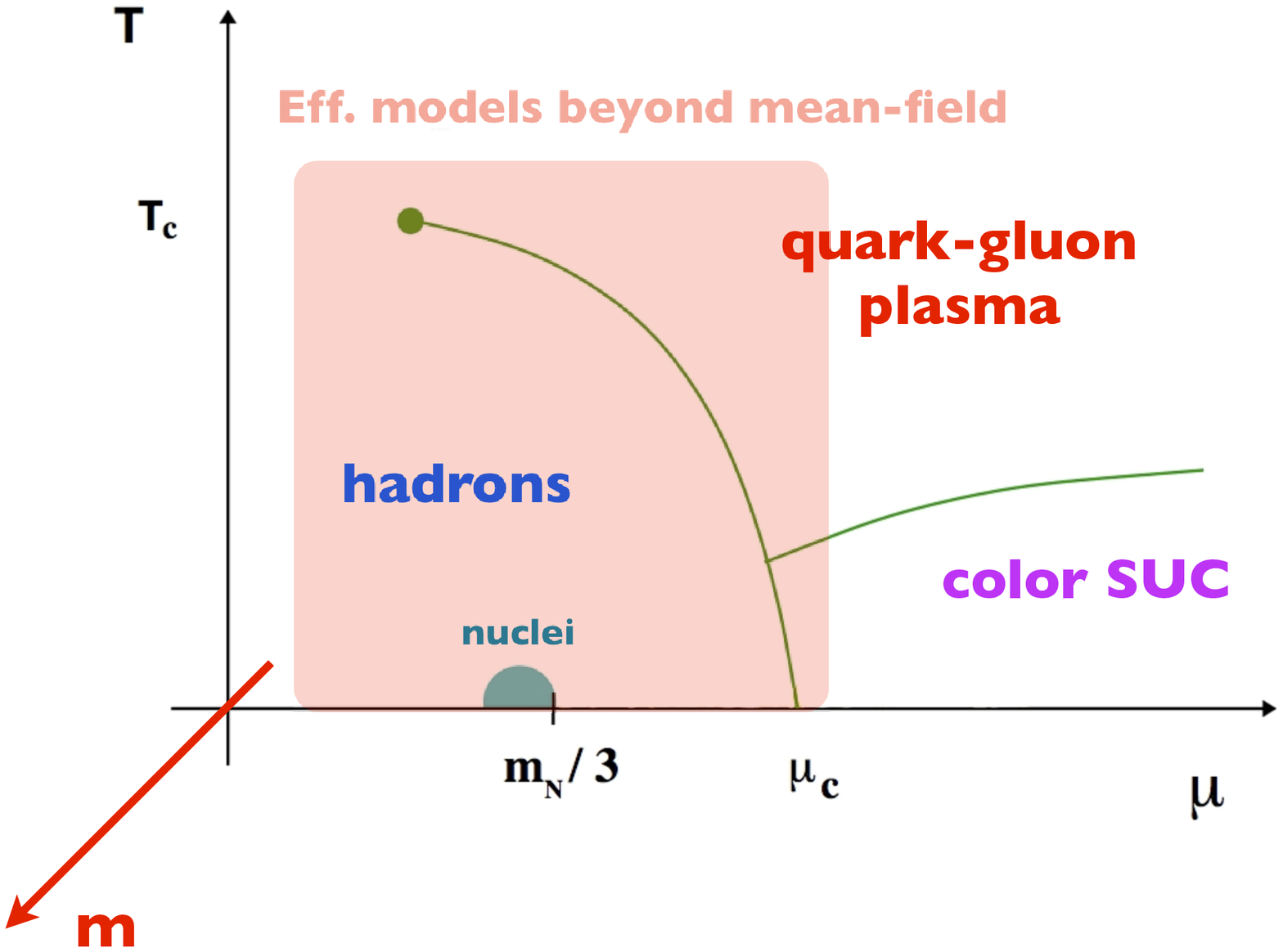}

\vspace{4.5cm}


Towards the goal of building a robust description of the phase diagram of Strong Interactions via effective models, one should pursue an approach that can accommodate finite temperature and chemical potentials, different masses, condensates and nonperturbative interactions. Since order parameters are often related to condensates, which in turn commonly appear in effective masses, it is desirable to have a framework that combines a nonperturbative method with the capability of investigating the role played by different massive fields.

This chapter presents a nonperturbative computation of the thermodynamic potential of a general Yukawa theory within Optimized Perturbation Theory to two loops, allowing also for the presence of a scalar condensate \cite{Fraga:2009pi}. We shall keep the discussion in general grounds, concentrating on the analysis of the nonperturbative method in question, as compared to na\"ive perturbation theory and mean-field approximation. The results are applicable to a wide range of interacting fermionic environments; not only Strong Interactions, but also the Higgs Sector of the Standard Model and Condensed Matter systems described by the Hubbard Model, as argued in the introduction.
%

\section{Introduction}

Matter under extreme conditions often requires the use of effective field theories in the 
description of its thermodynamical properties, independently of the energy scale under 
consideration. For instance, in the phenomenon of spontaneous symmetry breaking and 
temperature dependence of antiferromagnetic order in high $T_{c}$ superconductors described 
by a field-theoretic version of the Hubbard model \cite{hubbard} at relatively low energies, one has 
to incorporate the interaction of fermions with the crystal lattice via phonons, in a coarse 
graining of quantum electrodynamics. Moving up in energy scale, the phase structure of strong 
interactions and the quark-gluon plasma \cite{Rischke:2003mt}, investigated in high-energy 
heavy-ion collisions experiments \cite{QM} and in the observation of compact stars \cite{stars}, 
often demands simplifications of quantum chromodynamics, producing a variety of low-energy 
effective models \cite{Stephanov:2007fk}. 

Besides the major role it plays in the mechanisms of spontaneous symmetry breaking and mass 
generation in the Standard Model of particle physics, the Yukawa interaction stands out as one 
of the main ingredients in the construction of simplifying effective field theories to study the 
thermodynamics of systems under extreme conditions, especially if one imposes renormalizability. 
Although an effective theory does not require renormalizability to be consistent, a physically-motivated 
cutoff being usually more than satisfactory in this case, this feature can prove to be useful in the 
study of scale dependence and running via renormalization group methods \cite{Palhares:2008yq}. 

In this chapter we investigate the full nonperturbative thermodynamics of the Yukawa theory at finite 
temperature and density by computing its pressure using the optimized perturbation theory (OPT) framework.
In our evaluations we consider contributions up to two loops, which include direct (Hartree like)
as well as exchange (Fock like) terms, so that the Yukawa thermodynamics can be investigated
in the presence of condensates and also in their absence. In the first case, besides  
OPT and ordinary perturbation theory (PT) we shall also perform a mean field (MFT) evaluation.
This is an important step in establishing the OPT reliability since the reader will see 
that when exchange contributions are neglected OPT exactly reproduces MFT results 
which can be considered ``exact'' in this (large $N$) limit\footnote{
The reliability of the OPT framework applied to symmetric and broken phases was studied
previously in the context of $\phi^4$ scalar theories (see for instance Ref. \cite{Farias:2008fs}).
}.
We present results  which are 
valid for arbitrary fermion and scalar masses, temperature, chemical potential, and coupling. 
The region of large values of the coupling is particularly interesting and useful, since several effective 
field theory models in particle and nuclear physics exhibit Yukawa coupling constants that are much 
larger than one, as is the case, for instance, in the linear sigma 
model \cite{GellMann:1960np}, described in Appendix \ref{LSM} and frequently used in the description of the chiral phase 
transition (e.g. in Refs. \cite{quarks-chiral,Scavenius:2000qd,Scavenius:2001bb,ove,Scavenius:1999zc,Caldas:2000ic,Bilic:1997sh}
and in Chapters \ref{FS} and \ref{surften} of this thesis) 
and in pion-nucleon models extracted from chiral Lagrangians \cite{Weinberg:1978kz,vanKolck:1999mw}.

When condensates are present we show that, as expected, PT has a poor performance since it cannot 
resum direct (tadpole) contributions associated with symmetry breaking. On the other hand, OPT 
improves over MFT by also incorporating exchange terms in a nonperturbative fashion.

In the absence of condensates and in the regime of very small coupling our numerical findings can be verified in 
the limit of vanishing temperature by comparison to exact analytic two-loop perturbative results 
previously obtained for the equation of state of cold and dense Yukawa theory within 
the $\overline{\rm MS}$ scheme \cite{Palhares:2008yq} (cf. also Refs. \cite{thesis,procsYuk}). In Ref. \cite{Palhares:2008yq}, the 
two-loop momentum integrals were computed analytically for {\it arbitrary} fermion and scalar masses, 
the final result being expressed in terms of well-known special functions, which provides us with 
a solid and  clear reference in this limit. 

Furthermore, this comparison also provides another way of testing the idea 
that perturbation theory at high density and zero temperature in the symmetric phase 
is much better behaved than its 
converse \cite{Palhares:2008yq,Fraga:2001id,tony,andersen,Braaten:2002wi}, which 
has well-known severe infrared problems \cite{kapusta-gale}. Our analysis shows that, surprisingly, 
second-order PT agrees quite well with the OPT nonperturbative results up to values of the 
coupling of order one in the case of cold and dense as well as hot and dense Yukawa theory.

The framework of OPT \cite{OPT}, also known as the linear delta expansion, 
is an example of a variational method that implements the resummation of certain classes of 
Feynman diagrams, incorporating nonperturbative effects in the computation of the thermodynamic 
potential (for related methods, see Refs. \cite{OPT2}). 
It has been successfully applied to the study of many different physical situations, such as mapping 
the phase diagram of the 2+1 dimensional Gross-Neveu model \cite{Kneur:2007vj}, where a previously undetermined 
``liquid-gas'' phase has been located,  and determining the critical 
temperature for Bose-Einstein condensation in dilute interacting atomic gases \cite{BEC-OPT}.
Some early applications at finite temperature can be found in Refs. \cite{OPT3}.

The chapter is organized as follows. In Section \ref{Yuk} we present the in-medium Yukawa theory and set 
up the notation to be used in what follows. Section \ref{PertOm} contains a sketch of the derivation of the perturbative 
thermodynamic potential at finite temperature and chemical potential. In Section \ref{OPTOm} we apply the 
OPT machinery to the evaluation of the nonperturbative thermodynamic potential in the Yukawa 
theory. The thermodynamics of the Yukawa theory, in the presence of condensates, is
considered in Section \ref{OPTResCond} where the results from PT, OPT and MFT are contrasted. 
In Section \ref{OPTRes} we consider the symmetric case, where condensates are absent. 
Comparing the results from OPT and PT, we show that the latter performs well also for 
extreme temperature values up to two-loop order.
Section \ref{OPTFR} contains our conclusions. Technical details involved in the calculation of
the vacuum contributions and direct terms in the two-loop thermodynamic potential are left for appendices \ref{apOPTA} and \ref{apOPTB}\footnote{The technicalities concerning Matsubara sums, renormalization, and the analytic 
evaluation of in-medium momentum integrals in the calculation of the medium contributions up to two loops
were extensively addressed in the appendix of Ref. \cite{Palhares:2008yq} (cf. also \cite{thesis}).}.


\section{In-medium Yukawa theory\label{Yuk}}

In what follows, we consider a gas of $N_{F}$ flavors of massive spin-$1/2$ fermions 
whose interaction is mediated by a massive real scalar field, $\phi$, with an interaction 
term of the Yukawa type, so that the Lagrangian has the following general form:
\begin{equation}
\mathcal{L}_{Y}
= \mathcal{L}_{\psi}+
\mathcal{L}_{\phi}
+\mathcal{L}_{int} \, ,
\label{Lyukawa}
\end{equation}
where
\begin{eqnarray}
\mathcal{L}_{\psi} &=& \sum_{\alpha=1}^{N_F} \bpsi_{\alpha}\left( i\dslash -m \right)
\psi_{\alpha} \, ,
\\
\mathcal{L}_{\phi} &=& \frac{1}{2} (\partial_{\mu}\phi)(\partial^{\mu}\phi)-
\frac{1}{2} m_{\phi}^2 \phi^2
-\lambda_3\phi^3-\lambda\phi^4 \, ,
\\
\mathcal{L}_{int} &=& \sum_{\alpha=1}^{N_F} g~\bpsi_{\alpha}\psi_{\alpha}\phi \, . 
\label{Lint}
\end{eqnarray}
Here, $m$ and $m_{\phi}$ are the fermion and boson masses, respectively, assuming all the 
fermions have the same mass, for simplicity. The Yukawa coupling is represented by $g$; 
$\lambda_3$ and $\lambda$ are bosonic self-couplings allowed by renormalizability. 
Here we choose $\lambda_3=\lambda=0$,  disregarding  bosonic self-interactions.

We work in the imaginary-time Matsubara formalism of finite-temperature 
field theory, where the time dimension is compactified and associated with the inverse 
temperature $\beta=1/T$ \cite{kapusta-gale}. In this approach, one has to impose 
periodicity (anti-periodicity) for the bosonic (fermionic) fields in the imaginary time $\tau$, 
in order to satisfy the spin-statistics theorem. Therefore, only specific discrete Fourier modes 
are allowed, and integrals over the zeroth component of four-momentum are replaced by 
discrete sums over the Matsubara frequencies, denoted by $\omega^B_n=2n\pi T$ for bosons 
and $\omega^F_n=(2n+1)\pi T$ for fermions, with $n$ integer. Nonzero density effects are included 
by incorporating the constraint of conservation of the fermion number via a shift in the zeroth component 
of the fermionic four-momentum $p^0=i\omega_n^F\mapsto p^0=i\omega_n^F+\mu$, $\mu$ being 
the chemical potential.

\vspace{0.4cm}
\begin{figure}[htb]
\includegraphics[width=17cm]{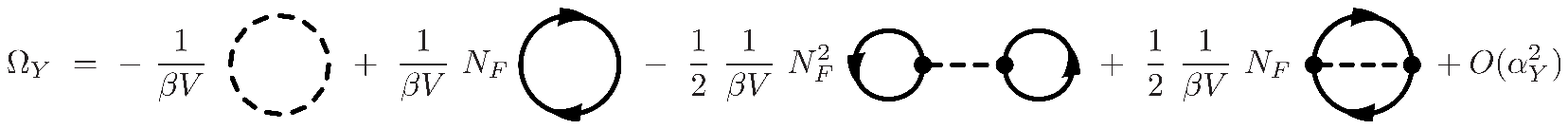}
\caption{Perturbative thermodynamic potential for the Yukawa theory to two loops. 
Solid lines represent fermions and dashed lines stand for  bosons. 
Here we omit the diagrams containing counterterms. The third (tadpole) contribution can
be neglected when we consider the symmetric case.}
\label{OmegaY-fig}
\end{figure}

From the partition function written in terms of the euclidean action for the lagrangian (\ref{Lyukawa}), 
 $Z_{Y}(T,\mu) = {\rm Tr} ~\exp(-S_Y)$, one derives the perturbative series for the thermodynamic 
 potential $\Omega_{Y}\equiv - (1/\beta V) \ln Z_Y$ \cite{kapusta-gale}:
\begin{equation}
\Omega_{Y}=
-\frac{1}{\beta V} \ln Z_0-
\frac{1}{\beta V}\ln \left[ 1+\sum_{\ell=1}^{\infty} \frac{(-1)^\ell}{\ell !}
\langle S_{int}^{\ell} \rangle_0 \right] \, ,
\end{equation}
where $V$ is the volume of the system, $Z_0$ is the partition function 
of the free theory and $S_{int}$ represents the euclidean interaction action. 
Notice that Wick's theorem implies that only even powers in the above 
expansion survive, yielding a power series in $\alpha_Y\equiv g^2/4\pi$. 
However, at finite temperature the perturbative expansion also contains 
odd powers of $g$ coming from resummed contributions of the zeroth 
Matsubara mode for bosons, such as in the case of the plasmon 
contribution \cite{kapusta-gale}. In the zero-temperature limit this is not 
the case and, even including hard-dense-loop corrections, only $g^{2\ell}$ terms 
are modified \cite{andersen}. Since we restrict our analysis to two 
loops (the resummed OPT calculation also departs from a two-loop perturbative 
setting), we are not concerned with this issue. Omitting the diagrams representing 
counterterms, the perturbative thermodynamic potential is shown, diagrammatically, in 
Fig. \ref{OmegaY-fig}. 
The first two diagrams correspond to the free gas, 
while the last two represent interaction terms: the third diagram is a contribution of
the direct type and the fourth is of the exchange type. Notice that the third diagram,
which contains tadpoles, belongs to the one-particle-reducible class, that
does not contribute in the absence of condensates. It will therefore
be neglected in Section \ref{OPTRes}, where only the symmetric case is considered.

\section{Perturbative Thermodynamic Potential\label{PertOm}}

Considering the thermodynamic potential only up to the first non trivial (two-loop) contribution, 
which corresponds to direct plus exchange terms, one must evaluate the following $O(g^2)$ quantity:
\begin{equation}
\Omega_Y =\frac{i}{2}  ~\sumint_{K}
\ln \left[K^2 - m^2_ \phi \right]
+i  ~\sumint_{P} {\rm Tr}\ln \left[\not \! P - m \right] 
- \frac{i}{2}  ~\sumint_{P}  {\rm Tr} \left[\frac {\Sigma_{\rm dir}(0)+\Sigma_{\rm exc}(P)}{(\not \! P - m)}\right] \,\,,
\label{PT}
\end{equation}
where the trace is to be performed over the Dirac structure, and the $4$-momenta are given 
in terms of the Matsubara frequencies and the $3$-momenta for fermions, 
$P=\left( p^0=i\omega_{n}^{F}+\mu \, ,\,  {\bf p}\right)$, and bosons, 
$K=\left( k^0=i\omega_{\ell}^{B}\, ,\, {\bf k} \right)$. We choose the metric tensor signature 
$g^{\mu\nu}={\rm diag}(+,-,-,-)$, and the following short-hand notation for sum-integrals:
\begin{equation}
\sumint_{P} = T \sum_{n} \int \frac{d^{3}{\bf p}}{(2\pi)^{3}} \, .
\end{equation}
The direct contribution is given by
\begin{equation}
\Sigma_{\rm dir}(0)= -i \left (\frac{g}{m_\phi} \right)^2 ~\sumint_{Q}
{\rm Tr}\frac{1}{ [\not \! Q - m]} \, ,
\end{equation}
and, the exchange self-energy is given by the following sum-integral over fermionic momenta:
\begin{equation}
\Sigma_{\rm exc}(P)= i g^2 ~\sumint_{Q}
\frac{1}{ [\not \! Q - m][(P-Q)^2 - m^2_\phi]} \, .
\end{equation}

Computing the traces at $\mu \ne 0$ and $T \ne 0$ (see Ref. \cite {Palhares:2008yq}
for details concerning the exchange term), one can write the 
thermodynamic potential as $\Omega_Y=\Omega_Y^{\rm vac}+\Omega_Y^{\rm med}(T,\mu)$, 
where the vacuum contribution prior to renormalization has the form
\begin{eqnarray}
\Omega_Y^{\rm vac}&=& \frac{1}{2} \int \frac {d^3 {\bf k}}{(2\pi)^3} 
\omega_{\bf k} - 2N_{\rm F} \int \frac {d^3 {\bf p}}{(2\pi)^3} E_{\bf p} 
-\frac{g^2N_F^2}{2m_{\phi}^2} \left[2m\int \frac{d^{3}{\bf p}}{(2\pi)^{3}}  \frac{1}{E_{\bf p}}\right ]^2
- \nonumber\\
&&-g^2 \frac {N_{\rm F}}{4} 
\int \frac {d^3 {\bf p}_1}{(2\pi)^3} \frac {d^3 {\bf p}_2}{(2\pi)^3} 
\frac { {\overline{\cal J}_-}(E_+-\omega_{12})}{\omega_{12}E_{{\bf p}_1}E_{{\bf p}_2}} 
\nonumber\\
&=&
-B(m_{\phi})+4N_F B(m)+\Omega^{{\rm dir}}_{{\rm vac}}(m,m_{\phi})+\Omega^{{\rm exc}}_{{\rm vac}}(m,m_{\phi})
\, , \label{OmegaY(0,0)}
\end{eqnarray}
\noindent where $\omega_{\bf k}=({\bf k}^{2}+m_{\phi}^{2})^{1/2}$,  
$E_{\bf p}=({\bf p}^{2}+m^{2})^{1/2}$, 
$E_{\pm} \equiv \Epu \pm \Epd$, 
$\omega_{12} \equiv ( |{\bf p}_1-{\bf p}_2|^2+m_{\phi}^2)^{1/2}$, 
and we have defined
\begin{equation}
\overline{\mathcal{J}}_{\pm} \equiv 
-2~\frac{m^2-{\bf p}_1\cdot {\bf p}_2\pm \Epu\Epd}{E_{\mp}^2-\omega_{12}^2}
= 1-\frac{4m^2-m_{\phi}^2}{E_{\mp}^2-\omega_{12}^2} \, .
\label{barJpm}
\end{equation}
Notice that Eq. (\ref{OmegaY(0,0)}) contains zero-point energy divergent terms 
which are $T$- and $\mu$- independent only within the PT approach in which case they can be conveniently absorbed by the usual vacuum subtraction which normalizes  the pressure so that it vanishes at $\mu=0$.
In practice this means that, as done in Ref. \cite {Palhares:2008yq}, one does not have to care to their explicit evaluation and renormalization. However, within the other approaches (OPT and MFT)  considered here  the PT bare mass, $m$, is replaced by effective ($T$- and $\mu$- dependent) masses within the same diagrams. Therefore, we must renormalize those contributions
appropriately. Within the $\overline {\rm MS}$ subtraction scheme, the fully renormalized vacuum term can be written 
as  (see Appendices \ref{apOPTA} and \ref{apOPTB} for details)
\begin{equation}
\Omega_Y^{\rm vac}=-B^{{\rm REN}}(m_{\phi})+4N_F B^{{\rm REN}}(m)+\Omega^{{\rm dir,REN}}_{{\rm vac}}(m,m_{\phi})+
\Omega^{{\rm exc,REN}}_{{\rm vac}}(m,m_{\phi}) \, ,
\end{equation}
where we have defined the following functions of the masses:
\begin{eqnarray}
B^{{\rm REN}}(M)&\equiv& \frac{M^4}{64\pi^2}\left[ \frac{3}{2}+\log\left( \frac{\Lambda^2}{M^2} \right) \right] \, ,
\\
\Omega^{{\rm dir,REN}}_{{\rm vac}}(m,m_{\phi})&\equiv&-\frac{g^2 N_F^2}{2 m_\phi^2} \left \{ \frac{m^3}{(2 \pi)^2} \left [  1+ \ln \left(\frac{\Lambda^2}{m^2} \right ) \right ] \right \}^2
\\
\Omega_{{\rm vac}}^{{\rm exc,REN}}(m,m_{\phi}) &\equiv&  
N_F \frac{g^2}{2} \frac{m^4}{64\pi^{4}}~
\Bigg\{ 
v_1\left( \frac{m_{\phi}^2}{4m^2} \right)
+\left[ \gamma +\log\left( \frac{\Lambda^2}{m^2}\right)
 \right]~v_2\left( \frac{m_{\phi}^2}{4m^2} \right)
+
%
\nonumber\\
&&\quad +\frac{1}{2}~\left[ \gamma +\log\left( \frac{\Lambda^2}{m^2}\right)\right]^2
~v_3\left( \frac{m_{\phi}^2}{4m^2} \right)
+
6~\alpha\left(m^2\right)-
\nonumber\\
&&\quad -
\frac{m_{\phi}^4}{m^4}\left( 1-6\frac{m^2}{m_{\phi}^2} \right)
\alpha\left(m_{\phi}^2\right)
\Bigg\} \, .
\end{eqnarray} 
Here $\gamma$ is the Euler constant, $\Lambda$ is the renormalization scale in the $\overline {\rm MS}$ 
scheme, and the functions $v_i(z)$ and $\alpha\left(m^2\right)$ are defined in Appendices \ref{apOPTA} and \ref{apOPTB}.

Adding the direct term to the free gas plus exchange medium-dependent terms evaluated in Ref. \cite {Palhares:2008yq} one 
obtains
\begin{eqnarray}
\Omega_Y^{\rm med}(T,\mu)
&=& \frac{T^4}{2\pi^2} \int_0^\infty z^2dz \log[1-e^{-{\omega_z}}] 
\nonumber \\
&-& T^4 \frac{N_{\rm F}}{\pi^2}\int_0^\infty z^2dz\left \{  \log[1+e^{-(E_z -\mu/T)}] +\log[1+e^{-(E_z +\mu/T)}] \right \}
\nonumber \\
&-& g^2 T^2 \frac{N_{\rm F}^2 m^4}{(4 \pi^4) m_\phi^2}\left [  1+ \ln \left(\frac{\Lambda^2}{m^2} \right ) \right ]\Bigg\{\int_0^{\infty} z^2 dz \frac{1}{[z^2 + m^2/T^2]^{1/2}} 
\times\nonumber
\\
&&\quad\times
\left[
\frac{1}{[1+e^{(E_{z} -\mu/T)}]} +\frac{1}{[1+e^{(E_{z} +\mu/T)}]} \right]\Bigg\} 
 \nonumber \\
&-& g^2 T^4 \frac{ N_{\rm F}^2 m^2}{(2 \pi^4) m_\phi^2}\left \{\int_0^{\infty} z^2 dz \frac{1}{[z^2 + m^2/T^2]^{1/2}} \left [
\frac{1}{[1+e^{(E_{z} -\mu/T)}]} +\frac{1}{[1+e^{(E_{z} +\mu/T)}]} \right]\right \}^2 
\nonumber \\
&-& g^2 T^2 m^2 \frac{N_{\rm F}}{(2\pi)^4} ~\alpha_1 \int_0^\infty z^2dz \left [ \frac{N_f(1)}{E_z}
 \right ] \nonumber \\
&-& g^2 T^2 \frac{N_{\rm F}}{(2\pi)^4}~( \alpha_2 + 3\alpha_3)
\int_0^\infty z^2dz \left [ \frac{n_b(\omega_z)}{\omega_z} \right ]
\nonumber \\
&+& g^2 T^4 \frac{N_{\rm F}}{2(2\pi)^4}\int_0^\infty z^2dz y^2 dy\int_{-1}^1du_{zy} 
\frac{1}{E_z E_y}\left [ {\tilde {\cal J}_+}\Sigma_1 +{\tilde {\cal J}_-}\Sigma_2 \right ]
\nonumber \\
&+& g^2 T^4 \frac{N_{\rm F}}{(2\pi)^4}\int_0^\infty z^2dz x^2 dx \int_{-1}^1du_{zx}
\frac{1}{\omega_{x} E_z E_{zx}} \left [{\tilde {\cal K}_-}{\tilde {E}_+} - 
{\tilde {\cal K}_+}{\tilde {E}_-} \right ] n_b(\omega_{x}) N_f(1) 
\,\,,\nonumber\\&&
\label{Omega(T,mu)}
\end{eqnarray}
%
where, in order to perform numerical investigations, we have defined the following dimensionless quantities: 
$\omega_z^2= z^2 + {m_{\phi}}^2/T^2$, $E_z^2= z^2+m^2/T^2$, ${\tilde {E}_\pm}=E_z \pm E_{zx}$, 
$E_{zx}^2= x^2+z^2+2xz ~u_{zx} + m^2/T^2$, and
\begin{eqnarray}
{\tilde {\cal J}_\pm}&=& 1 + \frac { 4(m/T)^2 - (m_\phi/T)^2} { (E_z\mp E_y)^2 - {\omega_{zy}}^2}\,\,,
\\
{\tilde {\cal K}_\pm}&=& 1 + \frac { 4(m/T)^2 - (m_\phi/T)^2} { {\tilde{E}_{\mp}}^2 - {\omega_{x}}^2}\,\,,
\\
N_f(1) &=& n_f(E_z+\mu/T ) +n_f(E_z-\mu/T ) \, ,
\label{Nfs}
\\  
\Sigma_1 &=& n_f(E_z+\mu/T )~n_f(E_y+\mu/T )+n_f(E_z-\mu/T )~n_f(E_y-\mu/T ) \, ,
\label{Sigma1}
\\
\Sigma_2 &=& n_f(E_z+\mu/T )~n_f(E_y-\mu/T )+n_f(E_z-\mu/T )~n_f(E_y+\mu/T ) \, ,
\label{Sigma2}
\\
\alpha_1&=&
-4 \frac{m_{\phi}}{m}\left( 1-\frac{m_{\phi}^2}{4m^2} \right)^{\frac{3}{2}}
\left\{ \tan^{-1}\left[ \sqrt{\frac{1}{\frac{4m^2}{m_{\phi}^2}-1}} \right]
+\tan^{-1}\left[ \frac{\frac{1}{2}-\frac{m_{\phi}^2}{4m^2}}{\sqrt{ \frac{m_{\phi}^2}{4m^2} }\sqrt{1-\frac{
m_{\phi}^2}{4m^2}}} \right] \right\}
\nonumber \\
&& +\frac{7}{2}-\frac{m_{\phi}^2}{2m^2}-\frac{3}{2}\log\left( 
\frac{m^2}{\Lambda^2} \right)+\frac{m_{\phi}^2}{m^2}\left( \frac{3}{2}
-\frac{m_{\phi}^2}{4m^2} \right)
\log\left( \frac{m^2}{m_{\phi}^2} \right)
\, ,
\label{alpha1res}
\\
\alpha_2&=& m^2- \frac{1}{6}m_{\phi}^2
\, ,
\\
\alpha_3&=& \frac{2}{3}\left[ 2m^2-\frac{5}{12}m_{\phi}^2 \right]-
\frac{1}{3}m_{\phi}^2\left( \frac{4m^2}{m_{\phi}^2}-1 \right)^{\frac{3}{2}}
\tan^{-1}\left[ \frac{1}{\sqrt{\frac{4m^2}{m_{\phi}^2}-1}} \right]
-\nonumber \\&&
-\left( m^2-\frac{m_{\phi}^2}{6} \right)\log\left( \frac{m^2}{\Lambda^2} \right)
\, ,\label{alpha3res}
\end{eqnarray}
with $\omega_{zy}^2=z^2+y^2+2zy ~u_{zy}+{m_{\phi}}^2/T^2$, $x=k/T$, $z=p_1/T$, $y=p_2/T$, $n_b(x)=[\exp(x)-1]^{-1}$, and 
$n_f(x)=[\exp(x)+1]^{-1}$.
%

%
Given the perturbative expressions obtained in this section, we can compute the thermodynamic 
potential $\Omega_{Y}$ in the OPT framework, which corresponds to resumming all dressed 
diagrams of the direct and exchange types, generating in practice an effective mass for the fermions.

\section{Nonperturbative thermodynamic potential\label{OPTOm}}

Following the standard procedure, the OPT framework \cite{OPT} can be implemented 
in the Yukawa theory as
\begin{equation}
\mathcal{L}_{\rm OPT} =  \bpsi_{\alpha}\left[ i\dslash -(m+ \eta^*) \right]
\psi_{\alpha} + \delta \left[\frac{1}{2} (\partial_{\mu}\phi)^2-
\frac{1}{2} m_{\phi}^2 \phi^2\right]+ \delta ~g\phi \bpsi_{\alpha}\psi_{\alpha} \, ,
\label{Lint}
\end{equation}
where a sum over flavors is implied, and $\eta^* = \eta(1-\delta)$. As one can easily notice from the 
deformed Lagrangian above, at $\delta=0$ the theory is an exactly solvable theory of massive fermions 
even in the case when, originally, $m=0$. In this particular case, the {\it arbitrary}  mass parameter, 
$\eta$, works as an infrared regulator which proves to be very useful in studies related to chiral symmetry 
breaking. We should remark that, due to our choice, the meson sector disappears at $\delta=0$ since 
there is no need for their mediation in this case. Finally, when $\delta=1$ the {\it original}, interacting 
theory is recovered, so that our $\mathcal{L}_{\rm OPT}$ interpolates between a free (exactly solvable) 
fermionic theory and the original one.

The Feynman rules generated by the interpolating theory are trivially obtained from the original ones: 
$m \mapsto m+\eta^*$, $g \mapsto \delta g$, and the bosonic propagator receives a $1/\delta$ factor. 
It is important to notice that the interpolation does not change the polynomial structure of the 
original theory and hence does not spoil renormalization, as proved in Ref. \cite {PRDMR}. Now, a 
physical quantity such as $\Omega_Y^{\rm OPT}$ is {\it perturbatively} evaluated in powers of the 
dummy parameter, $\delta$, which is formally treated as small during the intermediate steps of the 
evaluation. At the end one sets $\delta=1$, in a procedure which is analogous to the one used within 
the large-$N$ approximation, where $N$ is formally treated as a large number and finally set to its 
finite value at the end. The result, however, depends on the arbitrary parameter, $\eta$, which can be 
fixed by requiring that $\Omega_Y^{\rm OPT}$ be evaluated at the point where it is less sensitive to 
this parameter \cite{PMS} (Principle of Minimal Sensitivity, PMS). This can be accomplished by requiring
\begin{equation}
\frac{d \Omega_Y^{\rm OPT}}{d \eta}{\Big |}_{\eta={\bar \eta}, \delta=1}=0 \,\,.
\label{pms}
\end{equation}
In general, the optimum value $\bar\eta$ becomes a function of the original couplings via self-consistent 
equations, generating nonperturbative results. 

Now we can apply the method to the case of the two-loop perturbative $\Omega_Y$ given in 
Eq. (\ref {PT}). Using the OPT replacements for the Feynman rules and expanding $\eta^*$ to 
order $\delta$, one obtains the first nontrivial result
\begin{eqnarray}
\Omega_Y^{\rm OPT} &=& \delta ~\frac{i}{2}  \sumint_{K} 
\ln \left[ K^2 - m^2_ \phi \right]+i  \sumint_{P} 
{\rm Tr}\ln \left[\not \! P - (m+\eta) \right] 
+\delta~ i  \sumint_{P} 
{\rm Tr} \left[\frac {\eta}{[\not \! P - (m+\eta)]}\right] 
\nonumber \\
&& 
-\delta ~\frac{i}{2}  \sumint_{P} 
{\rm Tr} \left[\frac {\Sigma_{\rm dir}(0,\eta)+\Sigma_{\rm exc}(P,\eta)}{[\not \! P - (m+\eta)]}\right] 
\,, \nonumber \\
\label{TPopt}
\end{eqnarray}
where
\begin{equation}
\Sigma_{\rm dir}(0,\eta)= -i \left (\frac{g}{m_\phi} \right)^2 ~\sumint_{Q}
{\rm Tr}\frac{1}{ [\not \! Q - (m+\eta)]} \, 
\end{equation}
and
\begin{equation}
\Sigma_{\rm exc}(P,\eta)= i g^2 \sumint_{Q}
\frac{1}{ [\not \! Q - (m+\eta)][(P-Q)^2 - m^2_\phi]} \, .
\end{equation}
Notice that, when compared to the second-order perturbative result, Eq. (\ref {TPopt}) displays an extra contribution 
given by the third term on its right-hand side which represents a one-loop graph with a $\delta \eta$ insertion.
Now, using Eq. (\ref {pms}) one arrives at
\begin{eqnarray}
0&=&-i  \sumint_{P} {\rm Tr} \left[\frac {1}{[\not \! P - (m+{\bar \eta})]}\right]
+\delta~ i  \sumint_{P} {\rm Tr} \left[\frac {1}{[\not \! P - (m+{\bar \eta})]}\right]
+\delta~ i  \sumint_{P} {\rm Tr} \left[\frac {\bar \eta}{[\not \! P - (m+\eta)]^2}\right]
\nonumber \\
&& 
-\delta~ i   \sumint_{P} {\rm Tr} 
\left[\frac {\Sigma_{\rm dir}(0,{\bar \eta})+\Sigma_{\rm exc}(P,{\bar \eta})}{[\not \! P - (m+{\bar \eta})]^2}\right] \,, \nonumber \\
\end{eqnarray}
where, to obtain the last term, we have used a redefinition of momenta $p \mapsto q, 
q \mapsto p$ at an intermediate step. Setting $\delta=1$, one has a nontrivial, 
coupling-dependent self-consistent integral relation for the optimum mass parameter 
involving the self-energy given by
\begin{equation}
 i   \sumint_{P} {\rm Tr} 
\left[\frac {{\bar \eta}-\Sigma_{\rm dir}(0)-\Sigma_{\rm exc}(P,{\bar \eta})}{[\not \! P - (m+{\bar \eta})]^2}\right]
=0 \, .
\label{etapms}
\end{equation}
Since the OPT propagator has an infinite number of $\eta$ 
insertions, one can see that the optimization will resum exchange graphs in a nonperturbative way. 
It is very interesting to notice that when exchange terms are neglected 
${\bar \eta} = \Sigma_{\rm dir}(0)=-g\langle \phi \rangle_0 $, where $\langle \phi \rangle_0$
represents the scalar condensate, satisfying the MFT self-consistent relation for the 
effective mass (obviously, when $g\to 0$ the OPT results agree with the free gas case).
This type of result is consistent with applications of OPT to different
types of theories \cite {optmft} and  illustrates the way OPT works.
Notice that within the Hartree approximation one also adds and subtracts a mass 
term which is determined self-consistently. The basic difference is that the topology 
of this term is fixed from the start: direct terms in the Hartree approximation,
direct plus exchange in the Hartree-Fock approximation. Within OPT the effective
mass $m+\eta$ is {\it arbitrary} from the start, its optimum form being determined
by the topology of the contributions considered in the perturbative evaluation of a
physical quantity such as the thermodynamic potential considered here.

In the next two sections we shall study the pressure numerically, using different 
approximations in several different situations. With this aim we set 
$m=0.1 \LMS$, $m_\phi=m/2$ in our numerical routines\footnote{The value $m_\phi=0$,
which is of particular interest in situations motivated by QCD, will also be considered
in subsection \ref{mphi=0}.}. The temperature and chemical potential ranges considered
cover $0 \to 10m$ while the coupling values cover $0 \to \pi$.

\section{Results in the presence of a scalar condensate \label{OPTResCond}}

\begin{figure}[htb]
\center
\includegraphics[width=9cm]{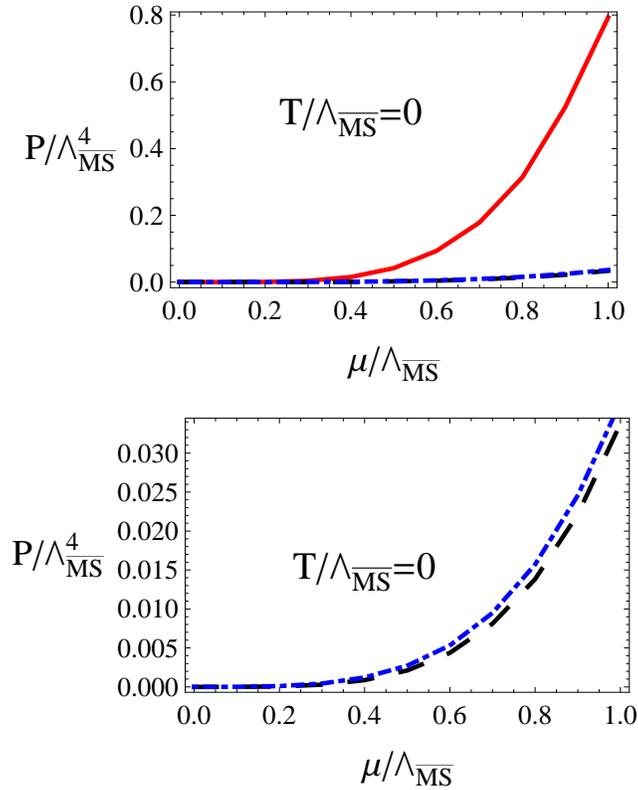}
\caption{
The pressure, $P/\LMS^4$,  as a function of $\mu/ \LMS$ at $T=0$. 
Top: results from PT (continuous line), MFT (dashed line) and OPT (dot-dashed line). 
Bottom: differences, due to exchange terms, between OPT and MFT. }
\label{DiExTzero}

\end{figure}

\begin{figure}[htb]
\center
\includegraphics[width=9cm]{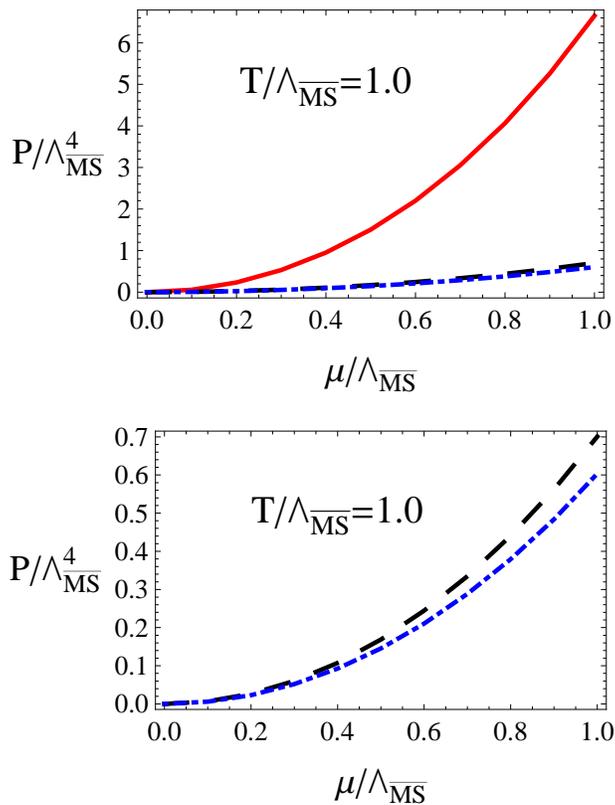}
\caption{The pressure, $P/\LMS^4$,  as a function of $\mu/ \LMS$ at $T=1.0 \LMS$. Top: results from PT (continuous line), MFT (dashed line) and OPT (dot-dashed line).
Bottom: differences, due to exchange terms, between the OPT and MFT. }
\label{DiExTum}
\end{figure}

In this section we compare the results generated by PT, OPT and MFT when the scalar
condensate represented by the direct (one-particle reducible) terms are considered. 
This case is also interesting because one can use the well-established MFT to 
analyze the results provided by OPT and PT. As already mentioned, when exchange
contributions are not considered in OPT, MFT results are {\it exactly}
reproduced since in this case both theories employ the same effective mass 
$M_{\rm eff}= m+\Sigma^{\rm MFT}_{\rm dir}(0)= m+ {\bar \eta}$. Each approximation
considers different two-loop contributions: PT takes all graphs shown in Fig. \ref {OmegaY-fig}
into account, while  OPT considers all plus the extra one-loop fermionic graph with the 
$\delta \eta$ insertion. In practice, as shown in the previous section, the MFT 
result is quickly recovered from the OPT one by neglecting the exchange term.
Figure \ref{DiExTzero} shows the pressure as a function of $\mu$ at $T=0$.
As one can see, PT predicts very high values for the pressure as $\mu$ increases, in disagreement
with MFT and OPT results. Since OPT agrees exactly with MFT when exchange
terms are neglected, one can also see in this figure the effects of resumming
exchange contributions: they yield slightly higher values of the pressure for increasing $\mu$.
Figure  \ref {DiExTum} shows the same situation but
at a high temperature. In this case, the OPT-predicted pressure values are
smaller than the MFT ones as $\mu$ increases.

\begin{figure}[htb]
\center
\includegraphics[width=9cm]{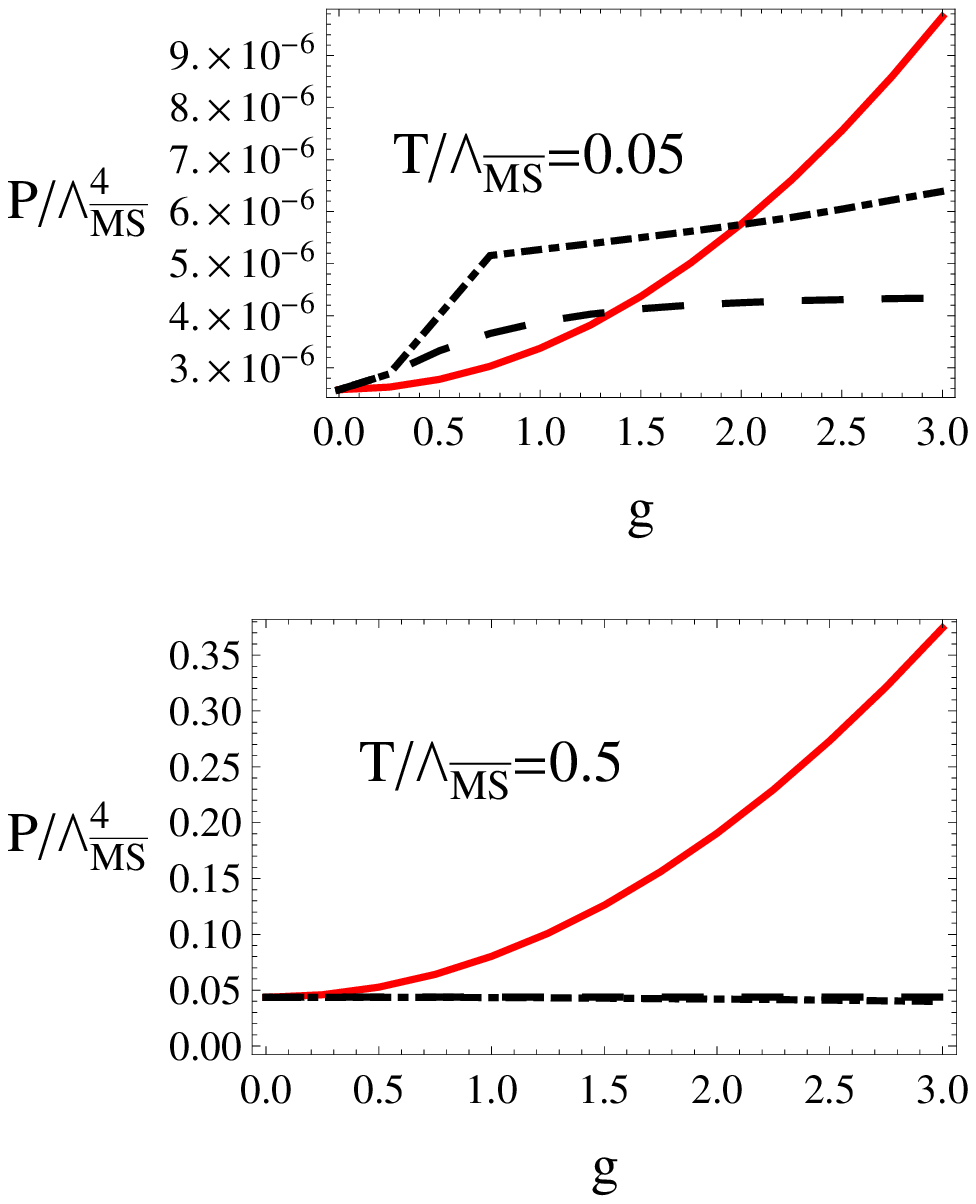}
\caption{The pressure, $P/\LMS^4$,  as a function of $g$ for $\mu=T=0.05 \LMS$ (top figure)  and $\mu=T=0.5 \LMS$ (bottom figure). PT corresponds to the continuous line, MFT to the dashed line and OPT to the dot-dashed line.}
\label{PvsgDEx}
\end{figure}

\begin{figure}[htb]
\center
\includegraphics[width=9cm]{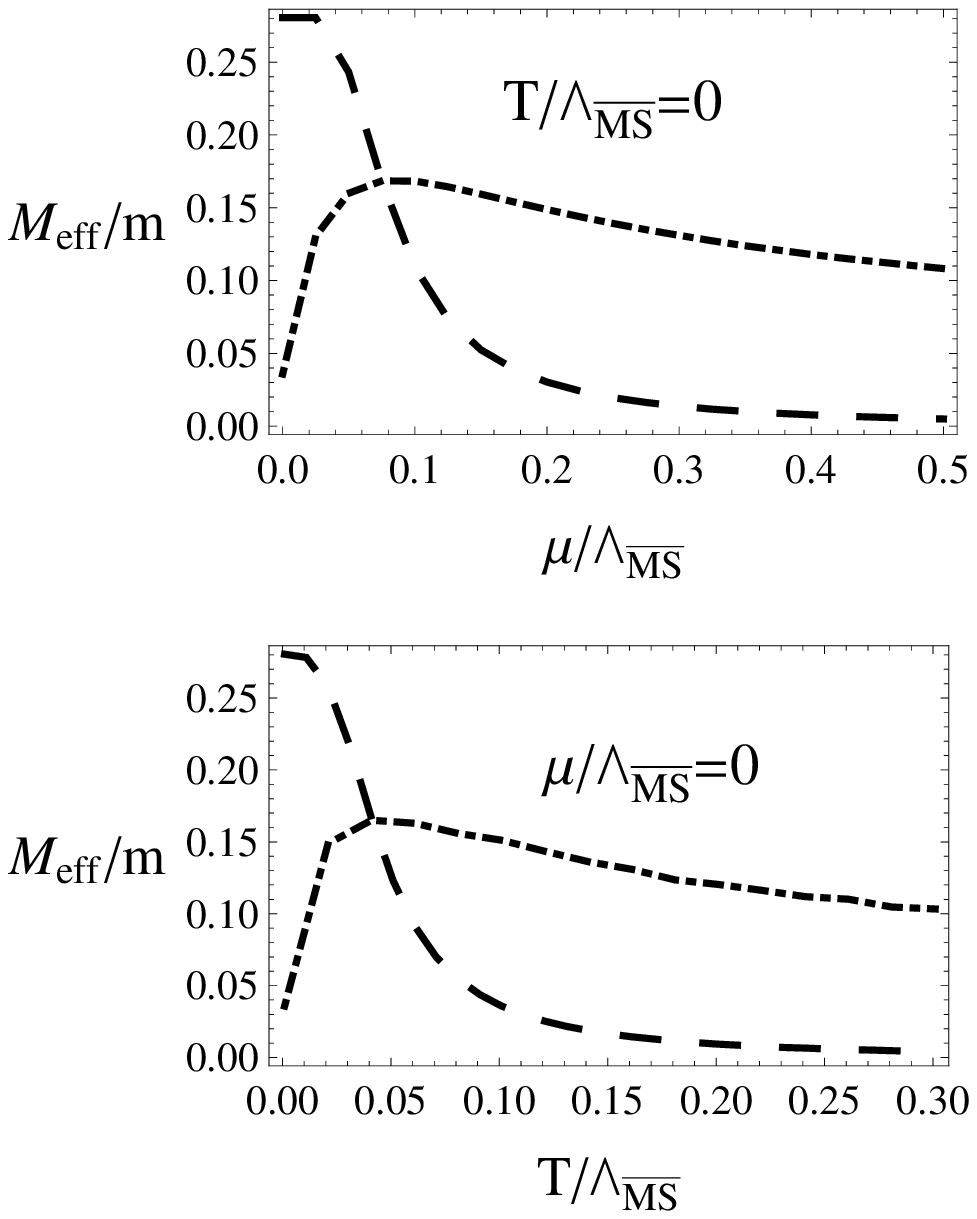}
\caption{The OPT (dot-dashed line) and MFT (dashed line) effective masses ($m + {\bar \eta}$ and $m+ \Sigma_{\rm dir}(0)$) in units of $m$ for $g=\pi$. 
Top: $M_{\rm eff}/m$ as a function of $\mu/ \LMS$ for $T=0$. Bottom: $M_{\rm eff}/m$ as a function of $T/ \LMS$ for $\mu=0$.}
\label{Meff}
\end{figure}

In Figure \ref {PvsgDEx} we analyze the pressure as a function of the coupling for low ($0.5 m$) and high ($5 m$) values
of $\mu$ and $T$. As expected, all methods agree with the free gas case when $g \to 0$. 
Also, at high $T,\mu$ values MFT and OPT results tend towards the free gas since one approaches the Stefan-Boltzmann limit,
while PT has a completely different behavior in this situation.
Finally, let us study the behavior of the OPT and MFT effective masses as functions of $T$ and $\mu$, as shown in Figure \ref{Meff} for $g=\pi$. 
Both quantities have a quantitatively as well as qualitatively different behavior  at small $T,\mu$ values but, as seen in the previous figures, 
this effect does not manifest itself in the pressure, probably being compensated by the presence (absence) of exchange terms within the OPT (MFT).
As $T$ or $\mu$ increases the qualitative behavior of both effective masses becomes the same although there are still quantitative differences.
These results clearly illustrate how PT is not appropriate to deal with this kind of situation since it cannot resum the condensate which arises from the interaction between the scalar field and fermions and which is related to symmetry breaking. Our application nicely illustrates the reliability of the OPT results since they exactly agree with MFT at large-$N$ (when exchange terms are neglected), while allows us to improve over this approximation by resumming exchange contributions.

\section{Results in the absence of a scalar condensate \label{OPTRes}}

Let us now follow Ref. \cite {Palhares:2008yq} and neglect the one-particle-reducible two-loop diagrams containing tadpoles. In this case, which is relevant for the situation where symmetry breaking is not present, one expects that PT will perform better than in the previous case where the tadpole contributions have been considered, especially in the zero-temperature limit. Clearly MFT is not applicable in this situation, and we will restrain our analysis to OPT and PT results. At this stage, the reader should be convinced that, as explicitly shown in the previous section, the former method is able to generate nonperturbative results which can be used to access the eventual breakdown of PT. 

\subsection{Cold and dense case with $m_\phi=0$\label{mphi=0}}

In this subsection, the order-$\delta$ OPT results will be compared to the second-order perturbative predictions of Ref. \cite {Palhares:2008yq}.
The case $m_\phi=0$ is of particular interest since in this situation one can write the thermodymanic potential 
$\Omega_Y=\Omega_Y^{{\rm vac}}+\Omega_Y^{{\rm med}}(T=0,\mu)$ in a simple and compact analytic form, in terms of the vacuum part
\begin{eqnarray}
\lim_{m_{\phi}\to 0}~\Omega_Y^{{\rm vac}}
&=&
4N_F B^{{\rm REN}}(m)+\lim_{m_{\phi}\to 0}~\Omega_{{\rm vac}}^{{\rm exc,REN}}(m,m_{\phi}) \, ,
\end{eqnarray}
with
\begin{eqnarray}
\lim_{m_{\phi}\to 0}~\Omega_{{\rm vac}}^{{\rm exc,REN}}(m,m_{\phi})
&=&
N_F \frac{g^2}{2} \frac{m^4}{64\pi^{4}}
\Bigg\{ 
v_1\left( 0 \right)
+\left[ \gamma +\log\left( \frac{\Lambda^2}{m^2}\right)
 \right]v_2\left( 0 \right)
+
 \nonumber\\
 &&
 \quad+ 
\frac{1}{2}\left[ \gamma +\log\left( \frac{\Lambda^2}{m^2}\right)\right]^2
v_3\left( 0 \right)
+
6~\alpha\left(m^2\right)
\Bigg\} \, ,
\end{eqnarray}
and the in-medium contribution
\begin{eqnarray}
\lim_{m_{\phi}\to 0}~\Omega_Y^{{\rm med}}(T=0,\mu)
&=&
- N_F\frac{1}{24\pi^2}
\left[
2~\mu p_f^3-3 m^2~u
\right]
-N_F\frac{g^2}{64\pi^4}
\bigg\{
3~u^2-4~p_f^4
+
\nonumber\\&&+
m^2~
u
\left[ 7-3\log\left( 
\frac{m^2}{\Lambda^2} \right) \right]
\bigg\}
\, , \label{OmegaYResT0mphi0}
\end{eqnarray}
where $p_f^2=\mu^2-m^2$ and $u=\mu p_f-m^2 \log\left(\frac{\mu+p_f}{m}\right)$.

\begin{figure}[htb]
\center
\includegraphics[width=9cm]{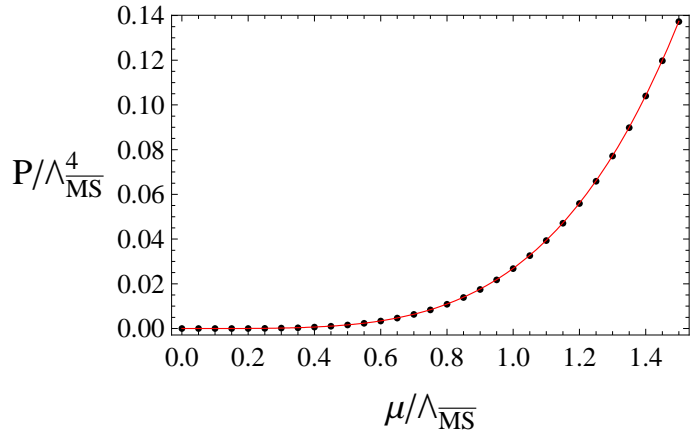}
\caption{Pressure in the absence of tadpoles normalized by $\LMS^{4}$ as a function of the chemical potential $\mu$ normalized 
by $\LMS$. Dots represent the OPT result and the line stands for PT. The fermion mass 
is fixed at $m=0.1\LMS$, and $g=\pi$.}
\label{fig2}
\end{figure}

Figure \ref{fig2} shows the pressure as a function of $\mu$ for a large value of the coupling, $g=\pi$.
As one can see both methods predict very similar results.

\vspace{0.4cm}
%
%
%

\begin{figure}[htb]
\center
\includegraphics[width=9cm]{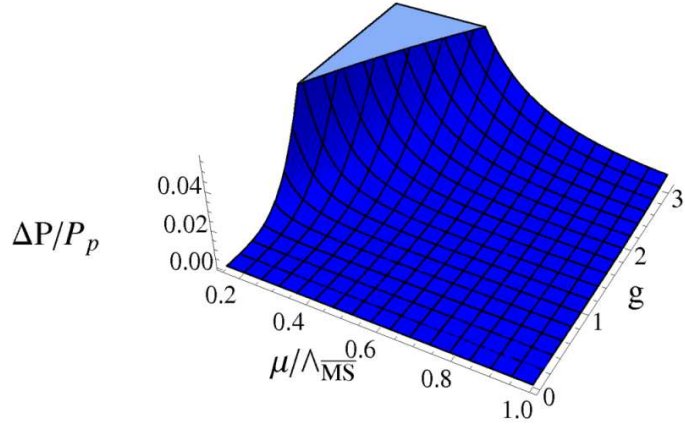}
\caption{Difference between the perturbative and the OPT pressures in the absence of tadpoles as a 
function of the coupling $g$ and the fermion chemical potential $\mu$ normalized by $\LMS$.
The fermion mass is fixed at $m=0.1\LMS$.}
\label{dif3dT0}
\end{figure}

To analyze the tiny differences, let us define the quantity $\Delta P/P_p= |(P_{\rm opt} - P_p)|/P_p$ 
where $P_p$ and $P_{\rm opt}$ are, respectively, the pressures predicted by PT and OPT.
Although the numerical discrepancies appear to be rather small, Figure \ref{dif3dT0} nicely illustrates that  $\Delta P/P_p$ increases with higher couplings and decreases with higher chemical potential values as opposed to the case where the scalar tadpole is  present. 
%
%

\vspace{0.4cm}

\hspace{2cm}

\begin{figure}
\center
\includegraphics[width=9cm]{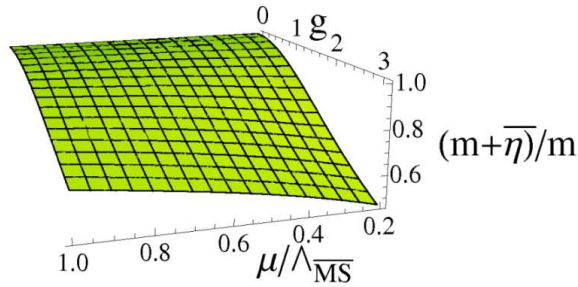}
\caption{The OPT effective mass $m+{\bar \eta}$ in units of $m$ as a 
function of the coupling $g$ and the fermion chemical potential $\mu$ normalized by $\LMS$ when tadpoles are absent.
The fermion mass is fixed at $m=0.1\LMS$.}
\label{mass3dT0}
\end{figure}

This behavior can be better understood if one analyzes how the OPT effective mass varies 
with $g$ and $\mu$. Figure \ref {mass3dT0} shows that the quantity $(m+{\bar \eta})/m$ deviates 
from $1$ as $g$ increases but, contrary to the case where tadpoles are present (see Fig. \ref {Meff}), 
approaches $1$ as $\mu$ increases.

%

Therefore, the nonperturbative OPT results support the PT results of Ref. \cite {Palhares:2008yq}, at $T=0$, when scalar condensates are not considered. Notice that even though we have considered in this subsection only the $m_\phi=0$ case our numerical simulations show that the agreement between PT and OPT remains valid for $m_\phi \ne 0$. For more details concerning the effects of the scalar mass the reader is referred to Ref. \cite {Palhares:2008yq}.

\subsection {Thermal effects\label{ThOPT}}

\begin{figure}[htb]
\center
\includegraphics[width=9cm]{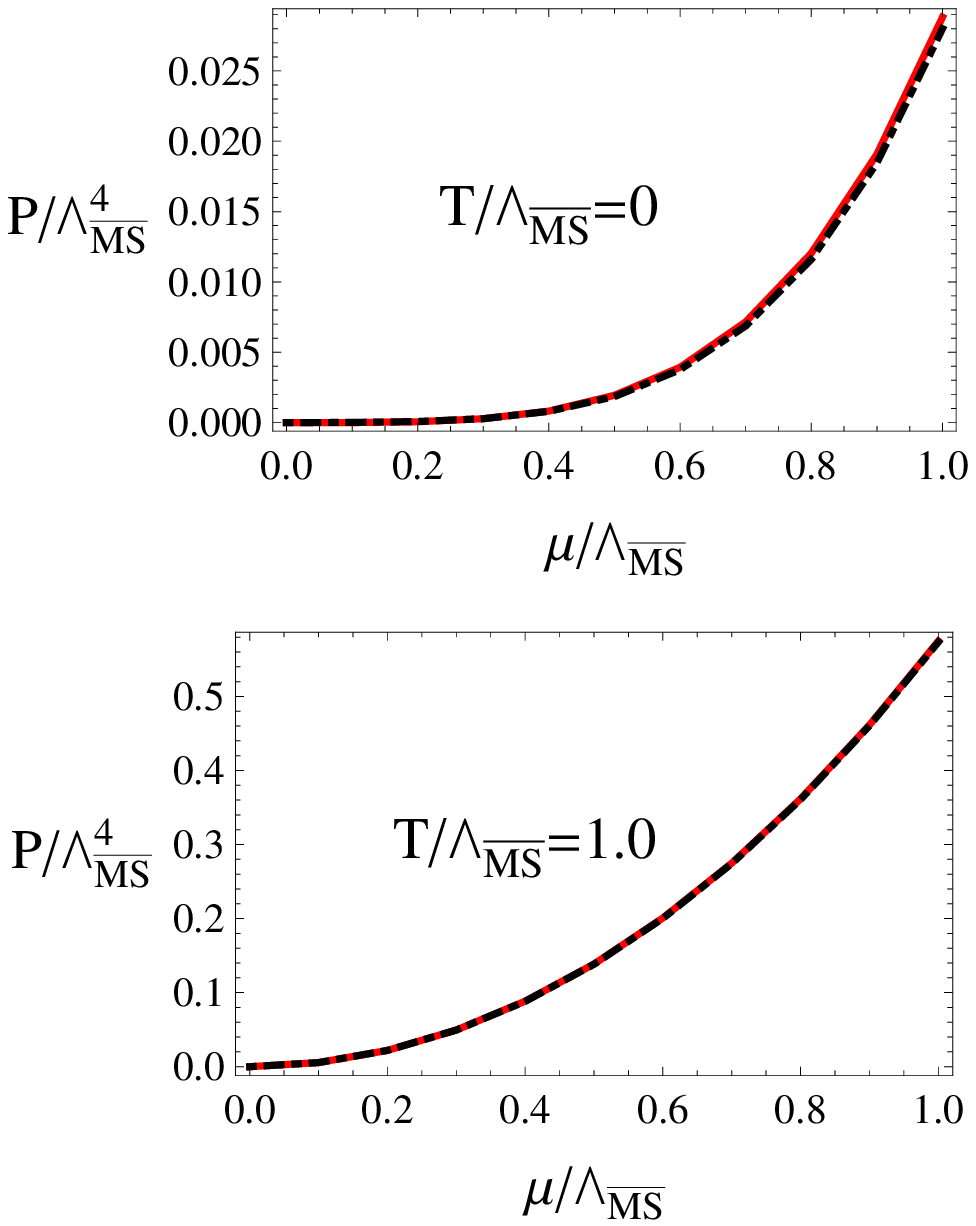}
\caption{The pressure, $P/\LMS^4$  as a function of $\mu/ \LMS$ at $T=0$ (top figure) and at $T=1.0 \LMS$  when tadpoles are absent.  PT  results are represented by the continuous lines while the OPT results are represented by the  dot-dashed lines. }
\label{SoEx}
\end{figure}

So far, we have seen through comparison with MFT and OPT that PT does not give reliable results when condensates are present but, as seen in the previous subsection, the situation improves in their absence, at least at $T=0$. In principle, it is not obvious that PT will furnish reliable results at high temperatures, even when only exchange contributions are considered. The aim of this subsection is to analyze this situation.
Figure \ref {SoEx} compares the OPT and PT results for the pressure as a function of $\mu$ in two extreme situations: $T=0$ and $T=10 m= 1.0 \LMS$. In the first case both methods agree well, as one should expect from the discussion performed in the previous section, but more surprisingly is the 
high-temperature result where OPT also supports PT.

\begin{figure}[htb]
\center
\includegraphics[width=9cm]{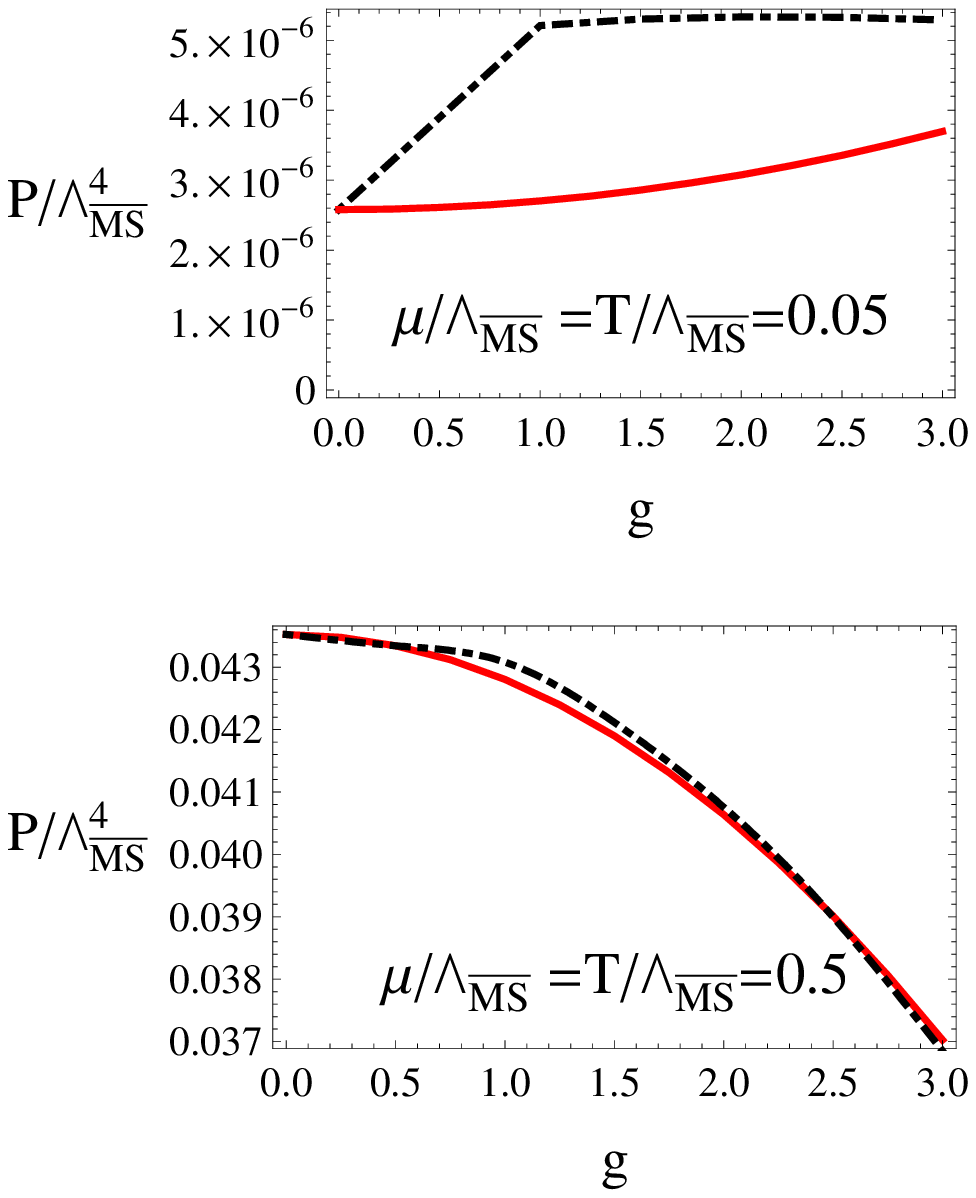}
\caption{The pressure, $P/\LMS^4$  as a function of $g$ for $\mu/=T= 0.05 \LMS$ (top figure) and at $\mu=T=0.5 \LMS$ when tadpoles are absent.  PT  results are represented by the continuous lines while the OPT results are represented by the  dot-dashed lines. }
\label{SoExg}
\end{figure}
%
%

Like in the previous $T=0$ case, the quantity $\Delta P/P_p$ assumes very small values which increase with high values of $g$ but decrease with high values of $T$ or $\mu$. The OPT effective mass in units of $m$, $(m+{\bar \eta})/m$, also behaves as in the previous case, deviating from $1$
as $g$ increases and, contrary to the case where tadpoles are present (see Fig. \ref {Meff}), approaching $1$ as $\mu$ and/or $T$ increase.
Therefore, it looks like the absence of direct terms means that $m \simeq m+{\bar \eta}$ at high $T$ and/or $\mu$, so that PT behaves like OPT.
Finally, Figure \ref {SoExg} suggests that even at relatively high $T$ and $\mu$ OPT and PT deviate from the free gas as the coupling increases, while in the case with condensates (see Fig. \ref {PvsgDEx} ) OPT (and MFT) had a better agreement with the free gas as opposed to PT. Therefore, the correct resummation of condensates seems to reduce the effects of interactions at high $T$ and $\mu$.

\section{Final Remarks\label{OPTFR}}

This chapter represents a first step towards a nonperturbative picture of the phase diagram of Strong Interactions via (chiral) effective models.
We have investigated the thermodynamics of the Yukawa model at finite temperature and chemical potential by evaluating its thermodynamic potential up to the two-loop level which includes direct (Hartree-like) as well as exchange (Fock-like) types of contributions. Three different methods have been considered: the usual perturbation theory (PT) with its bare mass, the optimized perturbation theory (OPT) with its effective mass given by the PMS variational criterion, and the well-known MFT with its self-consistent effective mass. 
Our comparative analysis shows that, as far as thermodynamic results are concerned, OPT is not only consistent with MFT when one-particle irreducible contributions are neglected, but also furnishes nontrivial corrections when the exchange class of diagrams is included. 

In this vein, OPT seems to be a well-defined framework that accommodates all the desirable ingredients to obtain a robust description of phase diagram of Strong Interactions via effective theories. It suffers, however, from its close connection to perturbative computation techniques and their renormalization procedure. In the case of the chiral phase transition described within the LSM, for instance, two-loop vacuum contributions apparently spoil the positivity of the effective action of the model (cf. e.g. \cite{thesis,Mocsy:2004ab}) \footnote{This issue is actually the analogous of the Higgs potential instability in the Standard Model of particle physics, which amounts to the requirement of a finetuning of the Standard Model Higgs mass. In the case of Strong Interactions, however, all the physical scales related to  spontaneous chiral symmetry breaking are known and a similar finetuning is therefore forbidden.} and further work to circumvent this obstacle is called for. 

Our results are nevertheless general, featuring coupling values ranging from $g=0$ to $g > > 1$ in situations where the temperature and chemical potential ranged from zero to ten times the highest mass value, which is kept fixed. As discussed in the Introduction, the Yukawa model usually emerges in the description
of various physical situations, ranging from low-energy condensed matter phenomena to 
extremely energetic QCD matter. In our approach, the characteristic features of each
physical system will be brought about essentially by the specific values of the coupling $g$ 
and the three energy scales: the fermion and
scalar masses $m$ and $m_{\phi}$, respectively, and the renormalization scale $\LMS$,
which normalizes all the quantities in our plots. As observed previously using 
the perturbative method \cite{Palhares:2008yq}, variations of the masses can significantly affect
the thermodynamic potential of the Yukawa theory, yielding
extremely different thermodynamical pictures. The renormalization scale $\LMS$ sets
the typical energy scale of the system of interest. In hadronic physics, for example,
it is reasonable to choose $\LMS\sim 1$ GeV, the confinement scale.
In this chapter, we kept the discussion in general grounds, normalizing physical quantities
by $\LMS$ and fixing the masses as $m=0.1~\LMS$ and $m_{\phi}=0 \, , \, 0.5~m$, and 
concentrating on the effects of Hartree- and Fock-like interactions and nonperturbative corrections.

First, we have analyzed the pressure with both direct and exchange contributions, with the former being associated to the presence of a scalar condensate driven by the interactions with fermions. We have shown that OPT and MFT are identical if one does not consider exchange terms, in agreement with many other applications \cite {optmft}, which is reassuring since in the limit of direct contributions only the MFT resummation can be considered as ``exact''. Moreover, nonperturbative effects of exchange  contributions are readily incorporated by considering OPT consistently up to two loops. In the light of the nonperturbative approaches OPT and MFT, our results show that, as expected, naive PT is inadequate to deal with this situation since it has no ability to resum tadpoles. 

As a byproduct of this application we could see how the resummation of exchange terms performed by OPT corrects the  MFT framework, 
which corresponds to the leading order of a $1/N$ type of approximation. Having established the reliability of the OPT, we have 
followed Ref. \cite {Palhares:2008yq} imposing the absence of tadpoles at $T=0$.  We have then shown that, in this case, the OPT results turn out to be very similar to the ones given by PT. Finally, still in the limit in which condensates are not present, we have investigated the high-temperature case where one could expect the breakdown of PT. However, our findings have shown that this is not the case and the numerical differences between the OPT and PT pressures are very small. The outcome of this analysis also suggests that, at high $T$ and $\mu$, the presence of condensates minimize the effects due to interactions when these contributions are properly resummed. 

One should notice that the direct application of these results as effective-model predictions 
in different physical contexts is restricted 
to definite energy regimes, within which the relevant physical degrees of freedom can be translated into
a Yukawa model. For instance, when discussing the thermodynamics of cold and dense baryonic matter,
the Yukawa model considered here might be a suitable effective theory only at small values
of the chemical potential, e.g. at $\mu\lesssim 350$ MeV. To describe the properties of baryonic matter
at higher values of $\mu$, a different framework is necessary to account for the phenomenon of color superconductivity
(for a review, see Ref. \cite{Alford:2007xm}). One alternative could be to extend this model
by adding boson fields from another, non-singlet, multiplet of the color $SU(3)$ group.
Nevertheless, since Yukawa-type interactions are almost ubiquitous in the description
of fermionic matter, this analysis provides an understanding
of the interplay between direct and exchange contributions to the thermodynamics as well
as the role played by nonperturbative effects in the scalar Yukawa sector of any given 
extended model.

%% file: relatBEC.tex
\chapter[Nonperturbative analysis of relativistic Bose-Einstein condensation of pions]{\label{relatBEC}}
\chaptermark{Nonperturbative analysis of relativistic...}

\vspace{1.5cm}

{\huge \sc Nonperturbative analysis of }
\vspace{0.3cm}

\noindent {\huge \sc  relativistic Bose-Einstein }
\vspace{0.3cm}

\noindent {\huge \sc   condensation of pions}


\vspace{-11.3cm}
\hspace{6cm}
\includegraphics[width=8.5cm,angle=90]{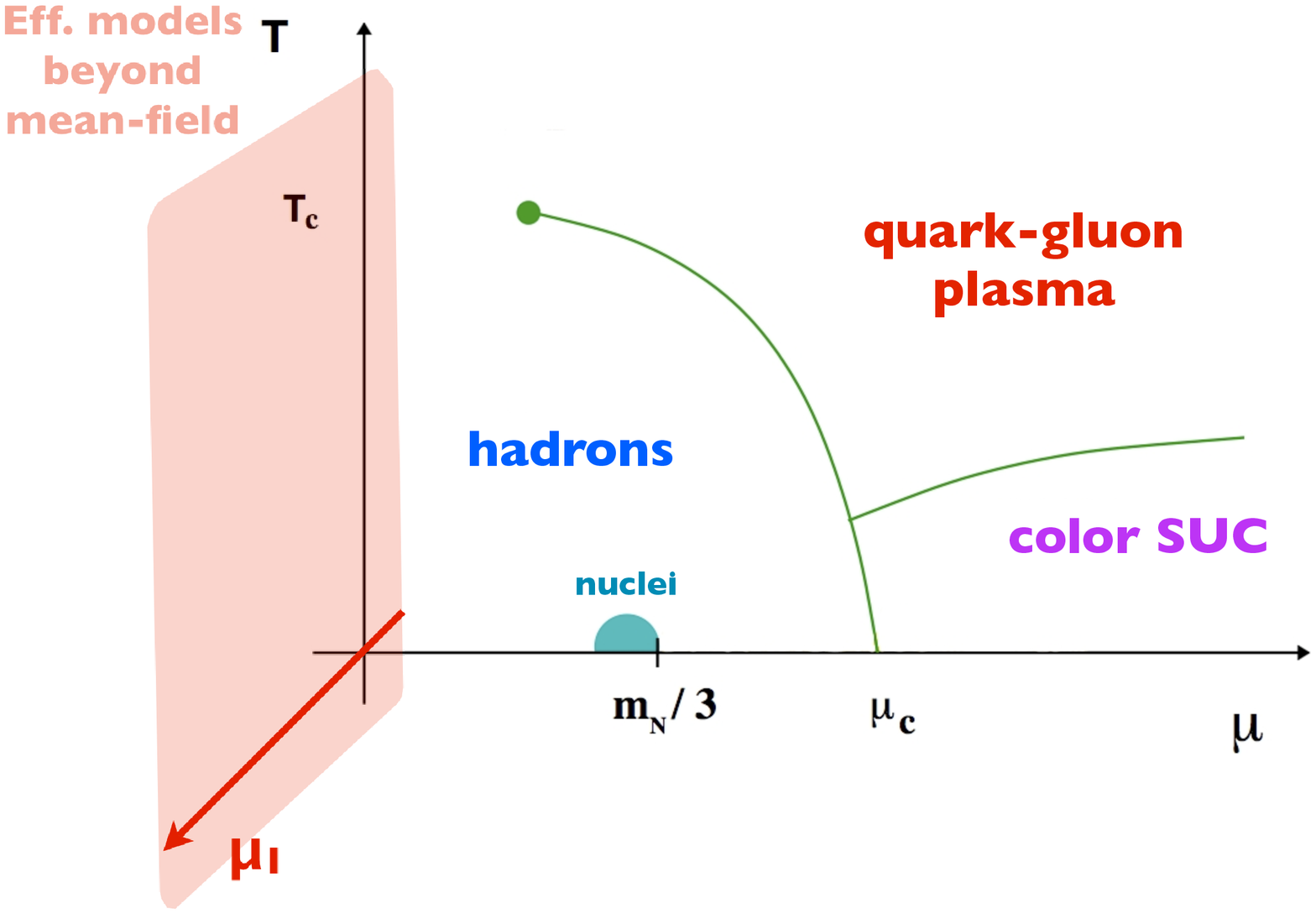}

\vspace{4.5cm}


In this chapter we will address the phase diagram of Strong Interactions in the temperature--isospin chemical potential plane through a nonperturbative approach. 

Instead of analyzing the deconfining phase transition in this type of media, as done in Chapter \ref{isospin}, we will neglect baryon density effects and concentrate on the phenomenon of relativistic Bose-Einstein condensation of pions to investigate the influence of (nonperturbative) interactions in this context \cite{RelatBEC-FRG-wip}. Besides its phenomenological motivation, the problem of relativistic BEC in isospin dense QCD represents a particularly interesting framework to develop efficient nonperturbative methods for dense systems: being Sign-Problem free, a robust reference from lattice simulations at finite isospin chemical potential is in principle available. Moreover, the role of interactions in the BEC phenomenon is a nontrivial fundamental question. Even for the nonrelativistic case, some issues were only understood quite recently, such as the shift of the critical temperature in cold atomic systems, that was successfully described using the Functional Renormalization Group framework \cite{Blaizot:2004qa}.
Our intent in this on-going work \cite{RelatBEC-FRG-wip}, centered in the relativistic version of BEC that should occur in QCD matter, is to analyze in detail how the Functional Renormalization Group technique and the different possible approximations that must complement it perform in this case.
%
%
%

%

\section{Basics of BEC in isospin dense QCD}

Proposals of pion condensation in Strongly interacting matter under extreme conditions were made back in the 70's by Migdal and others (for a review of these early works, cf. Ref. \cite{Migdal:1978az}). They were basically concerned with astrophysical environments and concentrated on electrically neutral media. The imposition of electric neutrality actually lead them to another (region of the) QCD phase diagram than the one to be addressed in this chapter: not only isospin density is present, but also baryonic. In this more complicated and phenomenological context, the pion condensate is found to be inhomogeneous and triggered by complicated interactions with nucleons and external fields.

Since we are interested in developing nonperturbative techniques for dense systems, we shall restrict our approach to isospin-dense, baryon-free matter for which the Sign Problem on the lattice is absent\footnote{Actually, one must consider two-flavor QCD with degenerate masses $m_u=m_d$, for which the imaginary part of the fermionic determinant cancels exactly.}. Pioneering work on this type of system was done by Son and Stephanov (cf. e.g. \cite{Son:2000xc}). Pion condensation in this case will be a clean relativistic realization of Bose-Einstein condensation and, differently from these early proposals, the condensate occurs as a constant and uniform state whose onset is triggered even for weak coupling.

Indeed, pion Bose-Einstein condensation in isospin dense environments is such a clear relativistic manifestation of thermodynamical properties of bosons that one can formulate a simple heuristic argument for its existence. Departing from the thermodynamical definition of chemical potential associated with a given net charge: the free energy the system {\it spends (gains)} when one {\it negative (positive)} charge is {\it taken out of (added to)} the system.  In the case of isospin-dense media, charged pions $\pi^{\pm}$ may play the role of $\pm 1$ isospin charges, but, being massive, the relativistic free energy spent when including one of them in the system must account for its mass $m_{\pi}$, at least. It is then straightforward to conclude that the free energy cost to add a zero-momentum $\pi^+$ decreases with increasing isospin chemical potential $\mu_I$, going to zero when it reaches the pion mass. At $\mu_I=m_{\pi}$ an instability sets in and the new thermodynamically stable ground state corresponds to a macroscopic occupation of $\pi^+$'s in the zero-momentum level, i.e. a Bose-Einstein condensate of $\pi^+$'s.

\subsection{Modeling in a chiral effective theory\label{chiralPionModel}}

Let us now build up an effective framework to address the issue of Bose-Einstein condensation of pions in isospin dense media. We will show that the heuristic argument presented above is indeed verified in a chiral model in the form of a nontrivial static and uniform pion condensate at high isospin densities.  

Departing from a standard chiral Hamiltonian,
the constraint of isospin number conservation is imposed via the method of Lagrange multipliers: $\mathcal{H}\mapsto \mathcal{H}-\mu_I n_I$, where the isospin chemical potential $\mu_I$ plays the mathematical role of the Lagrange multiplier for the conservation of the isospin number density $n_I$. Therefore, to describe an isospin dense system within a chiral effective theory, it is crucial to define isospin number $n_I$ in the context of chiral transformations and fields.

Isospin is the conserved charge associated with the invariance of Strong Interactions under a two-flavor subset of the global vectorial chiral transformations ${\rm exp}\left[-i\vec{\theta}\cdot\vec{\tau}_{\alpha\beta}/2\right]$, which are left unaffected by the spontaneous chiral symmetry breaking. Here, $\vec{\tau}=(\tau_1,\tau_2,\tau_3)$ are the Pauli matrices, $\alpha,\beta$ are indices in flavor space and $\vec{\theta}$ is the parameter of the transformation. The specific isospin subset is the one related to the light flavor sector of quarks up and down. We will be interested in the projection of the isospin on a specific coordinate axis, say $z$, so that the conserved isospin current can be obtained, through Noether's theorem, by transforming the Lagrangian locally using $\vec{\theta}=(0,0,\theta_3(x))$. In QCD, one derives in this way the conserved isospin current in terms of quarks ($\Psi$):
\be
\vec{j}_V^{\mu}
&=&
\overline{\Psi}_{\alpha}\gamma^{\mu}\frac{\vec{\tau}_{\alpha\beta}}{2}\Psi_{\beta}
\,,
\ee
while in a chiral effective theory involving mesons (such as the LSM or, equivalently, the $O(4)$ model), one obtains the isospin number $n_I\equiv n_3$ ($\vec{n}= \vec{j}_V^{0}$) related to the pion degrees of freedom $\vec{\pi}=(\pi_1,\pi_2,\pi_3)$:
\be
n_I&=&
\pi_1\partial^0\pi_2
-
\pi_2\partial^0\pi_1
=i\big[\pi^-\partial^0\pi^+-\pi^+\partial^0\pi^-\big]
\,,\label{nIdef}
\ee
where we have used the mass-eigenstate basis for the pions, $\pi^{\pm}=(\pi_1\mp i\pi_2)/\sqrt{2}$ and $\pi^0=\pi_3$, the latter being neutral under isospin rotations around the $z-axis$. A peculiar feature of the isospin number of pions is that it involves
not only the charged pion fields but also their time derivatives, i.e. the momenta canonically conjugated to these fields.

Since we are ultimately interested in thermodynamics and the phase transition to the pion condensed phase, we need to construct the partition function and the associated euclidean action for this isospin dense chiral model. For the LSM without constituent quarks, defined in the vacuum by the Lagrangian (the reader is referred to Appendix \ref{LSM} for an extended description and references):
\be
\mathcal{L}&=&
\frac{1}{2}(\partial_{\mu}\sigma)
(\partial^{\mu}\sigma)
+\frac{1}{2}(\partial_{\mu}\vec{\pi})
(\partial^{\mu}\vec{\pi})
- V\left(
[\sigma^2+\vec{\pi}^2],\sigma
\right)
\,,\label{LLSM}
\\
&&
 V_{\rm cl}\left(
[\sigma^2+\vec{\pi}^2],\sigma
\right)
=\frac{\lambda^2}{4}\left\{[\sigma^2+\vec{\pi}^2]-v^2\right\}^2-h\,\sigma
\,,
\label{VclLSM}
\ee
the result is (a detailed derivation is given in Appendix \ref{apSEisospin}; cf. also Ref. \cite{kapusta-gale}):
\be
Z&=&
\int [D\Phi]\; {\rm e}^{-S_E}
\,,
\ee
with the euclidean action for the isospin-dense LSM being given by:
\be
S_E&=& \int d^dx\bigg\{
\frac{1}{2}(\partial_{\mu}\sigma)
(\partial_{\mu}\sigma)
+\frac{1}{2}(\partial_{\mu}\vec{\pi})
(\partial_{\mu}\vec{\pi})
+\mu_I
\big[
\pi_1\partial_0\pi_2
-\pi_2\partial_0\pi_1
\big]
-\frac{1}{2}\mu_I^2\left(\pi_1^2+\pi_2^2\right)
+\nonumber\\
&&\quad
+V_{\rm cl}\left(
[\sigma^2+\vec{\pi}^2],\sigma
\right)
\bigg\}
\,.\label{SEis}
\ee

Within this model, pion Bose-Einstein condensation is realized in the form of a nontrivial pion vacuum expectation value at high isospin densities. To verify this and understand the quasiparticle degrees of freedom in both phases of this isospin-dense system, one should consider the theory in the presence of condensates and find the solution that minimizes the effective action as the isospin chemical potential $\mu_I$ is varied.

Adopting, as a first step, the classical (or mean-field or Landau-Ginzburg) approximation to obtain the condensate solutions for different $\mu_I$ domains, the effective action is assumed to have the same form as the euclidean action in Eq. (\ref{SEis}). The equilibrium expectation value of the fields is then given by the absolute minimum of this action.


Considering only static uniform condensate solutions $(\bs,\bvp)$ and small perturbations around it $(\xi,\vec{\eta})$,
\be
\sigma&=&\bs+\xi,
\\
\vec{\pi}&=&
\bvp+\vec{\eta}
\,,
\ee
the effective action in the classical approximation can be expanded as:
\be
S_E=
\int d^4x\left\{
\frac{1}{2}(\partial_{\mu}\xi)(\partial_{\mu}\xi)
+
\frac{1}{2}(\partial_{\mu}\vec{\eta})\cdot(\partial_{\mu}\vec{\eta})
+
\mu_I\left[
(\bp_2+\eta_2)\partial_0\eta_1-(\bp_1+\eta_1)\partial_0\eta_2
\right]
+
V_{\rm cl}^{\rm (L\sigma M)}(\sigma,\vec{\pi})
\right\}
\,,\label{SE-exp}
\ee
while the expansion of the potential, defined in Eq.(\ref{VclLSM}), reads
 (N.B.: $\vec{\pi}=(\pi_1,\pi_2,\pi_3)=\left(\frac{\pi^++\pi^-}{\sqrt{2}},i\frac{\pi^+-\pi^-}{\sqrt{2}},\pi^0\right)$):
\be
V_{\rm cl}^{\rm (L\sigma M)}(\sigma,\vec{\pi})
&=&
 \frac{\lambda^2}{4}
\Big\{
\bs^2+\bvp^2-v^2
\Big\}^2
-h\;\bs
-\frac{\mu_I^2}{2}
\left[
\bp_1^2+\bp_2^2
\right]
+\nonumber
\\
&&
+
\xi\Big[
\frac{\lambda^2}{4}\,4\bs(\bs^2+\bvp^2-v^2)
-h
\Big]
+\nonumber\\&&
+
\eta_i\Big[
\frac{\lambda^2}{4}
(
2\bs^2\,2\bp_i
+2\bvp^2\,2\bp_i
+4\bp_i(-v^2)
)
-\frac{\mu_I^2}{2}(1-\delta_{i3})
\left[
2\bp_i
\right]
\Big]
+\nonumber\\
&&+\xi^2
\Big[
\frac{\lambda^2}{4}
(4\bs^2
+2\bs^2
+2(\bvp^2-v^2)
)
\Big]
+\nonumber\\
&&
+\eta_i\eta_j
\Big[
\frac{\lambda^2}{4}
(
4\bp_i\bp_j
+
2\bs^2\delta_{ij}
+2\bvp^2\delta_{ij}
+2[-v^2]\delta_{ij}
)
-\frac{\mu_I^2}{2}\delta_{ij}(1-\delta_{i3})
\Big]
+\nonumber\\
&&
+\xi\eta_i
\Big[
\frac{\lambda^2}{4}
(
4\bs\,2\bp_i
)
\Big]
%
%
%
%
\nonumber\\&&
+\frac{\lambda^2}{4}
\Big\{
\xi^4
+\eta_i\eta_i\eta_j\eta_j
+4\bs\xi[\xi^2+\vec{\eta}^2
]
+2\xi^2[2\bvp\cdot\vec{\eta}+\vec{\eta}^2
]
+4\bvp\cdot\vec{\eta}
[\vec{\eta}^2
]
\Big\}
\, ,
\label{Vcl-exp}
\ee
where: $(i)$ the first line corresponds to the classical potential evaluated at the condensate; $(ii)$ the second and third lines (linear in the fluctuations) give the equilibrium conditions:
\be
\lambda^2\bs(\bs^2+\bvp^2-v^2)
-h&=&0
\label{EC-s}
\\
\left\{
\lambda^2
(
\bs^2
+\bvp^2
-v^2
)
-\mu_I^2(1-\delta_{i3})
\right\}\bp_i
&=&0
\quad\quad (i=1,2,3)
\,,\label{EC-pi}
\ee
$(iii)$ the terms quadratic in the fluctuations (4th, 5th and 6th lines) are related to the dispersion relations of the quasiparticles above in the condensed phase and define the stability of the condensate solution;
and $(iv)$ the last line corresponds to the interaction vertices for the fluctuations.

The dispersion relations for the quasiparticles in each phase are derived in Fourier space.
Using the following field expansion for each component of the fluctuation fields ($\varphi=\xi,\eta_i$):
\be
\varphi(x)&=&\sqrt{\frac{\beta}{V}}\, \beta V\sumint_Q\, {\rm e}^{iQ\cdot x}\varphi(Q)
\,,
\ee
Eq.(\ref{SE-exp}) then can be written as (up to quadratic order in the fluctuation fields and using the equilibrium conditions, Eqs. (\ref{EC-s}) and (\ref{EC-pi}), to cancel the terms which are linear in the fluctuations):
\be
S_E
&=&
\beta V\; V_{\rm cl}^{\rm (L\sigma M)}(\bs,\bvp)
+
\nonumber\\&&+
\beta V\sumint_Q
\left(
\begin{array}{cccc}
\xi(Q) & \eta_1(Q) & \eta_2(Q) & \eta_3(Q)\\
\end{array} 
\right)
\frac{\beta^2}{2}\mathcal{M}^2(Q)
\left(
\begin{array}{c}
\xi(-Q) \\ \eta_1(-Q) \\ \eta_2(-Q) \\ \eta_3(-Q)
\end{array} 
\right)
\nonumber\\
&&+O(3)
\,,
\ee
with:
\be
\mathcal{M}^2(Q)
&=&
\left( \begin{array}{cccc}
-Q_0^2+(\vec{q})^2+m_{\xi}^2 & 2\lambda^2\bs\bp_1 & 2\lambda^2\bs\bp_2 & 2\lambda^2\bs\bp_3 \\
2\lambda^2\bs\bp_1 & -Q_0^2+(\vec{q})^2+m_1^2 & 2\lambda^2\bp_1\bp_2+2i\mu_IQ_0 & 2\lambda^2\bp_1\bp_3\\
2\lambda^2\bs\bp_2 & 2\lambda^2\bp_1\bp_2-2i\mu_IQ_0 & -Q_0^2+(\vec{q})^2+m_2^2& 2\lambda^2\bp_2\bp_3\\
2\lambda^2\bs\bp_3 & 2\lambda^2\bp_1\bp_3 & 2\lambda^2\bp_2\bp_3 & -Q_0^2+(\vec{q})^2+m_3^2\\
\end{array} \right)\,,
\nonumber\\
\label{M2(Q)}
\ee
where we have defined:
\be
m_{\xi}^2&=&
\lambda^2[2\bs^2+\bvp^2+\bs^2-v^2]
\nonumber\\
m_i^2&=&
\lambda^2[2\bp_i^2+\bvp^2+\bs^2-v^2]-(1-\delta_{i3})\mu_I^2
\label{m_i's}
\ee
We have also used the notation $O(3)$ to represent the interaction between the perturbations, i.e. the terms that are cubic or quartic in $(\xi,\eta_i)$.

The spectrum of quasiparticles is then given by the energies $Q_0$ which correspond to a zero of an eigenvalue of the mass matrix $\mathcal{M}^2(Q)$ in the limit of zero three-momentum (i.e. a pole of the propagator). Therefore, they are solutions of the following equation:
\be
{\rm det}\left[\mathcal{M}^2(Q_0,{\vec{q}=0})\right]=0
\,.
\label{det}
\ee

\vspace{1cm}


 In what follows, from the equilibrium conditions, Eqs. (\ref{EC-s}) and (\ref{EC-pi}), the condensate solution is obtained as a function of the chemical potential: $\left(\bs(\mu_I),\bvp(\mu_I)\right)$. Then the quasiparticle spectrum is derived through Eqs. (\ref{M2(Q)})--(\ref{det}) in each phase.

Consider the (physical) case with explicit chiral symmetry breaking 
($h=f_{\pi}m_{\pi}^2\ne 0$).
Since $h\ne 0$, Eq. (\ref{EC-s}) implies that there exists a nonzero chiral condensate, in the $\sigma-$direction, i.e. $\bs\ne 0$, and also that the combination $(\bs^2+\bvp^2-v^2)$ must be nonvanishing. This result together with Eq. (\ref{EC-pi}) implies that:
\be
\bp_3\equiv\bp^0=0\,.
\ee

On the other hand, the condensates $\bp_1$ and $\bp_2$ are not constrained similarly and there will be solutions of the equilibrium conditions for $\bp_1$ and $\bp_2$ both vanishing or not:

\begin{itemize}
\item \underline{Normal phase, with vanishing pion condensates ($\bp_i=0$, for all $i$):}
The equilibrium conditions in the pion directions are trivially satisfied, while in the $\sigma-$direction we get the isospin-independent condition:
\be
\lambda^2\bs(\bs^2-v^2)
-h=0
\label{EC-s-pi0}
\,.
\ee
This equation has in general 3 solutions, which will be real and distinct for the physical values\footnote{The procedure of parameter fixing within the LSM is discussed in Appendix \ref{LSM}, Subsection \ref{ParFix}.} of $\lambda^2\approx 20$, $v^2\approx 7700~$MeV${}^2$ and $h=f_{\pi}m_{\pi}^2$. The true minimum is associated with the solution $\bs\gtrsim v$, while the others correspond to a metastable minimum and a maximum in between the minima.

The parameter fixing of the model implies that the true minimum of the potential in the vacuum ($\mu_I=0$, in this case) occurs at $\bs=f_{\pi}$ (cf. Appendix \ref{LSM}). Therefore, Eq. (\ref{EC-s-pi0}) becomes a condition\footnote{Strictly speaking, one should first verify that this condensation case (and not one with e.g. $\bvp^2\ne 0$) indeed minimizes the effective potential in the vacuum. This can be seen from the equilibrium conditions, Eqs. (\ref{EC-s}) and (\ref{EC-pi}), with $\mu_I=0$: they cannot be solved simultaneously if $h\ne 0$ and $\mu_I=0$ unless we consider $\bp_i=0$.} on the parameter $v$:
\be
\lambda^2f_{\pi}(f_{\pi}^2-v^2)
-h=0
\quad
\Rightarrow 
\quad 
v^2=f_{\pi}^2-\frac{h}{\lambda^2f_{\pi}}=
f_{\pi}^2-\frac{m_{\pi}^2}{\lambda^2}
\,.\label{v2-sol}
\ee

The value of the classical approximation for the true minimum of the effective potential in this case reads:
\be
\left.V_{\rm cl}^{\rm (L\sigma M)}\right|_{\rm min, \bvp^2=0}
&=&
\frac{\lambda^2}{4}
\left[
\bs^2-v^2
\right]^2
-h\bs
=\frac{1}{4\lambda^2}m_{\pi}^4-hf_{\pi}
\\
&=&
m_{\pi}^2\left[
\frac{m_{\pi}^2}{4\lambda^2}-f_{\pi}^2
\right]
\label{Vcl-pi0}
\ee

In this normal phase ($\mu_I<m_{\pi}$), the mass matrix in Eq.(\ref{M2(Q)}) becomes diagonal except for the terms $\sim\mu_IQ_0$, and the determinant equation for the spectrum becomes:
\be
&&
\hspace{-1.2cm}
(-Q_0^2+m_3^2)(-Q_0^2+m_{\xi}^2)\left\{
(-Q_0^2+m_1^2)(-Q_0^2+m_2^2)+4i^2\mu_I^2Q_0^2
\right\}
\stackrel{!}{=}0
\\
\Rightarrow
&&Q_0=|m_3|\,,\quad  Q_0=|m_{\xi}|
\quad {\rm or}\quad Q_0^4-Q_0^2(m_1^2+m_2^2+4\mu_I^2)+m_1^2m_2^2=0
\,.\ee
Using Eqs. (\ref{m_i's}) and the condensate solution $(\bs=f_{\pi},\bvp=0)$, the resulting spectrum has the following masses:
\be
Q_0^{(\sigma)}&=&|m_{\xi}|=\sqrt{2\lambda^2f_{\pi}^2+m_{\pi}^2}=m_{\sigma}\,,
\nonumber
\\
Q_0^{(\pi^0)}&=&m_{\pi}\,,
\nonumber
\\
{\left[Q_0^{(\pi^{\pm})}\right]}^2&=&
\frac{1}{2}\left(m_1^2+m_2^2+4\mu_I^2\right)\mp \frac{1}{2}\sqrt{
(m_1^2+m_2^2+4\mu_I^2)^2-4m_1^2m_2^2
}
\nonumber\\
&=&
m_{\pi}^2+\mu_I^2\mp \frac{1}{2}\sqrt{
4(m_{\pi}^2+\mu_I^2)^2-4(m_{\pi}^2-\mu_I^2)^2}
\nonumber\\
&=&(m_{\pi}\mp\mu_I)^2\,.
\label{spec-chiral}
\ee

\item \underline{Pion-condensed phase ($\bvp^2=(\bp_1^2+\bp_2^2)\ne 0$):}
Let us define radial and angular coordinates such that
\be
\bvp^2&=&(\bp_1^2+\bp_2^2)\stackrel{!}{=}\rho^2\ne 0
\\
\bp_1&=&\rho\cos\alpha
\\
\bp_2&=&\rho\sin\alpha
\,.
\ee
The equilibrium conditions then become:
\be
\lambda^2\bs(\bs^2+\rho^2-v^2)
-h&=&0
\label{EC-s-pin0}
\\
\left\{
\lambda^2
(
\bs^2
+\rho^2
-v^2
)
-\mu_I^2
\right\}\rho\cos\alpha
&=&0
\\
\left\{
\lambda^2
(
\bs^2
+\rho^2
-v^2
)
-\mu_I^2
\right\}\rho\sin\alpha
&=&0
\,,\label{EC-pi-pin0}
\ee
which are satisfied for $\bvp^2=\rho^2\ne 0$ if and only if:
\be
\lambda^2\bs(\bs^2+\rho^2-v^2)
-h&=&0
\\
\lambda^2(\bs^2
+\rho^2
-v^2
)
-\mu_I^2&=&0
\,,
\ee
so that
\be
\bs&=&\frac{h}{\mu_I^2}
\nonumber\\
\rho^2&=&\frac{\mu_I^2}{\lambda^2}+v^2-\frac{h^2}{\mu_I^4}
\nonumber\\
&\stackrel{(\ref{v2-sol})}{=}&
\frac{\mu_I^2-m_{\pi}^2}{\lambda^2}+f_{\pi}^2-\frac{h^2}{\mu_I^4}
\nonumber\\
&=&
\frac{\mu_I^2-m_{\pi}^2}{\lambda^2}+f_{\pi}^2\left[1-\frac{m_{\pi}^4}{\mu_I^4}\right]
\,.
\label{sols}
\ee
%

%
%

It should be noted that the equilibrium conditions are completely independent of the angle $\alpha$ (as the remaining symmetry under rotations around $\pi_3$ requires).
Physics does not depend on the direction of the condensate in the $\pi_1-\pi_2$ plane, which is defined by a choice of the coordinate system.
If we choose the direction of the pion condensate such that $\sin\alpha=0$ ($\cos\alpha=0$), the condensate occurs in a direction that is a linear combination of $\sigma$ and $\pi_1$ ($\pi_2$), while $\pi^0$ mass rises linearly with the chemical potential and the remaining pion stays massless at any $\mu_I$ for which this solution is the true equilibrium state of the theory. Independently of the value of $\alpha$ we follow the literature (\cite{Stephanov-book} and references therein) and name the quasiparticles in the condensed phase by a continuity criterium in the spectrum.

The minimum of the effective potential in the classical approximation in this case is:
\be
\left.V_{\rm cl}^{\rm (L\sigma M)}\right|_{\rm min, \bvp^2\ne 0}&=&
\frac{\lambda^2}{4}\left(\frac{\mu_I^2}{\lambda^2}\right)^2-h\frac{h}{\mu_I^2}-\frac{\mu_I^2}{2}\left\{
\frac{\mu_I^2-m_{\pi}^2}{\lambda^2}+f_{\pi}^2\left[1-\frac{m_{\pi}^4}{\mu_I^4}\right]
\right\}
\\
&=&
\frac{1}{4\lambda^2}\left[
\mu_I^4-2\mu_I^4+2m_{\pi}^2\mu_I^2
\right]
-f_{\pi}^2m_{\pi}^2\left[
\frac{m_{\pi}^2}{\mu_I^2}
+\frac{1}{2}\frac{\mu_I^2}{m_{\pi}^2}
-\frac{1}{2}\frac{m_{\pi}^2}{\mu_I^2}
\right]
\\
&=&
\frac{1}{4\lambda^2}\left[
-\mu_I^4+2m_{\pi}^2\mu_I^2
\right]
-f_{\pi}^2m_{\pi}^2\frac{1}{2}\left[
\frac{\mu_I^2}{m_{\pi}^2}
+\frac{m_{\pi}^2}{\mu_I^2}
\right]
\\
&\stackrel{(\ref{Vcl-pi0})}{=}&
\left.V_{\rm cl}^{\rm (L\sigma M)}\right|_{\rm min, \bvp^2=0}
-
\frac{1}{4\lambda^2}\left[m_{\pi}^2-\mu_I^2
\right]^2
-f_{\pi}^2m_{\pi}^2\frac{1}{2}\left[
\frac{\mu_I}{m_{\pi}}
-\frac{m_{\pi}}{\mu_I}
\right]^2
\,,
\ee
so that it is clear that it is smaller than $\left.V_{\rm cl}^{\rm (L\sigma M)}\right|_{\rm min, \bvp^2=0}$ for $\mu_I>m_{\pi}$. 
For $\mu_I<m_{\pi}$, $\rho^2<0$, the solution is unstable in the directions of the $\pi_1$ and $\pi_2$ fields, as can be verified by examining the quadratic terms in the expanded potential, Eq. (\ref{Vcl-exp}).

For this phase with pion condensation and $\sin\alpha=0$, the Eq. (\ref{det}) for obtaining the quasiparticle spectrum becomes:
\be
(Q_0^2-m_{3}^2)(Q_0^2)\left[
-m_{\xi}^2m_1^2+(2\lambda^2\bs\rho)^2-m_{\xi}^2(2\mu_I)^2
+Q_0^2\left(
m_{\xi}^2+m_1^2+(2\mu_I)^2
\right)-Q_0^4
\right]
&=&0
\,,
\nonumber\\
\ee
whose solutions are:
\be
Q_0&=&\sqrt{m_3^2}=|m_3|\,,
\\
Q_0&=&0\,
\,,
\ee
together with the solutions of the quadratic equation:
\be
-m_{\xi}^2m_1^2+(2\lambda^2\bs\rho)^2-m_{\xi}^2(2\mu_I)^2
+Q_0^2\left(
m_{\xi}^2+m_1^2+(2\mu_I)^2
\right)-Q_0^4&=&0
\,,
\ee
namely:
\be
Q_0^2&=&
\frac{1}{2}\left\{
m_{\xi}^2+m_1^2+(2\mu_I)^2\pm\sqrt{\left(-m_{\xi}^2+m_1^2+(2\mu_I)^2\right)^2+4(2\lambda^2\bs\rho)^2}
\right\}
\ee

Using the results for the condensate solutions, Eqs. (\ref{sols}) and (\ref{m_i's}), we arrive at:
\be
Q_0^{(a)}&=&|\mu_I|
\\
Q_0^{(b)}&=&0
\\
\left[Q_0^{(c,\pm)}\right]^2&=&
\frac{7}{2}\mu_I^2-m_{\pi}^2+\lambda^2f_{\pi}^2
\pm
\nonumber\\&&
\pm\sqrt{
\left[
\frac{5}{2}\mu_I^2-m_{\pi}^2+\lambda^2f_{\pi}^2\left(
1-2\frac{m_{\pi}^4}{\mu_I^4}
\right)
\right]^2
+4\lambda^2f_{\pi}^2
\frac{m_{\pi}^4}{\mu_I^4}
\left[
\mu_I^2-m_{\pi}^2+\lambda^2f_{\pi}^2\left(1-\frac{m_{\pi}^4}{\mu_I^4}\right)
\right]
}
\nonumber\\
\ee
At the transition ($\mu_I=m_{\pi}$), 
 these solutions continuously connect to the particle spectrum in the first phase (Eqs.(\ref{spec-chiral})). We will, for the sake of definiteness and connection with the literature, use the same names for the quasiparticles in the second phase, assuming continuity: $(a)\mapsto\pi^0$, $(b)\mapsto\pi^+$, $(c,+)\mapsto\sigma$ and $(c,-)\mapsto\pi^-$. Nevertheless, it should be emphasized that the states in the second phase are (apart from the neutral pion, which decouples) nontrivial mixings of the sigma and the charged pion fluctuation fields.

\end{itemize}


%
\begin{figure}[h!]
\centering\vspace{-1cm}\mbox{
\includegraphics[width=0.65\textwidth,angle=90]{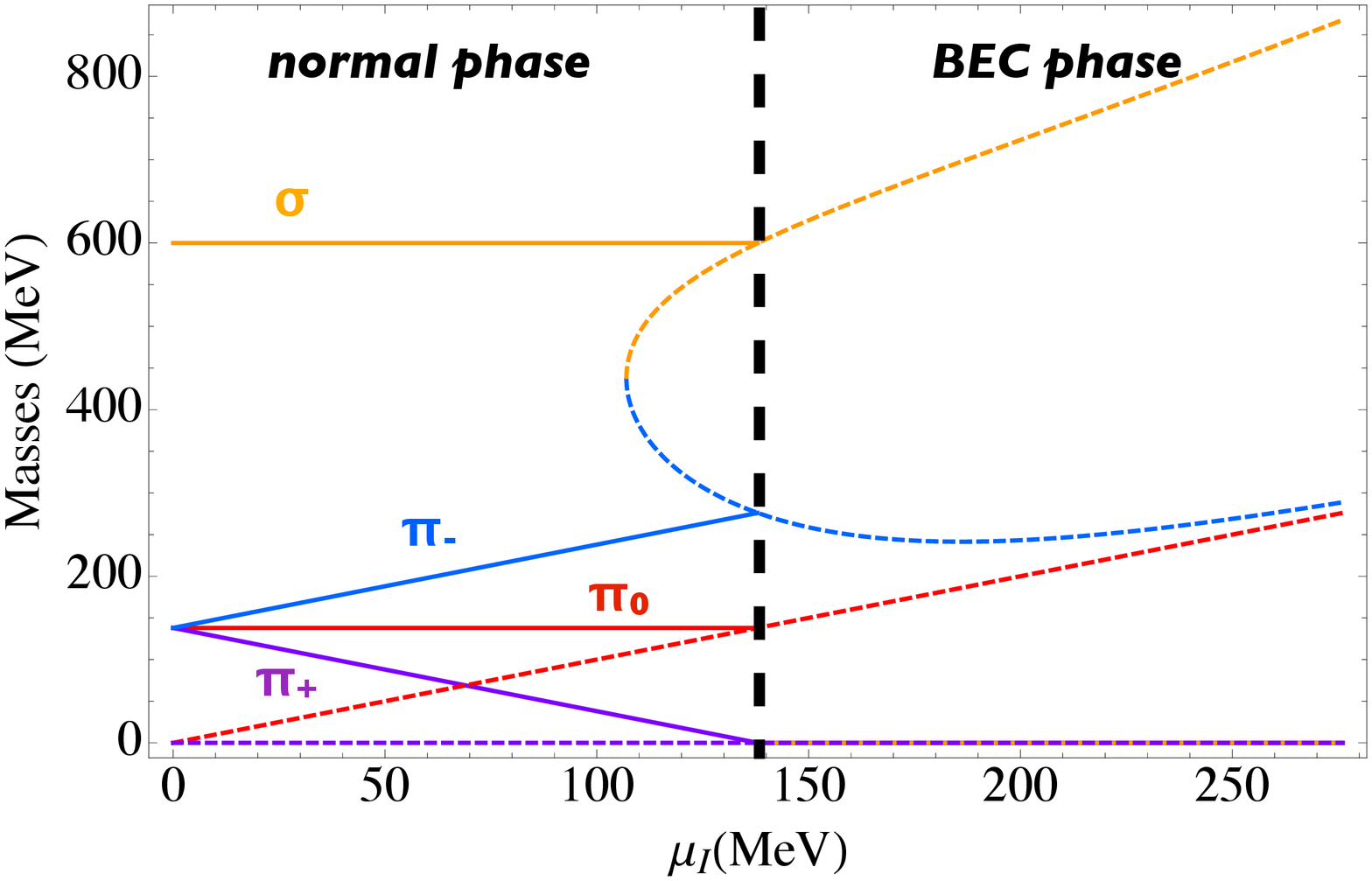}}
\vspace{-2cm}
\caption{Full lines are the masses in the normal phase, with standard chiral condensation, corresponding to, from top to bottom: $\sigma$, $\pi^-$, $\pi^0$ and $\pi^+$. The dashed lines are the masses of the quasi-particles of the phase with pion condensation.
}
\label{masses-twophases}
\end{figure}

The results for the mass spectrum in both chiral and pion condensed phases are summarized in Fig. \ref{masses-twophases}. In the normal phase, for low densities,
$\sigma$ and $\pi^0$ are independent of $\mu_I$, due to their invariance under isospin rotations around the $z$-axis. The free energy cost related to one charged pion, however, changes linearly with $\mu_I$. As discussed heuristically above, when $\mu_I=m_{\pi}=138~$MeV, the $\pi^{+}$ becomes effectively massless and its bosonic character causes a Bose-Einstein condensate to develop at higher densities. The phase transition is therefore of the second order. Due to the effective isospin rotation the ground state of the system suffers for $\mu_I>m_{\pi}$, originating the nonzero pion condensate,  all the fluctuations on top of it have clearly a nontrivial isospin dependence, as shown by the dashed lines.

\newpage

\section{The Functional Renormalization Group}

Before applying the Functional Renormalization Group (FRG)  to investigate nonperturbative aspects of relativistic Bose-Einstein condensation of pions, let us discuss in more fundamental grounds the formalism itself.
This section is not intended to be a thorough presentation of the subject, but rather a brief introduction to the main ideas and hypotheses that form the basis of FRG. Along the discussion, we shall demonstrate and/or justify the main results that will be used in what follows. For comprehensive texts the reader is referred to \cite{Berges:2000ew,Delamotte:2007pf,Blaizot:2008xx}.

The main aim of the formalism of the Functional Renormalization Group or non-Perturbative Renormalization Group is to provide a systematic framework for the inclusion of nonperturbative effects in field theory calculations. This is implemented via the construction of a set of effective actions $\Gamma_{\k}[\phi]$ which interpolate between the classical action $S[\phi]$ and the {\it full} effective action $\Gamma[\phi]$.

Within the FRG formalism, this set of interpolating effective actions is constructed in the form of a flow, i.e. a trajectory in the effective action space parameterized by the scale $\k$. It is determined via an exact renormalization group equation for the effective action $\Gamma_{\k}$, which actually encodes an infinite hierarchy of coupled exact renormalization group equations involving the $n-$point functions. 

The full solution of such a hierarchy is formally equivalent to solving the full interacting quantum field theory and as such is prohibitively complicated in the great majority of the physically interesting cases. An adequate truncation or approximation is needed to render the exact FRG flow a practical nonperturbative method. By adequate we mean an approximation that contains the relevant physical features of the system under investigation. Some of the most popular truncations and approximations are critically discussed in subsection \ref{FRGtool}.


In principle any $\k-$dependent deformation of a given theory defines an exact flow in the effective action space. Nevertheless, for any particular flow to be useful as a tool for computing nonperturbative effects some special features are required (or desired). Among them a smooth $\k\rightarrow 0$ limit and simplicity are relevant for pragmatic applications. The deformed theory at finite $\kappa$ should partially include the nonperturbative corrections, so that the $\k\rightarrow 0$ theory encodes the full nonperturbative result.  

In field theory, nonperturbative effects are usually connected to the extremes of momentum values: ultra-violet (UV) or infra-red (IR) modes. The UV regime is well-understood and the theory of renormalization provides a systematic method for taking these nonperturbative effects into account within the perturbative expansion of a renormalizable theory \cite{Peskin:1995ev}. The IR regime on the other hand seems to pose more severe difficulties for a pragmatic definition of a computable method, even at weak-coupling (as it happens, for instance, in thermal perturbation theory \cite{kapusta-gale}). It is reasonable therefore to define an exact flow that incorporates in a controlled and successive manner the IR modes.

With this in mind, the interesting $\k-$dependent deformation of the theory should affect differently UV and IR modes, including the effects of lower and lower momentum modes as the $\k\rightarrow 0$ limit is approached. The freedom of choice of the form of the deformation allows for choosing the simplest one, affecting only the quadratic part of the theory, or, equivalently, its propagator. 
The simplicity of the modification of the theory will translate into the exact FRG flow equation to be solved in the end. A quadratic deformation guarantees a one-loop structure in the final flow equation, as we shall verify in the next subsection.

Putting these features together, we conclude that one possible interesting definition of an exact flow is the one defined as the original theory with a modified propagator which suppresses IR modes with momentum $q\lesssim\k$.
Since the aim is that of interpolating between the classical action $S[\phi]$ and the full effective action $\Gamma[\phi]$, the trajectory of the FRG flow in effective action space should satisfy the following features at different scales $\k$:

\begin{center}
  \begin{tabular}{| c || c | }
    \hline
    $\k=0$ & all quantum fluctuations are included ($\Gamma_{\k=0}=\Gamma$) \\ \hline
     Intermediate $\k$ & IR modes ($q^2\lesssim \k^2$) suppressed; UV fluctuations ($q^2> \k^2$) included  \\ \hline
    $\k=\Lambda\to\infty$ &  all fluctuations are suppressed; physics is classical ($\Gamma_{\k=\Lambda}=S$) \\
    \hline
  \end{tabular}
\end{center}
%

%
%

Notice that the existence of such trajectory for a finite UV scale $\Lambda$ relies on a nontrivial hypothesis: the classical action must be the full effective action at some scale $\Lambda$, i.e., there exists a scale $\Lambda$ at (or above) which all fluctuations are indeed suppressed in the full theory.
Strictly speaking, this is in general not the case in quantum field theories.
From the experience with perturbative UV renormalization, we know that not even renormalizable theories present this exact property. When one renormalizes the theory,  integrating out the UV modes with wavelengths $\k\in [\Lambda,\infty\}$, it amounts to considering the same operators as in the original theory, but with dressed parameters.
Stated in a more precise way, the initial effective action $\Gamma_{\k=\Lambda}$ for the FRG flow is not the na\"ive classical theory, but rather an  action containing the same operators as the classical one with {\it dressed} coefficients.
Some remaining fluctuations with wavelengths $\k>\Lambda$ are encoded in the dressing of the parameters of the classical action.

All the subtleties and intricacies of the perturbative renormalization procedure are therefore implicitly included in the FRG formalism through its initial conditions. Once finite initial conditions are given and a sufficiently large initial scale $\Lambda$ is chosen, all results at the end of the flow ($\k=0$) are both finite and consistent.
%

\subsection{The exact flow equation\label{ExactFlowEq}}

Let us now illustrate the derivation of the exact flow equation that generates an interpolating trajectory as described above by considering a scalar field theory. The general guideline, however, is universal: starting from the classical deformation of the original theory through a $\k$-dependent quadratic term, one constructs (via a standard Legendre transform, including a convenient modification) the full effective action for the theory at the scale $\k$. The exact FRG equation is then obtained straightforwardly by taking an ordinary derivative of this definition with respect to the flow scale $\k$.

In euclidean space, a general scalar $O(N)$ model with $(\Phi^2)^2$ interaction is defined classically as:
\be
\mathcal{L}
&=&
\frac{1}{2}(\partial_{\mu}\Phi)(\partial_{\mu}\Phi)+\frac{1}{2}m^2\Phi^2+\frac{\lambda}{4!}(\Phi^2)^2
\,,
\ee
where the field is $\Phi=(\varphi_1,\dots,\varphi_N)$.

In momentum space,
\be
\Phi(x)&=&
\frac{1}{V^{1/d}}\,V\int_q {\rm e}^{i q \cdot x} \Phi(q)
\\
\delta(q)&=&\int d^dx \, {\rm e}^{iq\cdot x} 
\,,
\ee
so that the euclidean action reads:
\be
S[\Phi]&=&
\int d^dx \, \mathcal{L}
\nonumber\\
&=&
\frac{V^2}{V^{2/d}}\frac{1}{2}\int_{q}  \varphi_i(q)\left[q^2+m^2\right] \varphi_i(-q)
+\nonumber\\
&&
+\frac{V^4}{V^{4/d}}\frac{\lambda}{4!}\int_{q_1,\dots,q_4} \delta(q_1+q_2+q_3+q_4) 
\varphi_i(q_1)\varphi_i(q_2)\varphi_j(q_3)\varphi_j(q_4)
\,.
\ee

As discussed above, the modified theory is defined through the addition of a quadratic term to the action in order to suppress IR modes and leave UV modes unaffected, the threshold between these regimes being given by the flow parameter $\k$. These features are implemented by adding a term to the action $S\mapsto S_{\k}=S+\Delta S_{\k}$, with:
\be
\Delta S_{\k}[\Phi]&=&
\frac{V^2}{V^{2/d}}\frac{1}{2}\int_{q}  \varphi_i(q) R^{ij}_{\k}(q) \varphi_j(-q)
\,,\label{DeltaSk}
\ee
in which we wrote a general form without explicit $O(N)$ symmetry and the regulator $R^{ij}_{\k}(q)$ implements the desired suppression of IR modes with $q\lesssim \k$. Thus, typically, $R^{ij}_{\k}(q)\to \k^2$ when $q^2\ll \k^2$, and $R^{ij}_{\k}(q)\to 0$ when $q^2\gtrsim \k^2$. It is also usually convenient to use a symmetric regulator: $R^{ij}_{\k}(q)= \delta^{ij}R_{\k}(q)$.

The generating functional of connected Green's functions of the $\k-$modified theory is then given by:
\be
W_{\k}[J]&=&
\log
\int D\Phi\, \exp\left\{
-S[\Phi]-\Delta S_{\k}[\Phi]+\int d^4x\, \Phi(x)J(x)
\right\}
\,.\label{defWk}
\ee
Following the standard procedure, one can construct from $W_{\k}[J]$, via a Legendre Transform, the effective action of the $\k-$modified theory:
\be
\tilde{\Gamma}_{\k}[\phi]&=&-W_{\k}[J_{\phi}]+\int d^d x \, \phi\,J_{\phi}(x)
\label{tildeGammak}
\\
J_{\phi}(x) \quad&|&\quad \left.\frac{\delta W_{\k}[J]}{\delta J(x)}\right|_{J=J_{\phi}}
=\langle\Phi(x)\rangle_{\k,J_{\phi}}
\stackrel{!}{=}\phi
\,.\label{Jphi}
\ee
This functional has interesting properties, namely:
\begin{itemize}
\item in the absence of external currents, $\tilde{\Gamma}_{\k}[\phi]$ is extremized by the vacuum expectation value of the field $\phi=\langle\Phi\rangle$, where the average is defined using $W_{\k}$, i.e.:
\be
\left.\frac{\delta\tilde{\Gamma}_{\k}[\phi]}{\delta\phi}\right|_{J=0,\phi=\langle\Phi\rangle}=0
\,;
\ee
\item $\tilde{\Gamma}_{\k}[\phi]$ is the generating functional of one-particle irreducible correlation functions;
\item In the limiting case in which fluctuations are suppressed (classical limit; Landau-Ginzburg approximation), this effective action reduces to the original action of the $\k-$modified theory: $S+\Delta S_{\k}$.
\end{itemize}

However, as discussed above, the effective action we are interested in is actually one which interpolates between the full, exact theory at $\k=0$ and the original classical theory at large scales $\k=\Lambda$, i.e. $\Gamma_{\Lambda}[\phi]\to S[\phi]$.
While the first property is fulfilled by $\tilde{\Gamma}_{\k}$ (since $\Delta S_{\k=0}=0$), the second is not. In this vein, it is convenient to define a {\it Modified} Legendre Transform, subtracting the extra term $\Delta S_{\k}$ to reproduce the desired classical theory when fluctuations are suppressed without affecting the full quantum theory (the one at $\k=0$):
\be
\Gamma_{\k}[\phi]&=&-W_{\k}[J_{\phi}]+\int d^d x \, \phi\,J_{\phi}(x)-\Delta S_{\k}[\phi]\,,
\label{MLT}
\ee
and $J_{\phi}$ is still the source for which the vacuum expectation value of the field equals $\phi$, determined via Eq. (\ref{Jphi}).

The exact Renormalization Group equations satisfied by $\Gamma_{\k}$ can then be derived directly from this expression for the {\it Modified} Legendre Transform.

Taking the derivative of Eq. (\ref{MLT}) with respect to $\k$ we obtain:
\be
\partial_{\k}\Gamma_{\k}[\phi]
&=&
-\partial_{\k} \left\{W_{\k}[J_{\phi}]\right\}+\int d^d x \, \phi\,\partial_{\k}\left\{J_{\phi}(x)\right\}-\partial_{\k}\Delta S_{\k}[\phi]
\\
&=&
-\left.(\partial_{\k}W_{\k}[J])\right|_{J=J_{\phi}}
-\int d^dx\, \frac{\delta W_{\k}[J_{\phi}]}{\delta J_{\phi}}(\partial_{\k}J_{\phi})
+\int d^d x \, \phi\,\partial_{\k}\left\{J_{\phi}(x)\right\}-\partial_{\k}\Delta S_{\k}[\phi]
\nonumber \\
&\stackrel{(\ref{Jphi})}{=}&
-\left.(\partial_{\k}W_{\k}[J])\right|_{J=J_{\phi}}
-\partial_{\k}\Delta S_{\k}[\phi]
\,.
\ee
%

%
%

Using the definition of the generating functional of connected Green's functions $W_{\k}$, Eq. (\ref{defWk}), the right-hand side of the flow equation above can be rewritten as:
%
\begin{eqnarray}
\partial_{\k}\Gamma_{\k}[\phi]
&=&
-\,\frac{\int D\Phi\, 
\left(-\partial_{\k}\Delta S_{\k}[\Phi]\right)
\exp\left\{
-S[\Phi]-\Delta S_{\k}[\Phi]+\int d^4x\, \Phi(x)J_{\phi}(x)
\right\}}{{\rm e}^{W_{\k}[J_{\phi}]}}
-\partial_{\k}\Delta S_{\k}[\phi]
\nonumber\\
&=&
\frac{V^2}{V^{2/d}}\int_{q} \frac{\partial_{\k}R_{\k}(q)}{2}\delta_{ij} \Big[\langle\varphi_i(q)  \varphi_j(-q)\rangle_{J_{\phi}}
-\phi_i(q)\phi_j(-q)
\Big]\,,
\end{eqnarray}
where we have used the quadratic form of the (diagonal) regulator term (cf. Eq. (\ref{DeltaSk})) and the average represents:
\be
\langle\dots\rangle_{J_{\phi}}
&=&
\frac{\int D\Phi\, 
\left(\dots\right)
\exp\left\{
-S[\Phi]-\Delta S_{\k}[\Phi]+\int d^4x\, \Phi(x)J_{\phi}(x)
\right\}}{{\rm e}^{W_{\k}[J_{\phi}]}}
\,,
\ee
so that $\langle\varphi_i\rangle_{J_{\phi}}=\phi_i$ (cf. Eq. (\ref{Jphi})) and it becomes clear that:
\begin{eqnarray}
\partial_{\k}\Gamma_{\k}[\phi]
&=&
V^{\frac{d-2}{d}}\,V\int_{q} \frac{\partial_{\k}R_{\k}(q)}{2} \;\delta_{ij}\frac{\delta \phi_j}{\delta J_{\phi,i}}
\nonumber\\
&=&
V\int_{q} \frac{\partial_{\k}R_{\k}(q)\;\delta_{ij}}{2} \,V^{\frac{d-2}{d}}\,\left[\frac{\delta J_{\phi,i}}{\delta \phi_j}\right]^{-1}
\nonumber\\
&=&
{\rm Tr}\; V\int_{q} \frac{\partial_{\k}R^{ij}_{\k}(q)}{2} \;
\,V^{\frac{d-2}{d}}\,\left[\frac{\delta^2 (\Gamma_{\k}+\Delta S_{\k})}{\delta\phi_i\delta \phi_j}\right]^{-1}
\,,
\end{eqnarray}
where the full propagator of the $\k$-modified theory appears\footnote{The functional derivative notation is built in Fourier space, in which all fields are dimensionless by construction. Thus the overall dimensional factor $V^{\frac{d-2}{d}}=\sqrt{V}\;(d=4)$ that appears. At finite temperature, $V^{\frac{d-2}{d}}\mapsto \beta^2$ in our notation.}:
\be
G_{\k}^{ij}(q)&=&V^{\frac{d-2}{d}}\,\left[\frac{\delta^2 (\Gamma_{\k}+\Delta S_{\k})}{\delta\phi_i\delta \phi_j}\right]^{-1}
=
V^{\frac{d-2}{d}}\,\left[\frac{\delta^2 \Gamma_{\k}}{\delta\phi_i\delta \phi_j}
+\delta_{ij}R_{\k}(q)
\right]^{-1}
\,.\label{fullpropk}
\ee

The final form for the exact FRG flow equation for the effective action is therefore:
\be
\partial_{\k}\Gamma_{\k}[\phi]
&=&
\frac{1}{2}\, {\rm Tr}\;V
\int_{q} \;[\partial_{\k}R_{\k}(q)]
\, G_{\k}(q)
\,,
\label{FLOWEQ}
\ee
which can be viewed as a simple one-loop bubble diagram
with an insertion $\partial_{\k}R_{\k}(q)$ and the line representing the full propagator at the scale $\k$.


We have stated above that in the FRG framework renormalization is encoded in the initial conditions of the flow. Once initial parameters at the large scale $\k=\Lambda$ are given, the physical results, obtained in the end of the flow, are all finite. This is realized by the finiteness of the flow equation itself. Indeed, in Eq. (\ref{FLOWEQ}) both IR and UV divergences are absent. The good IR behavior is guaranteed by construction due to the presence of the regulator in the full propagator $G_{\k}$ maintaining the modes in the deep IR with a large mass $\sim\k$. Another property of the regulator is responsible for the taming of UV divergences in Eq. (\ref{FLOWEQ}). The insertion $\partial_{\k}R_{\k}(q)$ in the flow equation plays the role of an effective UV cutoff, due to the requirement that the regulator leaves the UV modes unaffected.

The simplistic appearance of the FRG flow equation in Eq. (\ref{FLOWEQ}) is deceiving. Fully solving this exact flow equation is equivalent to solving the whole interacting quantum field theory under consideration. This is of course a prohibitively complicated task in most interesting cases. To apprehend the complexity encoded in Eq. (\ref{FLOWEQ}) it is crucial to note that $\Gamma_{\k}$ and $G_{\k}$ are the exact effective action and propagator of the full quantum theory at the scale $\k$. Therefore to solve the equation for $\Gamma_{\k}$ one must first know the full propagator $G_{\k}$ exactly. One can also obtain a flow equation for $G_{\k}$ from Eq. (\ref{FLOWEQ}) by taking two functional derivatives with respect to the fields. This procedure generates an equation containing the full 3- and 4-point functions of the quantum theory on the right-hand side, which are of course not known {\it a priori}. Proceeding in this line, one concludes that the compact FRG flow equation shown in Eq. (\ref{FLOWEQ}) represents actually an infinite hierarchy of coupled flow equations for all $n$-point functions of the full quantum theory being analyzed.

%
%

\subsection{FRG as a practical tool: truncations and approximations\label{FRGtool}}

As it stands, the flow equation in Eq. (\ref{FLOWEQ}) is a formal result and cannot be considered a nonperturbative method to tackle interacting physical systems.
An adequate truncation or approximation is needed to render the exact FRG flow a powerful nonperturbative tool. In practice, this amounts to transforming this infinite hierarchy of complicated flow equations into a small set of sufficiently simple equations.
Such a drastic procedure should in principle be implemented through the knowledge of the relevant properties of each system under investigation, in order to obtain a sensible physical framework in the end.

There are two mainstreams (cf. e.g. \cite{Berges:2000ew,Delamotte:2007pf,Blaizot:2008xx}) of simple-minded approximations in the literature that are often combined: the derivative expansion and the truncation of the hierarchy of equations for the $n-$point functions by setting them to zero for $n>n_0$, with $n_0$ small (usually 4 or 6).

The derivative expansion corresponds to a specific hypothesis for the form of the $\k$-modified effective action $\Gamma_{\k}[\{\phi(x)\}]$. It assumes that an expansion on gradients of the fields is convergent and well-defined. The most popular particular case of derivative expansion is the so-called Local Potential Approximation (LPA), which correspond to the leading-order term. In LPA, the functional effective action is reduced to an effective potential, which is an ordinary function of constant and uniform classical fields: $\Gamma_{\k}[\{\phi(x)\}]\approx V_{\k}\left(\bar{\phi}\right)$. The flow equation therefore becomes an ordinary differential equation involving derivatives with respect to $\k$ and the values of the classical fields $\{\phi\}$.

In many applications in the literature one simplifies further the LPA ansatz by truncating the taylor expansion of the LPA effective potential with respect to the fields. In this way, the FRG flow is transformed into a finite small set of coupled ordinary differential equations for the coefficients of the potential.

Though seemingly a very drastic simplification of the full FRG flow, the LPA ansatz has been shown to encode different nontrivial contributions, coming from the resummation of entire classes of diagrams in the perturbative framework. For example, it can be demonstrated \cite{Blaizot:2008xx,D'Attanasio:1997ej} that
 the exact diagrammatic computation in the large-N limit of n-point functions of the $\k-$modified $O(N)$ field theory satisfies the corresponding FRG flow equations in LPA.
The large-N approach is well-know as an alternative to the weak-coupling perturbative expansion, bringing about nonperturbative effects which are not present in the latter.
The same is therefore true for the LPA version of FRG.

Nevertheless, for physical phenomena in which interactions with nontrivial momentum dependence plays an important role, LPA may not be a good approximation. This proved to be the case for nonrelativistic Bose-Einstein condensation of cold atoms. The shift of the condensation critical temperature has been successfully  described \cite{Blaizot:2004qa}
 within a FRG analysis with a more sophisticated approximation scheme for the flow known as the BMW approximation \cite{Blaizot:2005xy}, whose main feature is the inclusion of momentum dependence for the $n-$point functions.

\section{FRG analysis of pion BEC}

The specific physical problem we are interested in is relativistic Bose-Einstein condensation of pions in isospin dense media. In analogy with what was done in the nonrelativistic case in cold-atoms physics, we would like to understand the role played by (nonperturbative) interactions in the second-order transition between the normal and condensed phases.

Being a renormalization group based technique, FRG should describe well the physics of a second order phase transition. The IR phenomena intrinsically present near criticality (due to the massless zero mode that condenses) is exactly the type of nonperturbativeness 
that the flow was constructed to deal with.

The fact that this method was successfully applied to the nonrelativistic case (where extensive and well-controlled experiments are available) suggests that FRG should be an adequate approach to this relativistic version. The usage of the same semi-analytic technique in both cases will allow us to identify similarities and differences between the nonrelativistic and relativistic BEC, so that in principle a lot can be learned about the fundamental features of BEC.


\subsection{Results within the Local Potential Approximation\label{FRGpion}}

\begin{figure}[h!]
\centering\mbox{
\includegraphics[width=0.6\textwidth]{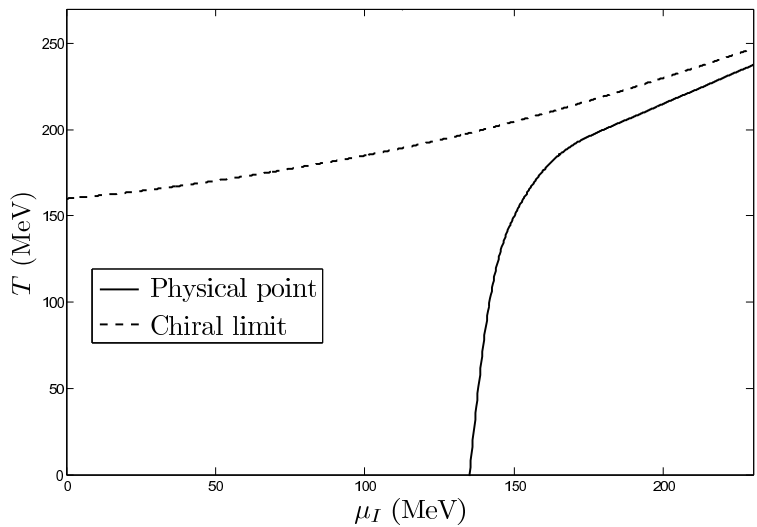}}
\caption{[extracted from Ref. \cite{Svanes:2010we}] T-$\mu_I$ phase diagram within LPA.}
\label{SvanesPhDiag}
\end{figure}

Adopting the pion chiral effective model discussed in Subsection \ref{chiralPionModel}, a FRG analysis using LPA was implemented to describe the BEC transition in the temperature {\it versus}  isospin chemical potential plane \cite{Svanes:2010we}.  The result is shown in Figure \ref{SvanesPhDiag}.


%

Consistently with the heuristic argument in the beginning of this chapter, BEC condensation of pions occurs for any nonzero isospin chemical potential in the chiral limit. At the physical point, the pion mass is nonzero $m_{\pi}=138~$MeV, so that condensation takes place only when $\mu_I>m_{\pi}$. Actually, as stated in the paper \cite{Svanes:2010we}, the second-order phase transition does not occur at $\mu_I=m_{\pi}$, but slightly below it. Another puzzling result is the transition line in the limit of $T\to 0$: it does not appear to be perpendicular to the $\mu_I$ axis as commonly happens.

The purpose of our work is to understand these issues and, in particular, find out whether they are unphysical artefacts of the nonperturbative method and approximations used.

\section{Going beyond LPA in a toy model}

Let us analyze the same phenomenon of relativistic Bose-Einstein condensation, but in a simple field-theory model: a complex scalar field $\pi=(\pi^1+i \pi^2)/\sqrt{2}$ with real mass ($m^2>0$) and $U(1)$ symmetry at finite density. 
Our goal is to study the issues encountered in the LPA computation of the phase diagram in the $T-\mu_I$ plane in a simpler theory that contains most of the features of interest. 
In what follows, physical similarities of this toy model and the chiral model presented above are made explicit, especially in what concerns the phase transition due to condensation and the fluctuations that should play an important role in the vicinity of criticality. This toy model will allow us to go beyond LPA in a more controlled fashion and address the questions concerning the quality of the nonperturbative method and its approximations.

The euclidean action $S_E$ (with the partition function given by $Z={\rm Tr}\exp (-S_E)$) including the $U(1)$ charge conservation constraint is:
\be
S_E&=&
\int d^4x
\left\{
-(\partial_0+i \mu_I)\pi^*(\partial_0-i \mu_I)\pi
+(\nabla \pi^*)\cdot(\nabla \pi)+
m^2\pi^*\pi+\lambda^2(\pi^*\pi)^2
\right\}\, ,
\ee
which can be rewritten as:
\be
S_E&=&
\int d^4x
\left\{
\partial_{\nu}\pi^*\partial_{\nu}\pi
-i\mu_I\left[\pi^*\partial_0\pi-\pi\partial_0\pi^*\right]
+(m^2-\mu_I^2)\pi^*\pi
+\lambda^2(\pi^*\pi)^2
\right\}\, ,
\ee
or, in terms of the real ($\pi^1/\sqrt{2}$) and imaginary ($\pi^2/\sqrt{2}$) parts of the field,
\be
S_E&=&
\int d^4x
\Big\{
\frac{1}{2}\partial_{\nu}\pi^1\partial_{\nu}\pi^1
+\frac{1}{2}\partial_{\nu}\pi^2\partial_{\nu}\pi^2
+\mu_I\left[\pi^1\partial_0\pi^2-\pi^2\partial_0\pi^1\right]
+\nonumber\\
&&
+\frac{(m^2-\mu_I^2)}{2}[(\pi^1)^2+(\pi^2)^2]
+\frac{\lambda^2}{4}[(\pi^1)^2+(\pi^2)^2]^2
\Big\}\, .
\label{SE}
\ee
Comparing the latter form with the problem of the linear sigma model at finite isospin density, one can see that this is a simplified version of it in which we consider only the charged pion directions and $-\lambda^2 v^2\mapsto m^2$.

\subsection{Condensates and fluctuation spectrum -- classical results}

With the aim of understanding the phase structure of the model and the quasiparticle spectrum at finite density, we can rewrite further the effective action as an expansion around (constant and uniform) condensates, parameterized by the {\it real} parameters $\rho$ and $\theta$ as follows,
\be
\pi^1(x)&=&\rho\cos\theta+\eta^1(x)\,,
\\
\pi^2(x)&=&\rho\sin\theta+\eta^2(x)\,,
\ee
where $\eta_i$ are the fluctuation fields around the condensates. One obtains [def.: $\eta^{\theta}\equiv (\eta^1\cos\theta+\eta^2\sin\theta)$]:
%
%
\be
S_E&=&
\beta V\Bigg(
\frac{(m^2-\mu_I^2)}{2}\rho^2 
+\frac{\lambda^2}{4}\rho^4
\Bigg)
+
\nonumber\\
&&
+
\int d^4x
\Bigg\{
\Big[
(m^2-\mu_I^2)\rho
+\lambda^2\rho^3
\Big]\eta^{\theta}
+
\nonumber\\
&&\quad+
\frac{1}{2}\partial_{\nu}\eta^1\partial_{\nu}\eta^1
+\frac{1}{2}\partial_{\nu}\eta^2\partial_{\nu}\eta^2
+\mu_I\left[\eta^1\partial_0\eta^2-\eta^2\partial_0\eta^1\right]
+\nonumber\\
&&
+\frac{(m^2-\mu_I^2)}{2}[
(\eta^1)^2+(\eta^2)^2]
+\frac{\lambda^2}{4}\Big[
4\rho^2(\eta^{\theta})^2+2\rho^2[(\eta^1)^2+(\eta^2)^2]\Big]
+\nonumber\\
&&
+\lambda^2\rho\eta^{\theta}[(\eta^1)^2+(\eta^2)^2]
+\frac{\lambda^2}{4}[(\eta^1)^2+(\eta^2)^2]^2
\Bigg\}\, .
\label{SE-exp2}
\ee
where we have omitted a term of the form $\propto\int d^4x\; \partial_0\eta$, which gives only a surface contribution. The second line yields the equilibrium condition for the condensate solution:
\be
&&\bar\rho(m^2-\mu_I^2)+\lambda^2\bar\rho^3=0
\quad \Rightarrow
\quad
\bar\rho=0 \quad {\textrm or} \quad \bar\rho^2=\frac{\mu_I^2-m^2}{\lambda^2}
\,.
\label{condensatesToy}
\ee
Since $\bar\rho \in \mathbb{R}$ by construction (and $\lambda^2>0$ to guarantee the stability of the potential), the nonzero condensate solution is only possible when $(\mu_I^2-m^2)>0$. Notice that in the absence of self-interactions ($\lambda^2=0$) condensation is only possible at $\mu_I^2=m^2$ (this reproduces Kapusta's result shown in the $\mu-$axis of Fig. 1 of Ref. \cite{Kapusta:1981aa}) and the theory becomes badly defined for chemical potentials above it (since the positivity of the action is violated).

In Fourier space, 
the effective action becomes:
\be
\frac{S_E}{\beta V}&=&
\frac{(m^2-\mu_I^2)}{2}\rho^2 
+\frac{\lambda^2}{4}\rho^4
+
\nonumber\\
&&
+
\frac{1}{\beta V}\int d^4x
\Bigg\{
\Big[
(m^2-\mu_I^2)\rho
+\lambda^2\rho^3
\Big]\eta^{\theta}\Bigg\}
+
\nonumber\\
&&
+\sumint_q\frac{\beta^2}{2}
\left(
\begin{array}{cccc}
 \eta_1(Q) & \eta_2(Q)\\
\end{array} 
\right)
\mathcal{G}^{-1}(Q)
\left(
\begin{array}{c}
\eta_1(-Q) \\ \eta_2(-Q) 
\end{array} 
\right)
+\nonumber\\
&&
+\frac{1}{\beta V}\int d^4x\, \Bigg\{\lambda^2\rho\eta^{\theta}[(\eta^1)^2+(\eta^2)^2]
+\frac{\lambda^2}{4}[(\eta^1)^2+(\eta^2)^2]^2
\Bigg\}
\, ,
\label{SE-expFT}
\ee
where we have defined the matrix of the inverse propagator (def.: $u_c\equiv \lambda^2\rho^2(1+2\cos^2\theta)$ and $u_s\equiv \lambda^2\rho^2(1+2\sin^2\theta)$) for the fluctuations around the condensate solution:
\be
\mathcal{G}^{-1}
=
\left( \begin{array}{cc}
-Q_0^2+(\vec{q})^2+m^2-\mu_I^2+u_c& \lambda^2\rho^2\sin(2\theta)-2i\mu_IQ_0 \\
\lambda^2\rho^2\sin(2\theta)+2i\mu_IQ_0 & -Q_0^2+(\vec{q})^2+m^2-\mu_I^2+u_s\\
\end{array} \right)\,.
\nonumber\\
\label{G-1(Q)}
\ee
The eigenvalues of the inverse propagator are directly obtained:
\be
g^{-1}_{\pm}&=&-\left(\sqrt{Q_0^2+\frac{\lambda^4\rho^4}{4\mu_I^2}}\mp \mu_I\right)^2+(\vec{q})^2+m^2+2\lambda^2\rho^2\left(1+\frac{\lambda^2\rho^2}{8\mu_I^2}\right)
\,.
\ee
%


The dispersion relations for the quasiparticles are then given by the zeros of these eigenvalues:
\be
\omega^2_{\pm}(\vec{q})&=&(\vec{q})^2+m^2+\mu_I^2+2\lambda^2\rho^2
\pm\sqrt{\lambda^4\rho^4+4\mu_I^2[(\vec{q})^2+m^2+2\lambda^2\rho^2]}
\,.
\ee
In particular, the standard result for the dispersion relations in the absence of condensates ($\rho=0$) is consistently reproduced:
\be
\lim_{\rho\to 0}
\omega^2_{\pm}(\vec{q})&=&
\left[ \sqrt{(\vec{q})^2+m^2} \pm \mu_I\right]^2
\,.
\ee

When $\mu_I>m$, the equilibrium condition for the true minimum yields $\rho^2=\bar\rho^2=(\mu_I^2-m^2)/\lambda^2$ (cf. Eq. (\ref{condensatesToy})) and the dispersion relations in this condensed phase become:
\be
\lim_{\rho\to \bar\rho\ne 0}
\omega^2_{\pm}(\vec{q})&=&
(\vec{q})^2+3\mu_I^2-m^2
\pm\sqrt{(\mu_I^2-m^2)^2+4\mu_I^2[(\vec{q})^2+2\mu_I^2-m^2]}
\nonumber\\
&=&
(\vec{q})^2+3\mu_I^2-m^2
\pm\sqrt{(3\mu_I^2-m^2)^2+4\mu_I^2(\vec{q})^2}
\,,
\ee
whose associated mass spectrum, obtained in the limit $\vec{q}\to 0$, is: $m_a^2=2(3\mu_I^2-m^2)$ and $m_b^2=0$. The mode $b$ is clearly the massless Goldstone boson in the condensed phase. 

\begin{figure}[htb]
\centering
\vspace{-1cm}\mbox{
\includegraphics[width=0.6\textwidth,angle=90]{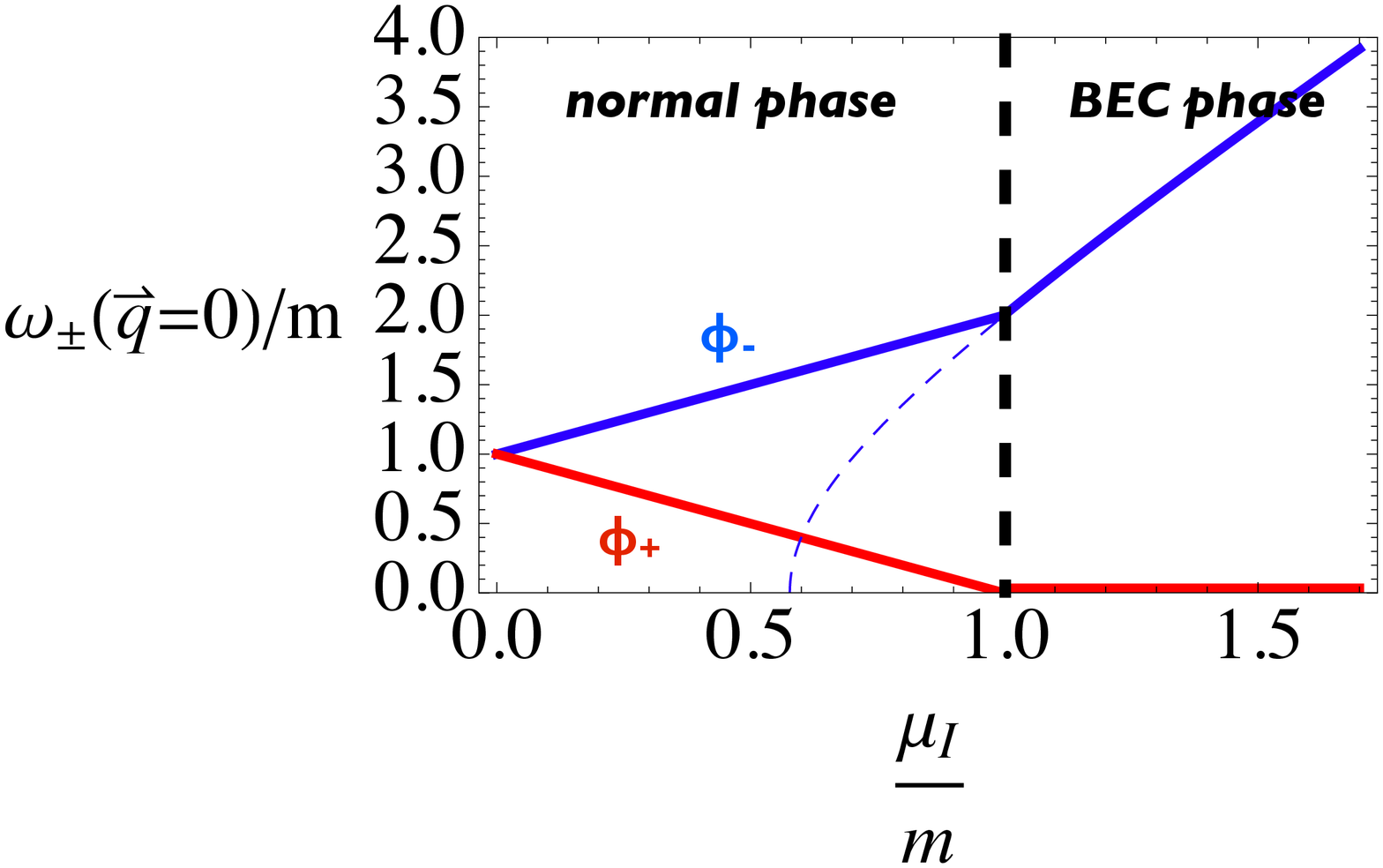}}
\vspace{-1.5cm}
\caption{Mass spectrum as a function of the chemical potential $\mu_I$. Units of mass $m$.}
\label{spec}
\end{figure}

Fig. \ref{spec} displays the mass spectrum as a function of the chemical potential.
Comparison with the spectrum of quasiparticles obtained for the chiral pion model, Figure 
\ref{masses-twophases}, shows that not only the BEC transition appears in this simplified model as a second-order one, but also the spectrum of fluctuations is similar to that of the charged pions. Therefore, this toy model contains the relevant features of relativistic BEC, including the spectrum of quasiparticles directly involved in it. Moreover, the extra complication regarding chiral spontaneous symmetry breaking is absent, allowing for a clean study of BEC physics and the role of interactions in it.

\subsection{FRG analysis within LPA}

Now, in order to study nonperturbative effects, we will implement a FRG flow. As discussed in detail in subsection \ref{ExactFlowEq}, a standard procedure of deforming the theory yields the FRG flow equation, Eq. (\ref{FLOWEQ}), for the $\kappa-$dependent effective action:
\be
\partial_{\kappa}\Gamma_{\kappa}[\pi^i]=\frac{1}{2}{\rm Tr}\left[
\partial_{\kappa}R_{\kappa} \left[\mathcal{G}^{-1}+R_{\kappa}\right]^{-1}
\right]
\,,
\label{flow-general}
\ee
where $\mathcal{G}$ is the {\it full} propagator matrix,
\be
\left[\mathcal{G}^{-1}(Q)\right]^{ij}&=&\frac{1}{\beta^2}\;\frac{\delta^2 \Gamma_{\kappa}}{\delta\pi^i(Q)\delta\pi^j(-Q)}
\ee
and $R_{\kappa}$ is the regulator, which we choose to be of the Litim form \cite{Berges:2000ew}: $R_{\kappa}(\vec{q}^2)=(\kappa^2-\vec{q}^2)\theta(\kappa^2-\vec{q}^2)$.

To proceed  further and obtain a closed, explicit form for the flow equation, one has to assume a form for the $\kappa-$dependent effective action $\Gamma_{\k}$, from which one can write both the left hand side and the full propagator itself. In practice, one must select an approximation or truncation scheme to work with.


First, it is interesting to analyze the toy model within the Local Potential Approximation. This serves two main purposes: $(i)$ to compare further the findings within the toy model with the ones from the chiral model for pion condensation \cite{Svanes:2010we} (cf. subsection \ref{FRGpion}) and $(ii)$ to provide a reference for comparison with the on-going investigations beyond LPA \cite{RelatBEC-FRG-wip}.

Let us therefore consider the LPA ansatz for the flowing effective action, i.e. in the form of the classical euclidean action as described above, but with $\kappa-$dependent parameters instead of $m^2,\lambda^2,\mu_I$ as follows:
\be
\Gamma_{\kappa}[\pi^i]&=&
\int d^4x
\Big\{
\frac{1}{2}\partial_{\nu}\pi^1\partial_{\nu}\pi^1
+\frac{1}{2}\partial_{\nu}\pi^2\partial_{\nu}\pi^2
+\mu_{\kappa}\left[\pi^1\partial_0\pi^2-\pi^2\partial_0\pi^1\right]
+\nonumber\\
&&
+\frac{(m_{\kappa}^2-\mu_{\kappa}^2)}{2}[(\pi^1)^2+(\pi^2)^2]
+\frac{\lambda_{\kappa}^2}{4}[(\pi^1)^2+(\pi^2)^2]^2
\Big\}
\nonumber\\
&\stackrel{!}{=}&
\int d^4x
\Big\{
\frac{1}{2}\partial_{\nu}\pi^1\partial_{\nu}\pi^1
+\frac{1}{2}\partial_{\nu}\pi^2\partial_{\nu}\pi^2
+\mu_{\kappa}\left[\pi^1\partial_0\pi^2-\pi^2\partial_0\pi^1\right]
+ V_{\kappa}\left(\alpha\right)
\, ,
\label{Gammakappa}
\ee
where we have defined $\alpha=(\pi^1)^2+(\pi^2)^2$ and
\be
V_{\kappa}\left(\alpha\right)&=&
\mathcal{V}_{\kappa}
+
\frac{(m_{\kappa}^2-\mu_{\kappa}^2)}{2}\, \alpha +\frac{\lambda_{\kappa}^2}{4}\, \alpha^2
\,.
\ee
Notice that the value of effective potential at zero field ($\mathcal{V}_{\kappa}$) will be in general a nontrivial function of temperature T and chemical potential $\mu$, being directly related to the pressure of the system (and its derivative with respect to $\mu$, to the density).

Partially transforming to Fourier space, we have:
\be
\Gamma_{\kappa}[\pi^i]&=&
\sumint_q\frac{\beta^2}{2}
\left(
\begin{array}{cccc}
 \pi^1(Q) & \pi^2(Q)\\
\end{array} 
\right)
\mathcal{K}(Q)
\left(
\begin{array}{c}
\pi^1(-Q) \\ \pi^2(-Q) 
\end{array} 
\right)
+
\int d^4x
\; V_{\kappa}\left(\alpha\right)
\, ,
\label{Gammakappa2}
\ee
where
\be
\mathcal{K}(Q)
&=&
\left( \begin{array}{cc}
-Q_0^2+(\vec{q})^2& -2i\mu_{\kappa}Q_0 \\
2i\mu_{\kappa}Q_0 & -Q_0^2+(\vec{q})^2\\
\end{array} \right)\,.
\ee

The full inverse propagator associated with this effective action is then:
\be
\mathcal{G}^{-1}(Q)&=&\frac{1}{\beta^2}\frac{\delta^2\Gamma_{\kappa}}{\delta\pi^a(Q)\delta\pi^b(-Q)}
\nonumber\\
&=&\mathcal{K}(Q)+
\int \frac{d^4x}{\beta V}
\Bigg\{
4 V''_{\kappa}(\alpha)
\pi^a(x)\pi^b(x)
+2V'_{\kappa}(\alpha)
\delta^{ab}
\Bigg\}
\,.
\ee
In the case of constant fields (consistently with the LPA ansatz), it simplifies to:
\be
\mathcal{G}^{-1}(Q)
=
\left( \begin{array}{cc}
-Q_0^2+(\vec{q})^2+4 V''_{\kappa}(\alpha)
(\pi^1)^2+2V'_{\kappa}(\alpha) & 4 V''_{\kappa}(\alpha)
\pi^1\pi^2-2i\mu_{\kappa}Q_0 \\
4 V''_{\kappa}(\alpha)
\pi^1\pi^2+2i\mu_{\kappa}Q_0 & -Q_0^2+(\vec{q})^2+4 V''_{\kappa}(\alpha)
(\pi^2)^2
+2V'_{\kappa}(\alpha)\\
\end{array} \right)\,,
\nonumber\\
\label{G-1LPA}
\ee
while the effective action reduces to the potential (note that one of the chemical potential terms involves derivatives and goes to zero in this approximation):
\be
\frac{\Gamma_{\kappa}[\pi^i]}{\beta V}
&=&
V_{\kappa}(\alpha)
\,.
\ee

Using the Litim regulator $R_{\kappa}(\vec{q}^2)=(\kappa^2-\vec{q}^2)\theta(\kappa^2-\vec{q}^2)$, we obtain:

\be
\partial_{\kappa}R_{\kappa}&=&
2\kappa\theta(\kappa^2-\vec{q}^2)
\,,
\\
&&\nonumber
\\
\left[\mathcal{G}^{-1}+R_{\kappa}\right]^{-1}
&=&
\left[\mathcal{G}^{-1}\Big(Q_0,\vec{q}^2\mapsto \vec{q}^2+(\kappa^2-\vec{q}^2)\theta(\kappa^2-\vec{q}^2)\Big)\right]^{-1}
\,,
\ee

\noindent so that the flow equation becomes:

\be
\partial_{\kappa}V_{\kappa}(\alpha)
&=&
\frac{1}{2}\sumint_Q
~2\kappa\theta(\kappa^2-\vec{q}^2)~
{\rm Tr}\left\{
\left[\mathcal{G}^{-1}\Big(Q_0,\vec{q}^2\mapsto \vec{q}^2+(\kappa^2-\vec{q}^2)\theta(\kappa^2-\vec{q}^2)\Big)\right]^{-1}
\right\}
\\
&=&
\sum_{Q_0}
\frac{4\pi}{(2\pi)^3} 
\kappa
\int_{0}^{\kappa} dq q^2
{\rm Tr}\left\{
\left[\mathcal{G}^{-1}\Big(Q_0,\vec{q}^2\mapsto \kappa^2\Big)\right]^{-1}
\right\}
\\
&=&
\frac{4\pi}{(2\pi)^3} 
\kappa
~\frac{\kappa^3}{3}
\sum_{Q_0}
{\rm Tr}\left\{
\left[\mathcal{G}^{-1}\Big(Q_0,\vec{q}^2\mapsto \kappa^2\Big)\right]^{-1}
\right\}
\,,
\label{flowVkappa}
\ee

\noindent and we are left with the Matsubara sum of the trace of the full propagator. The trace, in turn, can be computed in a straightforward way from the expression for $\mathcal{G}^{-1}$, Eq. (\ref{G-1LPA}). The result is [defs.: $v_{\kappa}\equiv \kappa^2+2 \alpha V''_{\kappa}
+2V'_{\kappa} $ and $w_{\kappa}\equiv \sqrt{ \mu_{\kappa}^2(\kappa^2+2V'_{\kappa})
+(\mu_{\kappa}^2+\alpha V''_{\kappa})^2}$]:
\be
{\rm Tr}\left\{
\mathcal{G}\Big(Q_0,\kappa^2\Big)
\right\}
&=&
2\,\frac{-Q_0^2+v_{\kappa}}{[-Q_0^2+v_{\kappa}]^2-4\alpha^2 [ V''_{\kappa}]^2-4\mu_{\kappa}^2Q_0^2}
\\
&=&
2\, \frac{-Q_0^2+v_{\kappa}}{[Q_0^2-(v_{\kappa}+2\mu_{\kappa}^2
+
2w_{\kappa})][Q_0^2-(v_{\kappa}+2\mu_{\kappa}^2
-
2w_{\kappa})]}
\nonumber\\
&=&
\left(\frac{\mu_{\kappa}^2}{w_{\kappa}}
+1\right)\frac{1
}{-Q_0^2+E_+^2}
+\left(-\frac{\mu_{\kappa}^2}{w_{\kappa}}
+
1\right)\frac{1}{-Q_0^2+E_-^2}
\,,
\ee
where we have defined:
\be
E_{\pm}^2
&=&
v_{\kappa}+2\mu_{\kappa}^2
\pm
2w_{\kappa}
\\
&=&
\kappa^2+2 \alpha V''_{\kappa}
+2V'_{\kappa}+2\mu_{\kappa}^2
\pm
2\sqrt{ \mu_{\kappa}^2(\kappa^2+2V'_{\kappa})
+(\mu_{\kappa}^2+\alpha V''_{\kappa})^2}
\,.
\ee

Explicitly, the flow equation for $V_{\kappa}$, Eq. (\ref{flowVkappa}) becomes:
\be
\partial_{\kappa}V_{\kappa}
&=&
\frac{\kappa^4}{6\pi^2} 
\left\{
\left(\frac{\mu_{\kappa}^2}{w_{\kappa}}
+1\right)
\sum_{Q_0}
\frac{1
}{-Q_0^2+E_+^2}
+\left(-\frac{\mu_{\kappa}^2}{w_{\kappa}}
+
1\right)
\sum_{Q_0}\frac{1}{-Q_0^2+E_-^2}
\right\}
\label{flowVkappa-Mat}
\,.
\ee
It is therefore clear that the Matsubara sums to be evaluated in the flow equations for $V_{\kappa}$ and its derivatives $V^{(n)}_{\kappa}$ will all be of the form
\be
\sum_{Q_0}
f_n(E^2)
&=&\sum_{Q_0}\frac{1}{(-Q_0^2+E^2)^n}
\,,
\ee
with
\be
\sum_{Q_0}
f_1(E^2)
&=&
\sum_{Q_0}
\frac{1}{-Q_0^2+E^2}
\nonumber\\
&=&
\frac{1}{2\sqrt{E^2}}\Big[
n_b(\sqrt{E^2})
-n_b(-\sqrt{E^2})
\Big]
=
\frac{1}{2\sqrt{E^2}}\Big[
1+2\,n_b(\sqrt{E^2})
\Big]
\,,
\ee
and:
\be
\sum_{Q_0} f_n(E^2)
&=&
(-1)^{n-1} \frac{1}{(n-1)!}~ [\partial_{E^2}]^{n-1} \,
\sum_{Q_0} f_1(E^2)
\,,
\ee
where $n_b(\omega)=[1-\exp(\beta \omega)]^{-1}$ is the Bose-Einstein distribution.

\vspace{.5cm}

The final closed form for the flow of the effective potential $V_{\kappa}$ in LPA is:
\be
\partial_{\kappa}V_{\kappa}
&=&
\frac{\kappa^4}{12\pi^2} 
\bigg\{
\left(\frac{\mu_{\kappa}^2}{w_{\kappa}}
+1\right)
\frac{1}{\sqrt{E_+^2}}\Big[
1+2\,n_b\left(\sqrt{E_+^2}\right)
\Big]
%
+\nonumber\\&&\quad\quad\quad
+\left(-\frac{\mu_{\kappa}^2}{w_{\kappa}}
+
1\right)
\frac{1}{\sqrt{E_-^2}}\Big[
1+2\,n_b\left(\sqrt{E_-^2}\right)
\Big]
\bigg\}
\label{flowVkappa-closed}
\,,
\ee
with:
\be
w_{\kappa}&=& \sqrt{ \mu_{\kappa}^2(\kappa^2+2V'_{\kappa})
+(\mu_{\kappa}^2+\alpha V''_{\kappa})^2}\,,
\\
E_{\pm}^2
&=&
v_{\kappa}+2\mu_{\kappa}^2
\pm
2w_{\kappa}
\\
&=&
\kappa^2+2 \alpha V''_{\kappa}
+2V'_{\kappa}+2\mu_{\kappa}^2
\pm
2\sqrt{ \mu_{\kappa}^2(\kappa^2+2V'_{\kappa})
+(\mu_{\kappa}^2+\alpha V''_{\kappa})^2}
\,.
\ee
The potential $V_{\kappa}$ in our ansatz is:
\be
V_{\kappa}(\alpha)
&=&
\mathcal{V}_{\kappa}+
\frac{m^2_{\kappa}-\mu_{\kappa}^2}{2}\;\alpha
+\frac{\lambda_{\kappa}^2}{4}\;\alpha^2
\,,
\ee
so that its derivatives give:
\be
V'_{\kappa}(\alpha)
&=&
\frac{m^2_{\kappa}-\mu_{\kappa}^2}{2}
+\frac{\lambda_{\kappa}^2}{2}\;\alpha
\,,
\\
V''_{\kappa}(\alpha)
&=&
\lambda_{\kappa}^2\,.
\ee
In this case, the parameters $w_{\kappa}$ and $E_{\pm}^2$ above become:
\be
w_{\kappa}&=& \sqrt{ \mu_{\kappa}^2(\kappa^2+
m^2_{\kappa}-\mu_{\kappa}^2
+\lambda_{\kappa}^2\;\alpha
)
+(\mu_{\kappa}^2+
\alpha 
\lambda^2_{\kappa}
)^2}
\\
&=&
\sqrt{ \mu_{\kappa}^2(\kappa^2+
m^2_{\kappa}
+3\alpha\lambda_{\kappa}^2
)
+
\alpha^2 
\lambda^4_{\kappa}
}
\,,
\\
E_{\pm}^2
&=&
\kappa^2
+
m^2_{\kappa}
+3\alpha\lambda_{\kappa}^2
+\mu_{\kappa}^2
\pm
2\sqrt{ \mu_{\kappa}^2(\kappa^2+
m^2_{\kappa}
+3\alpha\lambda_{\kappa}^2
)
+
\alpha^2 
\lambda^4_{\kappa}
}
\,.
\ee

\vspace{1.5cm}

Some simple limits are verified:
\begin{itemize}
\item 
{Free ($\lambda_{\kappa}=0$) or noncondensed ($\alpha=0$) theory}:
\be
w_{\kappa}
&=&
\sqrt{ \mu_{\kappa}^2(\kappa^2+
m^2_{\kappa}
)
}
\,,
\\
E_{\pm}^2
&=&
\kappa^2
+
m^2_{\kappa}
+\mu_{\kappa}^2
\pm
2\sqrt{ \mu_{\kappa}^2(\kappa^2+
m^2_{\kappa}
)
}
\nonumber\\
&=&
\left(
\sqrt{\kappa^2+m_{\kappa}^2}
\pm
\mu_{\kappa}
\right)^2
\,.
\ee

The free flow equation reduces to:
\be
\partial_{\kappa}V_{\kappa}
=
\frac{\kappa^4}{12\pi^2} 
\frac{1}{\sqrt{\kappa^2+m_{\kappa}^2}}
\Big[
2+2\,n_b\left(\sqrt{\kappa^2+m_{\kappa}^2}+\mu_{\kappa}\right)
+
2\,n_b\left(\sqrt{\kappa^2+m_{\kappa}^2}-\mu_{\kappa}\right)
\Big]
\label{flowVkappa-closed-free}
\,.
\ee

It is interesting to note that in general the field-independent part of the effective potential will satisfy a flow equation of the form of the {\it free} flow equation above. This is demonstrated straightforwardly by setting $\alpha=0$ on both sides of Eq. (\ref{flowVkappa-closed}):
\be
\partial_{\kappa}\mathcal{V}_{\kappa}
=
\frac{\kappa^4}{12\pi^2} 
\frac{1}{\sqrt{\kappa^2+m_{\kappa}^2}}
\Big[
2+2\,n_b\left(\sqrt{\kappa^2+m_{\kappa}^2}+\mu_{\kappa}\right)
+
2\,n_b\left(\sqrt{\kappa^2+m_{\kappa}^2}-\mu_{\kappa}\right)
\Big]
\label{flowvarVkappa}
\,.
\ee

\item 
{Vanishing chemical potential ($\mu_{\kappa}=0$)}:
\be
w_{\kappa}&=& 
\alpha 
\lambda^2_{\kappa}
\,,
\\
E_{\pm}^2
&=&
\kappa^2
+
m^2_{\kappa}
+(3\pm 2)\, \alpha\lambda_{\kappa}^2
\,.
\ee

The flow at $\mu_{\kappa}=0$ yields:
\be
\partial_{\kappa}V_{\kappa}
=
\frac{\kappa^4}{12\pi^2} 
\left\{
\frac{1}{\sqrt{E_+^2}}\Big[
1+2\,n_b\left(\sqrt{E_+^2}\right)
\Big]
%
+
\frac{1}{\sqrt{E_-^2}}\Big[
1+2\,n_b\left(\sqrt{E_-^2}\right)
\Big]
\right\}
\label{flowVkappa-closed-mu0}
\,.
\ee

\end{itemize}

The full flow equation for the effective potential $V_{\kappa}$, 
Eq. (\ref{flowVkappa-closed}), can be translated into a set of coupled equations for the parameters $\mathcal{V}_{\kappa}$, $m_{\kappa}$, $\mu_{\kappa}$ and $\lambda_{\kappa}$ of our ansatz (def.: $\bar m_{\kappa}^2\equiv m_{\kappa}^2-\mu_{\kappa}^2 $):
\be
\partial_{\kappa}V_{\kappa}(\alpha=0)&=&\partial_{\kappa}\mathcal{V}_{\kappa}
\,,
\label{varVkappa}\\
\partial_{\kappa}V'_{\kappa}(\alpha=0)&=&\partial_{\kappa}\bar m_{\kappa}^2/2
\,,
\nonumber\\
\partial_{\kappa}V''_{\kappa}(\alpha=0)&=&\partial_{\kappa}\lambda_{\kappa}^2\,,
\label{SetEqs}
\ee
where the left-hand side is then written in terms of (derivatives with respect to $\alpha$ of) the right-hand side of the full flow equation, Eq. (\ref{flowVkappa-closed}) evaluated at $\alpha=0$. 

Notice that with this ansatz for the effective potential and the assumption of constant fields we cannot fix both the flow of the chemical potential and that of the mass independently. Pragmatically, we assume a fixed $\mu_{\kappa}=\mu$.

Since $\mathcal{V}_{\kappa}$ does not appear at all in the right-hand side of Eq. (\ref{flowVkappa-closed}), its flow equation (explicitly given in Eq. (\ref{flowvarVkappa})) can be solved independently, after the set of two coupled equations (\ref{SetEqs}) is solved.

\vspace{1cm}

\noindent { \bf \underline{Numerical results at zero temperature}}

In what follows we investigate numerically the solutions of the set of coupled flow equations in Eq. (\ref{SetEqs}), with their left-hand side being given by the $\alpha-$derivatives of Eq. (\ref{flowVkappa-closed}). 

We use the following procedure to fix the parameters of the model (notice that, as discussed above, $\mu_{\kappa}$ is fixed at $\mu$, which will play the role of the control parameter to investigate the phase structure of the theory at zero temperature):
\begin{enumerate}
\item The initial conditions of the flow will be given at a scale $\kappa=\Lambda$: $\bar m_{\Lambda}^2= m^2-\mu^2$ and $\lambda_{\Lambda}^2=g$. We choose:
\be
\Lambda&=&1000\; {\rm a.u.}
\nonumber\\
m^2&=&-1000\; {\rm a.u.}
\nonumber\\
g&=&2
\ee
%
\item Consider the flow at $\mu=0$. Using the initial conditions above, the physical values, obtained in the end of the flow (at $\kappa=0$), are (cf. full $\kappa$ dependence in Figs. 
\ref{l2-T0mu0}):
\be
\bar m_{0}^2(\mu=0)&=& 38487.1\; {\rm a.u.}
\\
\lambda_{0}^2(\mu=0) &=&1.47035
\,.
\ee

\begin{figure}[h]
\center
\begin{minipage}{70mm}
\includegraphics[width=7.4cm]{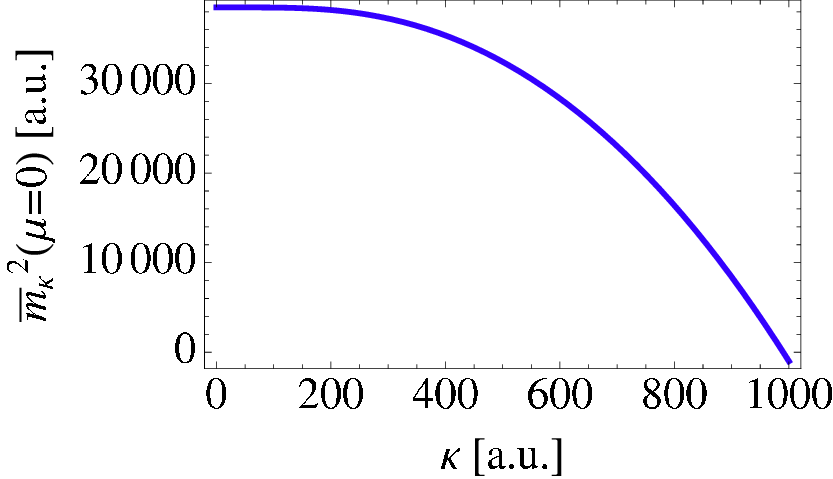}
\end{minipage}
\hspace{0.8cm}
\begin{minipage}{70mm}
\includegraphics[width=7cm]{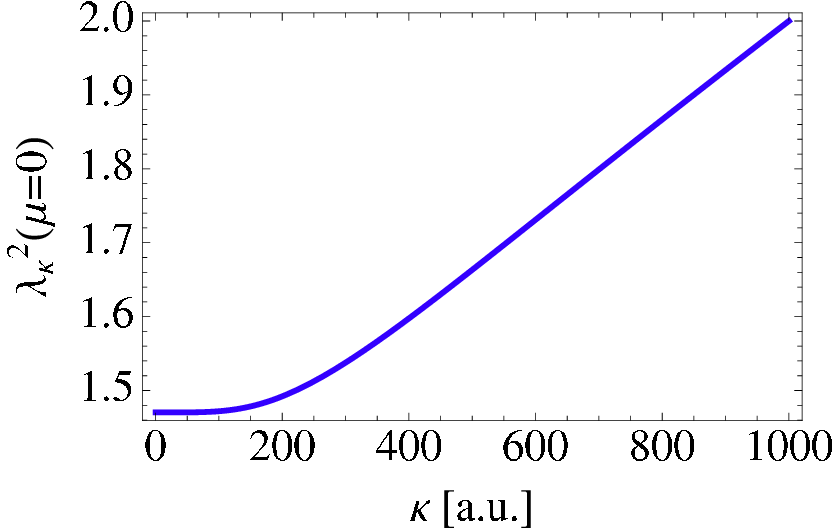}
\end{minipage}
\caption{Flow in the vacuum ($T=\mu=0$):  $\bar m_{\kappa}^2$ versus $\kappa$ (left) and $\lambda_{\kappa}^2$ versus $\kappa$ (right).}
\label{l2-T0mu0}%
\end{figure}
%
%

We use $\bar m_{0}(\mu=0)\equiv M$ to set the unit of energy in the problem. As shown in the analysis of the spectrum of excitations in the classical approximation above, we expect condensation to manifest at $\mu=m_{\rm phys}(\mu=0)$, i.e. in our case: $\mu\stackrel{!}{=}\bar m_{0}(\mu=0)\equiv M$.

It is interesting to note the saturation of the flow when $\k<M$. This is a generic feature of the FRG flow that reflects the fact the below $\k=M$ the IR modes of the theory are naturally suppressed and the regulator $R_{\k}$ plays no role in that anymore.

\end{enumerate}

\begin{figure}[h!]
\centering\mbox{
\includegraphics[width=0.9\textwidth]{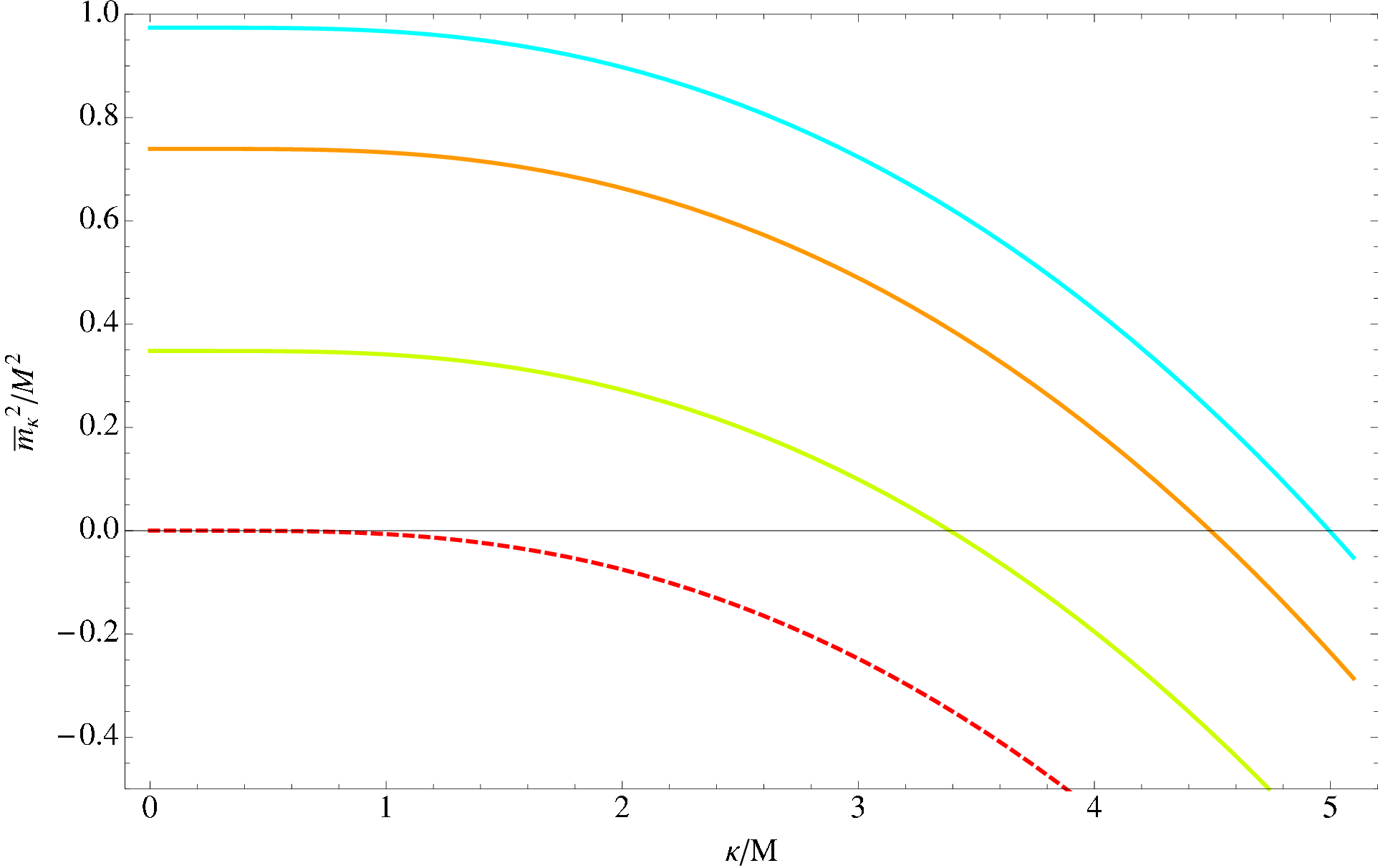}}
\caption{Flow for different $\mu$'s ($T=0$): $\bar m_{\kappa}^2$ versus $\kappa$ for $\mu^2/M^2=\big\{0.026\, ({\rm top}),\, 0.260,$ $0.650,\, 0.996 \, ({\rm bottom, dashed, critical})\big\}$.}
\label{barm2-mucrit}%
\end{figure}

 With the scale set, we turn on the chemical potential in the flow equations and look for the critical value $\mu_{\rm crit}$ at which $\bar m_{0}(\mu=\mu_{\rm crit})=0$. The vanishing effective mass characterizes the second-order phase transition corresponding to Bose-Einstein condensation. 

\begin{figure}[h!]
\centering\mbox{
\includegraphics[width=0.7\textwidth]{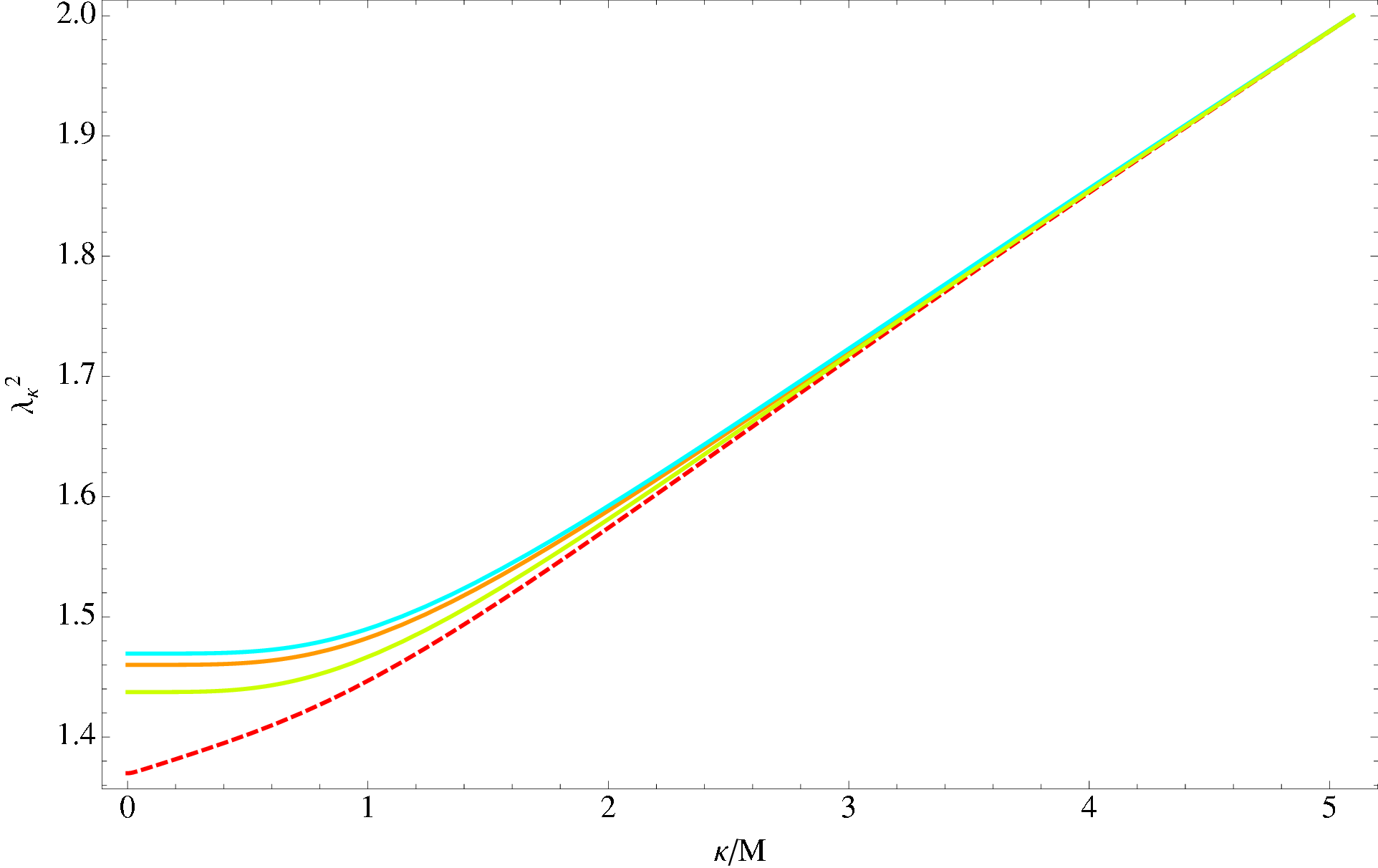}}
\caption{Flow for different $\mu$'s ($T=0$): $\lambda_{\kappa}^2$ versus $\kappa$ for $\mu^2/M^2=\big\{0.026\, ({\rm top}),\, 0.260,$ $0.650,\, 0.996 \, ({\rm bottom, dashed, critical})\big\}$.}
\label{l2-mucrit}%
\end{figure}

The flow for different values of chemical potential is shown in Figs. \ref{barm2-mucrit} and \ref{l2-mucrit}.
The effective mass $\bar m_{0}^2(\mu)$ decreases as expected when one increases the chemical potential $\mu$ and goes to zero when $\mu\to M$. We find:
\be
\frac{\bar m_{0}^2\Big(\mu^2=0.996\,M^2\Big)}{M^2}
&=&0.0000693186\,.
\ee
However, above $\mu^2\approx 0.996\,M^2$, numerical instabilities appear, signaling the criticality before the expected value $\mu_{\rm crit}^{\rm exp}=M$. Notice that this feature was already encountered in the LPA FRG analysis within the chiral model \cite{Svanes:2010we} (cf. Fig. \ref{SvanesPhDiag}).

\begin{figure}[h!]
\centering\mbox{
\includegraphics[width=0.5\textwidth]{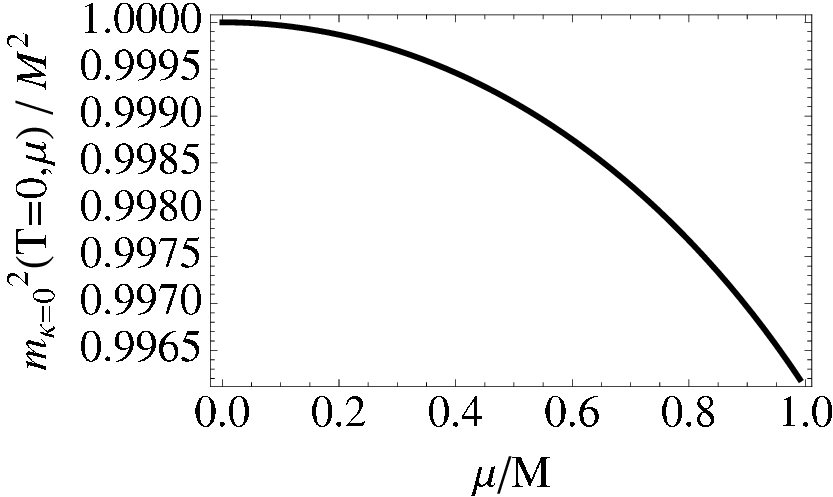}}
\caption{Physical mass at finite $\mu$, obtained at the end of the flow ($\kappa\to 0$).}
\label{m0zeroT-mu}%
\end{figure}

Actually, this can be understood as the effect of the running of the physical mass ($m_{\kappa}^2\equiv \bar m_{\kappa}^2+\mu^2$) at zero temperature and finite $\mu$, as illustrated by Fig. \ref{m0zeroT-mu}.

\newpage

\vspace{50cm}

\newpage

\vspace{50cm}

This running mass at zero-temperature and finite isospin chemical potential is also responsible for generating a small but nonzero net density in the normal phase.
This can be verified as follows.
The LPA result for the pressure and the density of the system can be obtained from the flow of $\mathcal{V}_{\kappa}(T,\mu)=-\mathcal{P}_{\kappa}(T,\mu)$, given by Eq. (\ref{flowvarVkappa}). At zero temperature and low $\mu$ (below the condensation point), we have:
\be
\partial_{\kappa}\mathcal{V}_{\kappa}(0,\mu)
&=&
\frac{\kappa^4}{6\pi^2} 
\frac{1}{\sqrt{\kappa^2+m_{\kappa}^2(0,\mu)}}
\label{flowvarVkappaZEROT}
\,.
\ee
In this case, we can fix the initial condition for the flow by setting the pressure to zero in the vacuum ($T=\mu=0$):
\be
\mathcal{V}_{\kappa=\Lambda}
&=&
\int_{0}^{\Lambda} d\kappa'
\frac{\kappa'^4}{6\pi^2} 
\frac{1}{\sqrt{\kappa'^2+m_{\kappa'}^2(0,0)}}
\,,
\ee
and assuming that, at the scale $\kappa=\Lambda$, fluctuations which are $T-,\mu-$dependent are totally suppressed and the potential $\mathcal{V}_{\kappa=\Lambda}$ is independent of $T$ and $\mu$. Using this result, the solution of the flow equation (\ref{flowvarVkappaZEROT}) can be written as:
\be
\mathcal{V}_{\kappa}(0,\mu)
&=&
\mathcal{V}_{\kappa=\Lambda}-\int_{\kappa}^{\Lambda}
d\kappa'
\frac{\kappa'^4}{6\pi^2} 
\frac{1}{\sqrt{\kappa'^2+m_{\kappa'}^2(0,\mu)}}
\label{flowvarVkappaZEROTSOL}
\,,
\ee
which can also be done with the zero-point energy term in the thermal case.

\begin{figure}[h!]
\center
\begin{minipage}{70mm}
\includegraphics[width=7.4cm]{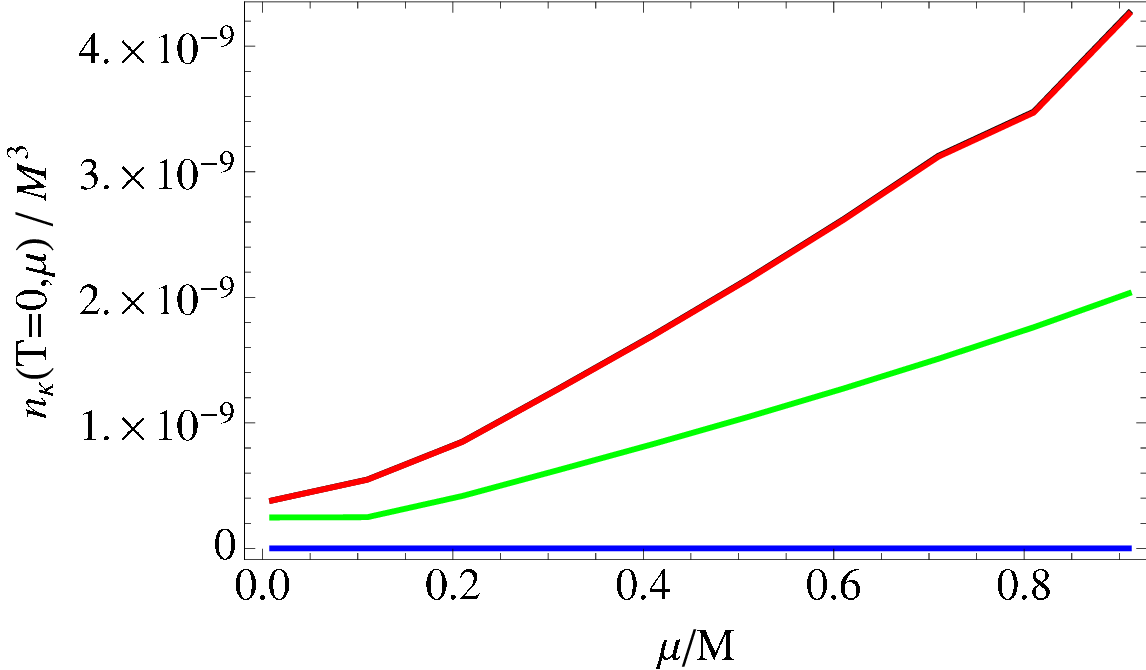}
\end{minipage}
\hspace{0.8cm}
\begin{minipage}{70mm}
\includegraphics[width=7cm]{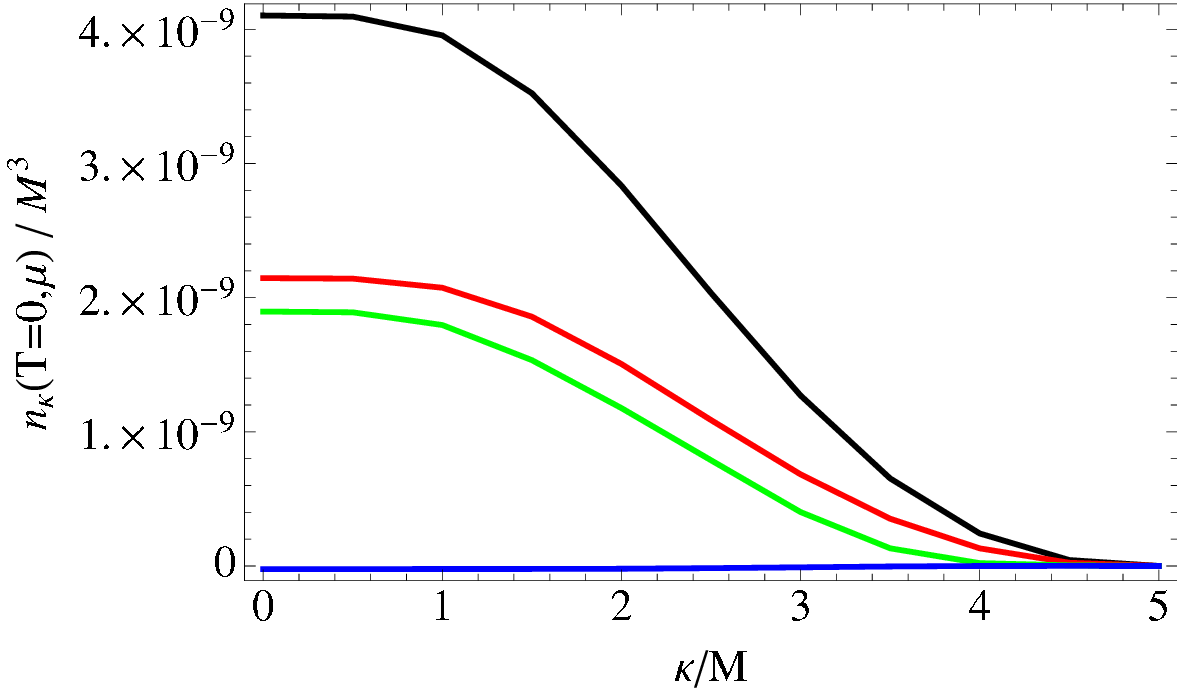}
\end{minipage}
\caption{Density {\it versus} chemical potential $\mu$ [left; $\kappa/\Lambda= 0$(top), $0.1,\; 0.5,\; 1$ (bottom)] and flow variable $\kappa$ [right; $\mu/M= 0.88$(top), $ 0.51,\; 0.16,\; 0.0005$ (bottom)].}
\label{densityZEROT}%
\end{figure}

The density is then given by the derivative with respect to $\mu$: $n_{\kappa}(T,\mu)=\frac{\partial \mathcal{P}_{\kappa}(T,\mu)}{\partial\mu}=\,-\,\frac{\partial \mathcal{V}_{\kappa}(T,\mu)}{\partial\mu}$. At zero temperature this is directly related to the $\mu-$derivative of the mass (which is clearly nonzero in Fig. \ref{m0zeroT-mu}):
\be
n_{\kappa}(0,\mu)
&=&
\,-\,\frac{\partial \mathcal{V}_{\kappa}(T,\mu)}{\partial\mu}
\nonumber\\
&=&
\int_{\kappa}^{\Lambda}
d\kappa'
\frac{\kappa'^4}{12\pi^2} 
\frac{1 }{(\sqrt{\kappa'^2+m_{\kappa'}^2(0,\mu)})^3}
\left(-\frac{ \partial m_{\kappa'}^2(0,\mu)}{\partial\mu}\right)
\,.
\ee
The numerical results are given in Fig. \ref{densityZEROT}.

%

\newpage

\noindent { \bf \underline{Numerical results at finite temperature}}

\begin{figure}[h!]
\center
\begin{minipage}{70mm}
\includegraphics[width=7.4cm]{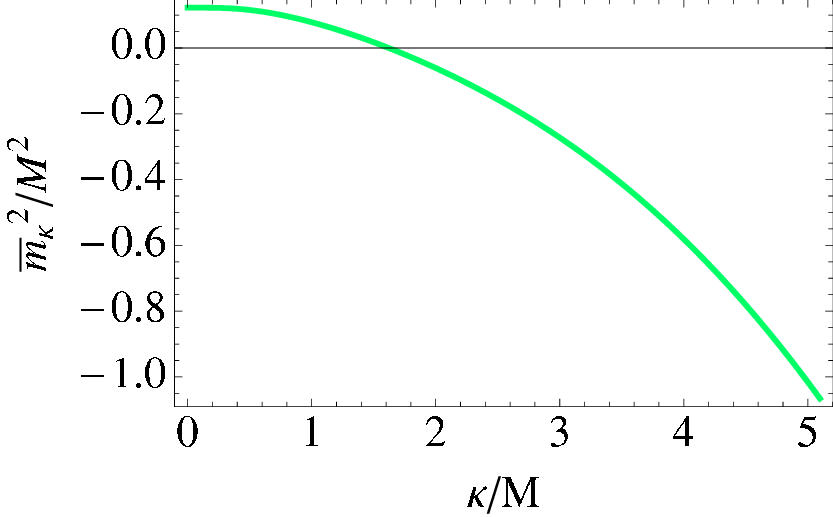}
\end{minipage}
\hspace{0.8cm}
\begin{minipage}{70mm}
\includegraphics[width=7cm]{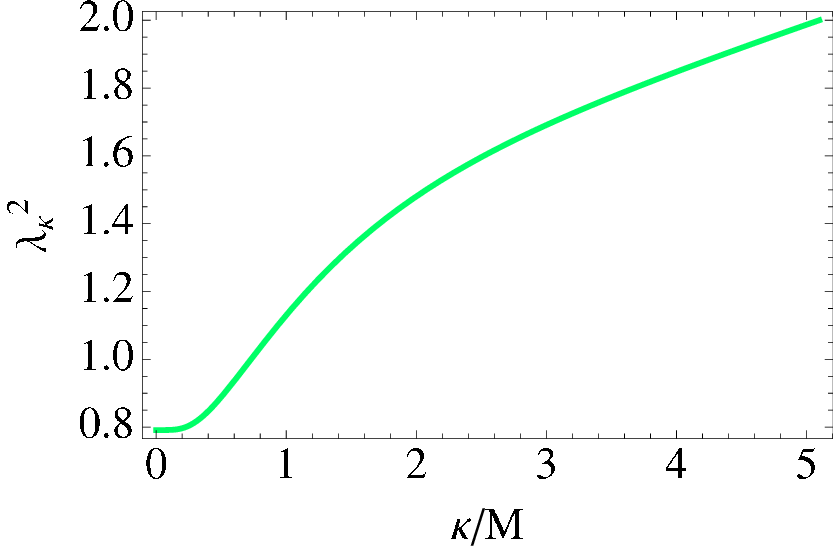}
\end{minipage}
\caption{Flow at $T/M \approx 0.51$ and $\mu^2/M^2 \approx 1.04$:  $\bar m_{\kappa}^2$ versus $\kappa$ (left) and $\lambda_{\kappa}^2$ versus $\kappa$ (right).}
\label{l2-Tmu}%
\end{figure}

Using the same procedure and initial conditions, we analyze temperature effects. Fig. \ref{l2-Tmu} shows the flow evolution of the parameters of the effective potential at finite temperature.

\begin{figure}[h!]
\centering\mbox{
\includegraphics[width=0.65\textwidth]{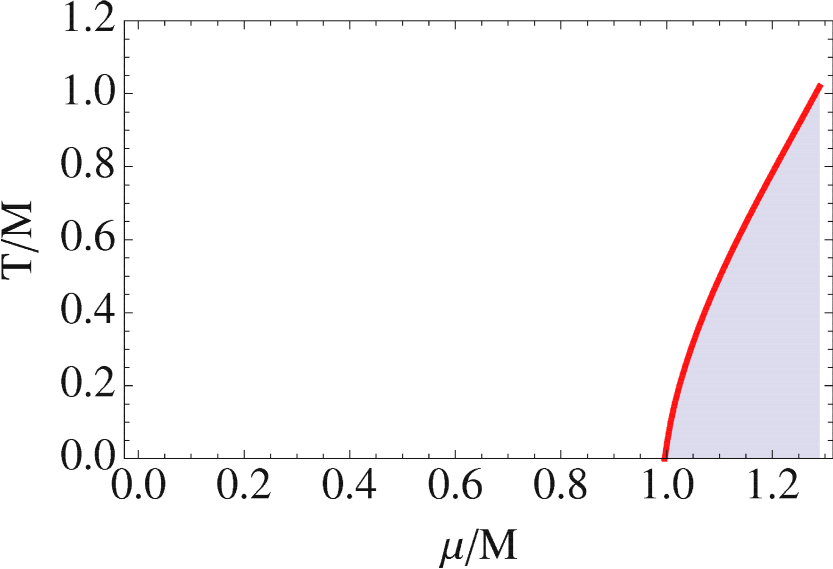}}
\caption{$T$ $\times$ $\mu$ phase diagram for BEC. The darker region is the condensed phase.}
\label{TmuLPA}%
\end{figure}

The critical line obtained for Bose-Einstein condensation is displayed in Fig. \ref{TmuLPA}.

A previous result for this line\footnote{In Ref. \cite{Andersen:2005yk}, the author obtains  $T_{\rm crit}(n)$, where $n$ is the density, within dimensional reduction (with lattice input), so that the comparison is less direct.} was obtained \cite{Kapusta:1981aa} within mean-field approximation (MFA),
\be
T_{\rm crit}^{\rm MF}&=&\sqrt{\frac{3}{\lambda^2}\left(\mu^2-m^2\right)}
\,.
\ee

\begin{figure}[h!]
\centering
\mbox{
\includegraphics[width=0.6\textwidth]{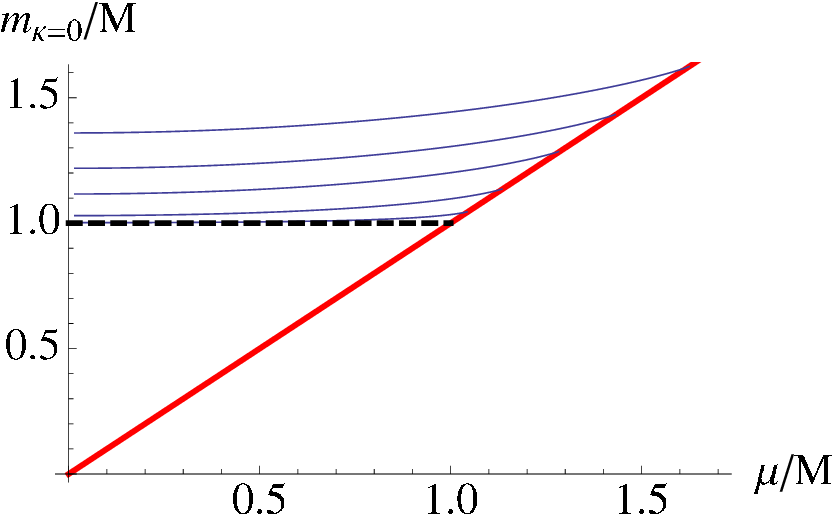}}
\caption{Mass spectrum as a function of the chemical potential $\mu$ for different temperatures: $T/M=\{ 0.30584 \textrm{(bottom)}, 0.611679, 1.01947, 1.42725, 2.03893~ \textrm{(top)}\}$. The thick red line is the ``critical'' curve $m_{\kappa=0}=\mu$, while the dashed one is the reference $m_{\k=0}=M$.}
\label{Thmasses}%
\end{figure}

As discussed for the results at zero-temperature, the shift of the critical line in LPA as compared to the MFA  is due to the effective running of the mass (i.e. the quadratic coefficient of the potential) in a medium. The hotter the environment the larger is this effect, as depicted in Fig. \ref{Thmasses}.

\begin{figure}[h!]
\centering
\vspace{-1cm}
\mbox{
\includegraphics[width=0.6\textwidth,angle=90]{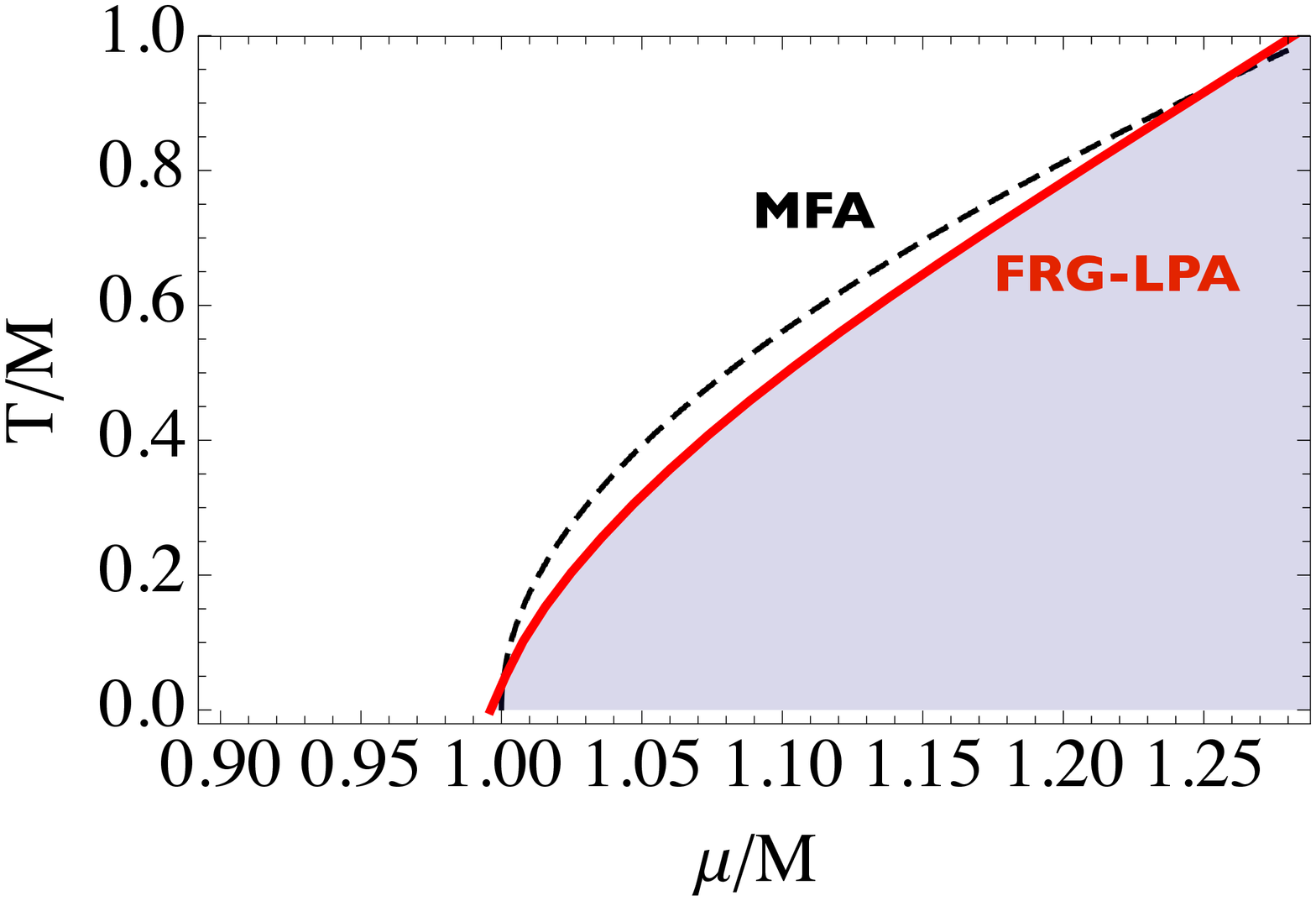}}
\vspace{-1cm}
\caption{Critical lines obtained within LPA (red, solid line) and MFA (black, dashed).
}
\label{TmuLPAvsMF}%
\end{figure}

The comparison between MFA (with $\lambda^2=2$ and $m=M$) and our LPA results is shown in Fig. \ref{TmuLPAvsMF}. This plot explicitates the fact that the critical line within the LPA FRG analysis is for most values of $\mu_I$, but not all, below the one within MFA.
Physically, however, one expects that, at fixed $\mu_I$, the treatment of fluctuations -- as implemented by FRG -- should  in principle always enlarge the disordered phase, i.e. the noncondensed one. This might be yet another peculiar feature of the LPA FRG result.

The second-order critical line obtained within LPA for this toy model is qualitatively very similar to the one in Fig. \ref{SvanesPhDiag}, computed within the pion chiral model, even though we have used a simplified LPA ansatz (neglecting $O([\phi^2]^3)$ terms). Once again, this simple field-theory appears to be a good toy model for the description of relativistic Bose-Einstein condensation. Moreover, the issues encountered in the analysis of Ref. \cite{Svanes:2010we} (namely, $\mu_I^{\rm crit}(T=0)<m_{\pi}$ and the transition line being nonperpendicular to the $\mu$-axis at zero temperature) are also present in this toy model computation and may be scrutinized in this cleaner framework.

\subsection{LPA': introducing the $Z-$factor and momentum dependence}

To investigate whether the issues encountered in the result for the phase diagram of relativistic Bose-Einstein condensation in isospin-dense systems within LPA (Figures \ref{SvanesPhDiag} and \ref{TmuLPAvsMF}) are indeed artefacts of this approximation, we would like to implement a FRG flow going beyond the zeroth order in the derivative expansion.

In order to incorporate the effect from momentum-dependent interactions in our computation of nonperturbative corrections, we go one step further in the derivative expansion for the effective action, including the $\kappa-$dependent wavefunction renormalization $Z_{\kappa}$ in our ansatz. This improved truncation scheme is the so-called LPA'.

Our LPA' ansatz for the effective action is therefore (recall that $\alpha\equiv (\pi^1)^2+(\pi^2)^2$):
\be
\Gamma_{\kappa}&=&
\int d^4x
\Bigg\{
\frac{Z_{\kappa}}{2}\partial_{\nu}\pi^1\partial_{\nu}\pi^1
+\frac{Z_{\kappa}}{2}\partial_{\nu}\pi^2\partial_{\nu}\pi^2
+\mu_{\kappa}\left[\pi^1\partial_0\pi^2-\pi^2\partial_0\pi^1\right]
+ V_{\kappa}\left(\alpha\right)
\Bigg\}
\, ,
\label{ansatzLPA'}
\ee

Following Ref. \cite{Wetterich:1991be} (section 5, in particular), one can obtain the flow equation for the wavefunction renormalization by expanding the field content in the flow equation for the effective action in the following way:
\be
\pi(x)&=& \varphi+\Delta(x)\,,
\label{BackField}
\ee
with $\varphi$ being a constant background field (that corresponds to the full $Q=0$ contribution to the field) and $\Delta(x)$, a small inhomogeneous fluctuation, i.e. satisfying $\Delta(x)\ll \varphi$ and $\Delta(Q=0)=0$. One can then show that the wave function renormalization is given by:
\be
Z_{\kappa}&=& \frac{1}{\beta^2}\lim_{Q^2\to 0}\frac{\partial}{\partial Q^2}\;\left.\frac{\delta^2 \Gamma_{\kappa}}{\delta \Delta^1(-Q)\delta\Delta^1(Q)}\right|_{\Delta=0}
\,.
\label{Zkappa}
\ee
Indeed, if we make the substitution (\ref{BackField}) in our LPA' ansatz for the effective action, Eq. (\ref{ansatzLPA'}), we obtain:
\be
\Gamma_{\kappa}&=&
\int d^4x
\Bigg\{
\frac{Z_{\kappa}}{2}\partial_{\nu}\Delta^1\partial_{\nu}\Delta^1
+\frac{Z_{\kappa}}{2}\partial_{\nu}\Delta^2\partial_{\nu}\Delta^2
+\mu_{\kappa}\left[(\varphi^1+\Delta^1)\partial_0\Delta^2-(\varphi^2+\Delta^2)\partial_0\Delta^1\right]
+
\nonumber\\&&
+ V_{\kappa}\Big(\alpha=\alpha_{\varphi}+\delta(x)
\Big)
\Bigg\}
\, ,
\label{ansatzLPA'-wf0}
\ee
where we have defined $\alpha_{\varphi}\equiv (\varphi^1)^2+(\varphi^1)^2$ and $\delta\equiv 2\varphi^1\Delta^1+2\varphi^2\Delta^2+(\Delta^1)^2+(\Delta^1)^2 = O(\Delta)$. Expanding up to quadratic order in the inhomogeneous fluctuation $\Delta$, we have:
\be
\Gamma_{\kappa}&=&
\int d^4x
\Bigg\{
\frac{1}{2}\Delta^i\Big[
\delta^{ij}\Big(-Z_{\kappa} \partial^2
+2V'_{\kappa}(\alpha_{\varphi})+4\alpha_{\varphi} V''_{\kappa}(\alpha_{\varphi})\Big)
+\delta^{i1}\delta^{j2}\mu_{\kappa}\partial_0-\delta^{i2}\delta^{j1}\mu_{\kappa}\partial_0
\Big]
\Delta^j
+\nonumber\\&&
+
\Big(
\mu_{\kappa}\varphi^{i-1}\partial_0
-\mu_{\kappa}\varphi^{i+1}\partial_0
+2\varphi^i V'_{\kappa}(\alpha_{\varphi})
\Big)\Delta^i 
+
V_{\kappa}(\alpha_{\varphi})
+ O(\Delta^3)
\Bigg\}
\, ,
\label{ansatzLPA'-wf1}
\ee
or, in momentum space,
%
\be
\Gamma_{\kappa}
&=&
\frac{\beta^2}{2}
(\beta V)\sumint_Q
\Delta^i(Q)\Big[
\delta^{ij}\Big(Z_{\kappa} [-Q_0^2+\vec{q}^2]
+2V'_{\kappa}(\alpha_{\varphi})+4\alpha_{\varphi} V''_{\kappa}(\alpha_{\varphi})\Big)
+\nonumber\\
&&
\quad+\delta^{i1}\delta^{j2}\mu_{\kappa}(-i Q_0)-\delta^{i2}\delta^{j1}\mu_{\kappa}(-i Q_0)
\Big]
\Delta^j(-Q)
+\nonumber\\&&
+
\beta V V_{\kappa}(\alpha_{\varphi})
+ O(\Delta^3)
\, ,
\label{ansatzLPA'-wf2}
\ee
where we have used the standard Fourier expansion for the fluctuation field:
\be
\Delta(x)&=&\sqrt{\frac{\beta}{V}}\, (\beta V)\sumint_Q {\rm e}^{iQ\cdot x}\Delta(Q)
\,,
\ee
and $\Delta(Q=0)=0$ in the last line. The functional derivatives of $\Gamma_{\kappa}$ with respect to the inhomogeneous fluctuation field $\Delta$ then give:
\be
\frac{\delta^2 \Gamma_{\kappa}}{\delta\Delta^b(-Q)\delta\Delta^a(Q)}
&=&
\beta^2
\Big[
\delta^{ab}\Big(Z_{\kappa} [-Q_0^2+\vec{q}^2]
+2V'_{\kappa}(\alpha_{\varphi})+
4\alpha_{\varphi} V''_{\kappa}(\alpha_{\varphi})\Big)
+
\nonumber\\&&\quad+
\delta^{a1}\delta^{b2}\mu_{\kappa}(-i Q_0)-\delta^{a2}\delta^{b1}\mu_{\kappa}(-i Q_0)
\Big]
+ O(\Delta)
\, ,
\label{ansatzLPA'-wf3}
\ee
so that (notice that $-Q_0^2+{\vec{q}}^2=Q^2$)
\be
\left.\frac{\delta^2 \Gamma_{\kappa}}{\delta \Delta^1(-Q)\delta\Delta^1(Q)}\right|_{\Delta=0}
&=&
\beta^2\left\{Z_{\kappa}\, Q^2+2V'_{\kappa}(\alpha_{\kappa})+4\alpha_{\kappa}V''_{\kappa}(\alpha_{\kappa})\right\}
\ee
and the expression (\ref{Zkappa}) for the $\kappa-$dependent wavefunction renormalization is verified.

In our case, we can similarly obtain the $\kappa-$dependent chemical potential (cf. Eq.(\ref{ansatzLPA'-wf3}) for $a=2$ and $b=1$):
\be
\mu_{\kappa}
&=&
\frac{-i}{\beta^2} \lim_{Q_0,|\vec{q}|\to 0}\frac{\partial}{\partial Q_0}\;\left.\frac{\delta^2 \Gamma_{\kappa}}{\delta \Delta^1(-Q)\delta\Delta^2(Q)}\right|_{\Delta=0}
\,,
\label{mukappa}
\ee
whose running is not accessible within LPA, as discussed above.

The results in Eqs. (\ref{Zkappa}) and (\ref{mukappa}) define special projectors\footnote{We followed the prescription by Wetterich in Ref. \cite{Wetterich:1991be}, but it seems to be completely equivalent not to expand the field. The same result is obtained for $\varphi\mapsto 0$, except that we would not be allowed in general to assume $\Delta(Q=0)=0$ and would have to evaluate $\pi=\varphi$ in the end, instead of $\Delta=0$.} that, if applied to the flow equation for the effective action (Eq. (\ref{flow-general})), yield flow equations for the $\kappa-$dependent wavefunction renormalization $Z_{\kappa}$ and chemical potential $\mu_{\kappa}$, respectively.
%
%
%
%

The derivation and solution of these projected equations for the LPA' ansatz is work in progress \cite{RelatBEC-FRG-wip}.

\section{Remarks and perspectives}

In this chapter a nonperturbative description of the phase diagram of Strong Interactions in the temperature {\it versus} isospin chemical potential plane was presented. We have discussed, using effective models, the physics concerning the formation of a new state of matter: the relativistic Bose-Einstein condensate of charged pions. Nonperturbative aspects were taken into account via an FRG flow.

Throughout the chapter it becomes clearer how one may render, through sensible approximations, the FRG formalism a powerful tool to address in a nonperturbative fashion the phase structure of in-medium field theories. The adequacy of approximations and truncations is, however, a subtle system-dependent issue. Our formal aim in this on-going work is to scrutinize the results provided by FRG and its approximations within a sufficiently simple theory containing a physically motivating phenomenon (BEC), gaining understanding of how this nonperturbative flow implements nontrivial contributions and what are the limitations of the approximations used.

In the near future, we expect to have results beyond the leading order in the derivative expansion \cite{RelatBEC-FRG-wip} that should eventually answer whether the puzzling features found in the description of the $T-\mu_I$ phase-diagram are physical or rather artefacts of the LPA ansatz.

%% file: conclusions.tex
\chapter[Conclusions]{
\label{conclusions}}

\vspace{1.5cm}

{\huge \sc Conclusions}
%

\vspace{2cm}

\vspace{-11.3cm}
\hspace{6cm}
\includegraphics[width=8.5cm,angle=90]{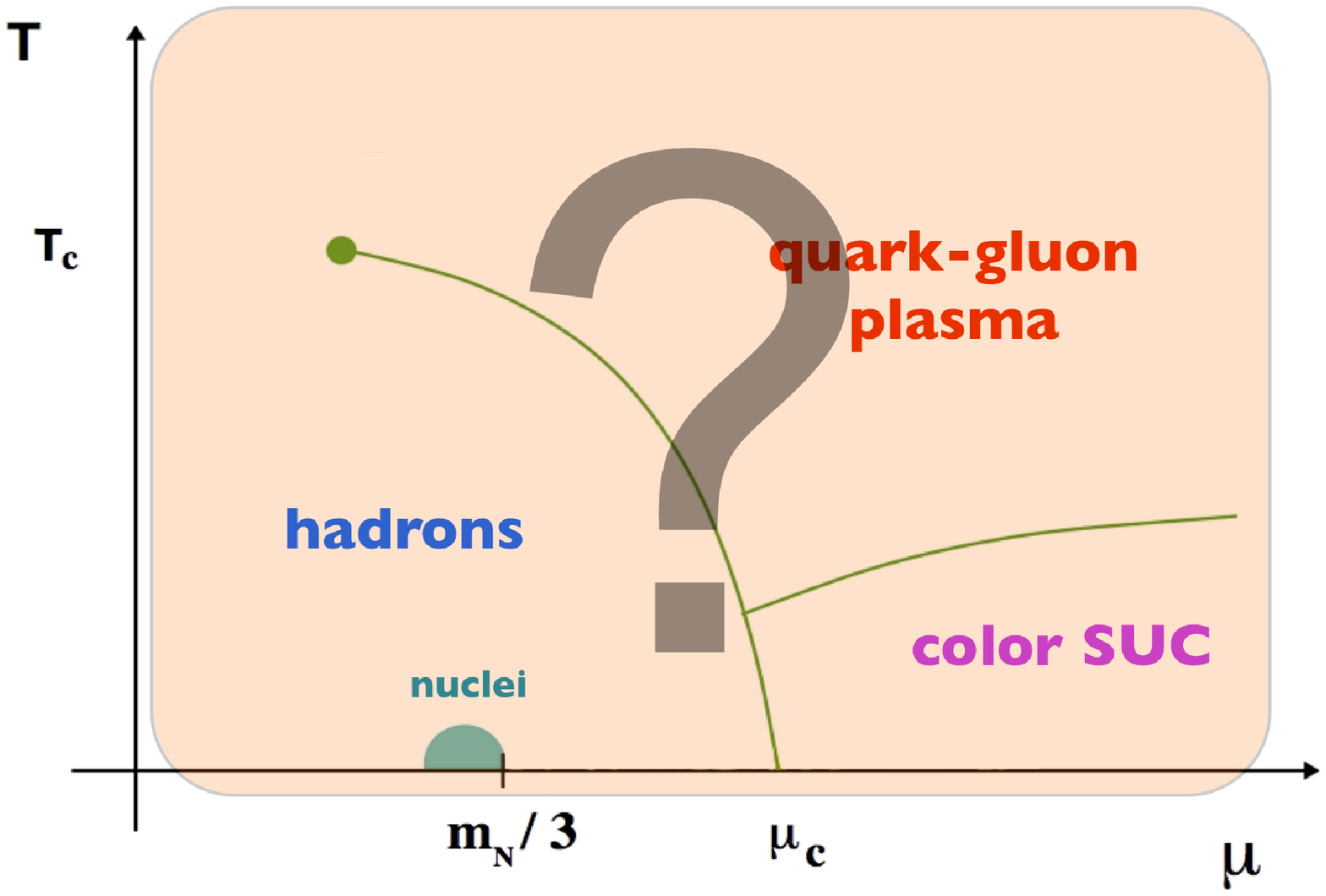}

\vspace{3.cm}

The study of the phase structure of Strong Interactions is an intricate enterprise. Being essentially an open problem, it generates an intense demand for theoretical development, since features of the different phases and the transitions between them influence the physics of observables in relevant contexts as, for instance, the evolution of the early universe, compact star structure and supernovae explosions. Moreover, the critical region of the QCD phase diagram is currently within the reach of ultra-relativistic heavy-ion collision experiments, especially via the on-going program Beam Energy Scan at RHIC-BNL and the facility FAIR at GSI which were designed to probe media with intermediate values of baryon chemical potential. This intense demand comes up against technical difficulties, so that it is inescapable to resort to various alternative/complementary methods, which range from nonperturbative, effective and out-of-equilibrium approaches to the study of cousin theories (with different masses and numbers of colors and flavors) and various macroscopic control parameters (temperature, chemical potentials, magnetic field, system size, etc). In this way the genesis of the {\it different} phase diagrams of Strong Interactions which are the subject of this thesis takes place.

The complexity of the problem is reflected in the plurality of projects and formalisms involved. The frameworks adopted included from universal statistical mechanics and simple-minded phenomenological models to intricate in-medium gauge theory and sophisticated nonperturbative methods. These were the tools used in different contexts to approach experimental and observational conditions or the domain of applicability of lattice QCD, with the aim of gathering complementary information about the physics of Strong Interactions in a medium.

The projects presented ranged from model-building with lattice input and HIC data analysis to the theoretical study of nonperturbative methods. The guideline underneath these various directions is unique, however: posing a well-defined question about the behavior of Strongly interacting matter, identifying the relevant ingredients in this context and adopting a sufficiently simple framework that still has the capability of accommodating most of these features, keeping close connection whenever possible to experiments, observations and lattice QCD.
Frequently drastic simplifications are needed, but in answering these, sometimes simple and isolated, questions, the hope is to advance in the puzzle of constructing the phase structure of Strong Interactions as it is realized in Nature.

Chapter \ref{isospin} illustrates how considering extra axes (mass and isospin chemical potential, in this case) in the phase diagram of Strong Interactions enhances sensibly the potential of constraining effective models. We have provided a unified description of the functional behavior of the deconfining critical temperature $T_c(m_{\pi},\mu_I)$ in reasonable agreement with available lattice QCD data and also predictions for the cold and dense critical line $\mu_B^{\rm crit}(m_{\pi},\mu_I)$. 
Even though our model is assumedly crude (it fails completely, for instance,  in predicting the crossover nature of the transition), the task of describing $T_c(m_{\pi},\mu_I)$, as far as we know, is currently not fulfilled by the most popular effective models for the QCD phase transitions. If not a striking coincidence, the results suggest that (i) the location of the critical region is reasonably well determined by the consistent description of high and low energy extreme domains and (ii) our crude model probably encodes the relevant ingredients in both high and low energy sectors. Moreover, since chiral models fail completely to describe $T_c(m_{\pi})$, it seems that the critical temperature is an observable driven dominantly by confinement rather than chiral symmetry.
Further analyses in this direction might help understanding the interplay between chiral and confinement features and hopefully building better models.

%
%


Approaching much closer the physics of HIC experiments, Chapter \ref{FS} presents an analysis of the pseudocritical phase diagram of Strong Interactions in a finite system. Finite-size effects on the experimental CEP search are extensively analyzed, both as a drawback of the known signatures and as the possibility of a new alternative signature based on FSS.
 First, it is shown that the sizes of the systems created at RHIC are comparable with the scale of the QCD critical region, causing the pseudocritical behavior to suffer significant shifts in the $T-\mu_B$ plane. Estimates for these shifts within the LSMq were obtained and consequences for the signatures of the second-order CEP were discussed, mainly representing a caveat in the search for nonmonotonic signals. The fact that HIC data can be organized as coming from an ensemble of systems of different sizes, however, brings a bright side of having significant finite-size corrections to the critical correlations: FSS, a universal statistical mechanics technique, can be used as a tool for the CEP search at HICs. Of course, a lot of caveats should be kept in mind, as discussed in the beginning of the chapter, due to the very complicated nature of the physical process occurring in a HIC. Nevertheless, we have pragmatically shown, in collaboration with an experimentalist from the STAR collaboration at RHIC, that a FSS analysis of data is feasible, has the capability of furnishing information about the location of the CEP and provides falsifiable predictions.


In Chapter \ref{surften} we turn our attention to the cold and baryon-dense region of the phase diagram. In this regime, some {\it observable} astrophysical phenomena are connected with the microscopic dynamics of Strong Interactions, via the possibility of nucleating deconfined bubbles in extremely dense systems such as compact stars and core-collapse supernovae. The key physical quantity coming from theory as an input to the modeling of astrophysical processes is the surface tension of the QCD phase transitions in the cold and dense regime. This is, however, not known via first-principle methods, due to its genuine nonperturbative nature and the Sign Problem on the lattice. We have provided estimates within the LSMq for the nucleation parameters of the cold and dense chiral phase transition, showing that the values obtained are small enough to allow for the proposed observable signals of quark matter formation in ultra-compact astrophysical objects. Finally, it is interesting to notice that astronomical observations might be entering a precision era in which it will start constraining the physics of Strong Interactions in the cold and dense domain, so that further theoretical predictions in this direction are called for.


Dynamics of quark bubble nucleation in supernova explosions is an example among several of the importance of nonequilibrium approaches in connecting theory and observations. In this work, we have concentrated on estimating time and energy scales associated with the dynamics of phase conversion by assuming the stationary scenario of homogeneous nucleation. Applying full out-of-equilibrium quantum field theory in a Strongly interacting medium is a crucial but extremely hard task. One alternative direction is to try to describe the approximate Langevin dynamics of the order parameters of the QCD phase transitions, in which memory effects should play an important role, especially since we are dealing with a relativistic system (as occurs in relativistic hydrodynamics).


In Chapter \ref{magneticQCD} a hot QCD medium exposed to a constant and uniform external magnetic field is considered in an on-going project. 
The temperature {\it versus} external magnetic field plane of the QCD phase diagram
combines three desirable features: experimental relevance, capability of being investigated on the lattice and a rich phenomenology.
The extra magnetic axis opens, therefore, a new channel of comparison between theory and numerical approaches. Our aim in this on-going project is to make proper use of this opportunity by providing predictions within the fundamental gauge theory for the thermodynamics of a hot QCD environment immersed in an intense magnetic field. 
Questions like ``what are the adequate quasiparticles?'' and ``how does the non-Abelian interaction affects the pressure in this extreme regime?'' are investigated analytically within this well-defined context. 
The background classical field is treated nonperturbatively as a magnetic dressing of the quarks, resulting in an effective dimensional reduction of the dynamics for very large fields. We have constructed the building blocks of the framework and discussed  its subtleties at very large $B$. Preliminary results for the two-loop pressure of a gas of gluons and magnetically dressed quarks are presented. We are looking forward to comparing our findings to recent lattice QCD simulations and to gaining, in this way, further information on the quasiparticles in this regime of the QCD phase diagram, but also on the behavior of thermal perturbation theory in this domain.


Towards having a robust description of the phase diagrams of Strong Interactions within effective field theories, it is crucial to go beyond the mean-field approximation and test for consistency. As discussed in Chapters \ref{OPT} and \ref{relatBEC}, this is an involved quest, especially because effective descriptions of a QCD medium usually require several ingredients, such as finite temperature and various chemical potentials, finite masses and condensates.
The adequate nonperturbative treatment will ultimately be dependent on the specific system treated and its main ingredients and shall be in practice considerably restricted by technicalities. In this thesis, we have used two nonperturbative frameworks to address completely different physical systems: the Optimized Perturbation Theory was applied to describe the thermodynamics of a general massive Yukawa theory in the presence of condensates in Chapter \ref{OPT} and the Functional Renormalization Group is applied, in the on-going project of Chapter \ref{relatBEC}, to the relativistic condensation of pions in hot isospin-dense media.

We have shown that a two-loop implementation of OPT consistently reproduces the mean-field result when only Hartree-like (but not Fock-like) contributions are included and further improves over MFA when the class of exchange diagrams are resummed. Nevertheless, this is an intricate framework, especially due to the complications of renormalization at higher loops, and, as discussed in the final remarks of Chapter \ref{OPT}, seems to encounter  difficulties to describe a stable effective potential in the physical case of a chiral model.

Finally, in Chapter \ref{relatBEC} the formalism of the functional renormalization group was presented and preliminary results on the $T-\mu_I$ phase diagram of a scalar toy model for relativistic Bose-Einstein condensation are shown. We believe that within this simplified toy model, we may apprehend the physics driving the phase transition to the condensed phase and, in particular, the role played by nonperturbative interactions. Our expectation is, as suggested by the analogous nonrelativistic problem, that the LPA (i.e. the zeroth order of a derivative expansion) will prove not to be a good approximation to describe BEC physics, in which momentum-dependent interactions seem to be important.

Of course, the problems addressed in this thesis represent only a restricted exploration of a subset of the various phase diagrams of Strong Interactions. There is still much to be explored and learned in these planes and others.
Not to mention the, much less investigated, phases of in-medium Electroweak Interactions and Beyond the Standard Model theories. In any of these rich physical scenarios, we believe that the development may be accomplished through an optimized utilization of complementary frameworks, as done throughout this thesis.

%% file: apLSM.tex
\chapter[Effective theory for the chiral phase transition]{
\label{LSM}}
\chaptermark{Effective theory for the chiral phase transition}


{\huge \sc Effective theory for the }
\vspace{0.3cm}

\noindent {\huge \sc   chiral phase transition }

\vspace{2cm}


The Linear Sigma Model with constituent quarks (LSMq), also known as the quark-meson 
model, is very suitable for the study of the chiral transition \cite{lee-book}. As argued in 
Ref. \cite{Pisarski:1983ms}, QCD with two flavors of massless quarks belongs to the same 
universality class as the $O(4)$ Linear Sigma Model (LSM), exhibiting the same qualitative behavior at criticality. 
Besides, the LSM is renormalizable \cite{Lee:1968da} and reproduces correctly the phenomenology 
of QCD at low energies, such as the spontaneous (and small explicit) breaking of chiral 
symmetry, meson masses, etc. In spite of the fact that an effective theory does not require 
renormalizability to be well posed, this attribute is highly desirable if one is interested in investigating 
the behavior of physical quantities as the energy scale is modified, which can be accomplished via 
renormalization group methods.

Since its proposal \cite{GellMann:1960np}, the LSM has been investigated in different 
contexts, from the low-energy nuclear theory of nucleon-meson interactions to ultra-relativistic 
high-energy heavy-ion collisions, and also in different varieties, e.g. including explicitly 
constituent quark degrees of freedom or not. The thermodynamical aspects of the model 
were first considered in the very early days of finite-temperature field theory \cite{Kirzhnits:1972ut}, 
and systematic approximations for the study of the chiral transition started to be implemented 
soon afterwards \cite{Baym:1977qb}, producing an extensive literature.

When used to mimic and study the chiral transition of QCD, the emphasis was generally on 
thermal effects \cite{Scavenius:2000qd,Scavenius:2001bb,Bilic:1997sh,Baym:1977qb,Bochkarev:1995gi,Petropoulos:1998gt,Roder:2003uz,Marko:2010cd} 
rather than considering a cold and dense scenario, although chiral symmetry 
restoration at high densities was predicted quite early \cite{Lee:1974ma}. This choice was, 
of course, well justified by the stimulating experimental results coming from high-energy 
heavy-ion collisions \cite{QM}, and by the possibility to compare to numerical output 
from lattice QCD \cite{Laermann:2003cv}. Usually the gas of quarks is treated as a thermal 
bath in which the long-wavelength modes of the chiral field evolve, the latter playing the role 
of an order parameter in a Landau-Ginzburg approach. The standard procedure is then integrating 
over the fermionic degrees of freedom, using a classical approximation for the chiral field, to 
obtain a formal expression for the thermodynamic potential from which one can compute 
all the physical quantities of interest. In this case, the sigma field is approximated by the condensate, 
and the functional integral over sigma fluctuations is not performed. The fermionic determinant is 
usually calculated to one-loop order assuming a homogeneous and static background 
field \cite{kapusta-gale}.

In a theory with spontaneous symmetry breaking, the presence of a condensate will 
modify the masses. In particular, in the case of the LSMq the masses of quarks and 
mesons will incorporate corrections that are functions of the chiral condensate, which 
is a medium-dependent quantity. Therefore, contributions from vacuum diagrams 
can not be subtracted away as trivial zero-point energies since they contain 
medium-dependent pieces via effective masses. So, although the presence of 
the medium brings no new ultraviolet divergence, in principle one has to incorporate 
carefully finite vacuum contributions that survive the renormalization procedure. These 
contributions have been very often discarded in studies of the LSMq, but were shown 
to play an important role by the authors of Ref. \cite{Mocsy:2004ab} who 
incorporate scale effects phenomenologically. Vacuum contributions were considered 
at finite density in the perturbative massive Yukawa model with analytic exact results up to 
two loops in Ref. \cite{Palhares:2008yq,thesis} and, more specifically, in optimized perturbation 
theory at finite temperature and chemical potential in Ref. \cite{Fraga:2009pi}, 
also comparing to mean-field theory. 
More recently, this issue was discussed in a comparison 
with the Nambu-Jona--Lasinio model \cite{Boomsma:2009eh},
in the Polyakov-LSMq model in the presence of an external magnetic field \cite{Fraga:2008qn}, 
and in the quark-meson model, with special attention to the chiral limit \cite{Skokov:2010sf}. 

To describe the chiral phase structure of strong interactions at finite density and vanishing 
temperature we adopt the LSMq, defined by the following lagrangian
\begin{eqnarray}
{\cal L} &=&
 \overline{\psi}_f \left[i\gamma ^{\mu}\partial _{\mu} - m_f - g(\sigma +i\gamma _{5}
 \vec{\tau} \cdot \vec{\pi} )\right]\psi_f \nonumber\\
&+& \frac{1}{2}(\partial _{\mu}\sigma \partial ^{\mu}\sigma + \partial _{\mu}
\vec{\pi}  \cdot \partial ^{\mu}\vec{\pi} )
- U(\sigma ,\vec{\pi})\;,
\label{lagrangian}
\end{eqnarray}
where
\begin{equation} 
U(\sigma ,\vec{\pi})=\frac{\lambda}{4}(\sigma^{2}+\vec{\pi}^{2} -
{\it v}^2)^2-h\sigma
\label{bare_potential}
\end{equation}
is the self-interaction potential for the mesons, exhibiting both spontaneous 
and explicit breaking of chiral symmetry. The $N_f=2$ massive fermion fields 
$\psi_f$ represent the up and down constituent-quark fields $\psi_{f}=(u,d)$. For 
simplicity, we attribute the same mass, $m_f=m_q$, to both quarks. The scalar field 
$\sigma$ plays the role of an approximate order parameter for the chiral transition, 
being an exact order parameter for massless quarks and pions. The latter are 
represented by the pseudoscalar field $\vec{\pi}=(\pi_{1},\pi_{2},\pi_{3})$. It is 
customary to group together these meson fields into a $O(4)$ chiral field 
$\phi =(\sigma,\vec{\pi})$. For simplicity, we discard the pion dynamics from 
the outset, knowing that they do not affect appreciably the phase conversion 
process \cite{Scavenius:2001bb}, and focus our discussion in the quark-sigma sector 
of the theory. Nevertheless, pion vacuum properties will be needed to fix the 
parameters of the lagrangian later. 
As will be discussed below, the parameters of the lagrangian are chosen such that 
the effective model reproduces correctly the phenomenology of QCD at low energies 
and in the vacuum, such as the spontaneous (and small explicit) breaking of chiral 
symmetry and experimentally measured meson masses. 

Since chiral symmetry is spontaneously broken in the vacuum, it is convenient 
to expand the $\sigma$ field around the condensate, writing 
$\sigma(\vec{x},\tau) =\langle\sigma\rangle + \xi (\vec{x},\tau)$, where 
$\tau$ is the imaginary time in the Matsubara finite-temperature formalism. 
Although we assume the system to be at temperature $T=0$, it is convenient to 
compute the loop expansion at finite temperature, taking the limit 
$T\rightarrow 0$ at the end. For a dense system, the chiral condensate, 
$\langle\sigma\rangle$, will be a function of the quark chemical potential, 
$\mu$. Given the shift above, the fluctuation field $\xi$ is such that 
$\langle\xi\rangle =0$ and $\xi (\vec{p}=0,\omega=0)=0$. From the 
phenomenology, one expects that 
$\langle\sigma\rangle (\mu \rightarrow \infty) \approx 0$ (being equal in 
the case of massless quarks).

Keeping terms up to $O(\xi^2)$, one obtains the following effective 
lagrangian
\begin{eqnarray}
{\cal L}  &=&
\overline{\psi}_f \left[i\gamma ^{\mu}\partial _{\mu} - 
M_q - g\xi \right]\psi_f
+ \frac{1}{2}\partial _{\mu}\xi \partial ^{\mu}\xi 
-\frac{1}{2}M_\sigma^2\xi^2 - U(\langle\sigma\rangle)\;,
\label{effective_lagrangian}
\end{eqnarray}
where we have defined the $\mu$-dependent effective masses
\begin{equation}
M_q \equiv m_q + g \langle\sigma\rangle \;\;,\;\; 
M_\sigma^2\equiv 3\langle\sigma\rangle^2 -\lambda {\it v}^2 \; .
\label{masses}
\end{equation}
Linear terms in $\xi$ can be dropped in the action because of the 
condition $\xi (\vec{p}=0,\omega=0)=0$. 

In the effective lagrangian, the medium-dependent masses will entangle 
vacuum and medium contributions in the loop expansion, rendering the 
renormalization process more subtle. This is a consequence of the presence 
of a nonzero $\mu$-dependent condensate in the broken phase. Of course, the 
ultraviolet divergences are the ones of the theory in the vacuum, and 
renormalization is implemented in the standard fashion by adding 
medium-independent counterterms to the original lagrangian (\ref{lagrangian}). 
In our case, though, it is more convenient to introduce counterterms in 
the effective theory, so that there are contributions from pure-vacuum, 
pure-medium and vacuum-medium pieces in the cancellation of ultraviolet 
divergences. Renormalization is then implemented by using standard methods 
of finite-temperature field theory \cite{kapusta-gale}.

\section{Cold and dense 1-loop effective potential and vacuum corrections}

The effective potential can be computed exactly and in closed form following 
the steps detailed in Ref. \cite{Palhares:2008yq} (cf. also \cite{thesis}), a procedure that can be even 
generalized to optimized perturbation theory \cite{Fraga:2009pi}, as presented in Chapter \ref{OPT}. Keeping only 
contributions to one loop order, the effective potential is given by
\begin{equation}
V_{\rm eff}(\bar\sigma)=U(\bar\sigma)+\Omega^{\rm ren}_{\xi}(\bar\sigma) \,\label{Veff0}
\end{equation}
where $\Omega^{\rm ren}_{\xi}(\bar\sigma)$ is the fully-renormalized thermodynamic 
potential for the fluctuation effective theory. The latter corresponds to a Yukawa theory 
for massive fermions and a massive scalar, with masses given by (\ref{masses}) as 
functions of $\bar\sigma$. The thermodynamics of this theory was fully solved analytically 
to two loops in the cold and dense regime in Ref. \cite{Palhares:2008yq}, where all details 
can be found. In the $\overline{\rm MS}$ scheme, $\Omega^{\rm ren}_{\xi}(\bar\sigma)$ can be 
written as a sum of a medium contribution
\begin{eqnarray}
\Omega_{{\rm med}}^{(1)} &=&
- N_f N_c~\frac{1}{24\pi^2}
\left\{
2\mu p_f^3 
-3 M_{q}^2~\left[ \mu p_f-M_{q}^2\log\left( \frac{\mu+p_f}{M_{q}} \right) \right]
\right\}
 \label{Cold-OmegaMed1}
\end{eqnarray}
and a vacuum contribution
\begin{eqnarray}
\Omega_{{\rm vac}}^{(1)} &=&-\frac{M_{\sigma}^4}{64\pi^2}
\left[ \frac{3}{2}+\log\left( \frac{\Lambda^2}{M_{\sigma}^2} \right) \right] 
+N_f N_c~\frac{M_{q}^4}{64\pi^2}
\left[ \frac{3}{2}+\log\left( \frac{\Lambda^2}{M_{q}^2} \right) \right]
\, ,\label{OmVac1Res}
\end{eqnarray}
where $u=\mu p_f-M_{q}^2\log\left( \frac{\mu+p_f}{M_{q}} \right)$, $p_{f}$ is the Fermi momentum 
and $\Lambda$ is the $\overline{\rm MS}$ scale. The latter can also be fixed by vacuum 
properties, as will be discussed in subsection \ref{ParFix}. 

\section{Thermal effects}

We can also incorporate thermal effects in the calculation of the effective potential within the LSMq. 
The inclusion of the temperature dependence allows for testing the validity of the cold ($T=0$) 
approximation at low temperatures and to investigate if the thermal corrections to
nucleation parameters can play an important role in the phase conversion process.

At one loop, the well-known temperature- and density-dependent correction to the thermodynamic 
potential is that of an ideal gas of massive sigma particles and constituent quarks 
($\omega_{\sigma}^2\equiv {\bf k}^2+M_{\sigma}^2$ and $E_q^2\equiv{\bf p}^2+M_q^2$):
\begin{eqnarray}
\Omega_{{\rm med,Th}}^{(1)} &=&
T~\int\frac{d^3{\bf k}}{(2\pi)^3}
~\log\left[
1-{\rm e}^{-\omega_{\sigma}/T}
\right]
-\nonumber\\
&&
-2TN_f N_c\int\frac{d^3{\bf p}}{(2\pi)^3}
\left\{\log\left[
1+{\rm e}^{-(E_q-\mu)/T}
\right]+
\log\left[
1+{\rm e}^{-(E_q+\mu)/T}
\right]
\right\}
\, . \label{TH-OmegaMed1}
\end{eqnarray}
%

\section{Parameter fixing\label{ParFix}}

As stated above, the LSMq is adopted as an effective model for QCD at low energies, so that
the parameters
$g$, $\lambda$, $m_q$, ${\it v}$, $h$ and $\Lambda$
are fixed in order to reproduce QCD properties either measured in the vacuum or calculated numerically
via lattice QCD. Therefore, the conditions for fixing the parameters are imposed on the
vacuum effective potential. Since we aim at comparing results from cases with different vacuum effective potentials (namely
$U$ and $U+\Omega_{\rm vac}^{(1)}$), and even with different parameter sets\footnote{The case with vacuum logarithmic terms has an extra parameter: the $\overline{\textrm{MS}}$ scale $\Lambda$.}, it is useful to make explicit the parameter fixing procedure, consistently.

The conditions for fixing the model parameters are the following:

\begin{itemize}

\item The chiral condensate in the vacuum is the pion decay constant, $f_{\pi}=93~$MeV, or, in terms
of the minimum of the vacuum effective potential,
\begin{equation}
\left.\frac{\partial V_{\rm{eff}}^{\rm vac}}{\partial \langle \sigma \rangle}\right|_{
\langle \sigma \rangle=f_{\pi}} =0 \, ; \label{ddsigma}
\end{equation}

\item The partial conservation of the axial current yields
\begin{equation}
h = f_{\pi}~m_{\pi}^2 = (93 ~\textrm{MeV})~(138 ~\textrm{MeV})^2 \; , \label{h}
\end{equation}
where $m_{\pi}$ is the pion mass;

\item The current quark masses calculated in lattice QCD (cf. e.g. Ref. \cite{Chiu:2003iw}) provide:
\begin{equation}
m_q = 4.1 ~\textrm{MeV}  \; , \label{mq}
\end{equation}
which we neglect\footnote{The current quark mass also does not bring any extra qualitative feature to the model nor changes significantly the quantitative results concerning the chiral phase transition \cite{thesis}.},
setting $m_q=0$, for the sake of comparison with previous model calculations;

\item Using the constituent quark mass in the vacuum as $1/3$ of the nucleon mass 
($m_{N}=938 ~$MeV), we can fix the Yukawa coupling, $g$:
\begin{eqnarray}
&& M_q\left(\langle \sigma \rangle(p_f=0)\right)= M_q^{\rm vac} \quad\Rightarrow 
\quad  g = \frac{1}{f_{\pi}}~\left( \frac{1}{3}~m_{N}-m_q \right) =
 3.32
 \; ; \label{g}
\end{eqnarray}

\item The value of the dressed mass of the $\sigma$ field is given by the experimental value 
of the mass of the $\sigma$ meson:
\begin{equation}
\left.\frac{\partial^2 V_{\rm{eff}}^{\rm{vac}}}{\partial \langle \sigma \rangle^2}\right|_{
\langle \sigma \rangle=f_{\pi}} = \left(M^{\rm vac}_{\sigma}\right)^2 \approx (600~\textrm{MeV})^2
 \; ; \label{d2dsigma2}
\end{equation}

\item The quark condensate is fixed by the lattice result 
(including only quarks {\it up} and {\it down}, we have \cite{Chiu:2003iw}:
$\langle \overline{\psi}_{f}\psi_{f} \rangle_{\rm vac} = -2 ~(225 ~\textrm{MeV})^3$),
so that
\begin{equation}
\left.\frac{\partial V_{\rm{eff}}^{\rm{vac}}}{\partial m_q}\right|_{{\rm vac} ; ~
\langle \sigma \rangle=f_{\pi}} = \langle \overline{\psi}_{f}\psi_{f} \rangle_{\rm vac}
\; . \label{ddmq}
\end{equation}
\end{itemize}

The conditions (\ref{h})--(\ref{g}) above fix directly the parameters $h$, $m_q$ and $g$,
independently of the inclusion of quantum corrections to the vacuum thermodynamic potential.
On the other hand, the Eqs. (\ref{ddsigma}), (\ref{d2dsigma2}) and (\ref{ddmq}) are coupled equations 
to determine the parameters ${\it v}$ and $\lambda$ (and $\Lambda$, if quantum corrections 
to the vacuum are considered) and depend on the explicit form of the vacuum effective potential. 
In the case without quantum corrections, i.e. with the vacuum effective potential being 
purely the classical potential ($V_{\rm eff}^{\rm vac}=U$), we find $\lambda^2\approx 20$ and 
${\it v}^2\approx 7696.8~$MeV$^2$. With the addition of the 1-loop vacuum 
term $\Omega_{\rm vac}^{(1)}$ to the vacuum effective potential, the solution of Eqs. (\ref{ddsigma}),
(\ref{d2dsigma2}) and (\ref{ddmq}) yields 
$\lambda^2\approx 16.65$, ${\it v}^2\approx 3296.89~$MeV$^2$ and $\Lambda \approx 16.48~$MeV.

\section{Influence of interactions: higher-loop corrections}\label{Int}

So far, we have only included the effects of interactions indirectly in the construction of the effective model itself, 
through dressed masses and the presence of a nonzero quark condensate. However, the interaction of the sigma meson with 
the medium constituent quarks could in principle alter the predicted dynamics for the chiral transition. The incorporation 
of such corrections in the calculation of the effective potential can be implemented via the perturbative technique order by order.

The first interaction correction in the present case appears at the two-loop order. The contribution to the effective potential corresponds then to that of the thermodynamic potential of an interacting Yukawa theory with massive scalars and massive fermions, which was obtained and analyzed in detail in Ref. \cite{Fraga:2009pi} (cf. Chapter \ref{OPT}), including the vacuum contribution and nonperturbative effects within the optimized perturbation theory framework.
Therefore, one has in principle all the machinery to investigate the influence of interactions on the phase diagram of the LSMq and
the associated process of homogeneous nucleation. It should be noted that the full case, including quantum corrections in the vacuum 
effective potential, requires the (nontrivial) solution of the coupled equations (\ref{ddsigma}), (\ref{d2dsigma2}) and (\ref{ddmq})
with the 2-loop result plugged in. Being a more technical analysis, it is out of the scope of this Appendix, but some preliminary results and discussion can be found in Ref. \cite{thesis}).


%% file: apOPTA.tex
\chapter[Vacuum thermodynamic potential and renormalization at two loops]{
\label{apOPTA}}
\chaptermark{Vacuum thermodynamic potential and ...}


{\huge \sc Vacuum thermodynamic potential }
\vspace{0.3cm}

\noindent {\huge \sc   and renormalization at two loops }

\vspace{2cm}


In this appendix, we address the details involved in the explicit derivation
of the vacuum contributions to the two-loop thermodynamic potential of the Yukawa
theory. In particular, we concentrate on the calculation and renormalization of the 
1-loop bubble diagrams and the exchange diagram in the vacuum. The vacuum and in-medium
direct contributions (the third diagram in Fig. \ref{OmegaY-fig}), which were not considered
in Ref. \cite{Palhares:2008yq}, are left for the next appendix.

After computing the Matsubara sums (cf. Ref. \cite{Palhares:2008yq} and the next appendix), 
the pieces which are not explicitly dependent on $T$ and/or $\mu$ correspond to the vacuum 
contribution in Eq. (\ref{OmegaY(0,0)}).

The first two terms in Eq. (\ref{OmegaY(0,0)}) can be expressed in terms of the following
UV-divergent function
\begin{eqnarray} 
B(M) 
&=&-\frac{1}{2}\int dM~M \int\frac{d^3{\bf p}}{(2\pi)^3} 
\frac{1}{\sqrt{{\bf p}^2+M^2}}
\, .
\end{eqnarray}
Those divergences are cancelled by a field-independent counterterm in the Lagrangian,
commonly known as a vacuum expectation value subtraction or a cosmological constant.
Within the $\overline{\textrm{MS}}$ subtraction scheme, 
the tridimensional momentum integral above is renormalized to \cite{Caldas:2000ic}:
\begin{eqnarray} 
B^{\textrm{REN}}(M) &=& \frac{M^4}{64\pi^2}\left[\frac{3}{2}+\log\left(  
\frac{\Lambda^2}{M^2}\right)\right]
\, .
\end{eqnarray}

\begin{figure}[h]
\center
\includegraphics[width=8cm]{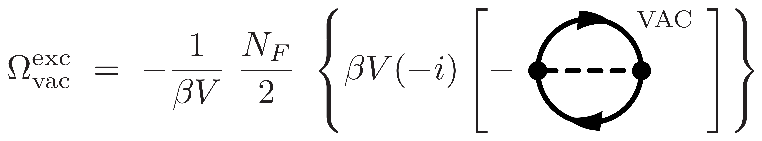}
\caption{Vacuum contribution of the exchange term written in terms of the associated vacuum diagram.}
\label{A1}
\end{figure}

The two-loop $T,\mu$-independent exchange contribution to the thermodynamic potential $\Omega_Y$, 
the last term in Eq. (\ref{OmegaY(0,0)}), can also be written in terms of UV-divergent vacuum integrals:
\begin{eqnarray}
\Omega_{{\rm vac}}^{{\rm exc}}
&=&  N_F\frac{g^2}{2} 
\int\frac{d^4Pd^4Q}{(2\pi)^8}~
\frac{4(m^2+P\cdot Q)}{\left( Q^2-m^2 \right) \left( P^2-m^2 \right) \left[ (Q-P)^2-m_{\phi}^2 \right]} 
\, ,\label{Omvac2fRF}
\end{eqnarray}
corresponding to the vacuum exchange diagram \footnote{
Throughout the appendices, whenever we refer to vacuum diagrams, we adopt the Feynman rules
from Ref. \cite{Peskin:1995ev}, with factors $(-N_F)$ associated with fermion loops excluded from the diagram definition.
}, as shown in Fig. \ref{A1}.

%

\begin{figure}[htb]
\includegraphics[width=17.5cm]{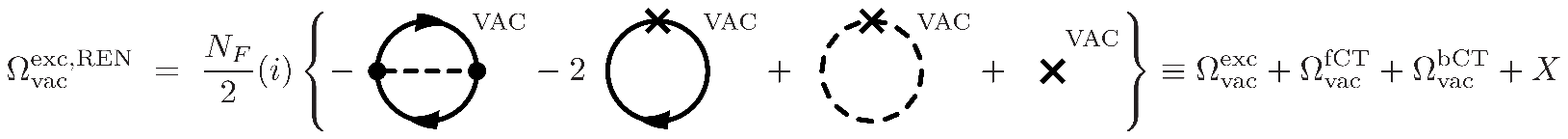}
\caption{Renormalized exchange contribution to the vacuum thermodynamic potential. The crosses indicate
counterterm vertices.}
\label{A2}
\end{figure}

The renormalization is then implemented through the usual procedure, with the addition of the
appropriate diagrams containing counterterms, as represented in Fig. \ref{A2}.
%

\begin{figure}[htb]
\center
\includegraphics[width=7cm]{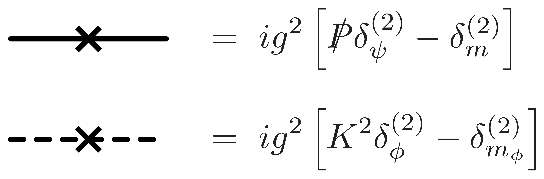}
\caption{Definition of counterterm vertices.}
\label{A3}
\end{figure}

The counterterm vertices are defined in Fig. \ref{A3} in terms of wavefunction and mass counterterms.
At this order within the $\overline{\textrm{MS}}$ subtraction scheme, these counterterm vertices
cancel exactly the 1-loop vacuum self-energy poles, yielding $(d=4-\epsilon)$:
\begin{eqnarray}
\delta_{\psi}^{(2)}=
-\frac{1}{2(4\pi)^2}~\frac{2}{\epsilon}
\quad &;& \quad
\delta_{m}^{(2)}
\frac{m}{(4\pi)^2}~\frac{2}{\epsilon}
\label{Adelpsim}
\, ,
\\
\delta_{\phi}^{(2)}=
-\frac{2}{(4\pi)^2}~\frac{2}{\epsilon}
\quad &;& \quad
\delta_{m_{\phi}}^{(2)}=
-\frac{12m^2}{(4\pi)^2}~\frac{2}{\epsilon}
\label{Adelphimphi}
\, .
\end{eqnarray}

%

%

Therefore, the second and third contributions in Fig. \ref{A2}, the vacuum bubble diagrams
with counterterm insertions, yield respectively:

\begin{eqnarray}
\Omega_{\textrm{vac}}^{\textrm{fCT}}
&=&
i\frac{N_F}{2}(-2) \int\frac{d^4P}{(2\pi)^4} ~\textrm{Tr}\left[
\left(ig^2[\slashed{P}\delta_{\psi}^{(2)}-\delta_{m}^{(2)}]
\right)
\frac{i}{\slashed{P}-m}
\right]
\nonumber\\
&=&
i\frac{N_F}{2}(-2)i(ig^2)~4\left[
m^2\delta_{\psi}^{(2)}-m\delta_{m}^{(2)}\right]\mathcal{I}(m^2)
\,,\label{AdiagfermionCTI}
\\
\Omega_{\textrm{vac}}^{\textrm{bCT}}
%
&=& 
i\frac{N_F}{2}\int\frac{d^4K}{(2\pi)^4} ~
\left(
ig^2(K^2\delta_{\phi}^{(2)}-\delta_{m_{\phi}}^{(2)})
\right)
\frac{i}{K^2-m_{\phi}^2}
\nonumber\\
&=& 
i\frac{N_F}{2}
i(ig^2)~
\left(
m_{\phi}^2\delta_{\phi}^{(2)}-\delta_{m_{\phi}}^{(2)}
\right)\mathcal{I}(m_{\phi}^2)
\, ,\label{AdiagbosonCTI}
\end{eqnarray}
where we have defined the following divergent integral
\begin{eqnarray}
\mathcal{I}(m^2)\equiv \int\frac{d^4P}{(2\pi)^4} \frac{1}{P^2-m^2}
\, ,
\end{eqnarray}
whose regularization yields
\begin{eqnarray}
\mathcal{I}^{\textrm{REG}}(m^2)
&=&
\frac{i m^2}{(4\pi)^{2}}
\left\{
\frac{2}{\epsilon}+1+
\log \left(
\frac{\Lambda^2}{m^2}
\right)
+\epsilon~
\alpha(m^2)
+O(\epsilon^2)
\right\}
\, ,\nonumber\\\label{AIreg}
\end{eqnarray}
with
\begin{eqnarray}
\alpha(m^2)\equiv \left[
\frac{1}{4}\left\{
1+
\log \left(
\frac{\Lambda^2}{m^2}
\right)
\right\}^2
+\frac{1}{4}\left(1+\frac{\pi^2}{6}\right)
\right]
\, .
\end{eqnarray}

Using the results (\ref{Adelpsim}), (\ref{Adelphimphi}) and (\ref{AIreg}) 
in (\ref{AdiagbosonCTI}) and (\ref{AdiagfermionCTI}), we arrive at the following final regularized 
expressions for the second and third terms in Fig. \ref{A2}, respectively:
\begin{eqnarray}
\Omega_{\textrm{vac}}^{\textrm{fCT}}
&=&
i\frac{N_F}{2}(-12)ig^2~\frac{m^4}{(4\pi)^{4}}
\left\{
\left(\frac{2}{\epsilon}\right)^2+\frac{2}{\epsilon}+
\frac{2}{\epsilon}\log \left(
\frac{\Lambda^2}{m^2}
\right)
+2~
\alpha(m^2)
+O(\epsilon)
\right\}
\,,\label{AdiagfermionCTF}
\\
\Omega_{\textrm{vac}}^{\textrm{bCT}}
&=& i\frac{N_F}{2}2ig^2~\frac{m_{\phi}^4}{(4\pi)^{4}}
\left[
1
-
\frac{6m^2}{m_{\phi}^2}~
\right]
\Bigg\{
\left(\frac{2}{\epsilon}\right)^2+\frac{2}{\epsilon}+
\frac{2}{\epsilon}\log \left(
\frac{\Lambda^2}{m_{\phi}^2}
\right)
+2~
\alpha(m_{\phi}^2)
+\nonumber\\&&\quad\quad
+O(\epsilon)
\Bigg\}
\, .\label{AdiagbosonCTF}
\end{eqnarray}
%

In Fig. \ref{A2}, the contribution still to be calculated explicitly in a regularized form is
$\Omega_{\textrm{vac}}^{\textrm{exc}}$.
Using the identity
\begin{eqnarray} 
m^2+P\cdot Q &=& 
2m^2-\frac{1}{2}m_{\phi}^2 -\frac{1}{2}\left[ (Q-P)^2-m_{\phi}^2 \right]
+\frac{1}{2}(P^2-m^2)+\frac{1}{2}(Q^2-m^2) \, ,
\nonumber\\&&
\end{eqnarray} 
we can rewrite $\Omega_{{\rm vac}}^{{\rm exc}}$, Eq. (\ref{Omvac2fRF}), as:
\begin{eqnarray} 
\Omega_{{\rm vac}}^{{\rm exc}} =  N_F 2g^2 \left\{ \left( 2m^2-\frac{1}{2}m_{\phi}^2  \right)
~I_1^{b} -\frac{1}{2} ~I_2^{b} +I_3^{b}\right\}
\, , \label{Omvac2fIs}
\end{eqnarray} 
in terms of the following integrals:
\begin{eqnarray} 
I_1^{b} &\equiv& \int\frac{d^4Pd^4Q}{(2\pi)^8}~
\frac{1}{\left( Q^2-m^2 \right) \left( P^2-m^2 \right) \left[ (Q-P)^2-m_{\phi}^2 \right]} 
\, ,
\label{i1b} \\ 
I_2^{b} &\equiv& \int\frac{d^4Pd^4Q}{(2\pi)^8}~
\frac{1}{\left( Q^2-m^2 \right) \left( P^2-m^2 \right) } 
\, ,
\label{i2b} \\
I_3^{b} &\equiv& \int\frac{d^4Pd^4Q}{(2\pi)^8}~\frac{1}{\left( Q^2-m^2 \right)  \left[ (Q-P)^2-m_{\phi}^2 \right]}
\, .
\label{i3b}
\end{eqnarray} 
Defining
\begin{eqnarray} 
J(a,b) &\equiv& \int \frac{d^4Pd^4Q}{(2\pi)^8} \frac{1}{(P^2-a)(Q^2-b)} \, ,
\end{eqnarray} 
we have $I_2^b=J(m^2,m^2)$ and $I_3^b=J(m^2,m_{\phi}^2)$. The dimensional
regularization of $J(a,b)$ is straightforward, yielding:
\begin{eqnarray} 
J^{\textrm{REG}}(a,b) &=& 
-\frac{1}{(4\pi)^d}~\left( \frac{e^{\gamma}\Lambda^2}{4\pi} 
\right)^{\epsilon} ~
\left[\Gamma\left( 1-\frac{d}{2} \right) \right]^2
~(ab)^{\frac{d}{2}-1} \, ,\label{JREGres}
\end{eqnarray}
where $d=4-\epsilon$.

On the other hand, the evaluation of the integral $I_1^b$ is extremely involved,
essentially due to the absence of factorization of terms containing different masses.
This calculation was performed in Ref. \cite{Davydychev:1992mt} and the result is
\begin{eqnarray} 
I_1^{b,\textrm{REG}} &=& \left( \frac{e^{\gamma}\Lambda^2}{4\pi} \right)^{\epsilon} ~
\frac{\pi^{4-\epsilon}}{(2\pi)^{2d}} 
\left(m^2\right)^{1-\epsilon}
A\left(\frac{\epsilon}{2}\right)
\Big\{ -\frac{4}{\epsilon^2}(1+2z)+\frac{2}{\epsilon}\left[ 4z~\log(4z) \right]-
\nonumber \\
&&-
2z~\left[ \log(4z) \right]^2+
2(1-z)\Phi(z) +O(\epsilon)
\Big\}  \, , \label{i1bREGres}
\end{eqnarray} 
with $z \equiv \frac{m_{\phi}^2}{4m^2}$,
\begin{eqnarray}
A\left(\frac{\epsilon}{2}\right)&\equiv& 
\frac{\left[ \Gamma(1+\epsilon/2) \right]^2}{(1-\epsilon/2)(1-\epsilon)}
= 1+\epsilon~\beta_1+\epsilon^2\beta_2+O(\epsilon^3) \label{A}\\
\beta_1 &\equiv& \frac{3}{2}-\gamma \\
\beta_2 &\equiv& \frac{7}{4}-\frac{3}{2}\gamma+\frac{1}{2}\gamma^2+\frac{\pi^2}{24}
\, ,
\\
\Phi(z) &\equiv& 4z~\Bigg\{ \left[ 2-\log(4z) \right]~{}_2F_1\left(1,1,\frac{3}{2};~z\right)
-\left.\left[ \frac{\partial}{\partial a}~{}_2F_1\left(a,1,\frac{3}{2};~z\right)
\right] \right|_{a=1}
-\nonumber\\ &&
-\left.\left[ \frac{\partial}{\partial c}~{}_2F_1 \left( 1,1,c;~z \right)
\right] \right|_{c=\frac{3}{2}} \Bigg\}
\, ,
\end{eqnarray}
and ${}_2F_1$ is the hypergeometric function, defined by:
\begin{equation}
{}_2F_1(a,b,c; ~z) \equiv \sum_{k=0}^{\infty}\frac{(a)_k(b)_k}{(c)_k}~\frac{z^k}{k!} 
\, ,
\end{equation}
where $(a)_k \equiv \frac{\Gamma(a+k)}{\Gamma(a)}$ is the Pochhammer symbol. 
In Ref. \cite{Davydychev:1992mt}, simplified expressions for $\Phi(z)$ valid
in the regions $z>1$ or $z\le 1$ were also derived.

Taking the results in Eq. (\ref{JREGres}) and in Eq. (\ref{i1bREGres}) into the
Eq. (\ref{Omvac2fIs}) and expanding around $\epsilon=0$, one obtains, after a long algebra:
\begin{eqnarray}
\Omega_{\textrm{vac}}^{{\rm exc}}
&=&  
N_F \frac{g^2}{2} \frac{m^4}{64\pi^{4}}~
\Bigg\{
\Big[ \textrm{poles in}~ \epsilon=0 \Big]+ 
v_1\left( \frac{m_{\phi}^2}{4m^2} \right)
+\left[ \gamma +\log\left( \frac{\Lambda^2}{m^2}\right)
 \right]~v_2\left( \frac{m_{\phi}^2}{4m^2} \right)
+\nonumber\\
&&
\quad+ \frac{1}{2}~\left[ \gamma +\log\left( \frac{\Lambda^2}{m^2}\right)\right]^2
~v_3\left( \frac{m_{\phi}^2}{4m^2} \right)
+O(\epsilon)
\Bigg\}
\, , \label{apOmvac2fResFinal}
\end{eqnarray}
where
\begin{eqnarray}
v_1(z) &=& 2(\gamma_0-4\beta_2)-8z~(2\gamma_0+\beta_2)+16\beta_2~z^2 +4(1-z)^2~\Phi(z)
+\nonumber \\
&&\quad
+ \log(4z)~\left\{ 8z~\left[ 2( 1-z )~\beta_1 + \gamma_{1} \right] \right\}+
\left[ \log(4z) \right]^2~\left\{ 4z^2-6z \right\} \, ,\label{apv1}
\\ 
v_2(z)&=& 2(\gamma_1-4\beta_1)-8z(2\gamma_1+\beta_1)+16\beta_1~z^2+
8z(3-2z)~\log(4z) \, , \\ 
v_3(z) &=& -6-24~z+16~z^2 \, ,\label{apv3}
\end{eqnarray}
with
\begin{eqnarray} 
\gamma_0 &\equiv& \frac{3}{4}+\frac{\gamma(\gamma-2)}{2}+\frac{\pi^2}{24}
\, ,
\\
\gamma_{1} &\equiv& 1-\gamma
\, .
\end{eqnarray}

Finally, the 2-loop contribution to the vacuum expectation value counterterm, $X$,
is then defined in order to cancel the poles in Eqs. (\ref{AdiagfermionCTF}),
(\ref{AdiagbosonCTF}) and (\ref{apOmvac2fResFinal}).

Therefore, collecting the finite parts of Eqs. (\ref{AdiagfermionCTF}), 
(\ref{AdiagbosonCTF}) and (\ref{apOmvac2fResFinal}), we obtain
the final expression for the renormalized exchange vacuum contribution
to the thermodynamic potential:
\begin{eqnarray}
\!\!\!\!\!\!\!\!
\Omega_{\textrm{vac}}^{\textrm{exc,REN}}
&=&
\frac{N_F}{2}(i)
\Bigg\{
- i g^2 \frac{m^4}{64\pi^{4}} 
\Bigg( 
v_1(z)
+\left[ \gamma +\log\left( \frac{\Lambda^2}{m^2}\right)
 \right]~
v_2(z)
+ \frac{1}{2}~\left[ \gamma +\log\left( \frac{\Lambda^2}{m^2}\right)\right]^2~
v_3(z)
\Bigg)
-\nonumber\\
&&\quad
-12ig^2~\frac{m^4}{(4\pi)^{4}}
\left[
2~
\alpha(m^2)
\right]
+ 2ig^2~\frac{m_{\phi}^4}{(4\pi)^{4}}
\left[
1
-
\frac{6m^2}{m_{\phi}^2}~
\right]
\left[
2~
\alpha(m_{\phi}^2)
\right]
\Bigg\}
\, .\label{OmegaRenF}
\end{eqnarray}

%% file: apOPTB.tex
\chapter[Vacuum and in-medium direct parts of the two-loop thermodynamic potential]{
\label{apOPTB}}
\chaptermark{Vacuum and in-medium direct parts of the ...}


{\huge \sc Vacuum and in-medium direct}
\vspace{0.3cm}

\noindent {\huge \sc     parts of the two-loop $\Omega_{Y}$}

\vspace{2cm}

\begin{figure}[htb]
\includegraphics[width=16cm]{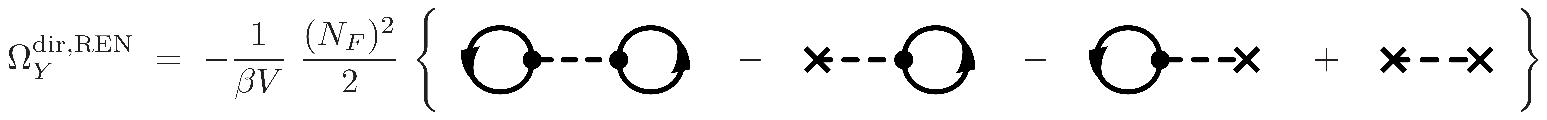}
\caption{Renormalized direct contribution to the thermodynamic potential, represented by
in-medium diagrams. As before, the crosses indicate (vacuum) counterterm vertices.}
\label{B1}
\end{figure}

The third term in the diagrammatic expansion of the perturbative thermodynamic potential,
shown in Fig \ref{OmegaY-fig}, is the direct term. In this appendix, we concentrate on the explicit evaluation
of both vacuum and in-medium parts of this contribution. As usual,
the full renormalized form of the direct term of the thermodynamic potential is obtained 
through the addition of the appropriate counterterms, as shown in Fig. \ref{B1}.

%
%
%
%
%
\begin{figure}[htb]
\center
\includegraphics[width=6cm]{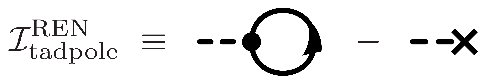}
\caption{Definition of the renormalized tadpole integral.}
\label{B2}
\end{figure}

Defining the renormalized tadpole integral, as in Fig. \ref{B2}, we can rewrite the renormalized
direct contribution to the thermodynamic potential as:
\begin{eqnarray}
\Omega^{\textrm{dir,REN}}_Y 
&=&
-\frac{1}{\beta V}~\frac{(N_F)^2}{2}~
\left\{\beta V \left[\frac{1}{m_{\phi}^2-K^2}\right]_{K^2=0} \left(\mathcal{I}_{\textrm{tadpole}}^{\textrm{REN}}\right)^2\right\}
=
-~\frac{(N_F)^2}{2m_{\phi}^2}~
\left(\mathcal{I}_{\textrm{tadpole}}^{\textrm{REN}}\right)^2
\label{Omegadir}
\, ,\nonumber\\
\end{eqnarray}

Following the $\overline{\textrm{MS}}$ prescription, the counterterm is defined through the renormalization
of the 1-loop fermionic self-energy, cancelling exactly the pole in $\epsilon=0$ of the tadpole integral, yielding:
\begin{eqnarray}
\mathcal{I}_{\textrm{tadpole}}^{\textrm{REN}}
&=&
\Bigg\{
(-g)~\sumint_P ~\textrm{Tr}\left[\frac{1}{\slashed{P}-m}\right]
\Bigg\}^{\textrm{REN}}
\nonumber\\
&=&
\Bigg\{
(-g)~4m\int\frac{d^3{\bf p}}{(2\pi)^3}\frac{1}{2E_{{\bf p}}}(-1)
\Bigg\}^{\textrm{REN}}
+
\nonumber\\
&&+
(-g)~4m\int\frac{d^3{\bf p}}{(2\pi)^3}\frac{1}{2E_{{\bf p}}}
\left[\frac{1}{\exp[E_{{\bf p}}/T-\mu/T]+1}+ \frac{1}{\exp[E_{{\bf p}}/T+\mu/T]+1} \right]
\nonumber\\
&\equiv&
I_{\textrm{tadpole}}^{\textrm{vac,REN}}+I_{\textrm{tadpole}}^{\textrm{med}}
\, ,
\end{eqnarray}

\noindent with $E_{{\bf p}}^2={\bf p}^2+m^2$,

\begin{eqnarray}
I_{\textrm{tadpole}}^{\textrm{med}}&=&-g~2m\int\frac{d^3{\bf p}}{(2\pi)^3}\frac{1}{E_{{\bf p}}}
\left[\frac{1}{\exp[E_{{\bf p}}/T-\mu/T]+1}+ \frac{1}{\exp[E_{{\bf p}}/T+\mu/T]+1} \right]
,
\end{eqnarray}

\noindent and

\begin{eqnarray}
I_{\textrm{tadpole}}^{\textrm{vac,REN}}
&=&
\Bigg\{
g~2m\int\frac{d^3{\bf p}}{(2\pi)^3}\frac{1}{E_{{\bf p}}}
\Bigg\}^{\textrm{REN}}
=
-4g~\frac{dB^{\textrm{REN}}(m)}{dm}
=-g~\frac{m^3}{(2\pi)^2}\left[
1+\log\left(
\frac{\Lambda^2}{m^2}
\right)\right]
\, .\nonumber\\
\end{eqnarray}

Finally, taking these results into Eq. (\ref{Omegadir}),
we obtain the following expression for the renormalized direct term of the 
thermodynamic potential:
\begin{eqnarray}
\Omega^{\textrm{dir,REN}}_Y &=&
-\,\frac{(N_F)^2}{2m_{\phi}^2}
\left(\mathcal{I}_{\textrm{tadpole}}^{\textrm{REN}}\right)^2
~=~
-\,\frac{(N_F)^2}{2m_{\phi}^2}
\left(I_{\textrm{tadpole}}^{\textrm{vac,REN}}+I_{\textrm{tadpole}}^{\textrm{med}}\right)^2
\equiv
\Omega^{\textrm{dir,REN}}_{\textrm{vac}}+ \Omega^{\textrm{dir}}_{\textrm{med}}
\label{OmegadirRES}
\, ,\nonumber\\
\end{eqnarray}
where
\begin{eqnarray}
\Omega^{\textrm{dir,REN}}_{\textrm{vac}}
&=&
-~\frac{(N_F)^2}{2m_{\phi}^2}~
\left(I_{\textrm{tadpole}}^{\textrm{vac,REN}}\right)^2
\\
&=&
-g^2~\frac{(N_F)^2}{2m_{\phi}^2}~
\left\{
\frac{m^3}{(2\pi)^2}\left[
1+\log\left(
\frac{\Lambda^2}{m^2}
\right)\right]\right\}^2
\, ,
\\
\nonumber
\\
\Omega^{\textrm{dir}}_{\textrm{med}}&=&
-~\frac{(N_F)^2}{2m_{\phi}^2}~
\left[2~I_{\textrm{tadpole}}^{\textrm{vac,REN}}~I_{\textrm{tadpole}}^{\textrm{med}}+
\left(I_{\textrm{tadpole}}^{\textrm{med}}\right)^2\right]
\\
&=&
-g^2T^2~\frac{(N_F)^2m^4}{(4\pi^4)m_{\phi}^2}~
\left[
1+\log\left(
\frac{\Lambda^2}{m^2}
\right)\right]
\nonumber\\&&\quad
\int~z^2~dz~\frac{1}{E_z}
\left[\frac{1}{\exp[E_{z}-\mu/T]+1}+ \frac{1}{\exp[E_{z}+\mu/T]+1} \right]
-
\nonumber\\
&&
-g^2T^4~\frac{(N_F)^2m^2}{(2\pi^4)m_{\phi}^2}~
\left\{\int~z^2~dz~\frac{1}{E_z}
\left[\frac{1}{\exp[E_z-\mu/T]+1}+ \frac{1}{\exp[E_z+\mu/T]+1} \right]
\right\}^2
\, ,\nonumber\\
\label{OmegadirRES}
\end{eqnarray}
in terms of the dimensionless quantities $E_z^2\equiv z^2+m^2/T^2$ and $z=p/T$.

%% file: apIsospin-Props.tex
\chapter[Quasiparticles in hot, isospin-dense media: modified pion propagators and physical masses]{
\label{ApIs}}
\chaptermark{Quasiparticles in hot, isospin-dense media: ...}


{\huge \sc Quasiparticles in hot, isospin-dense  }
\vspace{0.3cm}

\noindent {\huge \sc media: modified pion propagators }

\vspace{0.3cm}

\noindent {\huge \sc and physical masses }

\vspace{2cm}

In what follows we sketch the calculation of the physical masses in the dressed theory 
in the regime in which $\mu_{I}< m_{\pi}$ following 
Refs.  \cite{Loewe:2002tw}. The physical masses in the 
regime with $\mu_{I}\gtrsim m_{\pi}$ can be computed in a similar fashion, as was also 
discussed in Refs. \cite{Loewe:2002tw}, as well as the 
one in which $\mu_{I}\gg m_{\pi}$.

The tree-level {\it physical}
masses \footnote{Here we adopt the notation $m_{\pi}$ for the vacuum pion mass in the isospin 
symmetric case. In Refs. \cite{Loewe:2002tw}, this mass is denoted by $m$.}:
\be
m_{\pi^0}=m_{\pi} \quad , \quad
m_{\pi^{\pm}}=m_{\pi}\mp \mu_I
\, ,
\label{treelevelmasses}
\ee
which are associated with the following propagators (omitting the thermal piece)
\be
D_{DJ}(p)_{00} =
\frac{i}{p^2-m_{\pi}^2+i\epsilon} \quad , \quad
D_{DJ}(p)_{+-} =
\frac{i}{(p+\mu_I u)^2-m_{\pi}^2+i\epsilon}
\label{Props} \, ,
\ee
where $u^{\mu}=\delta^{\mu 0}$. On the other hand, the dressed propagators (denoted by $\bar{D}$) 
are given in terms of the self-energies as
\be
\bar{D}_{DJ}(p)_{00} &=&
\frac{i}{p^2-m_{\pi}^2-\Sigma_R(p)_{00}+i\epsilon} \, ,
\\
\bar{D}_{DJ}(p)_{+-} &=&
\frac{i}{(p+\mu_I u)^2-m_{\pi}^2-\Sigma_R(p)_{-+}+i\epsilon}
\, ,
\ee
whereas the self-energies up to one loop can be written as
\be
\Sigma_R(p)_{00} &=&
\alpha m_{\pi}^2 \left[  -\frac{1}{2}\bar{l}_3-32\pi^2\epsilon_{ud}^2l_7-2I_0+4I
\right]
\equiv m_{\pi}^2 \alpha \sigma_{00} \, ,
\\
\Sigma_R(p)_{-+} &=&
\alpha \left\{ (p_0+\mu_I)\epsilon(\mu_I)8m_{\pi}J+m_{\pi}^2\left[  
-\frac{1}{2}\bar{l}_3+2I_0
\right]
\right\}
\equiv \alpha(p_0+\mu_I)\sigma_0+\alpha m_{\pi}^2\sigma_1 \, ,
\nonumber\\
\ee
where
\be
\sigma_{00} &=& \left[  -\frac{1}{2}\bar{l}_3-32\pi^2\epsilon_{ud}^2l_7-2I_0+4I
\right] \, ,
\\
\sigma_0 &=& \epsilon(\mu_I)8m_{\pi}J \, ,
\\
\sigma_1 &=& \left[  -\frac{1}{2}\bar{l}_3+2I_0 \right] \, ,
\ee
with $\alpha\equiv (m_{\pi}/4\pi f_{\pi})^2$ , $\bar{l}_3=2.9\pm2.4$, $l_7=0.005$
(cf. Table B.1 of the last Ref. in \cite{Loewe:2002tw}, i.e. Ph.D. Thesis of C. Villavicencio), and the following additional 
convenient definitions:
\be
\epsilon_{ud}&\equiv&
\frac{m_u-m_d}{m_u+m_d}=-\frac{\delta{m}}{m} \, ,
\\
\epsilon(\mu_I)&\equiv& \frac{\mu_I}{|\mu_I|} \, ,
\\
I_0=I_0(m_{\pi},T)&\equiv&
\int_1^{\infty}dx~\sqrt{x^2-1}~2 n_B(m_{\pi}x) \, ,
\\
I=I(m_{\pi},T,\mu_I)&\equiv&
\int_1^{\infty}dx~\sqrt{x^2-1}\Big[ n_B(m_{\pi}x-|\mu_I|) + n_B(m_{\pi}x+|\mu_I|)\Big] \, ,
\\
J=J(m_{\pi},T,\mu_I)&\equiv&
\int_1^{\infty}dx~x\sqrt{x^2-1}\Big[ n_B(m_{\pi}x-|\mu_I|) - n_B(m_{\pi}x+|\mu_I|)\Big]
\, .
\ee

Therefore, we can rewrite the dressed propagators as:
\be
\bar{D}_{DJ}(p)_{00} &=& \frac{i}{p^2-m_{\pi}^2(1+\alpha\sigma_{00})+i\epsilon}
\equiv \frac{i}{p^2-\overline{m}_{0}^2+i\epsilon}
\label{dressedbarm}
\\
\bar{D}_{DJ}(p)_{+-} &=&
\frac{i}{(p_0+\mu_I-\frac{1}{2}\alpha\sigma_0)^2+O(\alpha^2)-{\bf p}^2
-m_{\pi}^2-\alpha m_{\pi}^2\sigma_1+i\epsilon}
\\
&\equiv&\frac{i}{(p_0+h)^2-{\bf p}^2-\overline{m}_{\pm}^2+i\epsilon +O(\alpha^2)}
\label{dressedhbarmpm}
\, ,
\ee
where we have defined the dressed masses and the parameter $h$:
\be
\overline{m}_0&\equiv& m_{\pi}\sqrt{1+\alpha\sigma_{00}} = m_{\pi}\left[ 1+\frac{1}{2}\alpha\sigma_{00}
+ O(\alpha^2) \right] \, ,
\\
\overline{m}_{\pm} &\equiv&
m_{\pi}\sqrt{1+\alpha \sigma_1} = m_{\pi}\left[ 1+\frac{1}{2}\alpha\sigma_1+O(\alpha^2) \right] \, ,
\\
h &\equiv&\mu_I-\frac{1}{2}\alpha\sigma_0 + O(\alpha^2)
\, ,
\ee
the latter playing the role of a dressed contribution to the zero-component of the $4$-momentum.
From Eqs. (\ref{dressedbarm}) and (\ref{dressedhbarmpm}), it is clear that the dressed propagators
have the same structure as the original ones, Eq. (\ref{Props}). Therefore, the {\it physical}
masses in the dressed theory are given by
\be
m_{\pi^0}&=& \overline{m}_0 
= m_{\pi}\left[ 1+\frac{1}{2}\alpha\sigma_{00} \right] \, ,
\\
m_{\pi^{\pm}} &=&
\overline{m}_{\pm}\mp h
= m_{\pi}\left[ 1+\frac{1}{2}\alpha\sigma_1 \pm\frac{1}{2}\alpha\frac{\sigma_0}{m_{\pi}} \right]\mp \mu_I
\ee
up to $O(\alpha)$ and with
\be
\sigma_{00}&=& \left[  -\frac{1}{2}\bar{l}_3-32\pi^2\epsilon_{ud}^2l_7-2I_0+4I \right] \, ,
\\
\sigma_0 &=& \epsilon(\mu_I)8m_{\pi} J \, ,
\\
\sigma_1 &=& \left[  
-\frac{1}{2}\bar{l}_3+2I_0
\right]
\, ,
\ee
which reproduces the results of Refs. \cite{Loewe:2002tw}.
In our effective theory, we use the dressed masses, $\overline{m}_{\pm}$ and $\overline{m}_0$,
and the function $h$ for the pion quasiparticles. 

Following a similar procedure for the second phase for $|\mu_{I}|\gtrsim m_{\pi}$, the equations remain the same as a first approach, but replacing $m_\pi$ by $|\mu_I|$.

%% file: apIntMag.tex
\chapter[Some typical integrals and traces in a magnetic background]{
\label{apIntMag}}
\chaptermark{Some typical integrals and traces in a ...}


{\huge \sc Some typical integrals and traces   }
\vspace{0.3cm}

\noindent {\huge \sc in a magnetic background }

\vspace{2cm}


In this appendix, we derive and collect some results for integrals and Dirac traces appearing in the analysis of QCD thermodynamics in a magnetic field.

\begin{itemize}
\item Properties of Dirac gamma matrices and the projector $\mathcal{P}_0=(1\!\!1+i\gamma^1\gamma^2)/2$ over the LLL degrees of freedom:
\begin{eqnarray}
(\mathcal{P}_0)^2&=&\mathcal{P}_0
\nonumber\\
\mathcal{P}_0\gamma^{\mu}\mathcal{P}_0
&\stackrel{\mu=1,2}{=}&0
\nonumber\\
\mathcal{P}_0\gamma^{\mu}
&\stackrel{\mu=0,3}{=}&\gamma^{\mu}\mathcal{P}_0
\nonumber\\
{\rm Tr}\left[\gamma^1\gamma^2\gamma^{\nu}\gamma^{\rho}\right]
&\stackrel{\nu,\rho=0,3}{=}&0
\nonumber\\
{\rm Tr}\left[\gamma^1\gamma^2\gamma^{\nu}\right]
&\stackrel{\nu=0,3}{=}&0
\nonumber\\
{\rm Tr}\left[\gamma^1\gamma^2\right]
&=&0
\label{P0Dirac}
\,,
\end{eqnarray}
so that, for any given matrix $\mathcal{M}$ that is a linear combination of $1\!\!1,\gamma^{\mu}$ and $\gamma^{\mu}\gamma^{\nu}$, with $\mu,\nu\ne 1,2$, we have:
\begin{eqnarray}
{\rm Tr}\left[\gamma^1\gamma^2\mathcal{M}\right]
&=&0
\\
{\rm Tr}\left[\mathcal{P}_0\mathcal{M}\right]
&=&\frac{1}{2}{\rm Tr}\left[\mathcal{M}\right]
\,.\label{TrP0}
\end{eqnarray}

From these properties, one concludes that the LLL traces appearing in Chapter 
\ref{magneticQCD} (e.g. in Eq.(\ref{excRF-LLL})) are essentially occuring in a (1+1)-dimensional subspace ($\mu=0,3$). This reduction shall be denoted by ${\rm Tr}_{\bar d=2}$.

\item Integrals in the phase $\phi_{x,y}$, defined in Eq. (\ref{defphi}):

Separating each of the 8 coordinate integrations, we have:
\be
\phi_{xy}&\equiv&
\int d^4xd^4y 
{\rm e}^{ik'\cdot x + ik\cdot y}
\nonumber\\
&&
{\rm exp}\Big\{
-i(p_0-q_0)(x_0-y_0)
+i(p_2-q_2)(x_2-y_2)
+i(p_3-q_3)(x_3-y_3)
\Big\}
\nonumber\\&&
\frac{eB}{\pi}{\rm exp}\Big\{
-\frac{eB}{2}\big[
\left(x_1-\frac{p_2}{eB}\right)^2
+\left(y_1-\frac{p_2}{eB}\right)^2
+\left(x_1-\frac{q_2}{eB}\right)^2
+\left(y_1-\frac{q_2}{eB}\right)^2
\big]
\Big\}
\nonumber\\
&=&
\frac{eB}{\pi}~
J_1(k'_1)\, J_1(k_1)~J_x(k'_0,p_0,q_0)\,J_x(k'_2,p_2,q_2)
\,J_x(k'_3,p_3,q_3)
\nonumber\\
&&
J_y(k_0,p_0,q_0)\,J_y(k_2,p_2,q_2)
\,J_y(k_3,p_3,q_3)
\,,
\ee
where the integrals over $x_1$ and $y_1$ are of the form
\be
J_1(k)&\equiv&
\int_{-\infty}^{\infty}
dx~
{\rm e}^{-ik x }
{\rm exp}\Big\{
-\frac{eB}{2}
\left[
\left(x-\frac{p_2}{eB}\right)^2
+
\left(x-\frac{q_2}{eB}\right)^2
\right]
\Big\}
\\
&=&
\sqrt{\frac{\pi}{eB}}
{\rm exp}\Big\{
-\frac{1}{4eB}
\left[
k^2+2i(p_2+q_2)k+(p_2-q_2)^2
\right]
\Big\}\,,
\ee
while the integrals over $x_i$ and $y_i$ for $i=0,2,3$ give directly momentum conservation delta functions in these directions and are respectively of the form:
\be
J_x(k,p,q)&\equiv&
\int_{-\infty}^{\infty}
dx~
{\rm e}^{i(k -p+q)x }
=
2\pi\;
\delta(k-p+q)
\,,
\\
J_y(k,p,q)&\equiv&
\int_{-\infty}^{\infty}
dx~
{\rm e}^{i(k +p-q)x }
=
2\pi\;
\delta(k+p-q)
\,.
\ee
Thus, the phase becomes:
\be
\phi_{xy}
&=&
(2\pi)^3\delta^{(0,2,3)}(k'-p+q)
(2\pi)^3\delta^{(0,2,3)}(k+p-q)
\nonumber\\&&
{\rm exp}\Big\{
-\frac{1}{4eB}
\left[
{k'}_1^2+k_1^2+2i(p_2+q_2)(k'_1+k_1)+2(p_2-q_2)^2
\right]
\Big\}
\,,
\label{resphixy}
\ee
where we have defined the 3-dimensional delta functions, excluding the $i=1$ direction.

\item Integral $I(k_1,k_2,eB)$, defined in Eq. (\ref{Ik}):
\begin{eqnarray}
I(k_1,k_2,eB)
&\equiv&
\int\frac{dx_1dy_1}{L_1}
{\rm e}^{ik_1(y_1-x_1)}
\frac{eB}{\pi}
\int\frac{dp_2dq_2}{(2\pi)^2}
(2\pi)\delta(p_2-q_2-k_2)
\nonumber\\
&\times&
{\rm exp}\Big\{
-\frac{eB}{2}\big[
\left(x_1-\frac{p_2}{eB}\right)^2
+\left(x_1-\frac{q_2}{eB}\right)^2
+\left(y_1-\frac{p_2}{eB}\right)^2
+\left(y_1-\frac{q_2}{eB}\right)^2
\big]
\Big\}\nonumber\\
\label{Ik-ap}
\end{eqnarray}

We can further simplify this expression as follows. Using (def.: $\bar x \equiv (x_1+x_2)/2$ and $\Delta=x_1-x_2$)
\begin{eqnarray}
\left(x_1-\frac{p_2}{eB}\right)^2
+\left(x_1-\frac{p_2}{eB}\right)^2
&=&
2\left[\frac{p_2}{eB}-\frac{x_1+y_1}{2}\right]^2-\frac{(x_1+y_1)^2}{2}+x_1^2+y_1^2
\nonumber\\
&=&
2\left[\frac{p_2}{eB}-\bar x\right]^2+\frac{\Delta^2}{2}
\\
-ik_1\Delta-\frac{eB}{2}\Delta^2&=&
-\frac{eB}{2}\Big[
\left(\Delta+i\frac{ik_1}{eB}\right)^2
+\frac{k_1^2}{eB}
\Big]
\\
dx_1dy_1 &=& d\bar x d\Delta
\,,
\end{eqnarray}
we get:
\begin{eqnarray}
I(k_1,k_2,eB)
&=&
\int\frac{d\bar x}{L_1}\frac{eB}{\pi}
\int d\Delta~{\rm e}^{
-\frac{eB}{2}\left[\left(\Delta+i\frac{ik_1}{eB}\right)^2+\frac{k_1^2}{eB}\right]}
\int\frac{dp_2dq_2}{(2\pi)^2}
(2\pi)\delta(p_2-q_2-k_2)
\nonumber\\
&\times&
{\rm exp}\Big\{
-eB\big[
\left(\frac{p_2}{eB}-\bar x\right)^2
+
\left(\frac{q_2}{eB}-\bar x\right)^2
\big]
\Big\}
\\
&=&
\int\frac{d\bar x}{L_1}\frac{eB}{\pi}
{\rm e}^{
-\frac{k_1^2}{2eB}}
\sqrt{\frac{2\pi}{eB}}
\int\frac{dp_2}{2\pi}
{\rm exp}\Big\{
-eB\big[
\left(\frac{p_2}{eB}\right)^2
+
\left(\frac{p_2-k_2}{eB}\right)^2
\big]
\Big\}
\,,\nonumber\\
\label{Ik-2}
\end{eqnarray}
where we have displaced $p_2\mapsto p_2+eB\bar x$ and $q_2\mapsto q_2+eB\bar x$ before using the $\delta$ function to evaluate the $q_2$ integral. Completing the square in the final integration, over $p_2$, one obtains [$p_2^2+(p_2-k_2)^2=2(p_2-k_2/2)^2+k_2^2/2$]:
\begin{eqnarray}
I(k_1,k_2,eB)
&=&
\int\frac{d\bar x}{L_1}\frac{eB}{\pi}
{\rm e}^{
-\frac{k_1^2+k_2^2}{2eB}}
\sqrt{\frac{2\pi}{eB}}
\frac{1}{2\pi}
\sqrt{\frac{eB\pi}{2}}
\\
&=&
\frac{eB}{2\pi}
{\rm e}^{
-\frac{k_1^2+k_2^2}{2eB}}
\,.
\label{Ik-3}
\end{eqnarray}

\end{itemize}

%% file: apPureGlue.tex
\chapter[Hot pure gauge QCD pressure at two-loops]{
\label{apPureGlue}}
\chaptermark{Hot pure gauge QCD pressure at ...}


{\huge \sc Hot pure gauge QCD pressure    }
\vspace{0.3cm}

\noindent {\huge \sc at two-loops }

\vspace{2cm}


In this appendix, we collect results for diagrams contained in 
the gluonic part of the thermodynamic potential, $\Omega_{QCD}^G$, defined in Eq. (\ref{OmegaQCDG}).

\begin{fmffile}{fmfapendicepureglue}
These contributions do not contain fermions, coming directly from hot Yang-Mills theory,
and were computed long ago (see, e.g., \cite{Kapusta:1979fh,Arnold:1994ps,kapusta-gale}). Picking the renormalized 
results for each piece from the appendix of Ref. \cite{Arnold:1994ps}, we have

\be
-~~
\parbox{10mm}{
\begin{fmfgraph*}(35,35)\fmfkeep{bolhagluon}
\fmfpen{0.8thick}
\fmfleft{i} \fmfright{o}
\fmf{gluon,left,tension=.08}{i,o}
\fmf{gluon,left,tension=.08}{o,i}
\end{fmfgraph*}}\quad^{\textrm{A-Z}}
&=& 
-~\frac{1}{\beta V}~~
\parbox{10mm}{
\fmfreuse{bolhagluon}
}
\quad
=
-4(N_c^2-1)\frac{\pi^2T^4}{90}
\\
\nonumber\\
-~~
\parbox{10mm}{
\begin{fmfgraph*}(35,35)\fmfkeep{bolhaghost}
\fmfpen{0.8thick}
\fmfleft{i} \fmfright{o}
\fmf{dashes,left,tension=.08}{i,o}
\fmf{dashes,left,tension=.08}{o,i}
\end{fmfgraph*}}
\quad^{\textrm{A-Z}}
&=&
\frac{1}{\beta V}~~
\parbox{10mm}{
\fmfreuse{bolhaghost}
}
\quad
=
2(N_c^2-1)\frac{\pi^2T^4}{90}
\\
\nonumber\\
-~~
\parbox{10mm}{
\begin{fmfgraph*}(35,35)\fmfkeep{exchange-ghost}
\fmfpen{0.8thick}
\fmfleft{i} \fmfright{o}
\fmf{dashes,left,tension=.08}{i,o,i}
\fmf{gluon}{i,o}
\fmfdot{i,o}
\end{fmfgraph*}}\quad^{\textrm{A-Z}}
&=&
\frac{1}{2}~\frac{1}{\beta V}~~
\parbox{10mm}{
\fmfreuse{exchange-ghost}
}
~~~ =
(N_c^2-1)N_c~g^2T^4~\frac{1}{4\times 144}
\ee

\be
-~~
\parbox{10mm}{
\begin{fmfgraph*}(35,35)\fmfkeep{exchange-3g}
\fmfpen{0.8thick}
\fmfleft{i} \fmfright{o}
\fmf{gluon,left,tension=.08}{i,o,i}
\fmf{gluon}{i,o}
\fmfdot{i,o}
\end{fmfgraph*}}\quad^{\textrm{A-Z}}
&=&
- ~\frac{1}{2}~\frac{1}{\beta V}\frac{1}{6}~~
\parbox{10mm}{
\fmfreuse{exchange-3g}
}
~~~ =
(N_c^2-1)N_c~g^2T^4~\frac{9}{4\times 144}
\nonumber\\
\nonumber\\
 - ~~~~~
\parbox{10mm}{
\begin{fmfgraph*}(35,35)\fmfkeep{db-4g}
\fmfpen{0.8thick}
\fmfbottom{i} \fmftop{o}
\fmf{phantom}{i,v}
\fmf{gluon,tension=.5}{v,v}
\fmf{phantom}{v,o}
\fmf{gluon,left=90,tension=.5}{v,v}
\fmfdot{v}
\end{fmfgraph*}}\quad\quad~^{\textrm{A-Z}}
&=&
 -~\frac{1}{2}~\frac{1}{\beta V}\frac{1}{8}~~~~~
\parbox{10mm}{
\fmfreuse{db-4g}
}\quad
~~~ =
(N_c^2-1)N_c~g^2T^4~\frac{3}{144}
\ee

\noindent where the superscript A-Z labels the diagrammatic representation adopted in
Ref. \cite{Arnold:1994ps}. Summing all the terms, we arrive at
\be
\Omega_{QCD}^G
&=&
-2(N_c^2-1)\frac{\pi^2T^4}{90}
+(N_c^2-1)N_c~g^2T^4~\frac{1}{144}
\, .
\ee
\end{fmffile}

%% file: apSEisospin.tex
\chapter[The euclidean action for isospin-dense chiral models]{
\label{apSEisospin}}
\chaptermark{The euclidean action for isospin-dense ...}


{\huge \sc The euclidean action for }
\vspace{0.3cm}

\noindent {\huge \sc isospin-dense chiral models}

\vspace{2cm}

In this appendix the derivation of the well-known result for the euclidean action
in the presence of a nonzero conserved charge involving momenta canonically conjugated to the fields is sketched, following the first Ref. in \cite{kapusta-gale}.

In the presence of a chemical potential, the partition function includes the constraint of charge conservation explicitly:
\be
Z&=&
\int [D\Pi][D\Phi]\;\exp\left\{
-\int d^dx\left[
-\Pi\, i \frac{\partial}{\partial\tau}\,\Phi
+\mathcal{H}[\Phi,\Pi]
-\mu n[\Phi,\Pi]
\right]
\right\}\,,
\label{PartFuncnI}
\ee
where $\Pi$ is the momentum conjugated to the field $\Phi$, being defined by ($\tau=it$):
\be
\Pi=\frac{\partial\mathcal{L}}{\partial(\partial_0\Phi)}
=
\frac{\partial\mathcal{L}}{\partial(i\partial_{\tau}\Phi)}
\,.
\ee
Note that the presence of a conserved charge that depends explicitly on the momentum $\Pi$ will modify the result of the functional integration over $\Pi$, rendering a modified euclidean action.

The isospin number $n_I$, Eq. (\ref{nIdef}), indeed depends on the momentum conjugated to the pion fields (let $\Phi=(\Phi_0,\Phi_1,\Phi_2,\Phi_3)=(\sigma,\pi_1,\pi_2,\pi_3)$):
\be
\Pi_i=\frac{\partial\mathcal{L}}{\partial(\partial_0\Phi_i)}=\partial_0\Phi_i
\,,
\ee
where we have used a Lagrangian of the form :
\be
\mathcal{L}&=&
\frac{1}{2}(\partial_{\mu}\Phi_i)
(\partial_{\mu}\Phi_i)
-V\left(
\Phi_i\Phi_i,\Phi_0
\right)
\,,
\label{Lag}
\ee
which is compatible with chiral models, since $\Phi_i\Phi_i$ has the form of a scalar under standard chiral rotations and the $\Phi_0$ ($\sigma$) dependence of the potential represents a possible explicit breaking, commonly included. For example, in the LSM (cf. Appendix \ref{LSM} neglecting constituent quarks) one has:
\be
V\left(
\Phi_i\Phi_i,\Phi_0
\right)
&=& V\left(
[\sigma^2+\vec{\pi}^2],\sigma
\right)
=\frac{\lambda^2}{4}\left\{[\sigma^2+\vec{\pi}^2]-v^2\right\}^2-h\,\sigma
\,.
\ee

The isospin number written in terms of the momenta $\Pi_i$ reads:
\be
n_I&=&
\Phi_1\partial^0\Phi_2
-
\Phi_1\partial^0\Phi_2=\Phi_1\Pi_2-\Phi_2\Pi_1\,.
\ee
The exponent in the integrand of the partition function, Eq. (\ref{PartFuncnI}), becomes then:
\be
-\int d^dx \Big[
-\Pi_i\, \partial_0\Phi_i+\mathcal{H}[\Phi,\Pi]-\mu_I\left(
\Phi_1\Pi_2-\Phi_2\Pi_1
\right)
\Big]
\,,
\ee
with
\be
\mathcal{H}[\Phi,\Pi]:=\Pi_i\partial_0\Phi_i-\mathcal{L}
\stackrel{(\ref{Lag})}{=}\frac{1}{2}\Pi_i^2+\frac{1}{2}
(\nabla \Phi_i)(\nabla \Phi_i)+V(\Phi_i\Phi_i,\phi_0)\,.
\ee
Although modified, the functional integration over the momenta $\Pi_i$ remains gaussian and can be done exactly once the terms linear in $\Pi_i$ are absorbed via completing the squares in the exponent above. Finally, the result for the partition function reads:
\be
Z&=&
\int [D\Phi]\; {\rm e}^{-S_E}
\,,
\ee
with the euclidean action for the isospin dense system being given by:
\be
S_E&=& \int d^dx\bigg\{
\frac{1}{2}(\partial_{\mu}\Phi_i)
(\partial_{\mu}\Phi_i)
+\mu_I
\big[
\Phi_1\partial_0\Phi_2
-\Phi_2\partial_0\Phi_1
\big]
-\frac{1}{2}\mu_I^2\left(\Phi_1^2+\Phi_2^2\right)
+\nonumber\\&&\quad\quad
+V\left(
\Phi_i\Phi_i,\Phi_0
\right)
\bigg\}\,.
\ee

\newpage

For the particular case of the LSM, we have $\Phi=(\Phi_0,\Phi_1,\Phi_2,\Phi_3)=(\sigma,\pi_1,\pi_2,\pi_3)$ and
\be
S_E&=& \int d^dx\bigg\{
\frac{1}{2}(\partial_{\mu}\sigma)
(\partial_{\mu}\sigma)
+\frac{1}{2}(\partial_{\mu}\vec{\pi})
(\partial_{\mu}\vec{\pi})
+\mu_I
\big[
\pi_1\partial_0\pi_2
-
\pi_2\partial_0\pi_1
\big]
-\frac{1}{2}\mu_I^2\left(\pi_1^2+\pi_2^2\right)
+\nonumber\\&&\quad\quad
+V\left(
[\sigma^2+\vec{\pi}^2],\sigma
\right)
\bigg\}\,.
\ee